\documentclass[]{aa} 

\usepackage{graphicx}
\usepackage{txfonts}
\usepackage{natbib}
\usepackage{longtable}
\usepackage{lscape}

\providecommand{\sorthelp}[1]{}

\begin{document}
 
\title{Galactic cold cores
IX. Column density structures and radiative-transfer modelling
\thanks{
{\it Planck} \emph{(http://www.esa.int/Planck)} is a project of the European Space
Agency -- ESA -- with instruments provided by two scientific consortia funded by ESA
member states (in particular the lead countries: France and Italy) with contributions
from NASA (USA), and telescope reflectors provided in a collaboration between ESA and a
scientific consortium led and funded by Denmark.
}
\thanks{{\it Herschel} is an ESA space observatory with science instruments provided
by European-led Principal Investigator consortia and with important
participation from NASA.}
}

\author{M.     Juvela\inst{1},
        J.     Malinen\inst{2},
        J.     Montillaud\inst{3},
        V.-M.  Pelkonen\inst{3},
        I.     Ristorcelli\inst{4,5},
       L.V.   T\'oth\inst{6},
        }

\institute{
Department of Physics, P.O.Box 64, FI-00014, University of Helsinki,
Finland, {\em mika.juvela@helsinki.fi}                                
\and
Institute of Physics I, University of Cologne, Germany
\and
Institut UTINAM, CNRS UMR 6213, OSU THETA, Universit\'e de Franche-Comt\'e, 
41 bis avenue de l'Observatoire, 25000 Besan\c{c}on, France
\and
Universit\'e de Toulouse, UPS-OMP, IRAP, F-31028 Toulouse cedex 4, France   
\and
CNRS, IRAP, 9 Av. colonel Roche, BP 44346, F-31028 Toulouse cedex 4, France  
\and
Lor\'and E\"otv\"os University, Department of Astronomy, P\'azm\'any P.s. 1/a,
1117 Budapest, Hungary (OTKA K62304)
}

\authorrunning{M. Juvela et al.}

\date{Received September 15, 1996; accepted March 16, 1997}

\abstract { 
The $Galactic$ $Cold$ $Cores$ (GCC) project has made $Herschel$
photometric observations of interstellar clouds where $Planck$
detected compact sources of cold dust emission. The fields are in
different environments and stages of star formation.
} 
{
Our aim is to characterise the structure of the clumps and their
parent clouds, and to study the connections between the environment and
the formation of gravitationally bound objects. We also examine the
accuracy to which the structure of dense clumps can be determined from
sub-millimetre data.
}
{
We use standard statistical methods to characterise the GCC fields.
Individual clumps are extracted using column density thresholding.
Based on sub-millimetre measurements, we construct a
three-dimensional radiative transfer (RT) model for each field. These are used to
estimate the relative radiation field intensities, to probe the clump
stability, and to examine the uncertainty of column density estimates.
We examine the structural parameters of the clumps, including their
radial column density profiles.
}
{
In the GCC fields, the structure noise follows the relations
previously established at larger scales and in lower-density clouds.
The fractal dimension has no significant dependence on column density
and the values $D_{\rm P} = 1.25 \pm 0.07$ are only slightly lower
than in typical molecular clouds. The column density 
probability density functions (PDFs) exhibit large variations, for
example, in the case of externally compressed clouds. At scales
$r>0.1$\,pc, the radial column density distributions of the clouds
follow an average relation of $N\sim r^{-1}$. In spite of a great
variety of clump morphologies (and a typical aspect ratio of 1.5),
clumps tend to follow a similar $N \sim r^{-1}$ relation below $r \sim
0.1$\,pc. RT calculations indicate only factor 2.5 variation in the
local radiation field intensity. The fraction of gravitationally bound
clumps increases significantly in regions with $A_{\rm V} > 5$\,mag
but most bound objects appear to be pressure-confined.
}
{ 
The host clouds of the cold clumps in the GCC sample have statistical
properties similar to general molecular clouds. The gravitational
stability, peak column density, and clump orientation are connected to
the cloud background while most other statistical clump properties 
(e.g. $D_{\rm P}$ and radial profiles) are insensitive to the
environment. The study of clump morphology should be continued with a
comparison with numerical simulations.
}


\maketitle

\begin{table*}
\caption[]{List of {\it Herschel} fields. The columns are: (1) name of the
field, (2)-(3) centre coordinates, (4) estimated distance, (5)-(6) size of
the modelled region in arcmin and in 9$\arcsec$ pixels, (8)-(9) Plummer parameters of
the default LOS density profile.} 
\centering
\begin{tabular}{lcccccccc}
\hline
\hline
Field  &     RA      & DEC        & Distance  &  \multicolumn{2}{c}{Model size} &  $A_{\rm V}^{\rm BG}$ & $R$ & $p$ \\ 
       &   (J2000.0) & (J2000.0)  &  (pc)     &  (arcmin) &  (pixels)   &  (mag)                & (arcmin) &  \\ 
(1)    &    (2)      &  (3)       &  (4)      &  (5)      &  (6)        &  (7)                  & (8) & (9) \\ 
\hline
    G0.02+18.02  &  16 40 56.7  &  -18 35 03.7  &  160   &   28.1   &  187$\times$187  & 0.5 &  2.9 & 2.4  \\ 
    G0.49+11.38  &  17 04 41.9  &  -22 13 52.3  &  160   &   35.1   &  234$\times$234  & 0.7 &  1.3 & 3.3  \\ 
     G1.94+6.07  &  17 27 50.7  &  -24 01 54.2  &  145   &   40.0   &  267$\times$267  & 1.6 &  11.5 & 8.1 \\ 
    G2.83+21.91  &  16 34 41.2  &  -14 09 27.4  &  300   &   38.1   &  254$\times$254  & 0.6 &  1.8 & 2.1  \\ 
     G3.08+9.38  &  17 17 29.6  &  -21 22 16.4  &  160   &   42.0   &  280$\times$280  & 0.7 &  1.1 & 1.8  \\ 
    G3.72+21.02  &  16 39 45.9  &  -14 02 09.3  &  160   &   40.0   &  267$\times$267  & 0.6 &  0.6 & 1.8  \\ 
    G4.18+35.79  &  15 53 43.4  &  -04 40 55.9  &  110   &   35.1   &  234$\times$234  & 0.5 &  2.7 & 3.6  \\ 
    G6.03+36.73  &  15 54 05.9  &  -02 54 39.7  &  110   &   35.1   &  234$\times$234  & 0.4 &  0.7 & 2.2  \\ 
    G9.45+18.85  &  17 00 22.2  &  -10 53 06.1  &  280   &   40.0   &  267$\times$267  & 0.6 &  4.7 & 2.9  \\ 
    G20.72+7.07  &  18 03 38.9  &  -07 30 25.0  &  260   &   30.0   &  200$\times$200  & 1.3 &  0.8 & 1.8  \\ 
   G21.26+12.11  &  17 46 47.7  &  -04 38 48.1  &  120   &   35.1   &  234$\times$234  & 0.9 &  -0.5 & 2.3 \\ 
    G24.40+4.68  &  18 19 21.5  &  -05 29 45.1  &  260   &   36.0   &  240$\times$240  & 1.7 &  0.8 & 2.1  \\ 
    G25.86+6.22  &  18 16 32.7  &  -03 23 49.4  &  260   &   36.0   &  240$\times$240  & 3.1 &  0.6 & 1.9  \\ 
    G26.34+8.65  &  18 08 37.9  &  -01 51 26.6  &  400   &   30.0   &  200$\times$200  & 1.3 &  1.0 & 3.6  \\ 
    G93.21+9.55  &  20 37 00.2  &  +56 58 46.8  &  440   &   30.0   &  200$\times$200  & 0.9 &  -0.4 & 2.5 \\ 
  G108.28+16.68  &  21 10 13.4  &  +72 52 58.5  &  300   &   35.1   &  234$\times$234  & 0.6 &  0.9 & 2.1  \\ 
  G110.62-12.49  &  23 37 39.8  &  +48 31 40.4  &  440   &   40.0   &  267$\times$267  & 0.2 &  0.6 & 2.6  \\ 
  G110.80+14.16  &  21 59 02.5  &  +72 52 56.0  &  400   &   40.0   &  267$\times$267  & 0.7 &  0.5 & 1.9  \\ 
   G116.08-2.40  &  23 57 06.7  &  +59 43 26.9  &  500   &   30.0   &  200$\times$200  & 1.3 &  0.8 & 2.3  \\ 
  G126.63+24.55  &  04 19 14.0  &  +85 52 06.5  &  125   &   30.0   &  200$\times$200  & 0.2 &  2.3 & 4.6  \\ 
  G141.25+34.37  &  08 48 58.2  &  +72 41 16.1  &  110   &   40.0   &  267$\times$267  & 0.1 &  1.2 & 1.8  \\ 
   G149.67+3.56  &  04 17 53.6  &  +55 15 05.2  &  170   &   40.0   &  267$\times$267  & 1.6 &  1.5 & 2.2  \\ 
   G150.47+3.93  &  04 24 37.8  &  +54 58 21.4  &  170   &   40.0   &  267$\times$267  & 1.7 &  -0.3 & 1.9 \\ 
   G151.45+3.95  &  04 29 53.9  &  +54 16 52.5  &  170   &   36.0   &  240$\times$240  & 1.2 &  3.5 & 2.6  \\ 
   G154.08+5.23  &  04 47 34.4  &  +53 05 02.4  &  170   &   34.0   &  227$\times$227  & 1.2 &  -0.5 & 2.4 \\ 
  G155.80-14.24  &  03 36 49.2  &  +37 42 31.1  &  350   &   40.0   &  267$\times$267  & 0.5 &  0.6 & 2.1  \\ 
   G157.08-8.68  &  04 01 55.7  &  +41 15 20.4  &  150   &   40.0   &  267$\times$267  & 1.1 &  0.5 & 2.1  \\ 
  G159.23-34.51  &  02 55 54.0  &  +19 37 10.2  &  325   &   40.0   &  267$\times$267  & 0.6 &  0.8 & 2.0  \\ 
   G161.55-9.30  &  04 16 06.1  &  +37 49 20.2  &  250   &   36.0   &  240$\times$240  & 0.8 &  0.8 & 2.0  \\ 
   G163.82-8.44  &  04 29 00.9  &  +36 43 21.1  &  420   &   50.1   &  334$\times$334  & 1.4 &  -0.4 & 2.1 \\ 
   G164.71-5.64  &  04 40 58.0  &  +37 59 06.6  &  330   &   42.0   &  280$\times$280  & 1.3 &  1.2 & 1.9  \\ 
   G167.20-8.69  &  04 36 34.8  &  +34 16 53.3  &  160   &   40.0   &  267$\times$267  & 0.9 &  0.5 & 1.8  \\ 
   G173.43-5.44  &  05 08 32.7  &  +31 26 38.8  &  150   &   45.0   &  300$\times$300  & 0.9 &  1.5 & 2.5  \\ 
  G181.84-18.46  &  04 44 00.3  &  +16 57 22.7  &  500   &   36.0   &  240$\times$240  & 0.7 &  1.4 & 2.8  \\ 
  G188.24-12.97  &  05 17 05.1  &  +14 59 33.2  &  445   &   40.0   &  267$\times$267  & 0.6 &  0.8 & 1.6  \\ 
  G189.51-10.41  &  05 29 55.2  &  +15 25 03.2  &  445   &   42.0   &  280$\times$280  & 0.6 &  0.5 & 2.1  \\ 
   G198.58-9.10  &  05 52 53.1  &  +08 22 34.1  &  450   &   35.1   &  234$\times$234  & 0.7 &  1.1 & 4.0  \\ 
   G203.42-8.29  &  06 04 47.6  &  +04 20 31.3  &  390   &   44.1   &  294$\times$294  & 0.7 &  1.6 & 3.6  \\ 
   G205.06-6.04  &  06 16 27.5  &  +04 07 44.0  &  400   &   44.1   &  294$\times$294  & 0.8 &  1.3 & 2.1  \\ 
  G206.33-25.94  &  05 07 01.2  &  -06 17 56.6  &  210   &   35.1   &  234$\times$234  & 0.1 &  0.8 & 2.6  \\ 
  G210.90-36.55  &  04 34 54.6  &  -14 23 35.1  &  140   &   50.1   &  334$\times$334  & 0.5 &  4.0 & 5.3  \\ 
  G212.07-15.21  &  05 55 49.7  &  -06 11 25.9  &  230   &   36.0   &  240$\times$240  & 0.6 &  1.3 & 2.0  \\ 
  G247.55-12.27  &  07 09 26.3  &  -36 16 39.6  &  170   &   44.1   &  294$\times$294  & 0.5 &  1.9 & 2.0  \\ 
  G298.31-13.05  &  11 39 22.1  &  -75 14 27.0  &  150   &   25.1   &  167$\times$167  & 0.5 &  1.1 & 2.6  \\ 
   G300.61-3.13  &  12 28 54.8  &  -65 47 40.5  &  200   &   36.0   &  240$\times$240  & 1.2 &  1.9 & 3.0  \\ 
   G300.86-9.00  &  12 25 17.0  &  -71 43 05.5  &  150   &   36.0   &  240$\times$240  & 0.7 &  1.3 & 3.1  \\ 
  G315.88-21.44  &  17 19 39.9  &  -76 55 17.2  &  250   &   35.1   &  234$\times$234  & 0.2 &  0.7 & 2.2  \\ 
   G341.18+6.51  &  16 25 05.1  &  -39 59 10.4  &  140   &   25.1   &  167$\times$167  & 1.4 &  3.2 & 6.2  \\ 
   G344.77+7.58  &  16 33 30.3  &  -36 39 02.1  &  240   &   40.0   &  267$\times$267  & 0.9 &  -0.5 & 2.4 \\ 
   G345.39-3.97  &  17 23 01.0  &  -43 26 24.7  &  225   &   33.0   &  220$\times$220  & 0.9 &  -0.5 & 3.0 \\ 
  G358.96+36.75  &  15 39 50.0  &  -07 12 09.0  &  110   &   35.1   &  234$\times$234  & 0.2 &  1.8 & 2.2  \\ 
\hline  
\end{tabular}  
\tablefoot{
List of clouds and masers associated to the $Herschel$ fields can be found in Table 1 in \cite{GCC-IV}.
}
\label{table:fields}  
\end{table*}

\section{Introduction} \label{sect:intro}

The all-sky survey of the {\it Planck} satellite \citep{Tauber2010} made it possible to
catalogue cold interstellar clouds at a Galactic scale. The angular resolution of
$\sim5\arcmin$ in the {\it Planck} sub-millimetre bands allowed the identification of 
compact sources that are associated with the early phases of star formation. Analysis of
{\it Planck} data led to the creation of the Cold Clump Catalogue of Planck Objects
\citep[PGCC, see][]{PGCC}, which lists basic properties of over 13000 sources. The low
colour temperatures (mostly $T\la14$\,K) are a direct indicator of the presence of high-column density structures. The objects are only partially resolved by the {\it Planck}
beam. Therefore, the catalogued objects, generally referred to as clumps, are likely to
contain substructure, including gravitationally bound prestellar and protostellar cores.

The {\it Herschel} Open Time Key Programme {\em Galactic Cold Cores} (GCC) carried out
continuum observations of 116 fields selected from an early version of the {\it Planck}
C3PO catalogue \citep{planck2011-7.7b}. The fields were mapped with {\it Herschel} PACS
and SPIRE instruments \citep{Pilbratt2010, Poglitsch2010, Griffin2010} between 100 and
500\,$\mu$m, the data enabling the study of the {\it Planck} clumps at up to $\sim$20
times higher spatial resolution \citep{Juvela2010, PlanckII, GCC-III, GCC-IV, GCC-VI}.
Typically the {\it Herschel} maps are some $40 \arcmin$ in size and contain a few
$Planck$ clumps and also observations of their environment. This is useful for studies
of dust property variations \citep{Juvela2011, GCC-III, GCC-V, GCC-VI} but also makes
it possible to trace density structures from gravitationally bound cores to clumps and
filaments, the parental molecular cloud, and sometimes even all the way to the
surrounding diffuse medium \citep{GCC-III, GCC-IV}. Studies have already been carried
out on the structure of filaments \citep{Rivera2016, Rivera2017} and high-latitude
clouds, especially the cloud LDN~1642 \citep{Malinen2014, Malinen2016}, and the
kinematics (e.g. the internal turbulence of the clumps) through molecular line
observations \citep{Parikka2015, Feher2017, Saajasto2017}.

Column densities can be estimated from dust emission but only assuming that the
properties of the emitting dust grains are known. The accuracy of the estimates is
limited by the uncertainty and the spatial variations of dust opacity, $\kappa$, and to a
lesser extent by the dust opacity spectral index, $\beta$. The sub-millimetre dust opacity
$\kappa$ is likely to vary between low- and high-density regions
\citep[e.g.][]{Stepnik2003, Lehtinen2007, Paradis2009, Martin2012, Ysard2013, Roy2013, GCC-V},
which increases the uncertainty of mass estimates and could cause systematic errors in
the estimated filament, clump, and core profiles. The dust opacity spectral index $\beta$
is also believed to increase towards the coldest regions \citep{Desert2008, Dupac2003,
Paradis2010, Planck2011b, planck2013-p06b, Schnee2014, GCC-VI}, possibly partially
cancelling out the bias caused by $\kappa$ variations. The true variations of $\kappa$
and $\beta$ are close to the sensitivity limit of current sub-millimetre observations.
Therefore, also in this paper, the potential effects of $\beta$ and $\kappa$ changes are
examined through a comparison of a couple of alternative scenarios.

Temperature variations along the line-of-sight (LOS) are another
notable source of error that cause the column densities to be 
underestimated towards cold clumps
\citep{Shetty2009a,Malinen2011,Juvela2012_Tmix}. The effect would thus
be opposite to that of $\kappa$ variations, if $\kappa$ increases in
the clumps above the value assumed in the analysis.
For externally heated
clouds with tens of magnitudes of visual extinction, this error
can also even  be a factor of several \citep{Juvela2013_colden, Pagani2015}.
The effect can be significantly reduced by internal heating sources
\citep{Malinen2011}, which makes it even more unpredictable. For
starless objects, it is possible to use radiative-transfer (RT) models
to derive an approximate correction \citep{Juvela2013_colden, GCC-V}.
It is clear that an accurate determination of real column densities
and of real density profiles is essential when observations are
compared to models of interstellar filaments or prestellar and
protostellar cores. 

In this paper, the emphasis is on density structures of clumps that are larger than the
protostellar cores. If the clumps are prestellar, they are particularly interesting as an
intermediate phase between the general turbulent density field and the later
gravity-dominated development. The shapes and density profiles of the clumps should 
reflect the processes that lead to the creation of gravitationally bound objects.
Depending on the region, the processes can include the random turbulent motions, larger-scale converging flows and cloud collisions, or more direct dynamic influences of stellar
populations in the form of radiation pressure, ionisation shocks, and supernova remnants.
Therefore, one can expect correlations between the properties of the clumps and their
environment. The GCC sample is well suited for such studies because it contains a
heterogeneous sample of objects from different Galactic environments.

In this paper, we examine the density structures of the GCC fields from individual clumps
to the extent of entire $Herschel$ maps. We characterise the basic statistical properties
of the identified clumps (e.g. aspect ratios and skewness) and extract their radial column
density profiles. We use RT modelling to quantify the uncertainties caused by the cloud
temperature structure and by the potential changes of dust-emission properties.  Models
also provide information on the differences in the local radiation field intensity. On
the other hand, we characterise the general properties of the {\it Herschel} fields with
standard statistical tools (e.g. structure noise, fractal dimensions, and column
density probability density functions). We also look for correlations between the
properties of the clumps and the large-scale cloud structure, for example, regarding the
clump orientations and the general anisotropy of the column density field.

The structure of the paper is as follows. The observations are described in
Sect.~\ref{sect:obs} and the main methods are listed in Sect.~\ref{sect:methods}. The
main results are presented in Sect.~\ref{sect:results}, including the basic statistical
parameters of the extracted structures, the clump radial profiles, and gravitational
stability of the clumps. We discuss the results in Sect.~\ref{sect:discussion} before
listing our main conclusions in Sect.~\ref{sect:conclusions}.

\section{Observations}  \label{sect:obs}

In the GCC project, the fields for {\it Herschel} observations
were selected based on {\em Planck} all-sky observations and
ancillary data, with the intention of covering a representative set of
objects in different Galactic environments. The target selection is
described in \citet{GCC-III} and an overview of the observations is
given in \citet{GCC-IV} and \citet{GCC-V}. The sample does not include
regions that were included in other {\it Herschel} key programmes
(e.g. the Galactic plane $|b|<1^{\degr}$ covered by the Hi-GAL
programme \citep{Molinari2010} and the individual clouds covered by
the Gould Belt survey \citep{Andre2010} and HOBYS \citep{Motte2010}
programmes).

The GCC observations cover 116 fields. The maps observed with the SPIRE instrument
correspond to wavelengths 250\,$\mu$m, 350\,$\mu$m and 500\,$\mu$m. The maps have an
average size of $\sim1800$\,arcmin$^2$. The fields are listed in \citet{GCC-VI} Table~3
and the {\it Herschel} observation identification numbers can be found in \citet{GCC-IV}.
We use data identical to those in \citet{GCC-VI}. The SPIRE observations were
reduced with the {\it Herschel} Interactive Processing Environment HIPE v.12.0,
using the official pipeline with iterative destriper, extended emission
calibration options, and naive map-making. The raw and pipeline-reduced data are
available via the {\it Herschel} Science Archive. The resolution of the maps is
approximately 18$\arcsec$, 25$\arcsec$, and 37$\arcsec$ for 250, 350, and
500\,$\mu$m, respectively.
The data were colour corrected and zero-point corrected as explained in \citet{GCC-VI}.
For SPIRE observations we adopt a 7\% uncertainty of absolute calibration and a 2\%
uncertainty of relative calibration\footnote{SPIRE Observer's manual, \\ {\em
http://herschel.esac.esa.int/Documentation.shtml}}. 
In this paper, we analyse 51 fields that have estimated distances below or equal to
500\,pc (see Table~\ref{table:fields}). The distance estimates are discussed in detail in
\citet{GCC-IV}.

For the RT models, we extracted from each SPIRE map a rectangular area
that was between $25\arcmin \times 25\arcmin$ and $50\arcmin \times
50\arcmin$ in size and covered the most interesting high-density
regions. To match the cell size of the RT models (see below), the data
were resampled onto 9.0$\arcsec$ pixels with the Montage
package\footnote{http://montage.ipac.caltech.edu/}. The 250\,$\mu$m
data remain at their original resolution. The 350\,$\mu$m and
500\,$\mu$m data also were convolved with a 10.0$\arcsec$ Gaussian as
they were resampled onto the pixels of the 250\,$\mu$m maps. 
Convolution was used to mitigate aliasing effects caused by the
different pixel locations of the different maps. 
We calculated dust optical depth maps directly via colour temperature,
assuming a fixed value of $\beta$=1.8, applicable as an approximate
average value over the fields \citep{GCC-VI}. The colour temperatures
(at a resolution of 40$\arcsec$) were used to colour correct the
surface-brightness data. Optical-depth data in the percentile range
0.5-6.0\% were used to select pixels that define the local background.
The average values of these pixels were then used to carry out
background subtraction for each surface-brightness map. The RT
models describe this emission and, thanks to background subtraction,
can be assumed to represent a finite volume in the LOS
direction. The background-subtracted maps were used to derive new
optical-depth maps via spectral energy distribution (SED) fitting and
they also serve as the basis of modelling, describing monochromatic
emission at the reference wavelengths and above diffuse emission.
These maps are also independent of the zero point corrections that in
the previous papers were calculated with the help of {\it Planck} data
\citep{GCC-V, GCC-VI}. Only the intermediate calculation of
dust-temperature maps (before colour corrections) relies on absolute
surface-brightness zero points that are the same as in \citet{GCC-VI}.

The effective point spread function (PSF) of SPIRE depends on the
source
spectrum\footnote{http://herschel.esac.esa.int/twiki/bin/view/Public/\\SpirePhotometerBeamProfileAnalysis}.
The RT models are intended to describe cloud structure at the
resolution of SPIRE 250\,$\mu$m data. Therefore, when models are
compared with observations, the model predictions are convolved from
the resolution of 250\,$\mu$m map down to the resolution of
350\,$\mu$m and 500\,$\mu$m maps. The kernels for these convolutions
were created adopting a temperature of 15.0\,K and a spectral index of
1.8. As noted in \citet{GCC-VI}, for parameter ranges of
$T\sim$10--20\,K and $\beta \sim$1.5-2.5, the variation of the PSF
shape is negligible.

\section{Methods} \label{sect:methods}

In this Section we describe the extraction of structures from column density maps. We
also describe the selection of clumps, their basic statistical analysis, and the
derivation of the clump radial column density profiles. We start with the general
properties of the fields before discussing the characterisation of individual clumps.
Finally, the RT models are discussed in Sect.~\ref{sect:RTmodels}.

\subsection{Characterisation of the fields} \label{sect:global}

To study the large-scale ($>0.1$\,pc) column density structures, we use a set of common
statistics that are used to analyse cloud observations and especially far-infrared and
sub-millimetre data. These include fractal dimensions, structure noise, and 
probability density functions (PDFs). Furthermore, we use template matching to quantify
the angular distribution of elongated structures.

The fractal dimension of interstellar clouds has been studied across a
wide range of size scales, down to the linear scales probed by our
$Herschel$ observations. The presence of a single fractal dimension
may be related to the scale-free nature of interstellar turbulence.
However, with the increasing role of self-gravity, the behaviour might
change at clump scales.
We estimate the fractal dimension $D_{\rm P}$
from the relation between the perimeter $P$ and the surface area $A$,
$P \propto A^{D_{\rm p}/2}$ \citep{Mandelbrot1983, Falgarone1991,
Stutzki1998}. Because $D_{\rm P}$ characterises the general cloud
structure, it is not sensitive to the spatial data resolution. We use
mainly the optical depth maps $\tau(250\mu{\rm})$ at 40$\arcsec$
resolution. The values of $D_{\rm P}$ are derived from 100 contours
between the peak value and the 10\% percentile of the pixel values.
All contours that touch the map boundaries or enclose an area of less
than 1.1\,arcmin$^2$ are ignored.

The structure function describes signal fluctuations as a function of the spatial scale,
in a way similar to the power spectrum analysis. Thus, it is also related to the forces
forming the cloud structure (e.g. turbulence). The second-order structure function
$S(\theta)$ is
\begin{equation}
S(\theta) = \langle |   Y(x) - Y(x+\theta) |^2 \rangle,
\end{equation}
where $\theta$ is the spatial (angular) separation and $Y$ are the data values (in our
case intensity or column density) and the averaging extends over all map pixels
\citep{Gautier1992}. Each value $Y(x)$ is an average over a measurement aperture with a
diameter $D$ or, in our case, over the beam ($D\sim$FWHM). The reference $|Y(x+\theta)|$
could be an average of values read from either side of the aperture,
$(|Y(x+\theta)|+|Y(x-\theta)|)/2$, or from a reference annulus \citep{Gautier1992,
Kiss2001, Martin2010}. We follow the first approach, using two reference positions.

The PDF analysis examines the probability distribution of surface
brightness or column density values \citep[e.g.][]{Kainulainen2009,
Schneider2015_tails}. Observations and simulations of the density and
column density of interstellar clouds have shown that the distribution
often resembles a log-normal function \citep[]{VazquezSemadeni1994,
Padoan1997}. A power-law tail, sometimes observed at high column
densities, has been attributed to the presence of gravitationally
bound objects, although there may be other contributing factors
\citep{Klessen2000,Kritsuk2011,Brunt2015}. We estimate the PDF
functions of the column density maps, within the limitations set by
the finite map sizes. For comparison, PDFs are also calculated for the
250\,$\mu$m surface brightness data, which is more weighted towards
the distribution of warm dust.

We use a template-matching method (TM) to measure the anisotropy of
column density structures at a given size scale $F$.  Template
matching can be seen as a subset of pattern recognition methods.
Instead of a single specific algorithm, one uses an image of the
search pattern that is compared to the data \citep{Brunelli2009}. Our
procedure is described in detail in \citet{Juvela2016TM}. The analysis
uses the difference between maps convolved with Gaussian beams with
FWHM values of (for example) $F$ and $2 \times F$, which separates
structures within a limited range of spatial frequencies. In the
second stage, data at each map position are compared to a pre-defined
template, which is rotated to find the best match to the data. To
match elongated structures, we use the same $3\times 3$ element
template as in \citet{Juvela2016TM}, which can be represented with a
matrix
\begin{equation}
\begin{pmatrix}
-1/3 & 2/3 & -1/3 \\
-1/3 & 2/3 & -1/3 \\
-1/3 & 2/3 & -1/3 \\
\end{pmatrix}.
\label{eq:TM}
\end{equation}
The significance of the match is calculated by multiplying the
template values with the corresponding data values (for the current
position and orientation of the template) and by summing the resulting
values. The significance (goodness-of-fit) and the position angle
giving the highest significance are registered as maps with dimensions
identical to the input map. We count position angles counter-clockwise
from the Equatorial north. 
The final result consists of position angle histograms.  We first
reject a certain percentage of pixels with the lowest significance
values. This is sufficient for our purposes (for a qualitative
description of the anisotropies) because the random position angles of
low-significance pixels simply add a flat pedestal to the histograms.

In the basic TM method, the computed significance depends on the
absolute pixel values and the result is weighted towards high-column density regions. We also use a variation where the data values under
the current template are normalised by their standard deviation. This
removes the dependence on absolute data values and enables the
characterisation of structures even in the most diffuse parts of the
fields \citep[see][]{Juvela2016TM}.

The TM method is not very sensitive to clumps although the
pattern of Eq.~(\ref{eq:TM}) will give a positive output at the
location of a spherically symmetric object. The signal could be strong
for very bright objects. However, with the above-mentioned
normalisation, even slightly elongated cores and clumps lead to a
smaller significance than faint but more elongated filaments.

The fields (and clumps) will also be characterised by the presence of
young stellar objects (YSOs). We use the compilation of YSO candidates
presented in \citet{GCC-IV}.

\subsection{Radiative-transfer models} \label{sect:RTmodels}

We construct for each field a three-dimensional (3D) radiative-transfer model whose
sub-millimetre emission matches the SPIRE observations. The main assumptions regarding
the dust model, the radiation field, and the density field are listed below. The
variations of the default models are summarised in Table~\ref{table:models}

\begin{table}
\caption[]{Parameters of the default radiative-transfer models and the assumption
differences in the alternative fits.} \label{table:models} \centering
\begin{tabular}{ll}
\hline
\hline
Model      & Assumptions \\
\hline
$D$        & default case, $\tau(250\mu{\rm m})/\tau(J)=1.0\times 10^{-3}$, $\beta=1.8$  \\
$P$        & $k_{\rm ISRF}$ fitted using pixels with highest 3\% of $N({\rm H}_2)$ only\\
$W$        & LOS cloud extent adjusted pixel by pixel \\
$\Delta A_{\rm V}$ & external field changed by $A_{\rm V}$=$\pm 1$\,mag \\
$K$        & $\tau(250\mu{\rm m})/\tau(J)=2.0\times 10^{-3}$, $\beta=2.1$  \\
$TD$       & dust changes with density from $D$ to $K$ assumptions \\
\hline
\end{tabular}
\end{table}

\subsubsection{Dust models} \label{sect:dust}

The essential dust parameters are the sub-millimetre dust opacity $\kappa$, the dust
opacity spectral index $\beta$, and the ratio of opacities at the wavelengths where
dust absorbs (ultraviolet, optical, and near-infrared) and emits (far-infrared,
sub-millimetre) most of energy. The dust opacity affects the dust temperature and, for
a given temperature, the scaling between the column density and the surface
brightness. In observations, $\beta$ is largely degenerate with dust colour
temperature. However, in RT models the temperatures are solved self-consistently and
the assumed value of $\beta$ has a more direct effect on the intensity ratios between
bands.

We use dust models loosely based on the \citet{Draine2003} dust model
of Milky Way dust, with a value of the total-to-selective
extinction ratio $R_{\rm V}=A_{\rm V}/E({\rm B}-{\rm V})=3.1$ and a
gas-to-dust ratio of 124. However, we modify this assuming a constant
value of $\beta$ at all wavelengths $\lambda \ge 30$\,$\mu$m and
explicitly specifying the optical depth ratio between 250\,$\mu$m and
the $J$ band. \citet{GCC-V} estimated that in the GCC fields the
optical depth ratio $\tau(250\mu{\rm m})/\tau(J)$ is higher than in
diffuse clouds, with a median value of $\sim 1.6 \times 10^{-3}$.
\citet{GCC-VI} concluded that the average spectral index increases
towards the coldest clumps of the GCC fields. The median far-infrared
spectral index was found to be $\sim 1.9$, the values sometimes
exceeding 2.0. Based on these findings, we use two alternative dust
models. The first one has $\tau(250\mu{\rm m})/\tau(J)=1.0 \times
10^{-3}$ and $\beta=1.8$ and the second one has $\tau(250\mu{\rm
m})/\tau(J)=2.0 \times 10^{-3}$ and $\beta=2.1$. These bracket most of
the observed parameter range and should show the quantitative effects
of dust property uncertainties. Conversely, this will give an idea of
the accuracy to which these parameters can be constrained by
observations.
We will also briefly experiment with spatial variations of dust properties, which is
implemented by varying, cell-by-cell, the relative abundance of the two dust species
described above.

\subsubsection{Radiation field} \label{sect:ISRF}

We start with the assumption that the model clouds are illuminated by
an isotropic external field with intensities given in
\citet{Mathis1983}. This is rarely sufficient to match the observed
dust colour temperatures. Therefore, the intensity of the radiation
field is scaled with a factor $k_{\rm ISRF}$, which varies from field
to field. In the first approximation, this is sufficient to match
observations in regions with different levels of heating. However,
also the spectral shape of the illuminating radiation is important.
For example, if the radiation field contains less short-wavelength
radiation, the temperature contrasts between low- and high-column
density regions becomes smaller. The question is particularly relevant
because we are modelling background-subtracted surface brightness
values, that is, an inner region of a cloud that may be surrounded by
an envelope with a non-negligible extinction. Because the modelled
surface brightness data are also background-subtracted, the external
radiation field should be attenuated by a dust layer that roughly corresponds to half
of the full LOS column density of the reference areas. This is only a crude
approximation and assumes that most of the material along LOS is actually around the
dense cloud. To examine the effects of the attenuation of the external field, we test cases where the \citet{Mathis1983} field is modified by a term
$exp(-\tau_{\nu})$, where the optical depth $\tau$ at the frequency $\nu$ corresponds
to a visual extinction of $A_{\rm V}=\pm 1$\,mag. Here the negative extinction simply
means a radiation field with a stronger short wavelength part.

\subsubsection{Model clouds}

Each cloud is modelled using a 3D density grid. The grid is uniform in the plane
of the sky (POS) with the cell size corresponding to the pixel size of the
resampled observations, $\Delta=$9.0$\arcsec$. In equatorial coordinates the
projected width and height of the maps, $N \times \Delta$, ranges from
20$\arcmin$ to 50$\arcmin$, depending on the SPIRE maps. The cell size was
selected as a compromise between the run times (increasing as $N^3$) and the
resolution required for comparison with observations. With $\Delta$=9.0$\arcsec$,
the 250\,$\mu$m observations with FWHM$\sim 18\arcsec$ (the highest-resolution
data used in the model fitting) are still Nyquist sampled. 

In the LOS direction the density distribution is unknown but we assume that it has a
Plummer-like profile,
\begin{equation}
     \rho = \frac{\rho_{\rm c}}{  (1+(z/R)^2)^{p/2} },
\label{eq:plummer}     
\end{equation}
where $\rho_{\rm c}$ is the maximum density and $z$ is the distance from the symmetry
plane that, in the LOS direction, is located half way through the cloud. The density
profile has two parameters, $R$ describing the extent of a central flattened part and
$p$ describing the steepness of the density drop at larger distances. These parameters
were defined for each field separately by fitting a column density cross-section of
the main clump (or a typical clump) along its minor axis with a corresponding Plummer
{\em column density} profile and converting this into a function of density
\citep[see][Eq. 1]{Arzoumanian2011}. Thus, the adopted LOS profile is most appropriate
for the densest regions, where the cloud shape has the largest effect on dust
temperatures. The procedure assumes that the LOS extent is similar to the smaller of
the clump dimensions seen in the POS. This is appropriate for spherical clumps and 
cylindrical filaments. However, it will underestimate the LOS extent, for example, when
a filament or an elongated clump is observed along its major axis. In the LOS
direction, the RT calculations employ a non-uniform discretisation where the number of
cells is 30\% lower than in the other two dimensions. The cell size decreases towards
the density peak ($z=0$) where it is equal to the resolution in the POS. This reduces
the time and memory used for calculations, at the same time retaining sufficient
resolution near the density peak.

Because the LOS extent of the cloud is poorly constrained, below we also examine models
where the LOS scale is a free parameter. This is implemented using a parameter map
$k_{\rm W}$ that, at the position of each map pixel, gives a linear scaling for the
LOS width of the density profile. Thus, in Eq.~(\ref{eq:plummer}), $R$ will be
replaced with $R \times k_{\rm W}$.

\subsubsection{Optimisation of the cloud models}

The RT calculations were performed with SOC, a new Monte Carlo program
\citep{Juvela_SOC} that has been compared with other RT
codes\footnote{See TRUST code comparison at
http://ipag.osug.fr/RT13/RTTRUST/} \citep{TRUST-I}. Given a density
distribution, a dust model, and an external radiation field, SOC
solves the dust temperature in each cell and computes surface-brightness maps at the requested wavelengths. Because we analyse
observations at wavelengths $\lambda > 100\mu$m, stochastically heated
grains have only a minor effect and our calculations assume that all
grains remain in thermal equilibrium with the radiation field.
The radiation field is estimated on a grid of 50 frequencies that
extend logarithmically from 10$^{11}$\,Hz to $3 \times 10^{15}$\,Hz.
Each frequency is simulated with about 10$^6$ photon packages. This
results in information about the absorbed energy within each model
cell and the integration over frequency provides the absorbed energy
that is then used to calculate dust temperatures. With the employed
number of photon packages, the Monte Carlo noise is below 0.1\,K in
terms of the dust temperature of individual cells. 
After the integration along the LOS and the convolution with the beam,
the noise of the synthetic surface-brightness maps is several times
below the uncertainty of the observed maps. Models give predictions
of the monochromatic surface brightness at 250\,$\mu$m, 350\,$\mu$m,
and 500\,$\mu$m at the resolution of the 250\,$\mu$m observations.
These data are convolved to the resolution of the corresponding
observed maps for the comparison with the measurements (see
Sect.~\ref{sect:obs}).

We optimise the models by scaling the model column densities, the
intensity of the external radiation field, and optionally some
additional parameters.

The intensity of the external radiation field is scaled by a single
factor $k_I$. We do not consider changes in either the spectrum or the
angular dependence of the incoming radiation. The parameter $k_I$ is
updated based on the ratio between the 250\,$\mu$m and 500\,$\mu$m
surface brightness,
\begin{equation}
     k_I =  
     \frac{I_{\rm Obs}(250\mu{\rm m}) I_{\rm Mod}(500\mu{\rm m})}{
     I_{\rm Obs}(500\mu{\rm m})
     I_{\rm Mod}(250\mu{\rm m})}.
     \label{eq:k_I}
\end{equation}
Because $k_I$ affects the whole model, we update it using average
surface-brightness values over a large area. The averaged pixels are
selected based on the optical depth $\tau_{\rm MBB}$  derived
from the modified blackbody (MBB) fits of the observations. By
default, we use pixels with $\tau_{\rm MBB}(250)$ between the 70\% and
98\% percentiles, thus concentrating more on the high-column density
regions. The upper limit of 98\% is used to reduce the effect of
point sources. We also carry out complementary calculations
where $k_I$ is determined by pixels in the 97.0-99.7\% range of
$\tau_{\rm MBB}(250)$. In this case, the radiation field is tuned to
give a good fit exclusively to the densest clumps.

Apart from the radiation field, the observed surface brightness
depends mainly on the line-of-sight column density. To match the
surface-brightness variations on the plane of the sky, the column
density is scaled pixel-by-pixel. The updates are done according to
the ratio of the observed and modelled 350\,$\mu$m surface
brightness,
\begin{equation}
k_N^i = I_{\rm Obs}^{i}(350\mu{\rm m})/I_{\rm Mod}^{i}(350\mu{\rm m}).
     \label{eq:k_N}
\end{equation}
The scaling applies equally to all cells along a LOS that correspond
to the same map pixel. The procedure is not optimal, because the model
column densities will mainly depend on a single frequency band. On the
other hand, this enables a cleaner separation between column density
and radiation field updates, which speeds up the optimisation.

Optionally, the LOS cloud profile can also be modified.  One
could use a single parameter to scale the cloud shape between oblate
and prolate geometries. However, our fields typically consist of a
number of separate clumps and the same scaling is not likely to work
for all substructures. Therefore we adjust the LOS cloud extent with
the parameter $k_I^{i}$  that is set for each map pixel $i$ separately.
The updates are based on the ratios
\begin{equation}
   k_W^i = \frac{I_{\rm Obs}^{i}(250\mu{\rm m}) I_{\rm Mod}^{i}(500\mu{\rm
   m})}{ I_{\rm Obs}^{i}(500\mu{\rm m})
   I_{\rm Mod}^{i}(250\mu{\rm m})}.
   \label{eq:k_W}
\end{equation}
If the LOS extent is increased ($k_W>1.0$), the medium receives more
radiation and the dust temperature increases.  Thus the ratio in
Eq.~(\ref{eq:k_W}) again traces temperature changes, however, unlike in
Eq.~(\ref{eq:k_I}), the updates are not global but affect each LOS
separately. There is some coupling between different LOS because of
the mutual shadowing of the volume elements. In our models, the
density always reaches its maximum in the central plane ($z$=0) of the
model volume. If the cloud consisted of separate clumps at different
$z$ locations, this would reduce the mutual shadowing of the clumps.
In the models, this can be compensated by having a larger $k_W$ value.

\begin{figure}
\includegraphics[width=8.8cm]{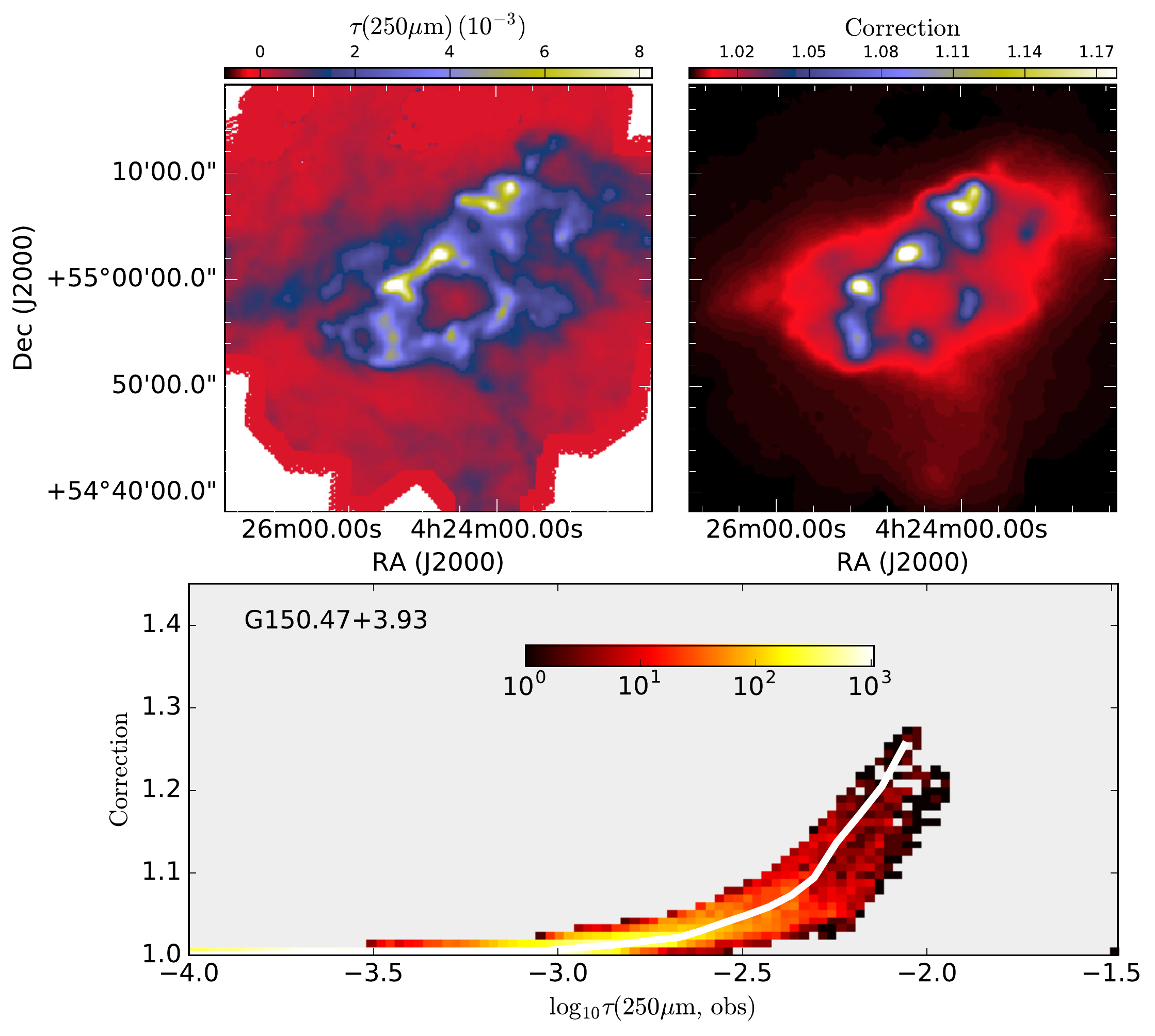}
\caption{
Example of probable column density biases (field G150.47+3.93). The first frame shows
the optical depth map derived from observations. Based on RT modelling, and the other
frames show the ratio $\xi_N$ between the true column density of the model cloud and
the column density derived from the model-predicted surface-brightness maps. The
bottom frame shows this ratio as a function of the 250\,$\mu$m optical depth. The
colours correspond to the number of map pixels per parameter area, as indicated by the
colour bar. The white curve shows the median relation. 
}
\label{fig:corr_G150}
\end{figure}

The models do not include details such as radiation field anisotropy
or internal sources. Even when these exist in the real data, the model
results are valuable by explicitly showing these effects in their
residuals. For partly the same reason, we do not directly use the
column density maps from the models to correct the MBB values. 
We use the ratio $\xi_N$ between the actual column density of the
model cloud and the column density that is estimated from the surface
brightness produced by the model. This gives a more robust estimate of
the {relative} bias in MBB analysis that is caused by LOS
temperature variations. Thus, in addition to the original column
density $N_{\rm obs}$ derived from observations via MBB fits, we will
use bias-corrected column density maps $N_{\rm cor}=\xi_N \, N_{\rm
obs}$. Figure~\ref{fig:corr_G150} shows the field G150.47+3.93 as an
example.

\subsection{Clump extraction} \label{sect:clumps}

\begin{figure}
\includegraphics[width=8.8cm]{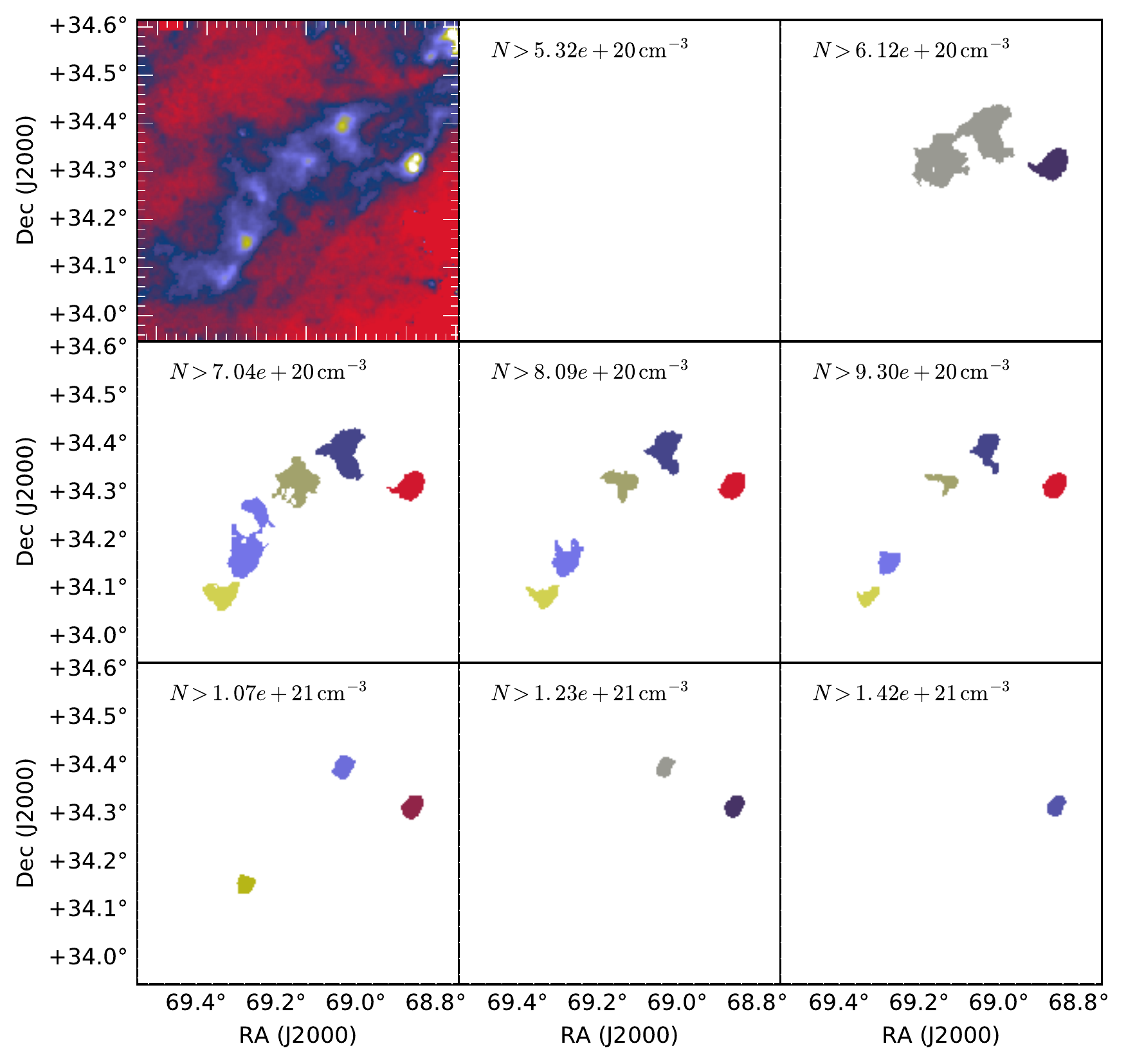}
\caption{
Example of the clump hierarchy. The first frame shows the column
density map of the field G167.20-8.69. The other frames show clumps
identified on the first seven column density thresholds, different
colours denoting different clumps. At the lowest column density
threshold, the structure extends beyond the map boundary and therefore
is not included in our sample.
}
\label{fig:sample_hierarchy}
\end{figure}

We extract from each field the densest structures that are
subsequently referred to as clumps. Our sample of fields is very
heterogeneous. Based on the estimates derived from SPIRE data, the
peak column densities vary by almost two orders of magnitude, from
$\sim 5\times 10^{20}$\,cm$^{2}$ up to $\sim 5\times
10^{22}$\,cm$^{2}$. The resolved linear scales also differ by up to a
factor of five because of the different distances. 
The main purpose of our clump selection is to locate density peaks,
most of which may also be relevant to star formation. We also use the
clump extraction for a more general characterisation of the density
structures. 
The term ``clump'' may thus refer to either (gravitationally bound)
cores, clumps, or even larger cloud structures. 

The two-dimensional (2D) selection is made by thresholding the column density maps,
which is a simple and objective way to characterise cloud
structure.
The column densities are estimated from SPIRE data with MBB fits,
assuming a fixed dust opacity spectral index $\beta=1.8$, a dust opacity
of $0.1 (\nu/1000\,{\rm GHz})^{\beta}\,{\rm cm}^2\,{\rm g}^{-1}$, and
a total mass of 2.8 atomic mass units per Hydrogen atom; see
\citet{GCC-VI}.  

We use 38 thresholds that are placed logarithmically between $0.2
\,10^{21}$\,cm$^{-2}$ and $3.5 \, 10^{22} \,{\rm cm}^{-2}$. Each
spatially connected region above a given threshold is counted as a
separate clump. Clumps that touch any of the map boundaries (over a
distance of five border pixels) or have an area smaller than
0.5\,arcmin$^2$ (corresponding to the 40$\arcsec$ resolution of the
column density maps) are rejected. Thresholding produces clump masks
that can be used for any 2D map, including those resulting from the
RT models. Figure~\ref{fig:sample_hierarchy} shows an example of 
extracted clumps.

Counting all the fields and clumps defined at all 38 column density
thresholds, we have 2998 detections. The number is large because of
clumps detected at a number of column density thresholds. Selecting in
each field the $N({\rm H}_2)$ threshold that results in the largest
number of objects in that field, the median number of clumps is 3 per
field. A typical clump is found at a column density level of $N({\rm
H}_2)=8 \times 10^{20}$\,cm$^{-2}$. 
The number of clumps per column density threshold is not
significantly different for fields with distances below and above
250\,pc, although more distant fields tend to have a slightly larger
fraction of their clumps below $N({\rm H}_2)=2 \times
10^{21}$\,cm$^{-2}$.

The 3D clumps are defined using volume-density isocontours and are
used only in connection with the 3D RT models. These are independent
of the 2D clumps but can also be selected to have a similar extent in
2D projection as the 2D clumps. These will be compared later in
Sect.~\ref{sect:stability} regarding their estimated stability, the 3D
clump analysis making use of the 3D clump density distributions as
they appear in the RT models.

\subsection{Characterisation of 2D clump structure} \label{sect:profile_fit}

We calculate first estimates of clump properties by finding the axis of maximum variance
for offsets weighted by column density. We can then proceed to calculate first moments
along the major and minor axes, to characterise the elongation (aspect ratio) based on the
second moments and asymmetry based on skewness. Similarly, kurtosis characterises the
column density profile, separating peaked (high kurtosis) and flat-topped (small kurtosis)
shapes.

We also fit 2D models to estimate the size, elongation, and radial profiles of the
clumps. We use a Gaussian model with seven parameters: the peak column density, two
components of the centre position, the position angle, and the FWHM values along the
minor and the major axes. The Gaussian models provide basic estimates of the size and
the orientation of the clumps.

To further characterise differences in the shape of the radial column density
profiles, we use 2D Plummer functions
\begin{equation}
N(x,y) = N_0  \left( 1 +  \frac{(x-x_0)^2}{R_x^2}  
 +  \frac{(y-y_0)^2}{R_y^2} 
\right)^{-p} + N_{\rm bg},
\label{eq:Plummer}
\end{equation}
where the parameters are the peak column density $N_0$, the centre coordinates ($x_0$,
$y_0$), and the exponent of the asymptotic powerlaw $p$. The coordinates ($x$, $y$)
are measured in a rotated coordinate system, making the position angle an additional
free parameter. We can either allow different distance scales through the $R_x$ and
$R_y$ parameters or assume $R_x \equiv R_y$ 
To take into account the data resolution, the models are always convolved to the map
resolution during the fitting.

\begin{table}
\caption[]{
Fractal dimensions $D_{\rm P}$ of column density maps, confusion noise estimates
$S(40\arcsec)$, and the parameters $A$ and $B$ of their dependence on angular separation.
The fields are in order of increasing distance.
} 
\centering
\begin{tabular}{lccccc}
\hline
\hline
Field &   $D_{\rm P}$  &   $S(40\arcsec)$ (mJy) &   $A^1$ (mJy) &    $B$  \\
\hline
   G159.23-34.51 &    1.25    &    52.6 &  78.00 &   0.54 \\
   G155.80-14.24 &    1.27    &    43.1 &  57.74 &   0.48 \\
    G157.08-8.68 &    1.31    &   128.3 & 143.12 &   0.35 \\
    G161.55-9.30 &    1.28    &    48.7 &  64.81 &   0.54 \\
    G149.67+3.56 &    1.31    &    47.2 &  61.85 &   0.62 \\
   G126.63+24.55 &    1.25    &    26.1 &  32.08 &   0.49 \\
    G150.47+3.93 &    1.34    &    66.7 &  98.81 &   0.55 \\
    G163.82-8.44 &    1.34    &    62.9 &  86.98 &   0.37 \\
    G151.45+3.95 &    1.15    &    50.8 &  58.04 &   0.86 \\
   G210.90-36.55 &    1.19    &    21.7 &  25.93 &   0.72 \\
    G167.20-8.69 &    1.25    &    30.0 &  40.13 &   0.44 \\
    G164.71-5.64 &    1.26    &    27.0 &  33.42 &   0.57 \\
   G181.84-18.46 &    1.21    &    28.3 &  42.74 &   0.69 \\
    G154.08+5.23 &    1.15    &    57.8 &  77.68 &   0.51 \\
   G206.33-25.94 &    1.17    &    44.8 &  66.11 &   0.73 \\
    G173.43-5.44 &    1.26    &    20.7 &  26.38 &   0.52 \\
   G188.24-12.97 &    1.34    &    25.4 &  30.83 &   0.43 \\
   G189.51-10.41 &    1.33    &    31.1 &  41.12 &   0.49 \\
    G198.58-9.10 &    1.23    &   147.7 & 163.51 &   0.33 \\
   G212.07-15.21 &    1.29    &    25.1 &  30.83 &   0.48 \\
    G203.42-8.29 &    1.29    &    42.8 &  55.33 &   0.51 \\
    G205.06-6.04 &    1.32    &    32.4 &  41.54 &   0.53 \\
   G247.55-12.27 &    1.25    &    26.2 &  37.33 &   0.60 \\
   G141.25+34.37 &    1.40    &    17.7 &  16.21 &   0.53 \\
   G298.31-13.05 &    1.16    &    29.8 &  42.45 &   0.64 \\
    G300.86-9.00 &    1.21    &    30.9 &  54.12 &   0.80 \\
    G300.61-3.13 &    1.07    &    34.2 &  49.54 &   0.51 \\
   G358.96+36.75 &    1.05    &    18.7 &  24.97 &   0.99 \\  
     G4.18+35.79 &    1.14    &    39.2 &  51.71 &   0.89 \\  
     G6.03+36.73 &    1.11    &    49.6 &  67.28 &   0.81 \\  
    G341.18+6.51 &    1.05    &    36.3 &  59.48 &   0.71 \\
    G344.77+7.58 &    1.30    &    37.3 &  49.40 &   0.38 \\
     G2.83+21.91 &    1.21    &    20.3 &  23.43 &   0.58 \\
     G3.72+21.02 &    1.26    &    20.1 &  24.26 &   0.50 \\
     G0.02+18.02 &    1.29    &    28.2 &  42.03 &   0.51 \\
     G9.45+18.85 &    1.26    &    20.0 &  20.20 &   0.57 \\
     G0.49+11.38 &    1.17    &    43.7 &  66.65 &   0.68 \\
      G3.08+9.38 &    1.29    &    34.6 &  47.19 &   0.55 \\
   G315.88-21.44 &    1.27    &    37.6 &  49.04 &   0.48 \\
    G345.39-3.97 &    1.27    &   134.8 & 177.72 &   0.37 \\
      G1.94+6.07 &    1.20    &    36.8 &  50.45 &   0.59 \\
    G21.26+12.11 &    1.17    &    53.2 &  70.25 &   0.58 \\
     G20.72+7.07 &    1.28    &    26.2 &  32.27 &   0.41 \\
     G26.34+8.65 &    1.18    &    31.5 &  48.08 &   0.45 \\
     G25.86+6.22 &    1.34    &    53.9 &  75.31 &   0.49 \\
     G24.40+4.68 &    1.24    &    36.8 &  50.57 &   0.48 \\
     G93.21+9.55 &    1.19    &    50.9 &  74.29 &   0.52 \\
   G108.28+16.68 &    1.40    &    32.3 &  37.75 &   0.44 \\
   G110.80+14.16 &    1.29    &    27.0 &  32.42 &   0.50 \\
   G110.62-12.49 &    1.10    &    24.1 &  36.99 &   0.70 \\
    G116.08-2.40 &    1.13    &    42.6 &  57.93 &   0.65 \\       
\hline
{$^{1} S(\theta)=A \times (\theta/1\arcmin)^B$}
\end{tabular}
\label{table:fieldstat}
\end{table}

\section{Results} \label{sect:results}

\subsection{Characterisation of the fields} \label{sect:chara}

Table~\ref{table:fieldstat} lists the fractal dimensions $D_{\rm P}$ 
calculated for the $\tau(250\mu{\rm m})$ maps. The mean value is 1.25,
equal to the median value, and with a scatter of $\sigma(D_{\rm
P})=0.07$. 
The use of 250\,$\mu$m surface-brightness data (instead of column
density) leads to a wider $D_{\rm P}$ distribution with $\sigma(D_{\rm
P})=0.13$. This scatter can be affected by noise, which is larger in a
single band than in the column densities derived from SED maps.
Surface-brightness maps are also more sensitive to dust temperature
variations and especially to the presence of warm point sources. The
median value is 1.29 and the mean is 1.33. These are higher than
the values estimated from column density data but the difference is
not  very statistically significant ($\sim1.5\,\sigma$).

The structure functions $S(\theta)$ were calculated for
$I_{\nu}(250\mu {\rm m})$ maps by using two symmetrically placed
reference positions (see Sect.\ref{sect:global}). To examine the
dependence on the angular scale, calculations were done at 
$40\arcsec$ steps up to a maximum scale of $\theta=600\arcsec$. The
value $\theta=40\arcsec$ is close to the resolution limit of the maps,
adopting $\theta > 2 \times D$. At the largest scales, the values
become biased, not only because of the finite map size but because the
maps are preferentially centred at column density maxima. This
selection effect tends to increase $S(\theta)$ when $\theta$
approaches the map radius.

\begin{figure}
\includegraphics[width=8.8cm]{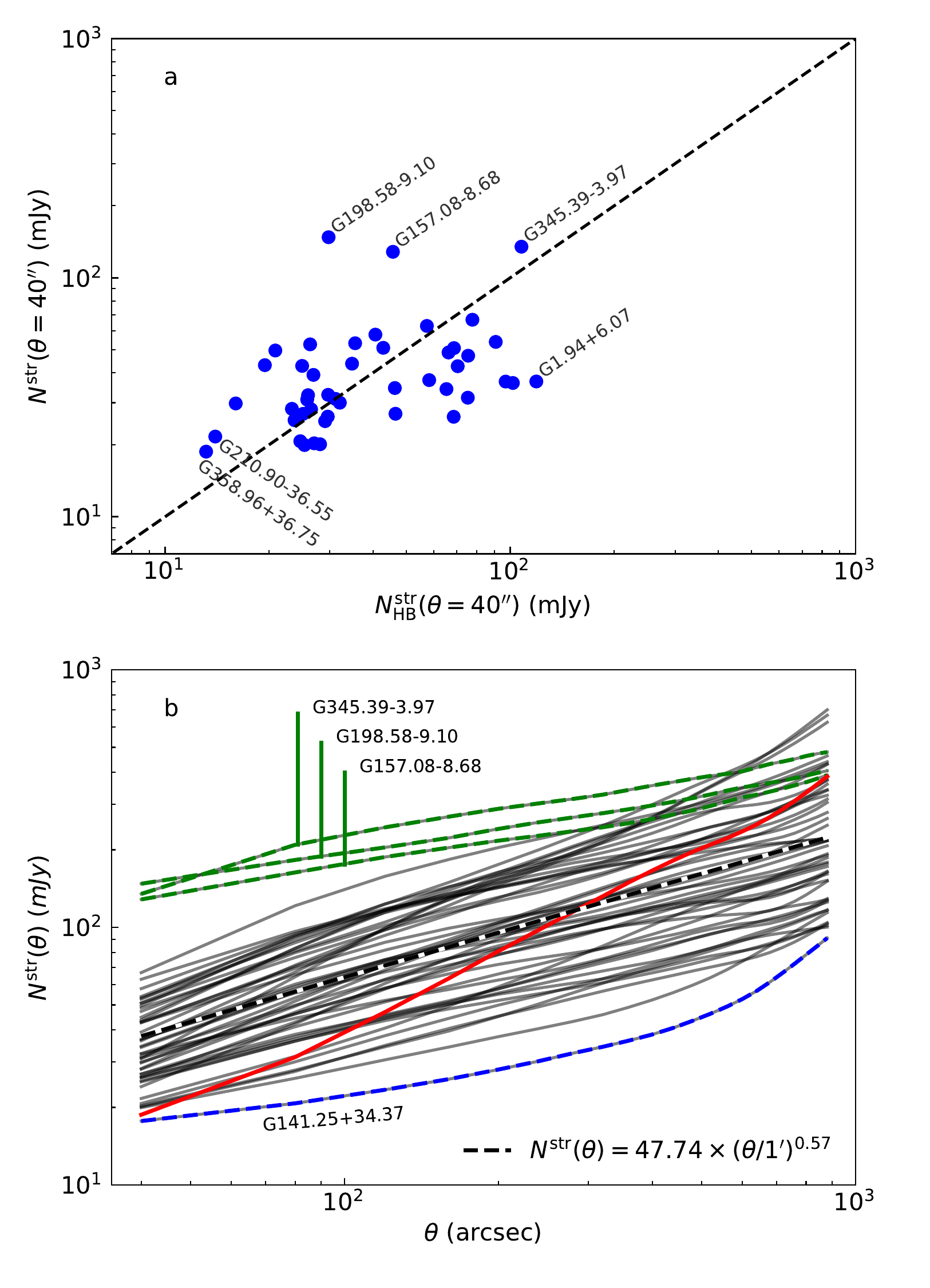}
\caption{
Structure noise $N^{\rm str}(\theta)$ in selected fields. The upper frame
shows $N^{\rm str}(\theta=40\arcsec)$ for the 250\,$\mu$m surface brightness 
(blue dots), compared to the surface-brightness dependence of
\citet{HelouBeichman1990} (dashed line).  The lower frame shows
$N^{\rm str}(\theta)$ in individual fields as a function of the angular scale
$\theta$. The black dashed line indicates the median relation. The
coloured lines highlight particular fields discussed in the text
(G141.25+34.37 as the bottom blue dashed line, G358.96+36.75 as the
red solid line, and the fields G345.39-3.97, G198.58-9.10, and
G157.08-8.68 as the uppermost dashed green lines).
}
\label{fig:sf}
\end{figure}

Structure function can be converted to structure noise $N^{\rm
str}$ via the relation $N^{\rm str}=\sqrt{S(\theta)}\, \Omega$, where
$\Omega$ is the solid angle of the measurement aperture
\citep{Kiss2001}.
In the upper frame of Fig.~\ref{fig:sf}, the $N^{\rm str}$ values are
compared to the analytical expression
\begin{equation}
N^{\rm str}_{\rm HB}(\theta) = 0.3\,{\rm mJy} \left( \frac{\lambda}{100\,\mu m} \right)
^{2.5} \left( \frac{D}{\rm 1\,m} \right) ^{-2.5} \left( \frac{\langle
B \rangle}{1\, \rm MJy/sr} \right) ^{1.5}
\label{eq:HB}
,\end{equation}
 presented by \citet{HelouBeichman1990}.
Here $\lambda$ is the wavelength, $D$ the telescope size (assuming
diffraction-limited observations that define the angular scale 
$\theta$), and $\langle B \rangle$ the average surface brightness. For
the ratio between the observed and predicted values, 
$N^{\rm str}(40\arcsec)/N^{\rm str}_{\rm HB}(40\arcsec)$.
Compared to \citet{Kiss2001}, our scatter is larger but we
observe a similar behaviour at low intensities, where the $N^{\rm
str}(\theta)$ values tend to rise above the Eq.~(\ref{eq:HB})
predictions.
Compared to \citet{Kiss2001}, our scatter is larger but we observe the
same behaviour where at low intensities the $N^{\rm str}(\theta)$
values tend to rise above the prediction of Eq.~(\ref{eq:HB}).

The lower frame in Fig.~\ref{fig:sf} shows the $N^{\rm str}(\theta)$
curves for all fields, with a median relation of $N^{\rm
str}(\theta)=48 \times (\theta/1\arcmin)^{0.57}$. The Figure
highlights some extreme fields. The highest $N^{\rm str}(40\arcsec)$
values are found for G198.58-9.10, G345.39-3.97, and G157.08-8.68.
Fields G198.58-9.10 and G157.08-8.68 are indeed filled with
significant small-scale structure while for G345.39-3.97 the result
can be explained by the compact central region where the surface
brightness exceeds 1000\,MJy\,sr$^{-1}$. In all three cases, $N^{\rm
str}$ increases only slowly with increasing $\theta$. Interestingly,
the smallest $N^{\rm str}(40\arcsec)$ values are found for
G141.25+34.37 and G358.96+36.75 (LDN~1780), the fields that
represented the two extremes of the fractal dimension distribution.
This suggests very little dependence between $D_{\rm P}$ and $N^{\rm
str}(40\arcsec)$, which is confirmed by a Pearson correlation
coefficient $r=$-0.02 for the whole sample. 
This conclusion does not depend on the scale at which structure
noise is evaluated (see Fig.~\ref{fig:sf}b) and remains true if
$N^{\rm str}(\theta)$ is estimated using column density instead of surface-brightness data.
Both fractal dimensions and structure noise appear to be
independent of the linear resolution of the data. The linear
correlation coefficient is 0.19 and 0.07 when $D_{\rm P}$ and
$N^{\rm str}(40\arcsec)$ are correlated with the field distance, respectively.

The PDFs of column density and $250\mu$m surface brightness are shown
in Appendix~\ref{sect:PDF}. PDFs show a range of shapes that are often
far from a log-normal distribution. Unlike the fractal dimension or
the structure noise, the PDFs also change if the analysis is done
with background-subtracted maps. The observations targeted local
column density peaks, which directly skews the statistics. The
occasional power-law tails towards high column densities are not
necessarily a sign of gravitationally bound structures. The PDF plots
are affected by the limited size of the fields and often reflect the
morphology of individual structures or even a single clump. For
example, the field G198.58-9.10 shows a well-defined power-law tail,
which is even more pronounced in column density than in surface
brightness. The cloud has a high-contrast boundary that is a clear
sign of external forcing, possibly by the nearby O star $\lambda$
Orionis or by previous generations of high-mass stars. Such structures
have a qualitatively similar effect on the PDF shape, irrespective of
gravitational stability of the region. The field G141.25+34.37 is 
again an outlier, showing a very distinct powerlaw tail towards {\em
smaller} column densities. In the 250\,$\mu$m data, the PDF extends
with a similar slope far below the range shown in Fig.~\ref{fig:pdf1},
which simply reflects the density profile of this diffuse cloud.

The log-normal function often results in a poor fit. To quantify the
asymmetry (e.g. in the case of a high-column density tail), we use the
skewness and a quantity calculated from the percentile values,
$[P(99\%)-P(50\%)]/[P(50\%)-P(10\%)]$. The correlation coefficients
between these quantities and the average field column density, the
cloud distance, and the Galactic latitude were all small, with
$r<0.15$.

\begin{figure}
\includegraphics[width=8.8cm]{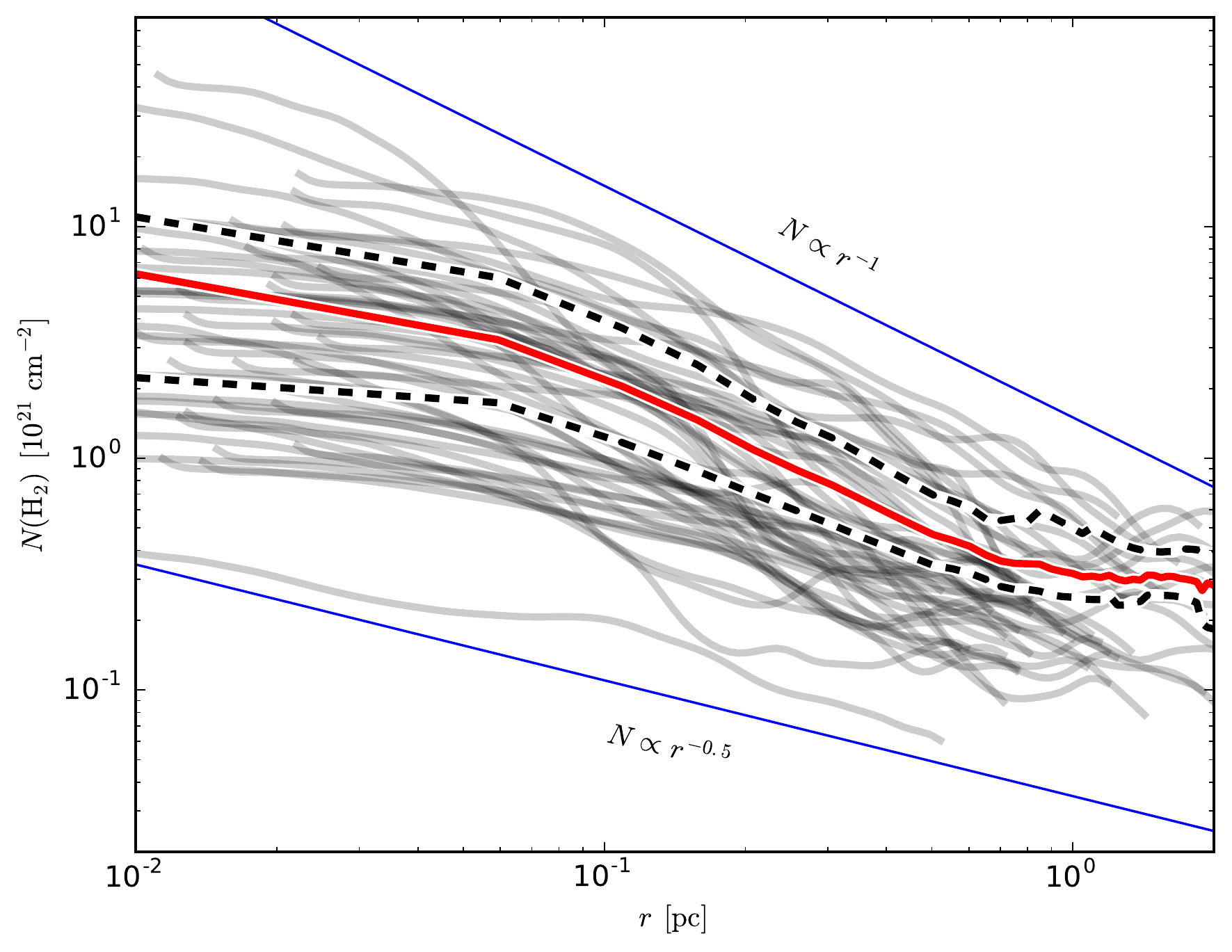}
\caption{
Average column density profiles as a function of the distance from the
highest column density peak in each field (grey lines). The
thick red line shows the mean column density profile and the dashed
lines the average profiles separately for fields with peak column
density below or above $5 \times 10^{21}$\,cm$^{-2}$. For comparison,
the solid blue lines show the $r^{-1.0}$ and $r^{-0.5}$ relations.
}
\label{fig:MvsR}
\end{figure}

To characterise the mass distribution in the fields, we examine the
column density profiles around the highest column density peak of each
field. The curves in Fig.~\ref{fig:MvsR} are obtained by averaging
over concentric circles. However, at each distance, the average is
evaluated only over those radial directions where the column density
is still a monotonously decreasing function of the distance. This
reduces the effects of other clumps and local column density peaks, in
an attempt to describe the underlying large-scale structure.
Beyond $r=$0.1\,pc, the radial profiles are well resolved even
for the most distant fields. Figure~\ref{fig:MvsR} shows that at
scales $r=0.1-1$\,pc the column density profiles are shallow and
typically correspond to $N(r)\propto r^{-1}$ or an even flatter
distribution. 
These describe the profiles of individual regions and are not to be
confused with size-mass relations of samples of distinct sources. In
the latter, the typical relation $M \sim r^2$ simply corresponds to a
constant average column density \citep{Larson1981, Friesen2016}.
\citet{Mueller2002} examined the radial profiles for a sample of
massive star-forming clouds. On average these corresponded to
$M(r)\sim r^{1.2}$. \citet{Kauffmann2010} found a similar relation
$M(r)\sim r^{1.27}$ for cluster-forming clouds \citep[see
also][]{Beuther2002, Zinchenko2005, Schneider2015_LOS, Lin2016}. These
results suggest a column density relation $N(r)\sim r^{-0.8}$.
The results of \citet{Shirley2000} on a sample of low-mass star-forming cores gave an average relation of $N(r)\sim r^{-1.1}$. That
result is dependent on assumptions of radial dust temperature profiles
and also partly relates to scales below our resolution. However, our
$N(r)$ profiles in Fig.~\ref{fig:MvsR} are compatible with the
above-quoted $N(r)\sim r^{-p}$ relations, with some preference for
values $p<1$.

\subsection{Basic statistics of the clumps} \label{sect:bstat}

\begin{figure*}
\includegraphics[width=17.0cm]{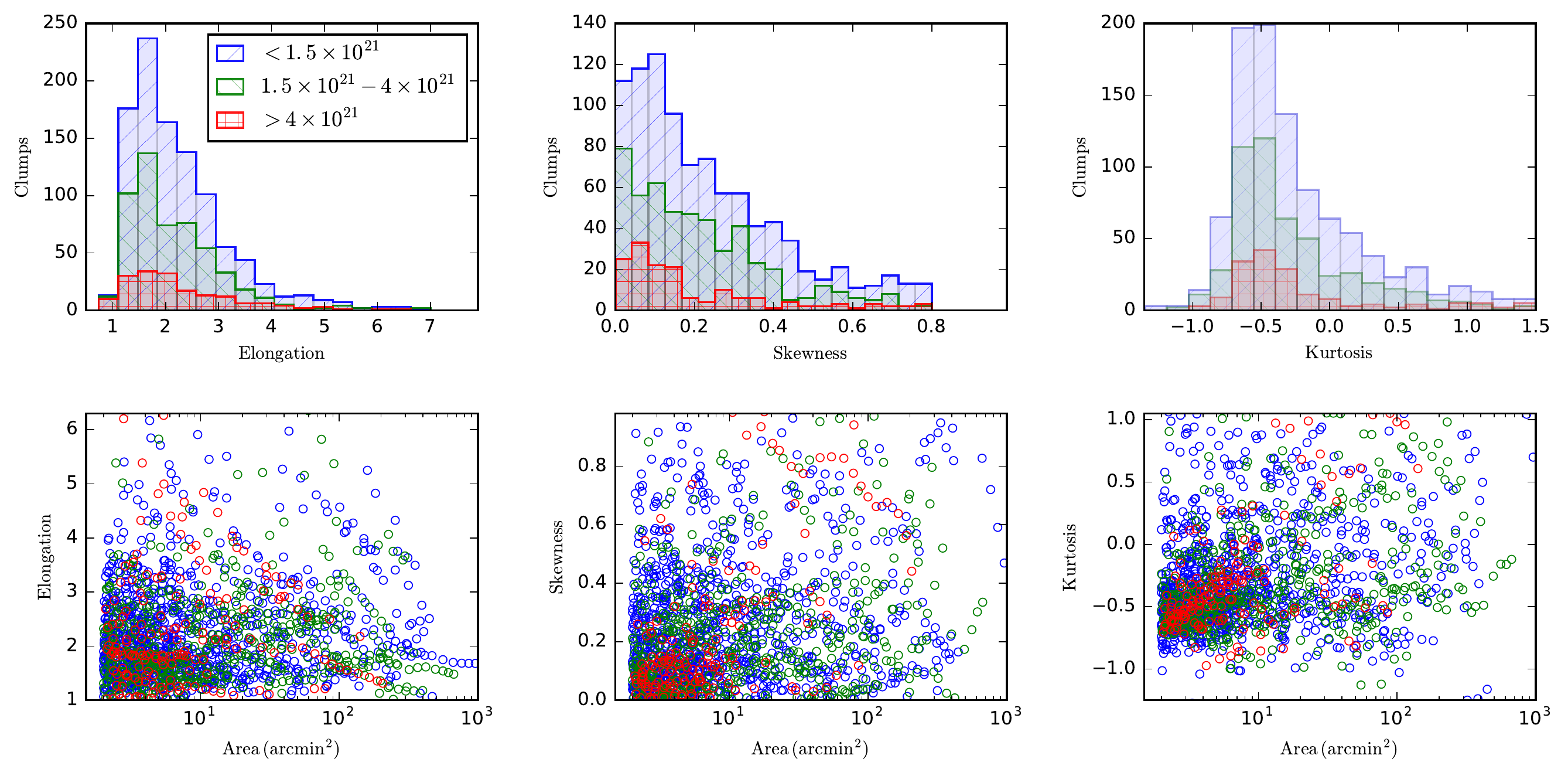}
\caption{
Basic statistics for selected clumps. The colours blue, green, and red correspond to
clumps with mean column density $N({\rm H}_2)< 2.1 \times 10^{21}$\,cm$^{-2}$, $N({\rm
H}_2)=2.1-5.0 \times 10^{21}$\,cm$^{-2}$, and $N({\rm H}_2)>5.0 \times
10^{21}$\,cm$^{-2}$, respectively. In the top row, skewness and excess kurtosis are
calculated for the minor axis direction. 
}
\label{fig:limited_clump_sample}
\end{figure*}

Clumps are identified through column density thresholding as explained
in Sect.~\ref{sect:clumps}.
After subtracting a threshold column density, we determine for each
clump the main axis as the direction of maximum
column-density-weighted standard deviation of pixel positions.
Skewness and kurtosis are calculated for this axis and for the
perpendicular direction. In Fig.~\ref{fig:limited_clump_sample}, we
have excluded clumps smaller than 2\,arcmin$^2$ and further divided
the clumps to three column density categories.
The overall median elongation is 1.5 and the values cover a broad
range of values but, surprisingly, are not significantly different for
different column density intervals. Even high-density clumps exhibit a
wide range of asymmetries, up to a skewness of $\sim$0.8. The third
frame of Fig.~\ref{fig:limited_clump_sample} shows the distribution of
kurtosis, or more precisely the excess kurtosis, which is defined to
be zero for the normal distribution. Negative values are suggestive of
flat-topped structures. Instead of very peaked isolated clumps, the
largest positive values are caused by compact structures seen on top
of extended diffuse emission.
Partly for the same reason, the only significant correlation is
observed between clump area and kurtosis. This becomes particularly
significant for the subsample of high-density clumps.

\begin{figure}
\includegraphics[width=8.8cm]{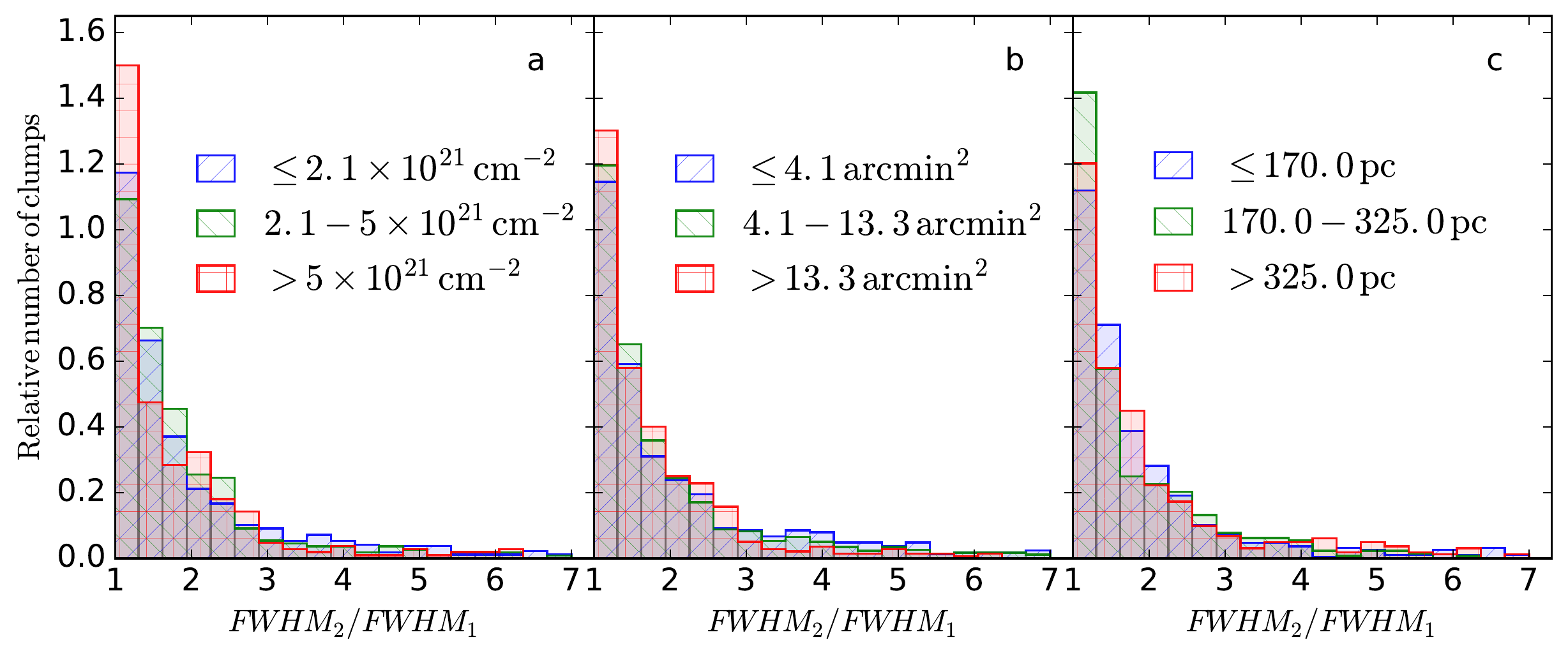}
\includegraphics[width=8.8cm]{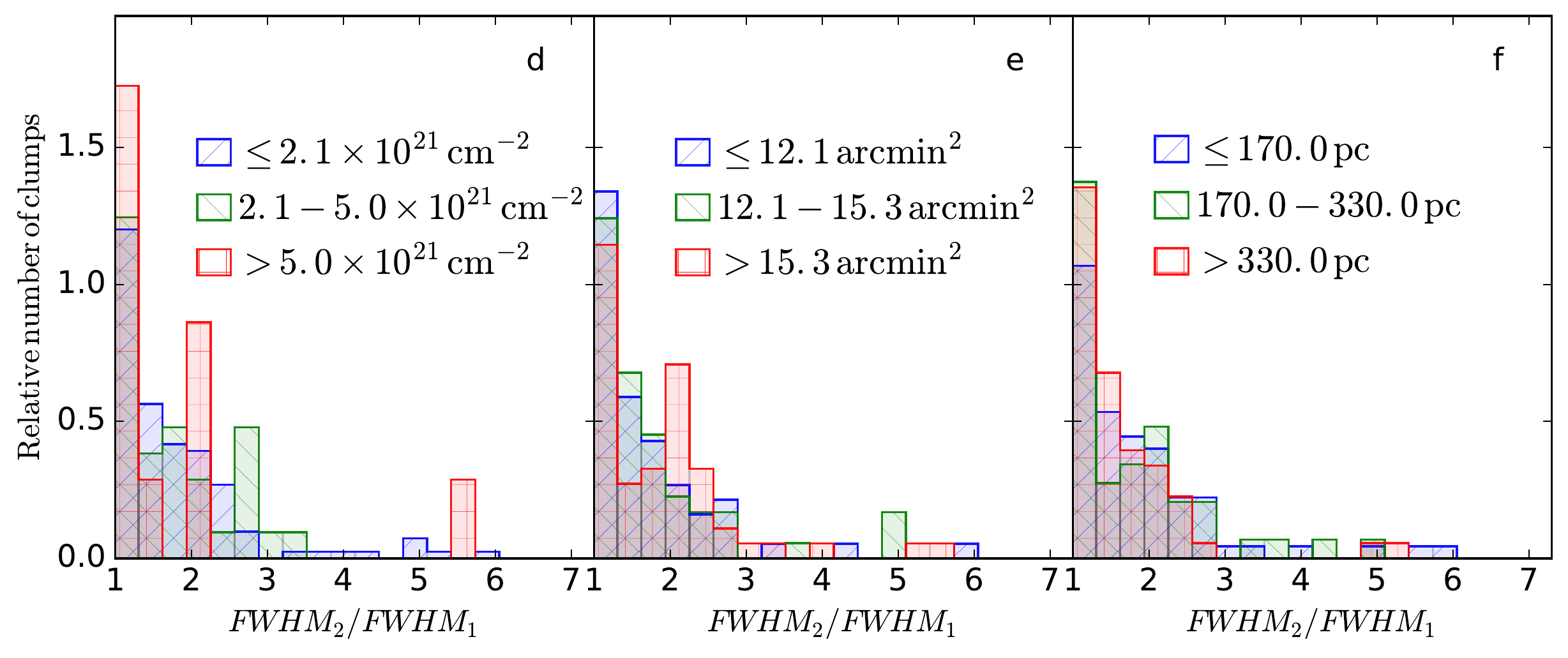}
\caption{
Aspect ratios of the fitted 2D Gaussians. The different histograms correspond to different
ranges of column density (frame a), clump area (frame b), and field distance (frame c), as
indicated in the frames. The lower frames d-f show the same data for a subset of clumps
with areas of 10\,arcmin$^2$.
}
\label{fig:plot_elongation}
\end{figure}

The aspect ratios and position angles of the clumps were also estimated by using the fits
of 2D Gaussians. Compared to direct dispersion measurements, these respond differently to
the presence of secondary peaks or extended, low-column density pedestals under dense
clumps. We exclude from the fitted sample clumps that are smaller than 2\,arcmin$^2$ or
for which the relative root mean square (rms) relative error of the fit exceeds 10\% (a
minor fraction of all clumps). Figure~\ref{fig:plot_elongation} shows the aspect ratios
(defined as the ratio of the FWHM values along the major and minor axes) for three column
density, clump size, and distance intervals. The statistics are again not strongly
dependent on any of these parameters. Symmetric clumps are slightly more likely to be
small in size and have high column densities, but this is partly a bias resulting from the
finite data resolution. In Fig.~\ref{fig:plot_elongation}, the lower frames show
corresponding distributions where, for each column density peak, we include only one clump
with an area of 10.0\,arcmin$^2$. The elimination of very extended clumps and clumps near
the resolution limit does not have a clear effect on the statistics.
Compared to Fig.~\ref{fig:limited_clump_sample}a, the axis ratio (elongation) of Gaussian
fits peaks closer to the value of one. This is probably in part a property of the methods
used, Fig.~\ref{fig:limited_clump_sample} being more easily affected by the diffuse
background above which more compact structures are seen.

\subsection{Structure orientation} \label{sect:orientation}

\begin{figure}
\includegraphics[width=8.8cm]{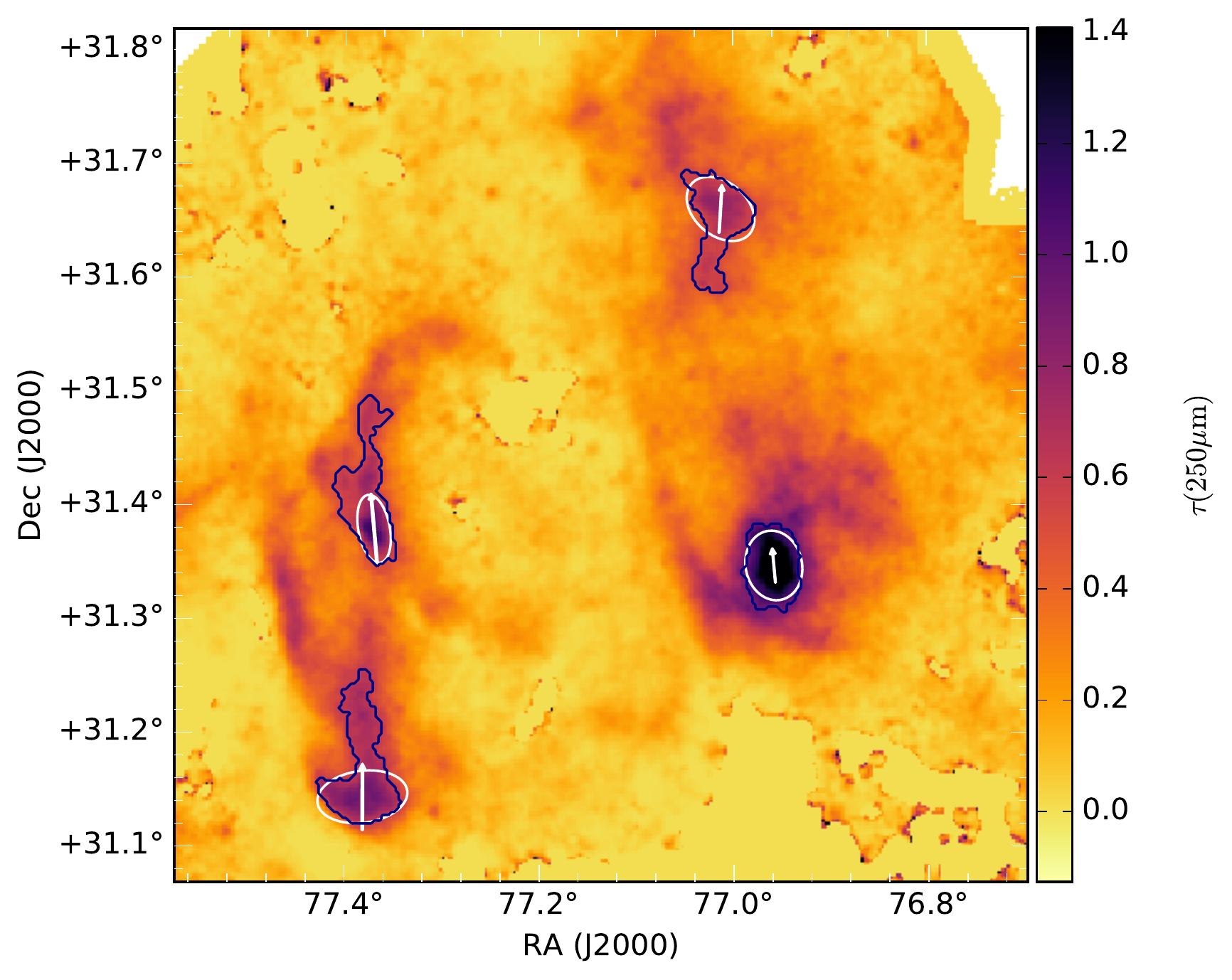}
\caption{
Clump orientations in the field G173.43-5.44. The black contours correspond to the
boundaries of the selected structures. The arrows show the direction of maximum variance,
with positive skewness along the arrow direction, and the ellipses show the orientation of
the fitted 2D Gaussians.
}
\label{fig:sample_orientations}
\end{figure}

The results of Sect.~\ref{sect:bstat} suggest that in some fields the clump orientations
may be correlated. As an example, Fig.~\ref{fig:sample_orientations} shows the
orientations of some 10\,arcmin$^2$ clumps in the field G173.43-5.44. The directions of
maximum variance are similar to each other, the result being slightly less clear for major
axis directions of Gaussian fits.

We selected the fields with more than one clump. Clumps were defined using column
density thresholds that resulted in clump areas $\sim$10\,arcmin$^2$. 
We calculated the quantities $|\theta_i-\langle \theta \rangle|$ for each field,
measuring the position angle difference between individual clumps and their median. The
average values of the individual fields are further grouped to samples with $n=$2, 3,
4, or 5 clumps per field. Above, $\langle \theta \rangle$ refers not to the average but
to the median value. Furthermore, when $n$ is an even number, we select as $\langle
\theta \rangle$ the value for the index $n/2$ (rather than calculating the average of
two elements in the vector).

\begin{figure}
\includegraphics[width=8.8cm]{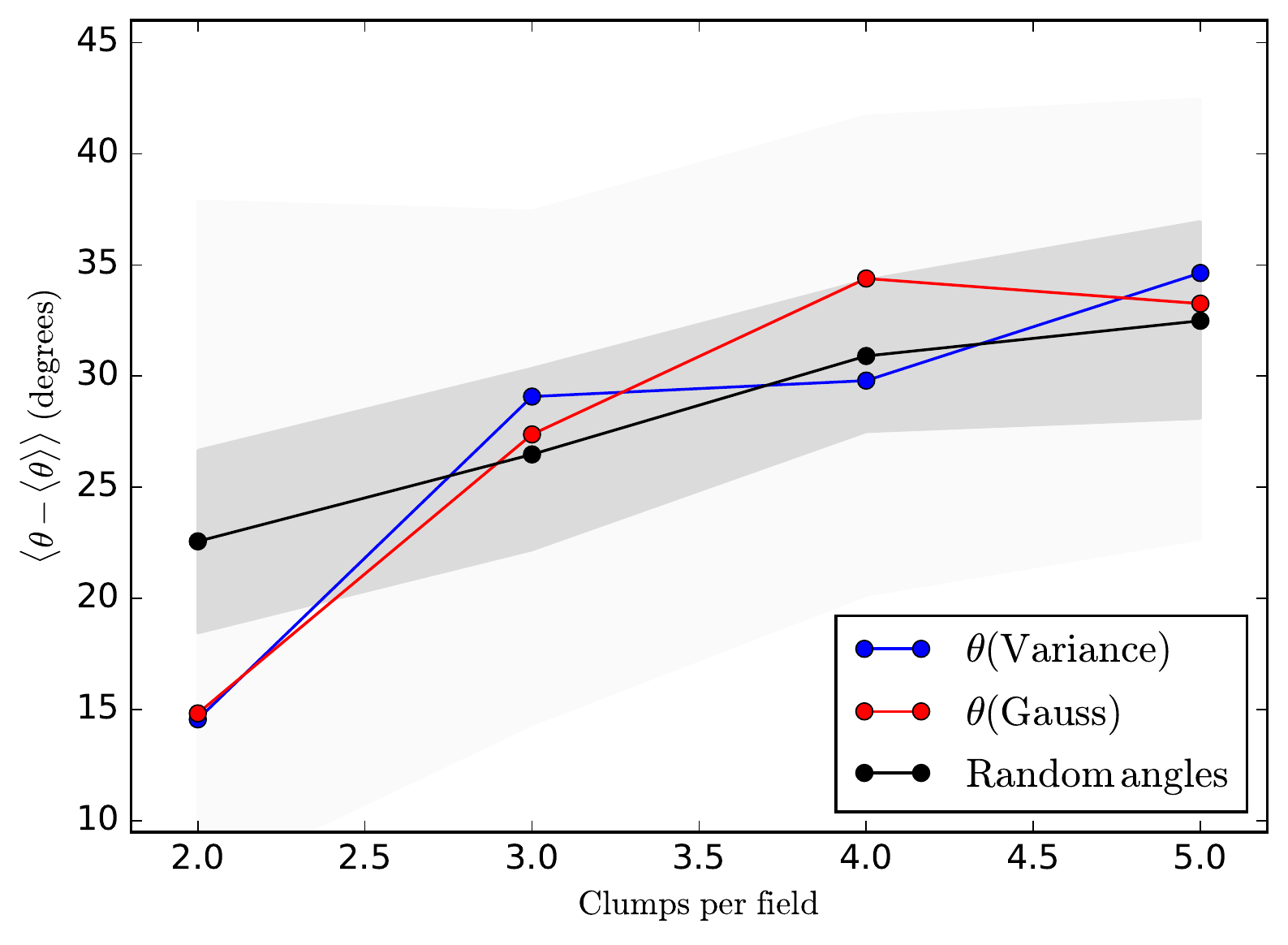}
\caption{
Average difference of the position angles of individual clumps and the average of all
clumps in a field. The relation is plotted as a function of the number of clumps in the
field, using the directions of maximum variance (blue curve) or the main axis of the
fitted 2D Gaussians (red curves). The black curve shows the expected relation for a
completely random distribution of position angles. The shaded region corresponds to 68\%
interval for a single field and the dark shaded region to the 68\% interval for the
average, taking into account the number of fields with the given number of clumps.
}
\label{fig:clump_ori}
\end{figure}

Figure~\ref{fig:clump_ori} compares the observed position angle
differences to the values expected for a completely random angle
distribution. In the fields with just two clumps, the relative
orientations of the clumps are significantly correlated. This sample
consists of 14 fields. If the angles in those fields were random, the
probability for their average of $\langle \theta - \langle \theta
\rangle \rangle$ to be below or equal to observed value of 15\% is
1.8\% (estimated with Monte Carlo simulations). For fields with three
or more clumps, the results are compatible with a random distribution.
{Many of these fields are at relatively large distances and thus have
a large physical separation between the clumps.} The median distances
in the four groups $n=$2-5 are 165, 245, 375, and 200\,pc,
respectively.

\begin{figure*}
\includegraphics[width=18.5cm]{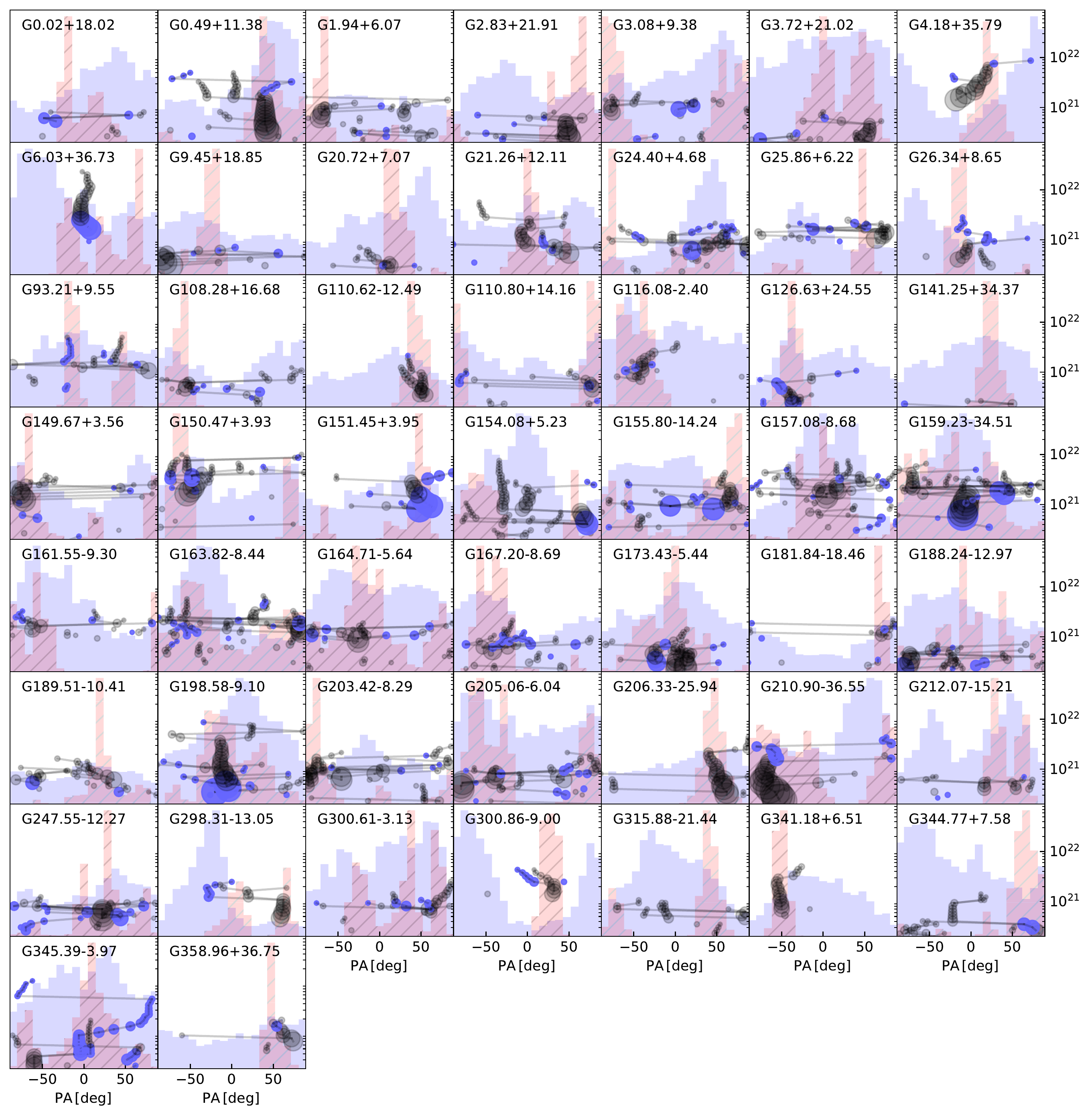}
\caption{
Orientation of structures in the 51 analysed fields. The
histograms show the distribution of position angles from the TM
analysis with ${\rm FWHM}=0.6\arcmin$ (blue histogram) and ${\rm FWHM}=5.0\arcmin$
(red hatched histograms). The histogram normalisation is arbitrary.
The circles show the position angles of the clumps at different column
density thresholds (right hand axis, in units of H$_2$ column density
cm$^{-2}$). The black and blue symbols are partly transparent and
correspond, respectively, to clumps with aspect ratios above and below
the value of 1.5. Lines connect clumps to the parent clumps at lower
column density levels. 
The radius of the plotted symbols is proportional to the
clump area raised to the power of 2/3. 
}
\label{fig:plot_directions}
\end{figure*}

We next compare the clump orientations to the large-scale anisotropy of the column
density structures over the entire field. Figure~\ref{fig:plot_directions} summarises
the results.
The histograms show the position angle distributions determined with the TM method (see
Sect.~\ref{sect:global}). To enable the examination of small spatial scales, the method is
applied to the 250\,$\mu$m surface-brightness maps with the original $18\arcsec$
resolution. The blue histograms correspond to structures extracted at the ${\rm FWHM}\sim
0.6\arcmin$ scale, using the normalisation that makes the method insensitive to the
absolute pixel values. The histogram shows the distribution of the position angles for
10\% of positions with the highest significance (see Sect.~\ref{sect:global}). The red
histogram is the corresponding distribution of position angles extracted at the larger
scale of ${\rm FWHM}=5.0\arcmin$. These correspond to the positions with 20\% of the highest
significance values, the calculations including the weighting by the local column density.
For examples of the TM extractions, see Fig.~\ref{fig:plot_TM_examples}.

In Fig.~\ref{fig:plot_directions}, the circles indicate the position
angles (the direction of maximum dispersion) of clumps extracted at
different column density levels (see Sect.~\ref{sect:clumps}). The
size of the symbols is related to the clump area and the lines
connect each clump to its parent/child clumps at the next lower/higher
column density level. Thus, Fig.~\ref{fig:plot_directions} shows the
change of structure orientations as function of both scale and column
density. 

Clump orientation is often similar to the directions probed by the red histograms, the
preferred large-scale orientation of the high-column-density structures. This is often
natural because the clumps themselves form a part of the high-column-density structures.
Clump orientation carries memory of the orientation of the parent structures, often from
column density levels lower by an order of magnitude.
We examined this separately for those clumps that include the main column density peak of
each field. Comparing the position angles of the clumps defined by the lowest- and the
highest-column-density contours, the average value of $\langle \theta - \langle \theta
\rangle \rangle$ over all fields is 17.2$\degr$. Comparison to Fig.~\ref{fig:clump_ori}
(for $n$=2) shows that this is still significantly smaller than expected for random
angles.
There are also exceptions and at the highest column densities the orientation may
rotate by up to full 90 degrees (e.g. G345.39-3.97). Significant changes of the
orientation are observed also for example in G4.18+35.79, G6.03+36.73, and
G167.20-20.89. Even in these cases, the position angles can be strongly correlated
between lower column density thresholds.

The Musca filament G300.86-9.00 is an example that shows a clear division between the
small- and large-scale structures. The main structure has a position angle of
$\theta\sim30$\,degrees and small-scale, lower-column-density striations are
preferentially perpendicular to the main filament \citep{PlanckXXXIII, Juvela2016TM,
Cox2016}. At high column densities, the field contains two main clumps that share the
general orientation of the main filament, which here, in fact, is mainly composed of those
two clumps. At the highest column densities, their position angles shift with respect to
the axis of the main filament. Similar dichotomy is observed, for example, in
G210.90-39.55, although this is a more complex object that contains subregions with
different preferred structure orientations \citep{Malinen2016}. The same caveat applies to
all data. Fields may locally have a more ordered structure than suggested by 
Fig.~\ref{fig:plot_directions}.

\begin{figure}
\includegraphics[width=8.8cm]{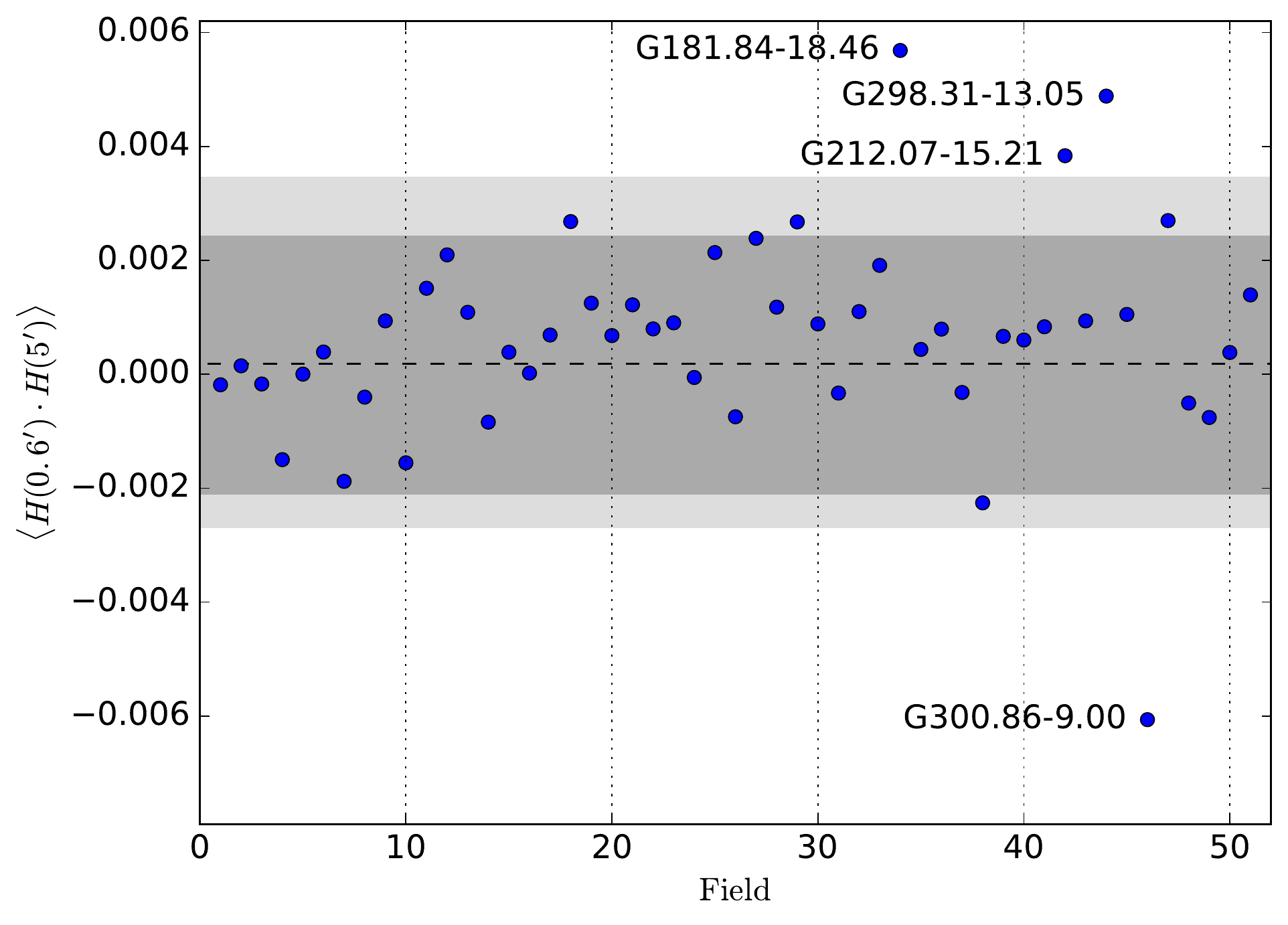}
\caption{
Correlations between 0.6$\arcmin$ and 5$\arcmin$ structures in TM position angle
histograms of Fig.~\ref{fig:plot_directions}, $\langle H(0.6\arcmin \cdot 5\arcmin)
\rangle$. The grey bands correspond to 10-90\% and 5-95\% percentile limits and the dashed
line to the median value derived from Monte Carlo simulations of fields with random
structure orientations. The fields are in order of increasing Galactic longitude. 
}
\label{fig:oricor1}
\end{figure}

The TM results can also be investigated in terms of the position angle correlations as a
function of the spatial separation or the scale $F$ used in the TM analysis.  The
differences between structures identified at the scales of 0.6$\arcmin$ and 5$\arcmin$ is
visible in Fig.~\ref{fig:plot_directions}. We quantify this further in
Fig.~\ref{fig:oricor1}. The Figure shows the quantity $\langle
H(0.6\arcmin) \cdot H(5\arcmin) \rangle$ for each field, where $H$ corresponds to the histograms of
Fig.~\ref{fig:plot_directions} that have been normalised by dividing by the histogram area
and by subtracting the mean histogram value. Thus, the plotted quantity is positive if the
0.6$\arcmin$ and 5$\arcmin$ histograms have a similar structure and negative if the
distributions are anticorrelated. The significance of the values was estimated with Monte
Carlo simulations of 1000\,arcmin$^2$ maps. The map size is relevant because it limits the
number of independent samples, especially at the 5$\arcmin$ resolution. The simulated maps
have Gaussian fluctuations, which follow a $k^{-2}$ power spectrum as the function of the
spatial scale $k$, have an average surface brightness of 50\,MJy\,sr$^{-1}$, and include
white noise with a standard deviation $\sigma=0.5$\,MJy\,sr$^{-1}$. The 10--90\% and
5--95\% confidence limits were determined from the analysis of 1000 simulated maps.
However, the resulting confidence limits should be considered only as rough estimates,
because of the varying size and surface-brightness level of the observed maps. According
to Fig.~\ref{fig:oricor1}, there are only three fields with a significant (at the 5\%
significance level) positive correlation between the small-scale and large-scale
structures. The Musca field G300.86-9.00 remains the only one with a significant negative
correlation. As an example, Appendix~\ref{sect:example_TM} shows the extracted structures
in the fields G181.84-18.46 and G300.86-9.00.

In principle, TM results could be used to investigate the spatial correlations of the
column density anisotropies as a function of the angular separation. This analysis is
complicated by the fact that TM only provides position angle estimates for a subset of all
map pixels. Thus, the correlations also depend on criteria used to select structures for
which the elongation is considered to be significant. Nevertheless,
Appendix~\ref{sect:ADF} shows some results on the angular dispersion functions. 
G300.86-9.00 is again a special case and has a particularly small dispersion of position
angles. At the opposite end can be found fields like G345.39-3.97, which was already found
to have one of the largest structure noise values (see
Fig.~\ref{fig:sf}).

\subsection{Radiative transfer models} \label{sect:RTresults}

RT models are used to investigate the uncertainty of the column density estimates and, for
example, variations of the radiation field intensity. In this Section, we describe the
main results of the RT models, before using these in the analysis of the radial profiles
(Sect.~\ref{sect:radial}) and the stability (Sect.~\ref{sect:stability}) of the clumps.

\subsubsection{The default model} \label{sect:default}

In the default RT models the dust properties are kept fixed and only the model column
density and the intensity of the external radiation field are optimised. Dust parameters are
characterised by the opacity and spectral index values listed in
Table~\ref{table:models}. 

Observations are fitted by adjusting the radiation field intensity and
column density. The models are constructed so that they reproduce the
350\,$\mu$m surface brightness and the average surface brightness
ratio between 250\,$\mu$m and 500\,$\mu$m (in practice, to an accuracy
of 1\% or better). At 250\,$\mu$m and 500\,$\mu$m, the residuals vary
from position to position but are typically a few percent and thus
smaller than the observational uncertainty.
Nevertheless, visual inspection shows several cases where the
residuals are significant. Some examples are given in
Appendix~\ref{sect:residuals}. First, the assumption of an isotropic
radiation field is not always fulfilled. This can be caused by local
heating source, the effect usually extending over a few Herschel
beams. However, in the field G110.62-12.49, a star (not visible in
submm maps) is located between the two main clumps and the effect is
more widespread (see Fig.~\ref{fig:G110_residuals}). 
The anisotropy may also be in the external field. The best examples are
G358.96+36.75, G4.18+35.79, and G6.03+36.73, where the asymmetry is
caused by the direction of the Galactic plane and the contribution of
the high-mass stars of the Sco-Cen OB association
\citep{Ridderstad2006}. The effect is particularly pronounced in the
case of the high-column-density fields G358.96+36.75 and G4.18+35.79.
Figure~\ref{fig:G4_residuals} shows data for G4.18+35.79 where the
maximum 250\,$\mu$m residuals are up to +15\% within the core and down
to -20\% on its shadow side. The residuals clearly show the presence
of a NW-SE gradient. The colour temperature map shows two cold
subclumps. The northern one is significantly colder than predicted by
the RT model while the southern one is warmer. Further quantitative
analysis would require modelling that explicitly includes the field
anisotropy. However, apart from the examples listed above, this is not
a significant factor and is not taken into account in this paper.

A potentially equally important effect is observed in some fields
where the densest clumps appear to reach much lower temperatures than
predicted by the models. As a result, the 250\,$\mu$m residuals are
negative with magnitudes up to $\sim10$\%. One example is shown in
Fig.~\ref{fig:G150_residuals}. This is a field with complex structure
where also the dense clumps (identifiable in the colour temperature
map) do not entirely coincide with the 250\,$\mu$m peaks. The RT model
overestimates the temperature of the main clumps and, because the
radiation field is adjusted based on average emission over a large
area, the model produces too little short wavelength emission for the
most diffuse material.

These discrepancies are interesting because they could indicate a
change in dust properties, an enhanced long wavelength emission that
leads to lower temperature (or a change in the opacity spectral
index). However, there are other possible explanations. First, the
ISRF level is adjusted according to the average spectrum over a large
area. In spite of background subtraction, this may include diffuse
material that may be subjected to a stronger radiation. The models
assume a cubic volume that, depending on the angular size and distance
of the field, can extend up to $\sim$6\,pc. The actual emission may
originate over a wide range of distances and in even completely
different radiation field environments. Second, the finite resolution
of the models (including the LOS density profile) may underestimate
the value of the peak column density. If the 350\,$\mu$m surface
brightness saturates because of an extreme density, a higher column
density may actually result in lower surface brightness
\citep[see][]{Juvela2013_colden}.

\subsubsection{Variations of the basic models} \label{sect:variations}

In addition to the default models of Sect.~\ref{sect:default} (model
version $D$), we carried out model fits with alternative sets of
assumptions that are listed in Table~\ref{table:models}. The versions
$P$ and $W$ are directly aimed at improving the fits of the dense
clumps by, respectively, concentrating on the higher-column-density
peaks or by including the LOS cloud size as additional free
parameters. A decrease (increase) of the external extinction layer
$\Delta A_{\rm V}$ could similarly help the fits by increasing
(decreasing) the temperature contrast between low- and high-density
regions. The versions $K$ and $TD$ are the same as $D$ except for
the use of different dust properties. In the $K$ version we use dust
with a higher sub-millimetre opacity and a higher opacity spectral
index (see Sect.~\ref{sect:dust}). In the $TD$ version the dust
properties transform smoothly from $D$ to $K$ dust as the density
increase from $n({\rm H_2})\sim10^3$\,cm$^{-3}$ to $n({\rm H_2})\sim
10^4$\,cm$^{-3}$. In practice, we modify the abundances of the two
dust components so that their sum remains constant and the relative
abundance of the default ($D$) dust is 
$0.5 \times (1+\tanh[2\,\log(n/1000\,{\rm cm}^{-3})])$.

\begin{figure*}
\includegraphics[width=16.8cm]{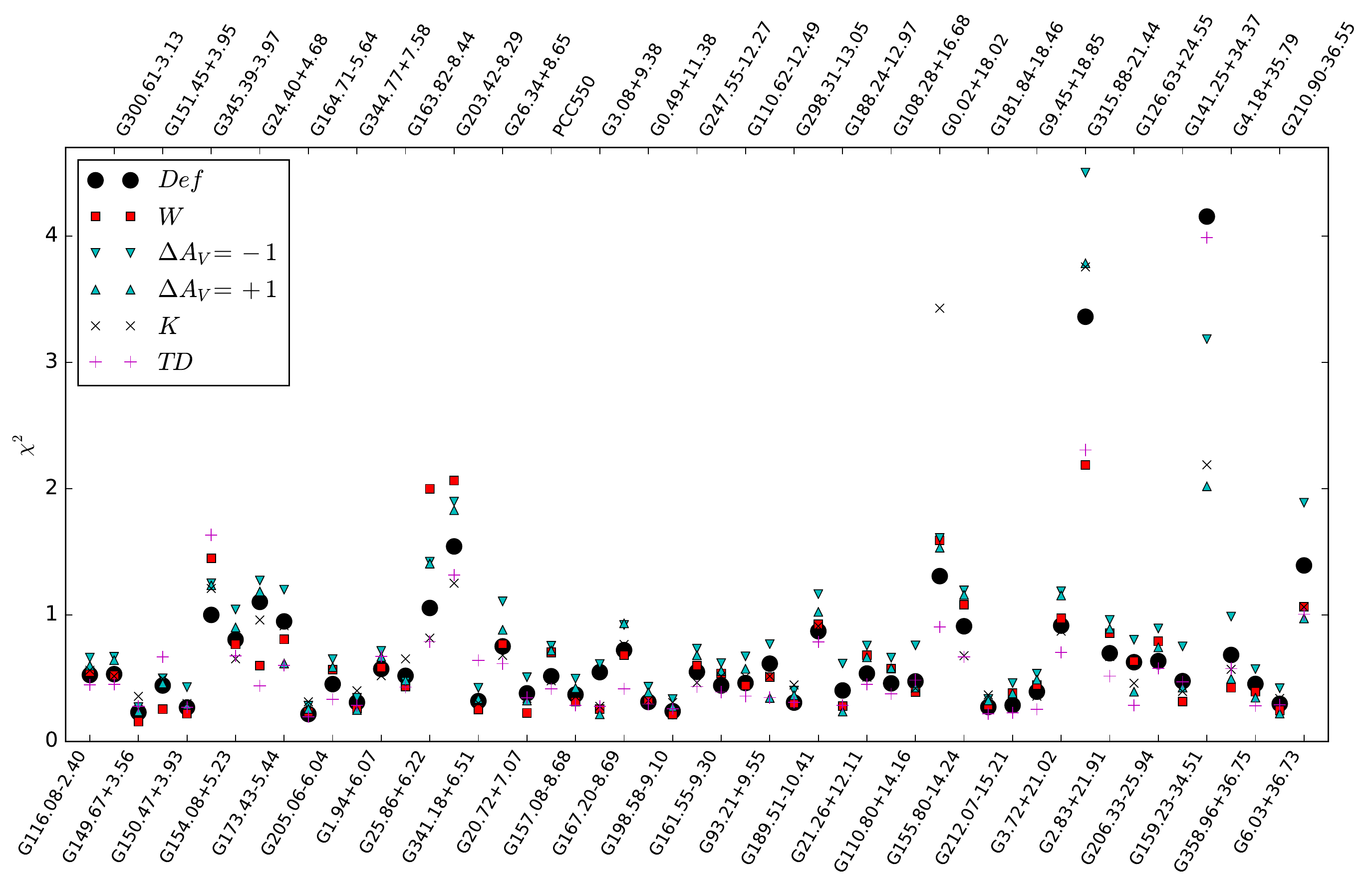}
\sidecaption
\caption{
$\chi^2$ values of the RT model fits. The reduced $\chi^2$ values are calculated over the
250, 350, and 500\,$\mu$m maps assuming an observational uncertainty of 7\%. The fields
are arranged in the order of increasing $|b|$. The legend refers to the model variations
listed in Table~\ref{table:models}.
}
\label{fig:model_chi2}
\end{figure*}

Figure~\ref{fig:model_chi2} gives a summary of the relative quality of
all fits. The only clear systematic effect is the somewhat higher
average $\chi^2$ value of $\Delta A_{\rm V}=-1$\,mag fits. The small
differences reflect the fact that most map pixels are located at
moderate column densities where the changes are not expected to have a
strong effect on the fit quality. The version $P$ fits (not shown)
concentrate on the small regions with high column densities and
therefore also show somewhat elevated $\chi^2$ values (similar to
those of version $\Delta A_{\rm V}=-1$\,mag) and give a better fit
only within the densest regions. The $\chi^2$ outliers G126.63+24.55
and G315.88-21.44 both have large areas with column density below
$N({\rm H}_2)=2\times 10^{20}$ (before background subtraction). In
these cases, the errors appear to be dominated by random noise rather
than systematic effects.

\begin{figure*}
\includegraphics[width=16.8cm]{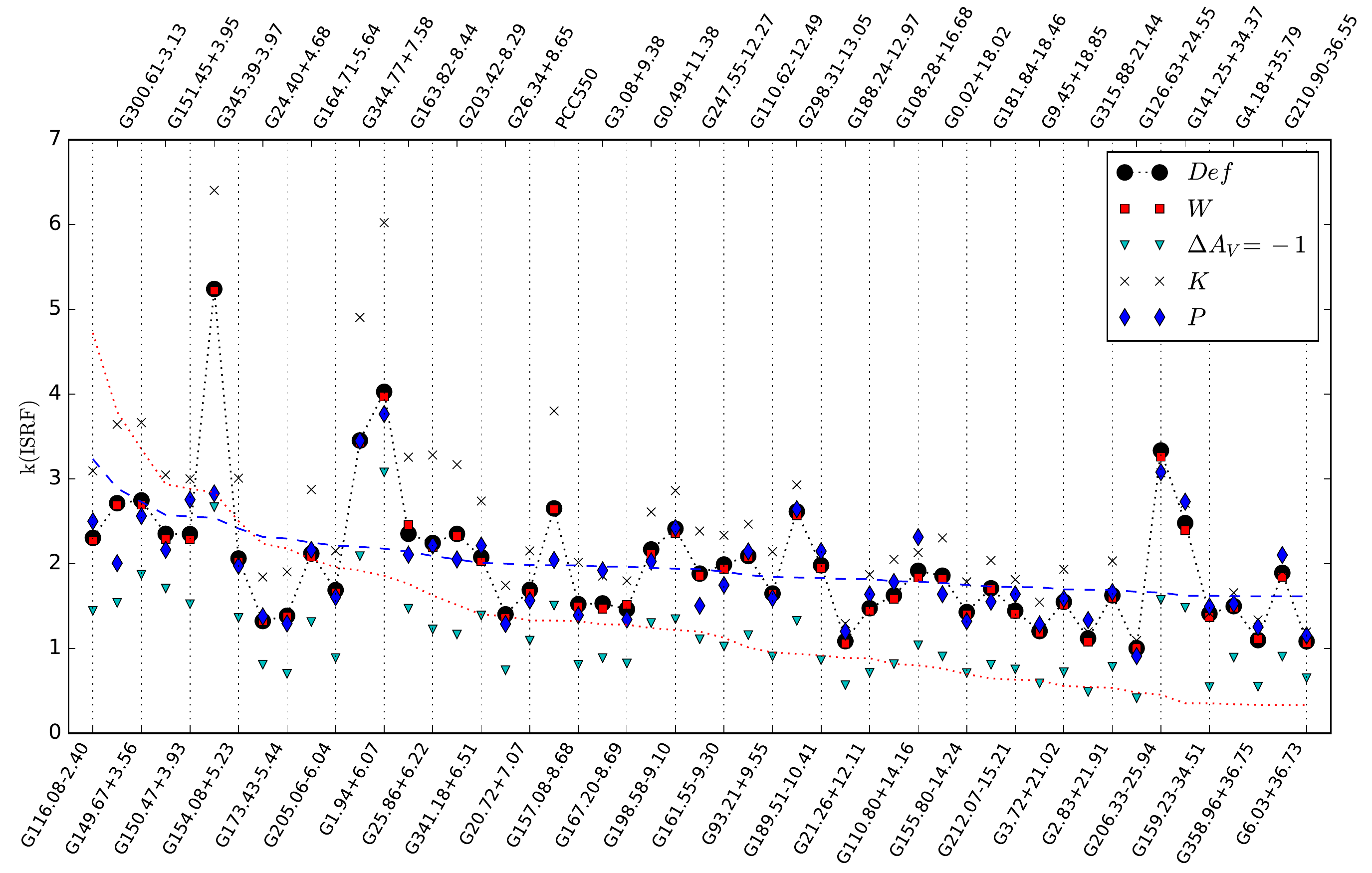}
\sidecaption
\caption{
Strength of the external radiation field, $k_{\rm ISRF}$, derived under various
assumptions of the RT models. The fields are in increasing order of
$|b|$. The blue dashed line shows the fitted least squares line $k_{\rm
ISRF}=(1.49\pm0.24) + (0.074\pm0.029)\times csc(|b|)$. For comparison, the red dotted
line indicates a pure cosecant law (with an arbitrary scaling).
}
\label{fig:plot_isrf}
\end{figure*}

Although the $\chi^2$ values are similar, different assumptions lead
to significantly different parameter values.
Figure~\ref{fig:plot_isrf} shows the estimated strength of the
radiation field $k_{\rm ISRF}$. As described in Sect.~\ref{sect:ISRF},
$k_{\rm ISRF}$=1 corresponds to a case where the incoming radiation is
assumed to be attenuated by an external layer that corresponds to the
amount of material in the reference area (used for background
subtraction). The case $\Delta A_{\rm V}=-1$\,mag is included in the
plot as the one resulting in the lowest $k_{\rm ISRF}$ values. The
magnitude of this drop is not trivial to predict because a change in
$\Delta A_{\rm V}$ also changes the shape of the incoming spectrum.
The plot shows the clear increase of $k_{\rm ISRF}$ in the case of a
higher sub-millimetre opacity.

The parameter $k_{\rm ISRF}$ does not show any systematic behaviour as
a function of Galactic longitude. The intensity tends to decrease with
increasing Galactic latitude, the overall trend in
Fig.~\ref{fig:plot_isrf} being significant at a $\sim$2.5\,$\sigma$
level. There is no similar dependence on distance. The correlation
between $k_{\rm ISRF}$ and the Galactic {\em height} is even slightly
(but not significant) negative.

\subsection{Clump radial density profiles} \label{sect:radial}

We examine in more detail the radial density structure of a subset of
clumps. The sample is selected by taking the clumps with areas close
to 10\,arcmin$^2$.
The value is well above the effective beam size but still corresponds
to relatively compact objects. Because the selection is based on
angular size, the physical sizes of the clumps vary depending on the
distance. The importance of this fact is examined at the end of this
Section.
To enable better fitting of the 2D surfaces mentioned in
Sect.~\ref{sect:profile_fit}, we exclude clumps that have strong
secondary peaks. This leaves a sample of 85 clumps.

\begin{figure}
\includegraphics[width=8.8cm]{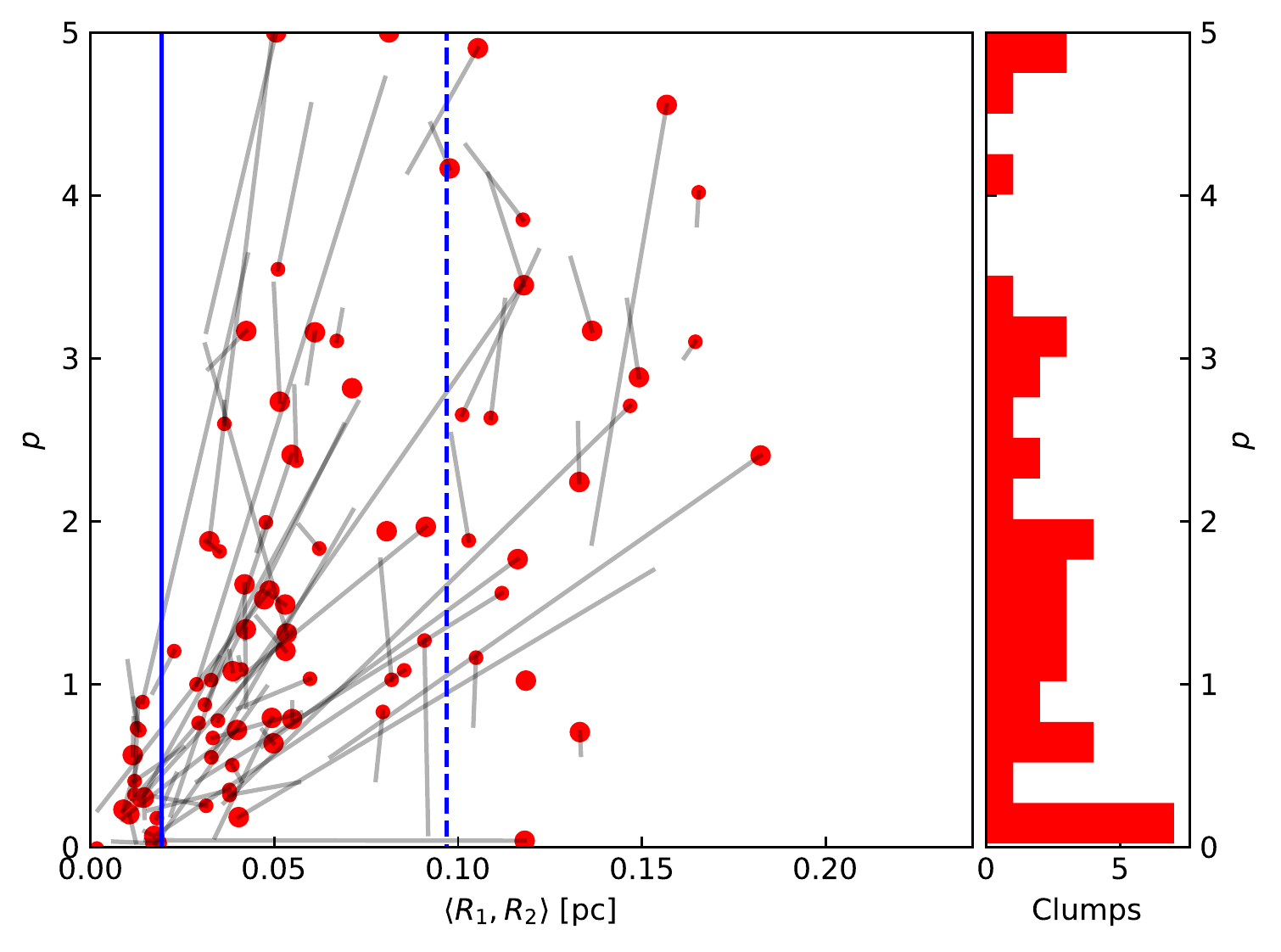}
\caption{
Results of 2D Plummer function fits. The circles correspond to the
parameters $p$ that are plotted against the geometrical mean of the
flat radii $R_1$ and $R_2$, which correspond to the clump major axis
and minor axis directions.
The solid lines are drawn from the plotted circles to the parameter
combination that is obtained when clumps are selected at a 15\% higher-column-density threshold. The solid and dashed vertical lines
correspond to 40$\arcsec$ at 100\,pc and 500\,pc distances,
respectively. The right panel shows the distribution of $p$ values ($0
\le p \le 5$ only) corresponding to the red symbols of the first
frame.
}
\label{fig:ppm_A}
\end{figure}

The clumps were fitted with 2D Gaussians and 2D Plummer functions. The statistics of all
fit parameters are shown in Appendix~\ref{sect:2dfits}. Even the simpler Gaussian model
gives relatively good fits with residuals below 10\%. The Plummer fits suffer from a large
number of free parameters (and degeneracy between $R$ and $p$ parameters). In particular,
the asymptotic power law exponent $p$ does not appear to be at all well constrained.
Figure~\ref{fig:ppm_A} shows the parameters $p$ and $R$ of the Plummer fits, assuming the
same value of $p$ for both the minor and major axes. The flat radius
$R$ is concentrated at values below 0.1\,pc but, depending on the distance of the clump,
this scale is only marginally resolved. The largest values near $R=$0.2\,pc are well below
the limit set by the selected 10\,arcmin$^2$ clump sizes. The values of the exponent $p$
are spread between zero and $p=5$, the maximum value allowed in the fits. Only the lack
of combinations of small $p$ and small $R$ values is related to the data resolution. In
total, at the scale of 10\,arcmin$^2$, the fit parameters scatter over a large parameter
range and, as far as characterised by the Plummer fits, the clump shapes do not show any
clear trends.

Because of the inconclusive results of the 2D fits, we made fits also to azimuthally
averaged column density profiles. Appendix~\ref{app:radial} shows the column density
maps and the radial optical depth profiles, also including the corrections derived from
the RT models. After subtracting the threshold column density, the median FWHM of the
clumps is 0.075\,pc. This differs only slightly between clumps below and above the
median column density, with 0.092\,pc and 0.071\,pc, respectively. The clumps with one
or more YSO candidates within their perimeter are more compact with FWHM=0.037\,pc
compared to FWHM=0.077\,pc for the remaining sources (median values).

The clump profiles were fitted with powerlaws and with one-dimensional (1D) Plummer
functions, the 1D versions of Eq.~(\ref{eq:Plummer}).
Both fits include a constant background as one of the free parameters.
In the Plummer fits, the beam convolution is also explicitly taken
into account. For powerlaw profile, $A r^{-p} + B$, to reduce the
effects of the finite beam, we simply limit the fits to angular
distances larger than 30$\arcsec$.

\begin{figure}
\includegraphics[width=8.5cm]{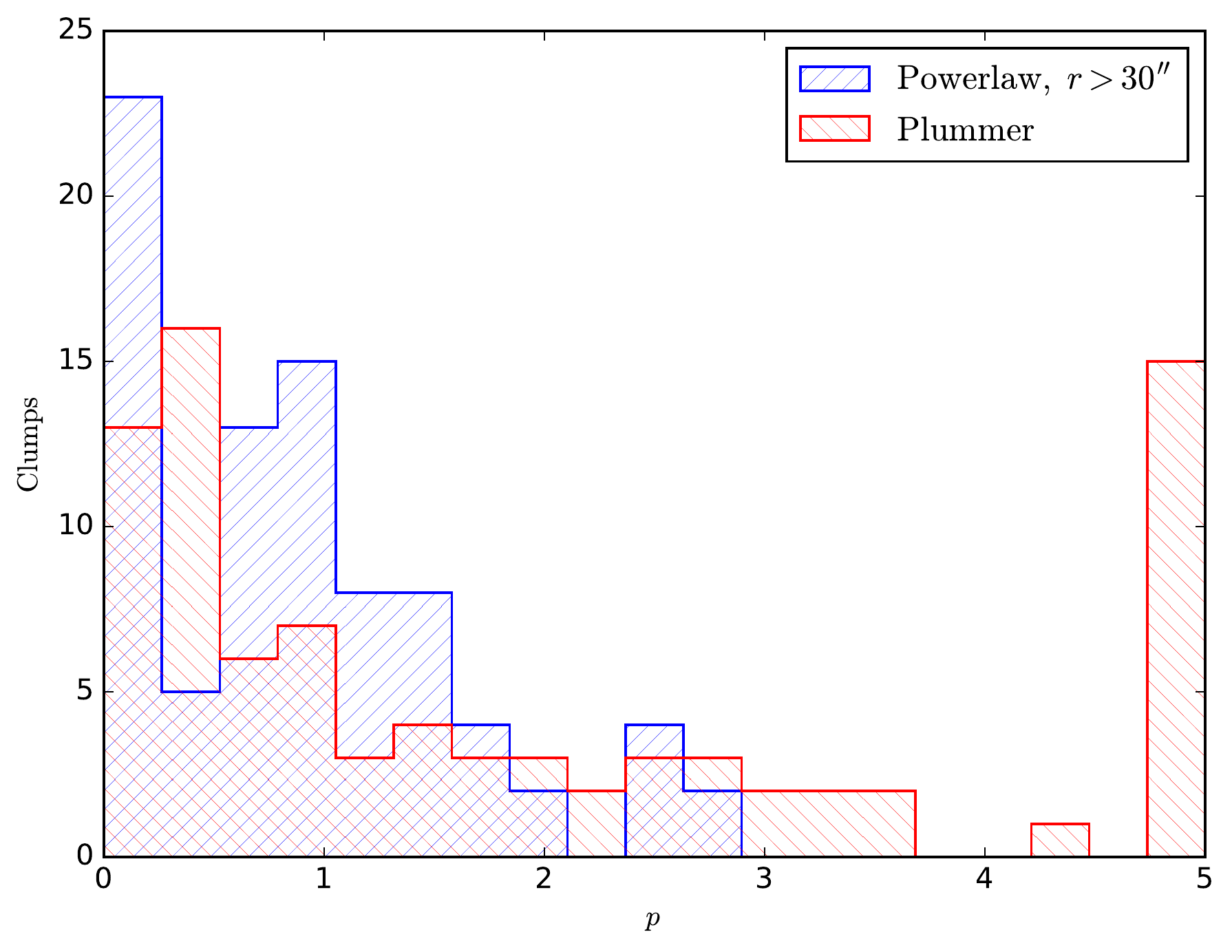}
\caption{
Distribution of the exponents of the azimuthally averaged optical depth profiles. The pure
powerlaw fits (blue histogram) are to data with $r>30\arcsec$. The Plummer fits (red
histogram for the parameter $p$) are to data $r<0.15$\,pc.
}
\label{fig:radial_expo}
\end{figure}

Figure.~\ref{fig:radial_expo} shows the distribution of the powerlaw
exponents. The plot is limited to a maximum value of $p=$5 (one
powerlaw fit resulted in a value above this limit). The Plummer
fits were directly constrained to values $p<5$. There is again a
large scatter in the Plummer parameters. For the pure powerlaw fits,
the exponent values are more concentrated below $2$ and there is a
clear local maximum around 1. The median value is +0.85 for the whole
sample but one in four clumps has a powerlaw exponent smaller than
0.2. 
Appendix~\ref{app:radial} shows that some of the very low $p$
values are associated with double-peaked column density structures or
poorly resolved clumps inside the 30$\arcsec$ radius. However, these
do not explain all the low $p$ values and generally the low $p$ values
are not associated with the particularly large fit residuals. 
There is some tendency for the profiles to be steeper in regions of
high column density. However, the correlation with the background
value is only marginal, both for the exponents of the pure powerlaw
fits ($r=$0.15) and for the Plummer $p$ parameter ($r=$0.30 for the
sample with fitted values $0<p<5$)

Because the distances of the fields range from $d$=110\,pc to
$d$=500\,pc, the fits apply to different physical scales. At
$d$=110\,pc the fitted radial range can be 0.016--0.08\,pc while for
$d$=500\,pc it could be 0.07--0.36\,pc. Here the calculated upper
limits correspond to a radial distance of 2.5$\arcmin$. Nevertheless,
the correlations between the fit parameters and the distance are weak.
The linear correlation coefficient is -0.02 in the case of the Plummer
$p$ parameter and +0.27 in the case of the powerlaw exponent. Even the
latter is only marginally significant, which suggests similarity in
the typical clump profiles across the $\sim$0.1\,pc scale.

\subsection{Clump stability} \label{sect:stability}

We estimate the clump stability using both the column densities
derived from observations and the 3D density distributions of the RT
models. We do not have uniform high-resolution line observations and
thus no precise knowledge of the thermal and turbulent support or the
external pressure. We make the assumptions that the gas kinetic
temperature is 10\,K inside the objects and 15\,K in their envelopes.
Following the compilation of observations in \citet{Kauffmann2013}, we
adopt a 1D, non-thermal velocity dispersion of
$\sigma_{1D, NT}$ = 0.7\,km\,s$^{-1}$ ($R_{\rm eff}$/1pc)$^{0.4}$, 
where $R_{\rm eff}$ is the radius of a circle with an area equal
to that of the clump. We assume that the velocity dispersion within
the source is smaller by 30\%, as justified below.
Line observations do exist for some of the selected clumps and for
some GCC fields that are not part of the present sample. The
observations are not sufficiently complete for the virial analysis
to be directly based on them. However, we can compare the
\citet{Kauffmann2013} relation with these data.
The relation is fully consistent with mean behaviour of the $^{13}$CO
data in \citet{Feher2017}, although observations show a 40\%
dispersion relative to the relation. \citet{Saajasto2017} investigated
the $Herschel$ field G82.65-2.00 that is not included in the present
paper because of its 650\,pc distance. In that field, which contains
a star-forming and strongly fragmented filamentary cloud, the large-scale velocity dispersion was found to be almost independent of the
linear scale. However, based on the $^{13}$CO data, the 1D velocity
dispersion of the main clumps was about 0.6\,km\,s$^{-1}$. With the
assumed cloud distance, the sizes of the clumps are about one parsec
and the values are again consistent with the Kauffmann et al.
relation.
\citet{Parikka2015} reported line widths that were based on C$^{18}$O
observations of compact objects with sizes below 0.5\,pc. The median
line width was some 30\% below the Kauffmann et al. relation.
We adopt $\sigma_{\rm 1D}$ values that are 30\% below the
\citep{Kauffmann2013} relation as an approximation of the internal
velocity dispersion of the clumps, as it would be seen in C$^{18}$O
observations. This is, of course, valid only statistically and should
not be used to infer the gravitational stability of any individual
object.

We calculate the virial parameter $\alpha_{2D} = M_{\rm vir}/M_{\rm
2D}$ using the clump masses integrated from column density maps, after
subtracting the background level that is estimated as the average
value within a one-arcmin-wide boundary just outside the clump
perimeter. The virial mass is obtained from
\begin{equation}
  M_{\rm vir} = 168.5 \, R_{\rm eff} \,  \sigma_{\rm 1D}^2 \, 8 \ln{2}
,\end{equation}
\citep{maclaren88}, which includes the assumption of an $r^{-1.5}$
density profile. In this form, $\sigma_{\rm 1D}$ includes both the
thermal and non-thermal velocity dispersions, which are added together
in squares. In the following, the $\alpha_{\rm 2D}$ parameters are
calculated by directly using the column densities derived from SED
fits, without the RT-derived corrections.

Alternatively, the clump stability can be estimated using the 3D models
and the direct balance of gravitational, kinetic, and external pressure
energies. The 3D models take into account the effects that temperature
gradients have on the observed surface brightness and may therefore
lead to different estimates of the gravitational energy. The 3D clumps
are defined by a density isocontour that has projected areas similar
to the selected 2D clumps. For comparison, we also consider smaller
clumps that are defined by density isocontours with 20\%, 40\%, and
60\% higher column density values (objects smaller than
1.1\,arcmin$^2$ are excluded).
We consider the gravitational potential energy $\Omega_{\rm G}$, the
internal (kinetic) energy $\Omega_{\rm K}=3 P V$, and the term
$\Omega_{\rm P}$ that is related to the external pressure.
The value of $\Omega_{\rm G}$ is calculated explicitly, using the
density distribution of the 3D model.
Because of the assumptions on $\sigma_{1D, NT}$ (see the beginning of
this Section), the pressure $P=\rho \sigma_{\rm 1D}^2$ only depends on
the scale at which it is estimated, the gas kinetic temperature, and
the density threshold $\rho$. Therefore, $\Omega_{\rm K}$ is
calculated with $\sigma_{\rm 1D}$ values that are estimated for
$R_{\rm eff}$ and include thermal motions at the assumed kinetic
temperature of 10\,K.
The energy related to the external pressure is $\Omega_P=-3 P_{\rm
ext} V$, where the volume is the actual volume of the 3D clump and the
pressure is again estimated at the scale $R_{\rm eff}$, assuming a
kinetic temperature of 15\,K.
We do not have observations of different molecules that could be
interpreted as direct measurements of the velocity dispersion inside
the clumps and at their boundary \citep[cf.][]{Pattle2015}. By using
the same $\sigma_{1D, NT}$ values for both $\Omega_{\rm K}$ and
$\Omega_{\rm P}$ (apart from the different kinetic temperature), we
may underestimate the importance of the outer pressure, provided that
turbulence is stronger outside the clump.

\begin{figure}
\includegraphics[width=8.8cm]{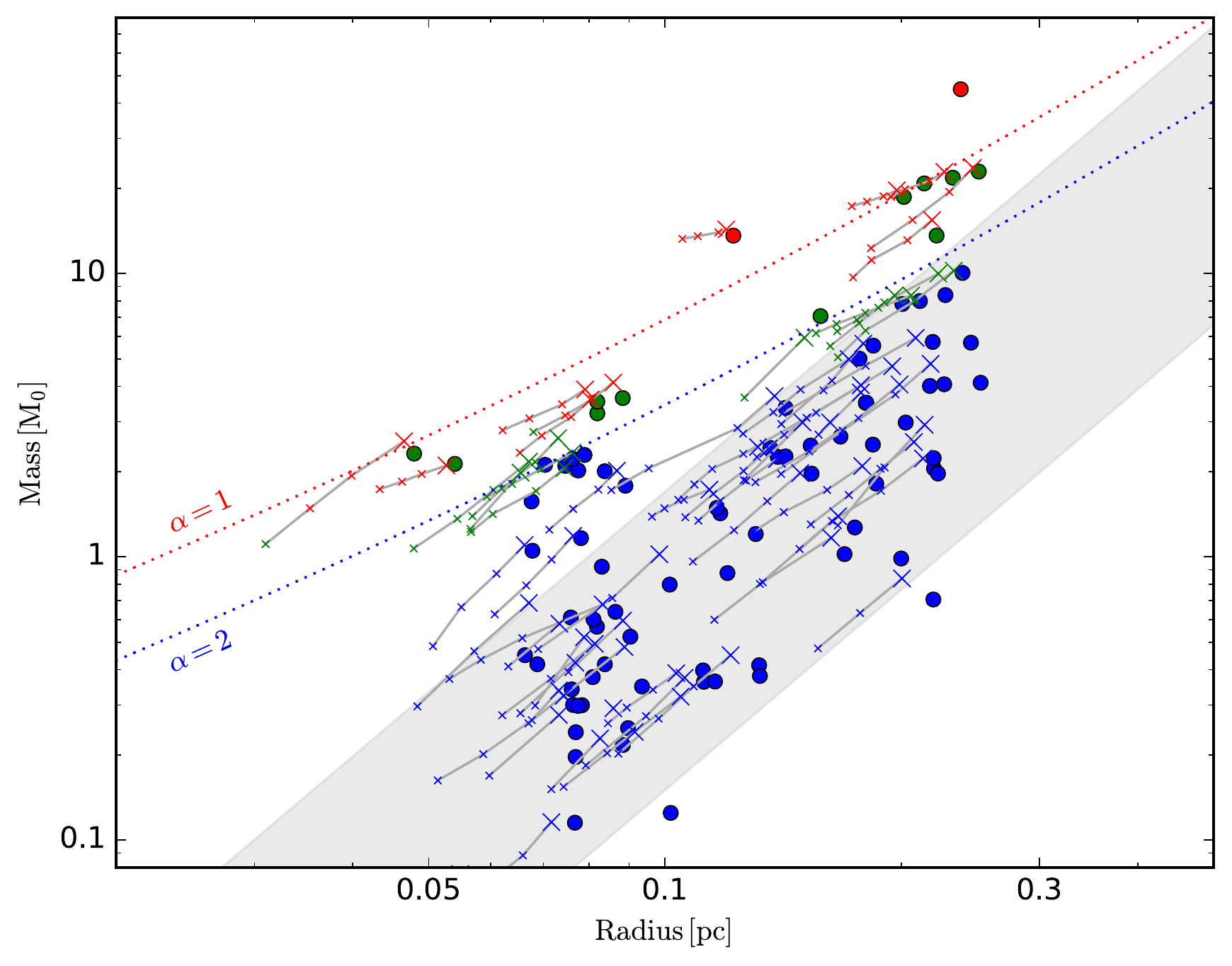}
\caption{
Clump masses as a function of the clump effective radius. The sample
consists of 10\,arcmin$^2$ clumps. The circles denote values derived
from the observed column density maps. The crosses are the values from
3D models, smaller symbols corresponding to progressively higher
volume density thresholds. The dotted lines correspond to the values
$\alpha=1$ and $\alpha=2$ of the virial parameter (3D clumps), with
the assumptions listed in the text. The red, green, and blue symbols
show the actual values with $\alpha<1$, $1<\alpha<2$, and $\alpha>2$,
respectively. The grey band corresponds to an empirical relation of CO
clumps \citep{Elmegreen1996, Pattle2015}.
}
\label{fig:virial_RM}
\end{figure}

We use the same sample as in Sect.~\ref{sect:orientation}, the clumps
with projected area of approximately 10\,arcmin$^2$.
Figure~\ref{fig:virial_RM} shows the 2D and 3D clump masses as a
function of the effective radius, $R_{\rm eff}$, which in the case of
3D models also corresponds to the projected area in the POS. The
results correspond to the default assumption of the dust properties
with $\tau(250\mu{\rm m})/\tau(J)=1.0\times 10^{-3}$. The use of a
higher emissivity would naturally lead to smaller clump masses and
higher values of the virial parameter.

\begin{figure}
\includegraphics[width=8.8cm]{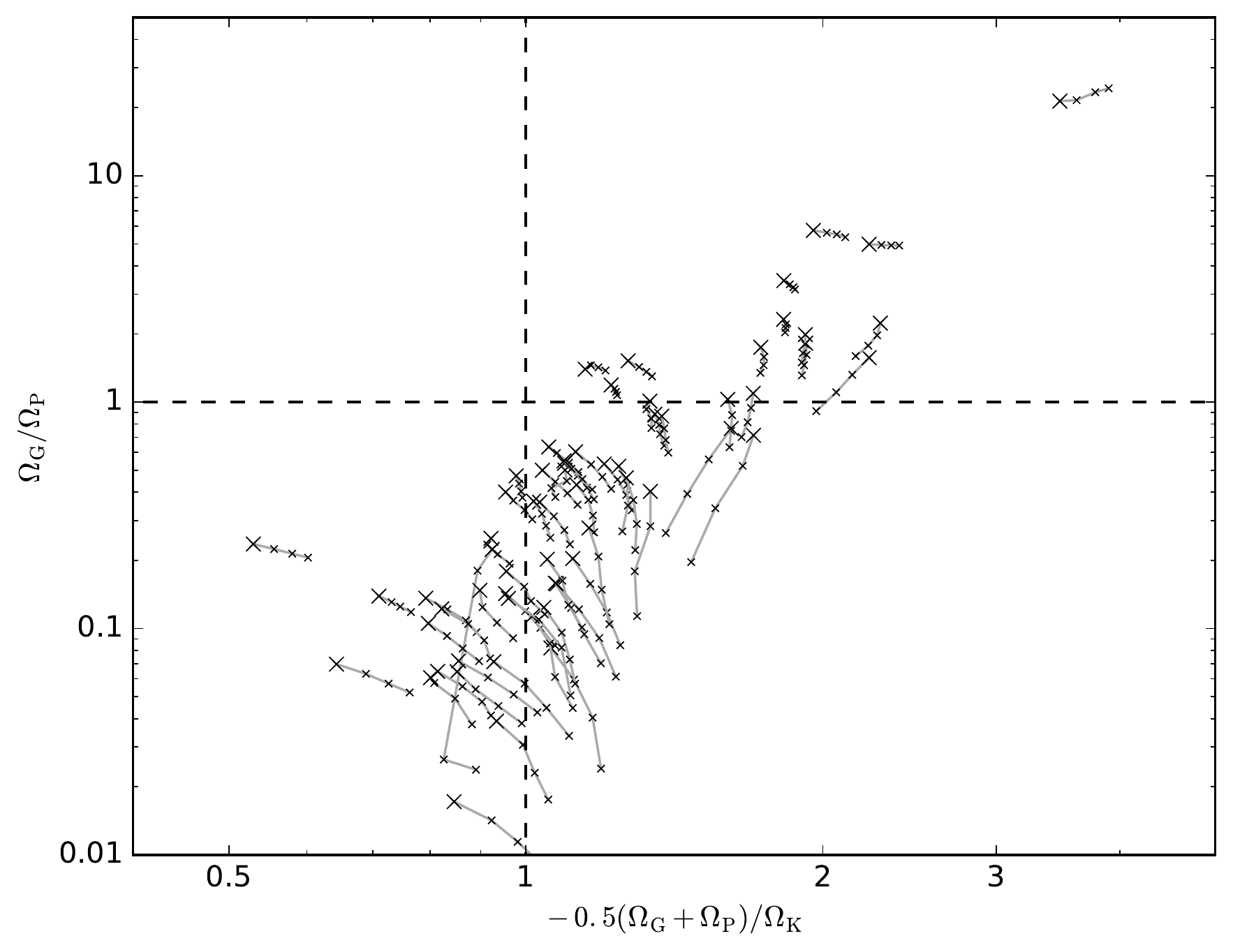}
\caption{
For the sample of 3D clumps in Fig.~\ref{fig:virial_RM}, energy ratios based on the terms
$\Omega_{\rm G}$, $\Omega_{\rm P}$, and $\Omega_{\rm K}$. The larger symbols refer to
clumps with a projected surface area of 10\,arcmin$^2$. The smaller symbols, joined by
lines, correspond to smaller clumps defined by, in steps of 20\%, higher volume density
thresholds. The clumps right of the vertical line are bound and the clumps above the
horizontal line are dominated by gravity.
}
\label{fig:virial_OO}
\end{figure}

Figure~\ref{fig:virial_OO} shows the same 3D data against the
different energy components. In this Figure, 25\% of the clumps
are gravitationally bound and 40\% are bound by the external pressure. It
is also clear that the selected density threshold has a non-negligible
effect on the virial parameter estimates. At a higher density
threshold, the $\Omega_{\rm G}/ \Omega_{\rm P}$ tend to be smaller
while several unbound objects also cross the $-0.5(\Omega_{\rm
G}-\Omega_{\rm P})/\Omega_{\rm K}$ boundary. Thus, by selecting a 40\%
higher density threshold, the fraction of gravitationally bound
objects decreases to 21\% while the number of pressure-bound clumps
increases to 53\%. Of course, the behaviour directly depends on
several assumptions, including the adopted velocity scaling and gas
temperatures.

\begin{figure}
\includegraphics[width=8.8cm]{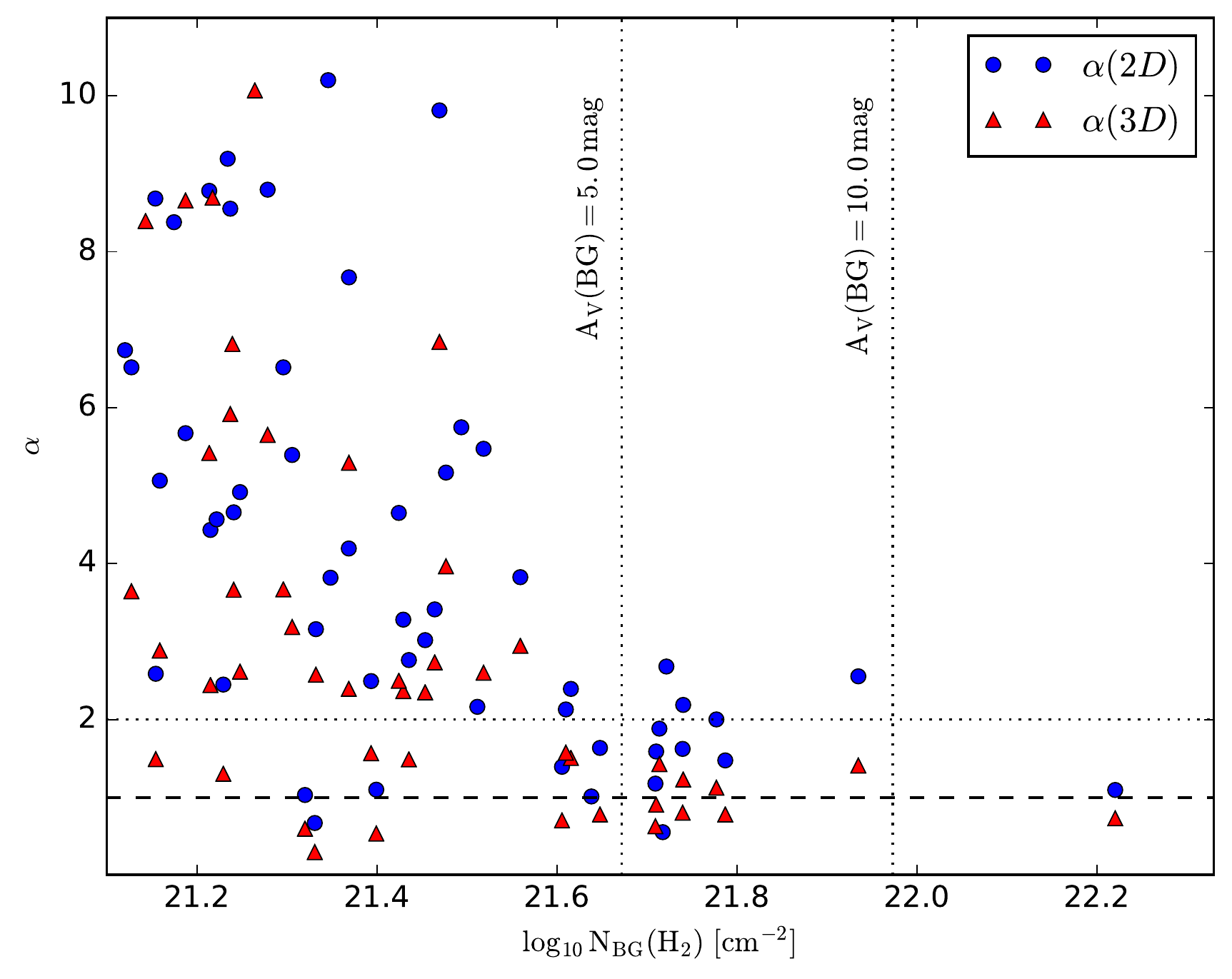}
\caption{
Virial parameter $\alpha$ as a function of the estimated background column density $N_{\rm
BG}$. The $\alpha$ values estimated from 2D and 3D data are, respectively, shown as blue
circles and red triangles. The vertical lines indicate the approximate values (depending
on the dust model) of $A_{\rm V}$ equal to 5 and 10 magnitudes. Horizontal lines are drawn
at $\alpha=1$ and $\alpha=2$.
}
\label{fig:virial_vs_BG}
\end{figure}

According to Fig.~\ref{fig:virial_vs_BG}, there is a clear correlation, at least for
the upper envelope of the $\alpha$ values, such that clumps with smaller virial
parameters are much more likely to be found in regions of higher column density. At $N_{\rm
BG}>5\times 10^{21}$\,cm$^{-2}$ ($A_{\rm V}>5\,$mag), most of the clumps are almost
bound ($\alpha \la 2$). Here the values $N_{\rm BG}$ consist of the immediate
surroundings of the clumps. It does not include the global background that for each
$Herschel$ field was subtracted when the column densities were estimated from
background-subtracted surface-brightness measurements. If those larger-scale
backgrounds were included, the correlation of Fig.~\ref{fig:virial_vs_BG} would remain
but also become much less pronounced. This is natural if the background subtraction has
removed mainly emission that is unrelated to the clumps and simply originates elsewhere
along the line of sight.

The virial parameter is not correlated with either skewness or kurtosis of the clumps. The
only potential dependence is seen with clump elongation. According to the Gaussian fits,
the average elongation (ratio of major and minor axes) is 1.63 for the $\alpha>2$ clumps
and 1.38 for the $\alpha<2$ clumps. Thus, more bound clumps also appear to be more
spherical. However, considering the standard deviations and counts within the two samples,
the difference is significant only at 1.8-$\sigma$ level.

The estimates of clump mass and of $M_{\rm vir}$ are affected by
several sources of uncertainty.
The separation of the clumps from the background may cause systematic
effects such as those associated with limb brightening at short FIR
wavelengths \citep{Menshchikov2016}. The average column density in
clump apertures is only $\sim$40\% higher than in the reference
annuli, which suggest that the uncertainties associated with
background fluctuations can be significant. An order-of-magnitude
estimate can be derived using Eq.~(\ref{eq:HB}). Assuming a dust
spectrum of $B_{\nu}(T=15{\rm K}) \nu^{1.8}$, the average clump
background level of 55\,MJy\,sr$^{-1}$ corresponds to a confusion
noise of $\sim$10\,MJy\,sr$^{-1}$ at the scale of 2$\arcmin$. This is
more than 40\% of the average clump signal relative to background. Of
course, this is error only so far as the fluctuation is not associated
with the main object. Because clumps are defined with column density
isocontours, a positive fluctuation at the clump location is likely to
increase both the mass and (via $R_{\rm eff}$) $M_{\rm vir}$ estimates,
with a smaller effect on the virial parameter.
The effect of the threshold used in the clump definition is directly
visible in Figs.~\ref{fig:virial_RM}-\ref{fig:virial_OO}.
The unknown LOS extent of the clumps acts as a similar error source,
especially for $\alpha_{\rm 3D}$. The difference between a sphere and
a bi-axial ellipsoid with an aspect ratio of 1:2 is 30\% in
gravitational energy. 
The distance estimates $d$ have a typical accuracy of 30\%
\citep{GCC-IV} but they affect the mass estimates in squares. $M_{\rm
vir}$ depends directly on $R_{\rm eff}$ while $\sigma_{\rm 1D}$ is in
our case proportional to $R_{\rm eff}^{0.8}$. Thus in our calculations
the distance dependence of the virial parameter $\alpha$ is only
$d^{-0.2}$.
The masses were derived for fixed values of dust opacity $\kappa$ and
opacity spectral index $\beta$ (Sect.~\ref{sect:clumps}). A $\beta$
uncertainty of $\sigma(\beta)=0.2$ corresponds to
30\% uncertainty in mass while the effect of the $\kappa$ uncertainty
is potentially even larger \citep{GCC-V}. These can increase the
scatter and introduce bias in
Figs.~\ref{fig:virial_RM}-\ref{fig:virial_vs_BG}.
$M_{\rm vir}$ is further affected by the uncertainty of the 
$\sigma_{\rm 1D}$ parameter. If C$^{18}$O line widths have a 30\%
scatter (normal distribution) around the adopted analytical relation,
the resulting noise in $M_{\rm vir}$ would be close to 60\% (standard
deviation). If C$^{18}$O does not accurately measure the relevant
velocity dispersion, the $M_{\rm vir}$ estimates can be 
correspondingly further biased.

\section{Discussion}  \label{sect:discussion}

We have investigated 51 fields that are within 500\,pc distances and
were originally targeted by $Herschel$ observations because they
contain $Planck$ detections of cold clumps. Low dust temperatures
imply large column densities that are able to produce large
temperature gradients, especially because PGCC, the catalogue of
$Planck$ detections, is based on the relative temperature of the
clumps and their environment. Thus, PGCC may also contain relatively
diffuse sources (sources with small internal temperature gradients) if
they happen to be seen against a much warmer background. The term
``background'' could mean other clouds along the LOS that either are
subjected to a stronger radiation field or have different dust
properties. However, previous studies of the GCC fields have already
confirmed the presence of cold clumps that truly dominate the LOS
column density. In this paper, we have tried to further characterise
both the global properties of the fields and the individual properties
of the selected clumps. This could shed some light on the clump
formation as an important intermediate step in the star formation
process.

\subsection{General properties of fields with Planck cold clumps}

We analysed the large-scale properties of the $Herschel$ maps using standard
statistical methods. The results show that the fields are in many respects typical of
interstellar clouds.

\subsubsection{Fractal dimension}

The fractal dimensions of the column density maps were $D_{\rm P}=1.25
\pm 0.07$. The nearby cloud G358.96+36.75 is one extreme and, with its
smooth radial density profile, appears almost non-fractal with $D_{\rm
P}=$1.05. The highest values were around 1.4 but these are sometimes
found in fields with very low column densities where the values are
affected by observational noise and even the extragalactic background.
Both act as noise, which tends to increase the $D_{\rm P}$ values
\citep{Vogelaar1994, Pant2013}. 

To an accuracy of 0.02 units, the median value of $D_{\rm P}$
over all fields remains the same if the calculations are done using
contours below or above the median of the original set of column
density thresholds. We also estimated $D_{\rm P}$ for the subset of
structures with surface area below or above the median structure size.
The median value is lower by 0.06 units for the smaller structures and
higher by 0.05 units for the larger structures. This could hint at the
presence of small, gravity-bound structures of low fractal dimensions.
However, because no significant correlation was found with the
distances, the dependence seems to be more on the angular rather than
the physical scale.

$D_{\rm P}$ values usually reported for interstellar clouds are in the
range of 1.2-1.5. \citet{Falgarone1991} obtained $D_{\rm P}$=1.36 from
$^{12}$CO data over a wide range of spatial scales.
\citet{Sanchez2005} and \citet{Sanchez2009} derived from CO
isotopologues estimates 1.30-1.35 for the Ophiuchus, Perseus, and
Orion A clouds. The IRAS 100\,$\mu$m data have given similar values
with $D_{\rm P}$=1.26 at high latitudes \citep{Bazell1988}, $D_{\rm
P}$=1.4 for the Taurus clouds \citep{Scalo1990}, and a range of values
1.2-1.3 for a sample of nearby molecular clouds \citep{Dickman1990}.
As noted by \citet{Sanchez2009}, values $D_{\rm P}\sim 1.35$ are more
consistent with a 3D fractal dimension $D_{\rm f}\sim 2.6$ than the
direct estimates of $D_{\rm P}$+1. Thus, the more recent values
reported by \citet{Elia2014} for column density maps of HiGal
$Herschel$ fields, $D_{\rm f}$=2.61--2.93, while being slightly
higher, are still relatively close to our values. \citet{Elia2014}
found that $D_{\rm f}$ decreases with the wavelength of the surface
brightness maps. Furthermore, $D_{\rm f}$ calculated from column
density was on average more than 0.1 units lower. The difference is
similar in our data, the 250\,$\mu$m surface brightness data resulting
in $D_{\rm P}$ values higher by 0.08 units (see
Sect.~\ref{sect:chara}). We saw no clear indication that the fractal
dimension would depend on the column density. If gravitation produces
more spherical dense structures, one could expect $D_{\rm P}$ to
decrease with column density. Because we exclude contours close to the
resolution limit, scales $<1 \arcmin$, we are not probing actual core
scales. Thus, the fractal dimensions are relatively constant between
the fields, between regions of different column density, and also
between small and large linear scales. Although our fields are outside
the Galactic plane, there is little difference even from the
\citep{Elia2014} data, which correspond to much longer sightlines
through the Galactic plane and thus could be expected to represent a
superposition of several clouds.

\subsubsection{Structure noise}

The structure noise $N^{\rm str}(\theta)$ of the 250\,$\mu$m surface
brightness (Fig.~\ref{fig:sf}) are consistent, within a factor of
$\sim 2$, with the intensity-dependence previously observed in
100\,$\mu$m IRAS data \citep{HelouBeichman1990, Gautier1992} and, for
example, in selected ISO satellite observations \citep{Kiss2001}. 
The structure noise appears to probe a generic property of the ISM and
the same laws apply to such different types of objects as cirrus
clouds and our sample of dense and in some cases actively star-forming
clouds. The $N^{\rm str}(\theta)$ values show variations as a function
of both the surface brightness and the angular scale. The average
angular dependence was found to be $N^{\rm str}(\theta) \sim
\theta^{0.57}$ but in individual fields the exponent could be as low
as 0.35. For nearby fields, the calculated $N^{\rm str}(\theta)$ is no
longer in the same sense a statistical description of the column
density fluctuations. When the map is dominated by a single clump, $N^{\rm str}(\theta)$ 
also mainly describes the radial profile of the
clump. As an extreme case, the smooth profile of G358.96+36.75
(LDN~1780) results in a very steep relation with $N^{\rm str}(\theta)
\sim \theta^{0.99}$.

\subsubsection{Column density PDFs} \label{disc:PDF}

The column density PDFs of clouds are often described as a combination
of a log-normal distribution and a power-law tail at large column
densities. These should be connected to key cloud properties; for example, the Mach number and the appearance of gravitationally
bound structures,
respectively \citep[e.g.][]{VazquezSemadeni1994, Padoan1997,
Kainulainen2009, Schneider2013, Ward_2014}. We show the column density
PDFs of the analysed fields in Appendix~\ref{sect:PDF}. Although the
average PDF is close to the log-normal shape expected of turbulent
clouds, the individual fields exhibit a wide range of shapes. Some
individual fields are examined further in Appendix~\ref{app:PDF}.

Because the 250\,$\mu$m surface brightness is usually a good proxy of
column density, the differences between surface brightness and column
density PDFs are small. When the differences are noticeable,
the column density PDF extends to higher values. This is caused by
high-column-density structures being colder and thus under-represented
in surface brightness data.  

Because observations target high-column-density structures, the
low-$N$ extent of the PDFs is dependent on the map size and on the low level to which column density observations extend
\citep{Ossenkopf_2016_PDF, Alves_2017_PDF}. Background subtraction
makes the distributions wider and, by construction, the PDF will
extend to zero column density (-$\infty$ on logarithmic scale). The
ambiguity of the background subtraction makes the interpretation of
the low-$N$ tail problematic \citep{Schneider2015_LOS,
Ossenkopf_2016_PDF,Alves_2017_PDF}. 
On the other hand, the high-$N$ side of the PDFs is, in first
approximation, only stretched as the mean column density decreases. The
background-subtracted data cover the regions selected for RT modelling
(see Table~\ref{table:fields}). The RT models indicate that the bias
in column density values (caused by LOS temperature variations) is
typically 10\% or less. This is small compared to the full dynamical
range and, after the renormalisation with the average column density,
the effect is not easily visible in logarithmic plots. However, the
correction can sometimes have a significant effect on the high-$N$
tail at $N({\rm H}_2)> 10^{22} {\rm cm}^{-2}$. In G6.03+36.74 the
correction makes the PDF tail a true powerlaw that extends down to the
resolution limit. This shows that as soon as optical depths are tens
of $A_{\rm V}$, the PDF derived from basic SED analysis can be
significantly biased.

A tail of the PDF towards high column densities is sometimes taken as
an indicator of the presence of dense cores or of a general tendency
to star formation. It is likely to be a more complex phenomenon,
involving the individual history of the cloud and the potential
effects of external forcing, magnetic fields, intermittent turbulence,
and feedback from the star-formation process \citep{Kainulainen2011,
Schneider2013, Schneider2015_LOS, Anathpindika_2017}. In our case, the
situation is further complicated by the fact that the observations
target distinct column density peaks with maps of limited size. Thus,
the extracted PDF can reflect the column density profile of a single
structure more than the relative statistics of low- and high-column-density objects (see Appendix~\ref{app:PDF}). A high-column-density
tail is also thus not automatically an indicator of gravitationally
bound substructures. 

We correlated the skewness and the asymmetry parameter
$[P(99\%)-P(50\%)]/[P(50\%)-P(10\%)]$ with several quantities. The
linear correlation coefficients with the column density were
compatible with zero ($r\sim0.015$). The surface density of YSO
candidates was estimated with the YSO catalogues of both
\citet{GCC-IV} and \citet{Marton2016}. The correlations were positive
but the correlation coefficients $r\sim 0.1$ are not significant even
at the 85\% confidence level. If the powerlaw tail is caused by the
presence of dense cores, these need to be at least partially resolved.
In their study of the Taurus molecular cloud, \citet{Pineda2010}
concluded that the powerlaw tail is noticeable at scales below
$\sim$0.4\,pc. These are in principle well resolved in all our fields
but the correlations between PDF asymmetry and distance were negative.
This may be the expected behaviour \citep{Alves_2017_PDF} but in our
study the correlation coefficients ($r=-0.11$ for skewness, $r=-0.06$
for the asymmetry parameter) were not statistically significant.

We do not find significant correlation between power-law tails at
high column density and the presence of gravitationally bound
structures or star-formation activity (YSOs), as reported in the
literature \citep{Kainulainen2009, Schneider2015_tails}. However, our
study is partly limited by the map sizes. The column density PDFs show
a wide range of shapes that often are far from a log-normal
distribution. The strongest asymmetries are related to the structure
of individual clumps or to dynamical interactions that lead to sharp
cloud boundaries.

\subsection{Radiative transfer models}

We carried out RT modelling of all the fields. This was done to
estimate the possible bias of the standard analysis through modified
blackbody
fits, to probe the systematic effects that result from
unknown dust properties and unknown LOS cloud shapes, and to derive
estimates of the relative strength of the radiation field. The 3D
model clouds were also used to examine the gravitational stability of
selected clumps. 

We compared the column densities of the optimised models and the
values derived from the surface-brightness maps predicted by the
models. This allowed us to estimate the bias of the normal SED
analysis and to derive multiplicative corrections. One example was
shown in Fig.~\ref{fig:corr_G150}, where the error rises to about 20\%
in the densest part of the field but is much smaller over most of the
field. The bias estimates themselves depend on assumptions about the
dust and the cloud properties. The bias systematically increases
with increasing column density. This trend could be changed only if
the observed surface brightness is affected by internal heating
sources \citep{Malinen2011}. Our models do not include embedded
sources but, according to surface brightness data, their effect is
constrained to small areas. The bias can be a significant source of
systematic error that should be considered in addition to the
(typically larger) uncertainty of the dust properties.

\begin{figure}
\includegraphics[width=8.8cm]{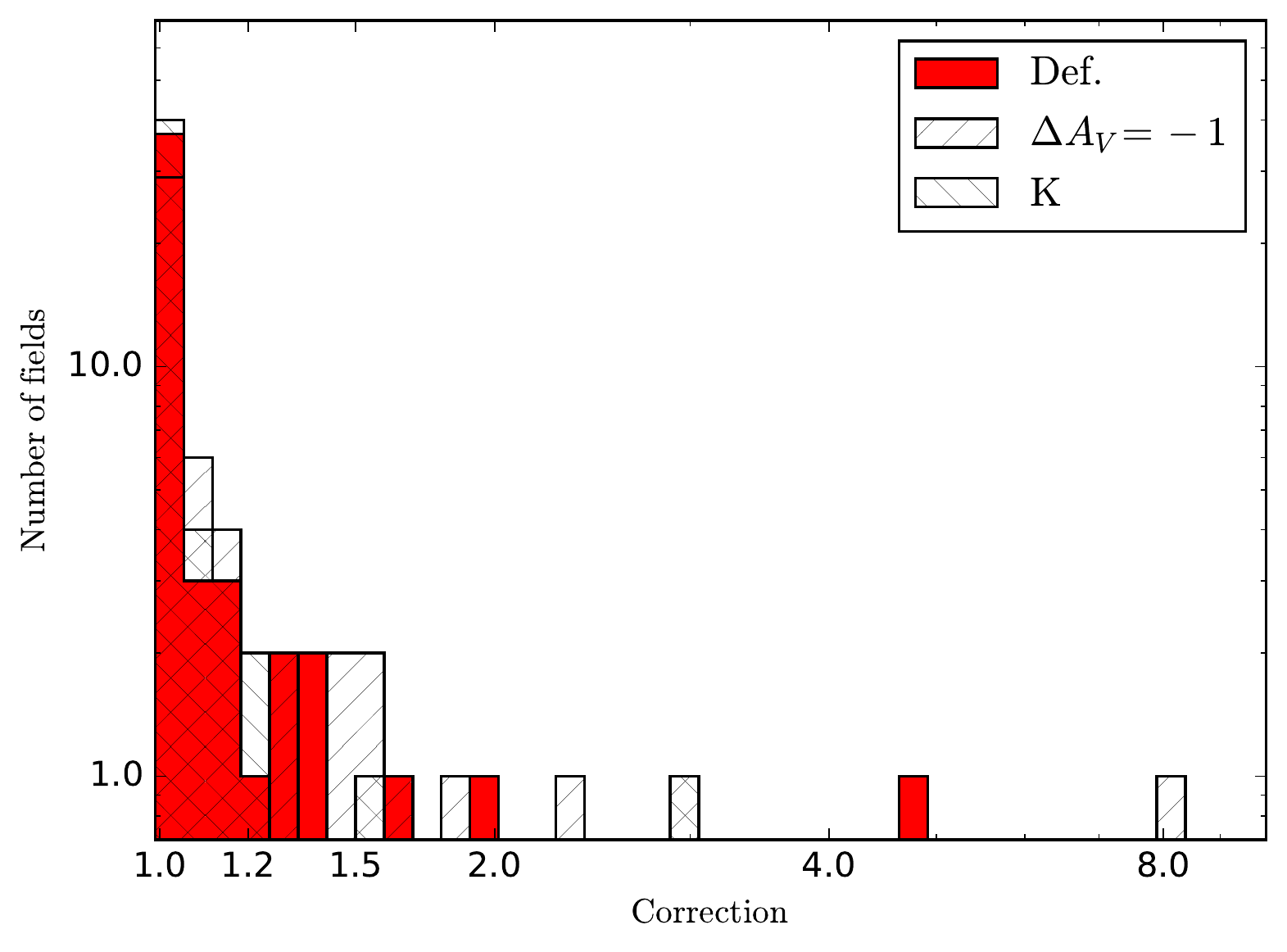}
\caption{
Bias in the column density estimates derived from SED fitting. The plotted quantity is
the ratio of the true column density and the column density estimated from SEDs. The
histograms represent the distribution of the maximum error in each field, estimated with
different versions of the RT models. The default model is shown as the red solid
histogram. The alternative model employs a harder radiation field parameterised as $\Delta
A_{\rm V}=-1$.
}
\label{fig:correction_stats}
\end{figure}

Figure~\ref{fig:correction_stats} shows the distribution of the
estimated maximum bias in each field. The errors are mostly below
20\%. In the cloud LDN\,183 (field G6.03+36.73) the estimated
error is a factor of five. In LDN\,183 molecular line data, 8\,$\mu$m
absorption, and millimetre dust emission are all consistent with a
maximum column density in excess of $N({\rm H}_2)=10^{23}$\,cm$^{-2}$
\citep{Pagani2015, Lefevre2016}. The bias estimate is roughly
consistent with the difference between these estimates and the values
derived from $Herschel$ SEDs. $Herschel$ bands are not sensitive to
very cold dust and the bias estimates from RT modelling are no longer
very reliable at such high column densities \citep{Juvela2013_colden,
Pagani2015}. Even after bias corrections, we may still underestimate
the true column density of the densest clumps.

The bias estimates depend on the assumptions of the radiation field
and the dust properties (Fig.~\ref{fig:correction_stats}). In the case
of $\Delta A_{\rm V}=-1$ the ISRF has a harder spectrum. This leads to
larger temperature gradients in the models and increases the bias
estimates typically by 14\% and in the case of LDN~183 by more than
60\%. On the other hand, the dust model with a higher ratio of
sub-millimetre and optical opacities results in models with lower
optical depth in the UV-optical regime. This reduces the average bias
by 7\% and the bias in LDN~183 by some 40\%.

RT models provide estimates of the radiation field intensity. The ISRF
strength decreases as a function of the Galactic latitude $b$
(Fig.~\ref{fig:plot_isrf}). Because there was no similar dependence on
the Galactic height, the trend is probably caused by factors other
than a true Galactic variation. 
Our sample is based on the $Planck$ PGCC survey where the central
detection parameter is the temperature contrast between the clumps and
their surroundings. Compared to high latitudes, where sources are seen
towards an empty sky (or the extragalactic background), the Galactic
plane is more confused. By boosting the temperature contrast, a warmer
background makes it more likely for a source to appear in PGCC. 
At low latitudes, a larger fraction of the background is likely to be
unrelated to the dense clumps. If it originates in diffuse regions
elsewhere along the line of sight, its colour temperature can be very
different. The trend in Fig.~\ref{fig:plot_isrf} may thus arise from
selection effects and from LOS confusion, rather than a Galactic trend
in the physical properties of the clouds that harbour cold clumps.

The RT results should be considered mainly as relative rather than
absolute measures of the radiation field intensity. Adoption of a
different dust model would affect on the $k_{\rm ISRF}$ values. Even
the assumption of grains in thermal equilibrium with the radiation
field has a small effect on the relation between the surface
brightness, column density, and radiation field \citep{TRUST-I}.

\begin{figure}
\includegraphics[width=8.8cm]{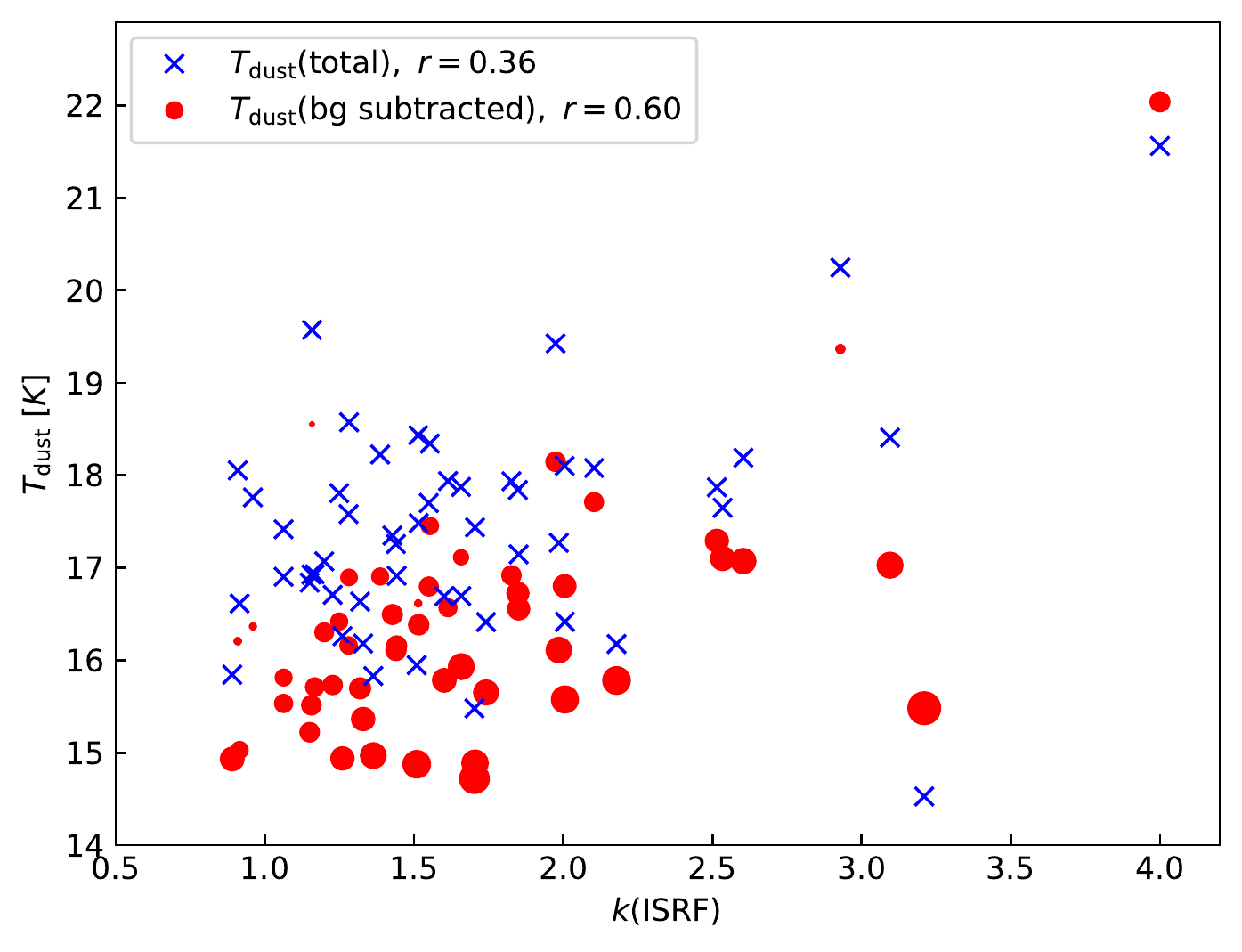}
\caption{
Comparison of the median dust colour temperature and the ISRF strength
obtained from RT models. Blue crosses correspond to $T_{\rm dust}$
values obtained from the full surface-brightness data. The red circles correspond
to values obtained from background-subtracted observations. The
$k({\rm ISRF})$ values are obtained by modelling the same
background-subtracted data. The diameter of the circles is
proportional to the logarithm of the median column density of a field.
These illustrates the effect of cloud opacity on the relation between
the derived radiation field and dust temperature estimates.
}
\label{fig:T_vs_kISRF}
\end{figure}

Figure~\ref{fig:T_vs_kISRF} compares $k_{\rm ISRF}$ to the median
$T_{\rm dust}$ values of the fields. Because the models are based on
background-subtracted observations, the correlation is naturally
better with $T_{\rm dust}$ that is derived from background-subtracted
data. Temperatures can be estimated only for those areas where the
residual surface brightness is clearly positive. In
Fig.~\ref{fig:T_vs_kISRF}, the median values of $T_{\rm dust}$
estimates are calculated for identical areas and can be thus directly
compared. The Figure shows that the background subtraction eliminates
a significant component of diffuse and warm dust emission. 

For these background-subtracted data, the linear correlation
coefficient between $k_{\rm ISRF}$ and $T_{\rm dust}$ is only $r=0.6$.
This is not entirely surprising because $T_{\rm dust}$ measures the
average radiation field inside the clouds while $k_{\rm ISRF}$
describes the field {outside} the modelled volume. In
Fig.~\ref{fig:T_vs_kISRF}, the effect of the column density is clear.
For a given value of $k_{\rm ISRF}$, dense fields have a significantly
lower median temperature. This shows that dust temperature is a good
proxy of the ISRF only outside optically thick regions. Even when the
column density along a particular LOS is low, $T_{\rm dust}$ can still
be affected by the shadowing caused by nearby dense clouds.

The $\chi^2$ values of Fig.~\ref{fig:model_chi2} show some trends.
First, the effect of the cloud extent in the LOS direction is not
significant. This indicates that the observations do not allow us to
make a clear distinction between (wrt LOS) oblate and prolate clouds.
Second, the use of modified dust with larger sub-millimetre opacity
tends to result in worse fits, either because of the resulting smaller
UV-optical cloud opacities or because of the larger opacity spectral
index. Because $\chi^2$ is calculated over large areas, it is not
particularly sensitive to the densest clumps. The models $TD$, where
the dust properties change as function of the density, might be
physically more justified. This complex model (but one without any
additional adjusted parameters) does sometimes lead to better fits but
the differences to the default model are usually not significant.
Third, the comparison between the $\Delta A_{\rm}$=+1\,mag and $\Delta
A_{\rm}$=-1\,mag models appears to favour larger $A_{\rm}$ values and
thus a radiation field with a softer spectrum. This seems
contradictory to the relatively bad performance of the alternative
dust model, which also should lead to smaller temperature gradients.
However, the main problem with the alternative dust model may be the
high spectral index, which may not be realistic outside the densest
clumps. Also, these two cases are fundamentally different, one having
lower extinction only at the cloud boundary, the other having lower
optical depths throughout the model volume. If the difference between
$\Delta A_{\rm}$=-1\,mag and $\Delta A_{\rm}$=+1\,mag cases were
significant, this would tell us something about the ISRF spectrum or
the cloud inhomogeneities. Alternatively, it could indicate that we
have underestimated the extinction of this external layer
(see~Sect.\ref{sect:ISRF}).

\subsection{Clump properties}

Figure~\ref{fig:MvsR} gave a crude overview of the mass distribution
around the highest column density peaks of each field. The Figure
covers distances up to about 1\,pc, or typical angular distances of
10$\arcmin$. At small scales, the plot represents the radial profile
of individual clumps, however, affected by the distance-dependent
linear resolution. Above $0.1$\,pc, one can see the resolved mass
distribution of the clouds. By only using radial lines with
monotonously decreasing column density values and by measuring
distances from a single position (not necessarily the geometric centre
of the cloud), the plot is by construction somewhat biased.
Nevertheless, the plots demonstrate the general similarity of the
fields, with an average profile $N(r) \sim r^{-1.0}$.

We investigated in more detail the clumps defined by column density
thresholding. The basic statistics revealed only weak trends, for
example, slightly lower average elongation at higher column densities.
The situation is complicated by the factor of five spread in distances
and the resulting selection biases. The median elongation of all
clumps was 1.5 so that most structures do have a well-defined position
angle. Examples like G173.43-5.44 suggested that there would be a
strong correlation between the orientation of nearby clumps (the
10\,arcmin$^2$ clump sample). Considering all the fields, the
correlation was statistically significant only in fields containing
two clumps. In several cases, those were also part of a single
filament.

We attempted to analyse the radial column density profiles of the
clumps using fits of 2D Gaussian and Plummer profiles. However,
especially in the case of the Plummer functions with many free parameters,
the scatter of parameter values was large and no clear systematic
trends could be discerned. Although azimuthal averaging is somewhat
questionable in the case of such elongated objects,
Appendix~\ref{app:radial} and Fig.~\ref{fig:radial_expo} show the
radial profiles and profile fits for 85 clumps with sizes close to
10\,arcmin$^2$. Even among resolved and single-peaked clumps, the
radial profiles do exhibit a fair amount of variation.  For the
resolved $r>30\arcsec$ parts, the power-law fits show a preference for
profiles close to $r^{-1}$. The median profile was $r^{-0.85}$,
although here the exponent is biased towards zero by some unresolved
clumps and clumps with secondary peaks. Nevertheless, the average
behaviour of individual clumps at scales close to 0.1\,pc is rather
similar to the behaviour of the entire clouds {above} 0.1\,pc
that, moreover, was not very dependent even on the column density
level (see Fig.~\ref{fig:MvsR}).
The $N(r)\sim r^{-0.85}$ relation is similar to earlier studies of
high-mass cores \citep{Beuther2002,Kauffmann2010,Lin2016}.
\citet{Shirley2000} found a slightly steeper profile $N(r)\sim
r^{-1.1}$ for a sample of low-mass cores but at scales that also are
partly below our spatial resolution.

The median FWHM size of the clumps (above the threshold column
density) is 0.075\,pc and thus smaller than many objects in early
molecular line studies \citep{BensonMyers1989, Myers1991}.
\citet{Arquilla1985} used $^{13}$CO observations of a set of clouds to
obtain a typical  density profile of $r^{-2}$ at scales of
several times 0.1\,pc. This is in approximate agreement with our
$r^{-1}$ {column density} relation\footnote{\citet{Arquilla1985}
adopt a finite cloud size and thus their column density for $r^{-2}$
density profile drops faster to zero at the cloud boundary; see their
Table~2.}.

The gravitational stability of clumps was estimated using both column
density maps and effective clump radii ($\alpha_{\rm 2D}$) and by
using the 3D density distributions of the RT models ($\alpha_{\rm
3D}$). The latter has, in principle, two advantages. First, the
radiative transfer modelling automatically takes into account
temperature gradients, thus resulting in more accurate estimates of
the clump mass. However, it is difficult to construct models that
describe the cloud exactly. Therefore, for analysis that only involves
column densities, it may be better to only extract relative
corrections from RT models (e.g. Fig~\ref{fig:corr_G150}) that are then
applied to the $N$ values obtained from regular SED fits. Second, we
can calculate $\Omega_{\rm G}$ directly from the mass distribution in
the 3D model, without further approximations on the general shape or
radial profile. Again, this may not be a significant advantage,
considering the overall uncertainty of the LOS extent of the
structures that we call clumps. For the sample of 85 clumps, the 3D
models indicates that only one in four objects was bound by gravity
while 40\% were bound by external pressure. Given the inhomogeneity of
the sample (and the lack of direct measurements of the turbulent
velocity dispersion), we can only conclude that most of the bound
objects appear to be pressure-confined. There was a clear dependence
on the background column density. Expressed in units of visual
extinction, most of the clumps below $A_{\rm V}=$5\,mag were clearly
unbound while most clumps above this threshold had virial parameters
below $\alpha=2$. At high column densities, $\alpha_{\rm 3D}$ values
tend to be almost a factor of two lower than the $\alpha_{\rm 2D}$
values. Most of the difference can be attributed to the higher column
densities (effects of temperature gradients). However, the 3D models
were also constructed so that their LOS extent matches the POS extent
along the clump minor axis (instead of a geometric mean of the sizes
along the minor and major axes), thus leading to systematically larger
values of $\Omega_{\rm G}$.

\section{Conclusions}  \label{sect:conclusions}

We have used {\em Herschel} observations and radiative transfer modelling to examine the
density structure of selected clumps and their cloud environment. 
The study led to the following conclusions.

On average, the structure noise $N^{\rm str}(\theta)$ of the fields
matches the surface-brightness dependence predicted by
\citet{HelouBeichman1990}, with a scatter of less than a factor of
two.
The fractal dimensions of the fields are relatively constant with
$D_{\rm P}=$1.25$\pm$0.07. The values are only slightly smaller than
typically found for general interstellar clouds. There was no clear
dependence on either the size or the column density of the structures.

The column density PDFs show a wide range of shapes. The strongest
asymmetries are often related to the density structure of individual
clumps or dynamical interactions leading to sharp cloud boundaries.
Column density bias, estimated with RT modelling, has a noticeable
effect on the high-$N$ tail ($N>10^{22}$\,cm$^{-2}$) of the PDFs in a
couple of fields. The low-column-density side of the PDFs is very
sensitive to LOS contamination or the details of background
subtraction.

The radiative transfer models suggest that, in our sample, the
standard SED analysis underestimates the peak column densities usually
by less than 20\% but the maximum errors can be a factor of several.
The strength of the radiation field is on average higher than the
\citet{Mathis1983} model and increases with decreasing Galactic
latitude. However, the absolute values depend on the assumed dust
properties and the dependence on $b$ is likely to be affected by
selection effects.

Both large-scale mass distribution ($r>0.1$\,pc) of the target fields
and the average azimuthally averaged column density profiles of
individual clumps ($r\sim 0.1$\,pc) follow an average relation
$N(r)\sim r^{-1}$. This is in agreement with previous studies of
star-forming clouds.
Clump orientation is often similar to the preferred orientation of
large-scale structures. The correlation can persist over more than one
order of magnitude in column density.

Clump stability was studied using both the projected column density
maps (standard SED analysis) and the 3D radial transfer models. For a
sample of well-resolved clumps, 25\% appeared to be gravitationally
bound and 40\% confined by external pressure. Above a background level
of $A_{\rm V}\sim$5\,mag, most clumps appear to be close to virial
equilibrium.

Our results are consistent with a picture where the clumps are created
by the universal turbulence but sometimes aided by specific converging
flows or direct external forcing. The objects are mainly
pressure-confined but, given sufficient ambient density and mass
reservoir, can evolve towards gravitational instability. The clumps
retain close links to the large-scale cloud environment and, for
example, often inherit their orientation from the filamentary
structure of the parent clouds. At these stages and in spite of the
large variety of cloud environments, the clump regions share many
statistical properties as reflected in the structure functions and
large-scale column density profiles.

\begin{acknowledgements}
This research made use of Montage, funded by the National Aeronautics
and Space Administration's Earth Science Technology Office,
Computational Technologies Project, under Cooperative Agreement Number
NCC5-626 between NASA and the California Institute of Technology. The
code is maintained by the NASA/IPAC Infrared Science Archive.
MJ and VMP acknowledge the support of the Academy of Finland Grant No.
285769 and VMP also acknowledges the financial support from the
European Research Council, Advanced Grant No. 320773.
JMa acknowledges the support of ERC-2015-STG No. 679852 RADFEEDBACK
\end{acknowledgements}

\bibliography{Juvela}

\appendix

\section{PDF analysis}  \label{sect:PDF}

Figures~\ref{fig:pdf1}-\ref{fig:pdf2} show the PDFs of the logarithmic column density and
of the logarithmic 250$\mu$m surface brightness.

The red and the blue curves correspond, respectively, to the surface brightness and
column density data over the full $Herschel$ map coverage and without background
subtraction. In RT modelling, we excluded some of the map boundaries and, furthermore,
carried out background subtraction based on the average surface brightness in regions
within a certain range of low column density values (see Sect.~\ref{sect:obs}). The black
histograms show the PDFs for the column densities calculated via SED analysis and using
these somewhat smaller and background-subtracted surface-brightness maps. Finally, the
grey lines are the corresponding histograms once the RT-derived column density
corrections are taken into account.

\begin{figure*}
\includegraphics[width=17cm]{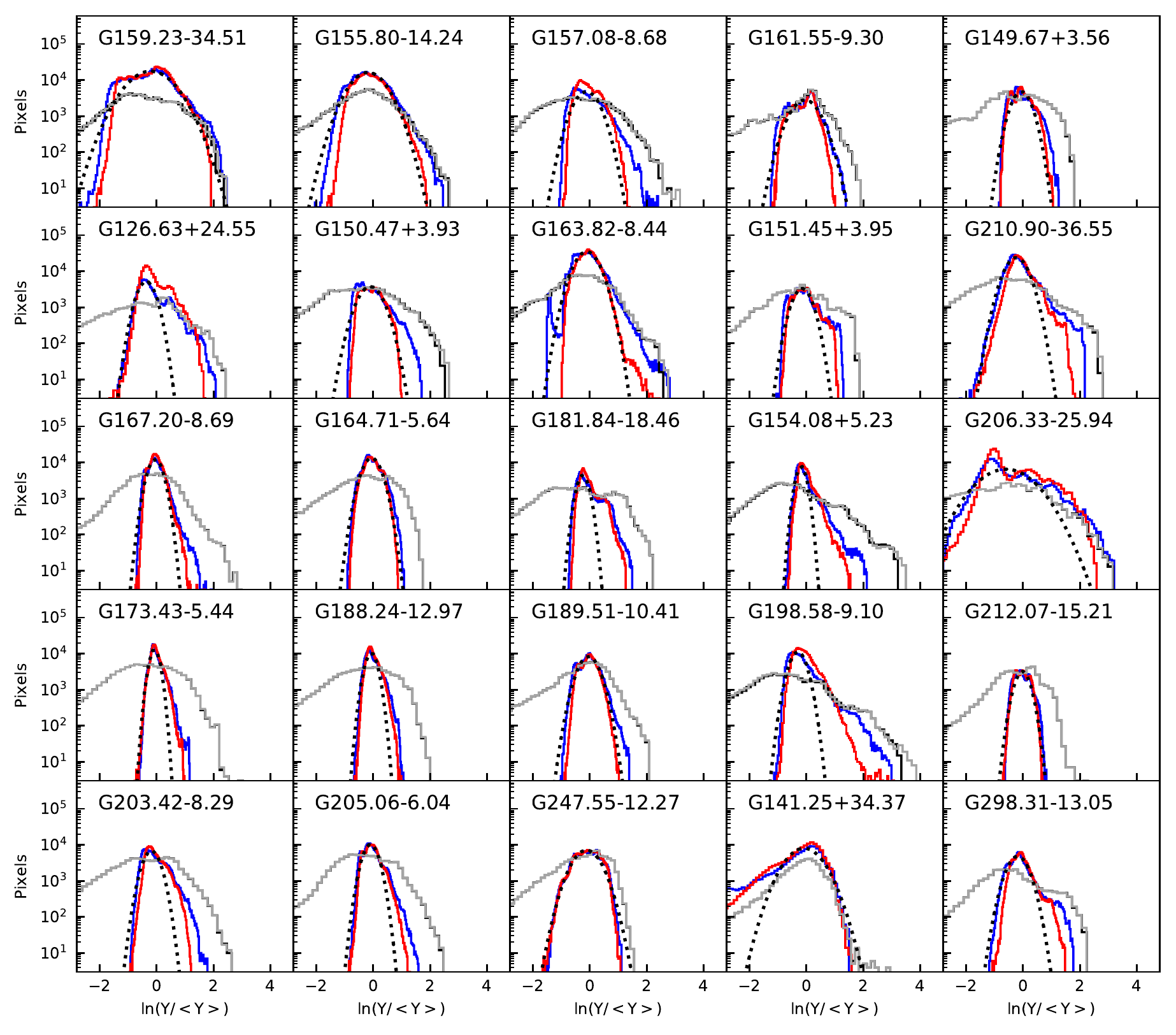}
\caption{
PDFs of the logarithm of the relative values of column density ($N/
\langle N \rangle)$; blue lines) and 250$\mu$m surface brightness
($S(250\mu{\rm m})/ \langle S(250\mu{\rm m}) \rangle)$; red lines).
Gaussian fits to the column density histograms are shown as dotted
black lines.
The black histograms are derived from column density that is estimated
based on background subtracted surface-brightness data. The grey
histograms additionally (mostly on top of the black histograms)
include the bias corrections derived from RT models.
The fields are arranged in the order of increasing distance (see
Table~\ref{table:fields}).
}
\label{fig:pdf1}
\end{figure*}

\setcounter{figure}{0}

\begin{figure*}
\includegraphics[width=17cm]{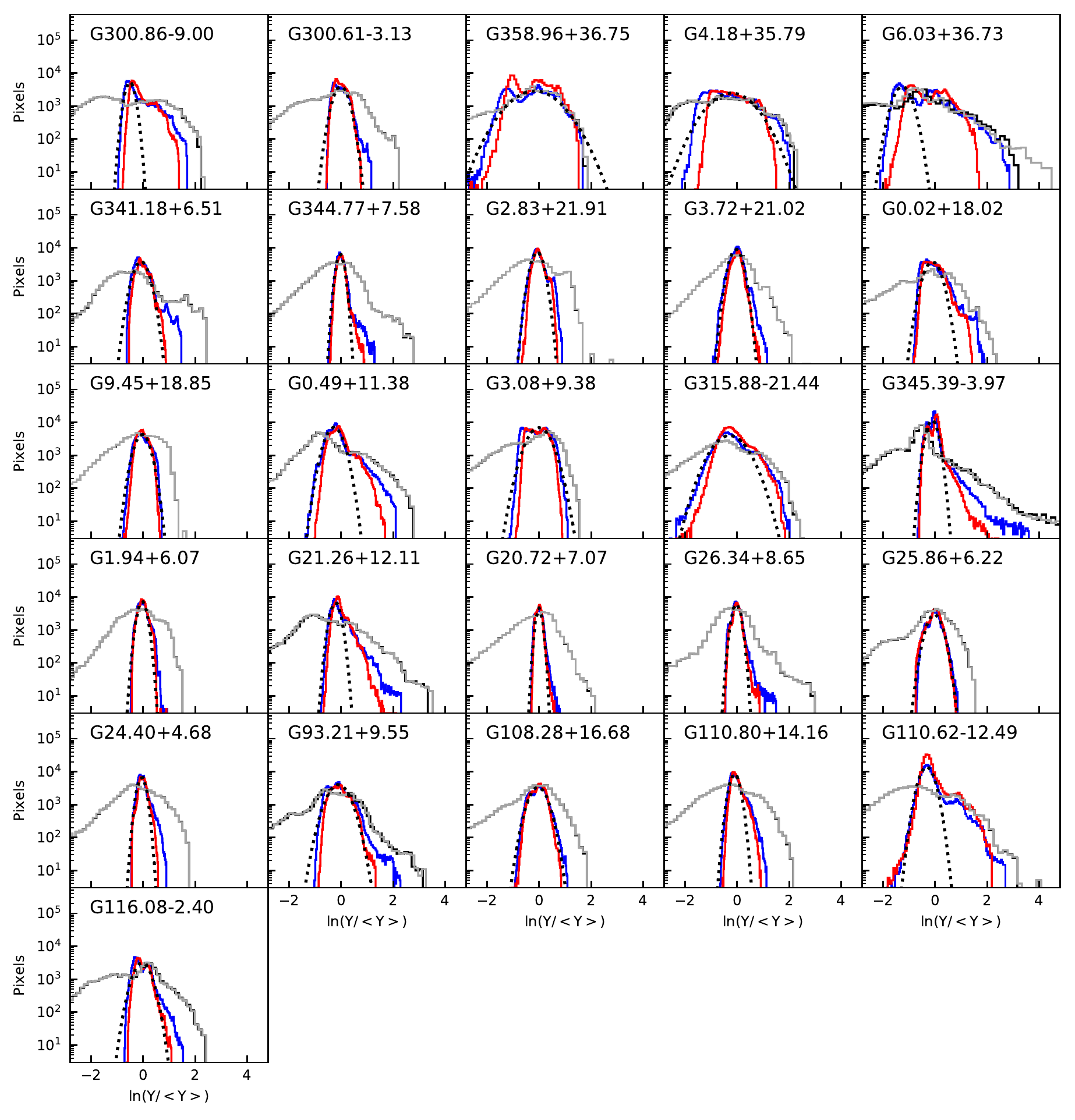}
\caption{
continued.
}
\label{fig:pdf2}
\end{figure*}

\section{Examples of residuals in radiative transfer models} \label{sect:residuals}

\begin{figure}
\includegraphics[width=8.8cm]{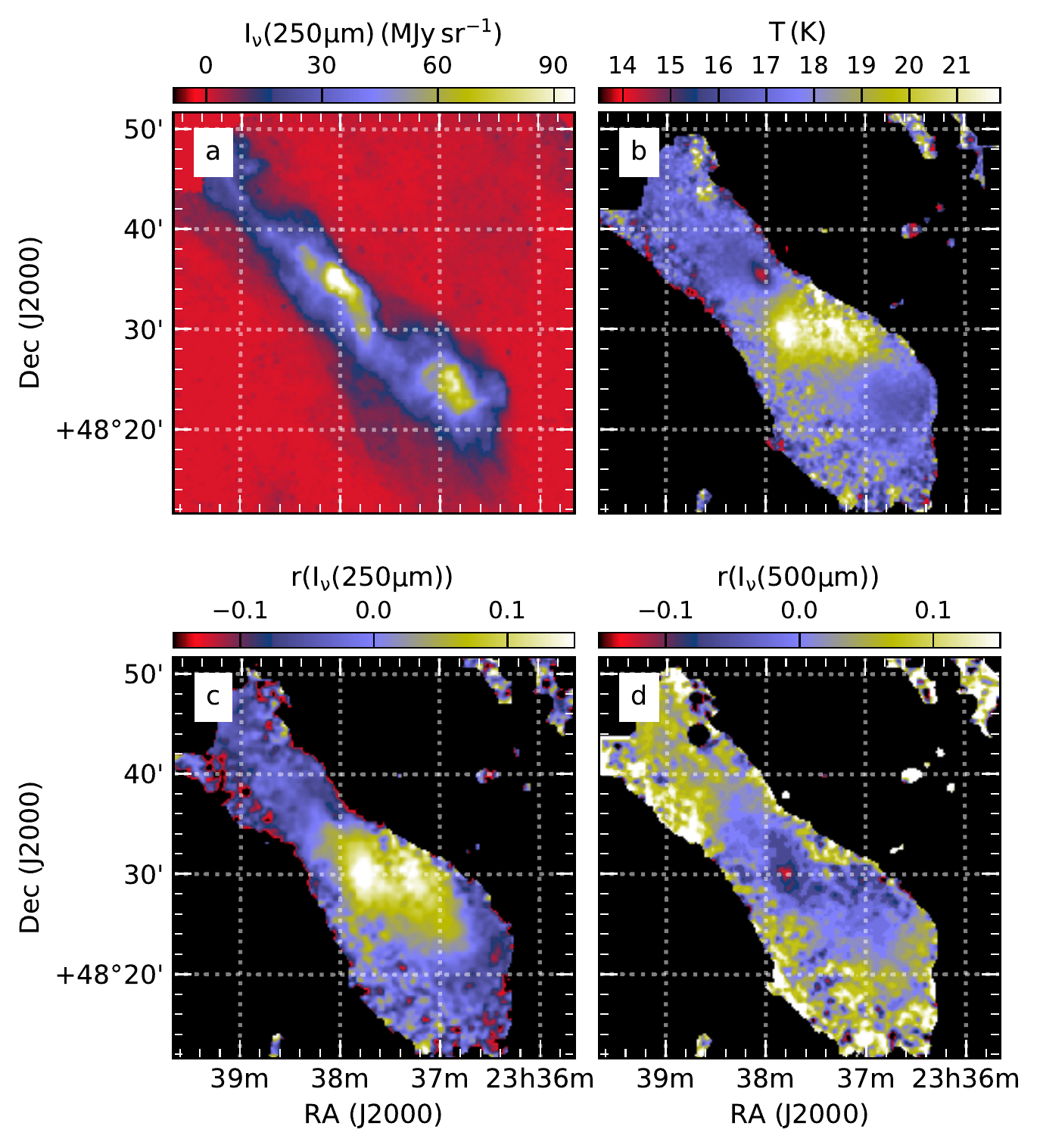}
\caption{
Observed 250\,$\mu$m surface brightness $I_{\rm nu}(250\mu{\rm
m})$ (frame a), colour temperature $T$ (frame b), and the 250\,$\mu$m
(frame c) and 500\,$\mu$m (frame c) residuals in the RT model of the
field G110.62-12.49. Colour temperature is estimated from observations
using $\beta=1.8$. The residuals are calculated as $r= (I_{\rm
Obs}-I_{\rm Mod})/I_{\rm Obs}$, where Obs and Mod refer to observed
and modelled values, respectively. All maps are at 40$\arcsec$
resolution.
}
\label{fig:G110_residuals}
\end{figure}

\begin{figure}
\includegraphics[width=8.8cm]{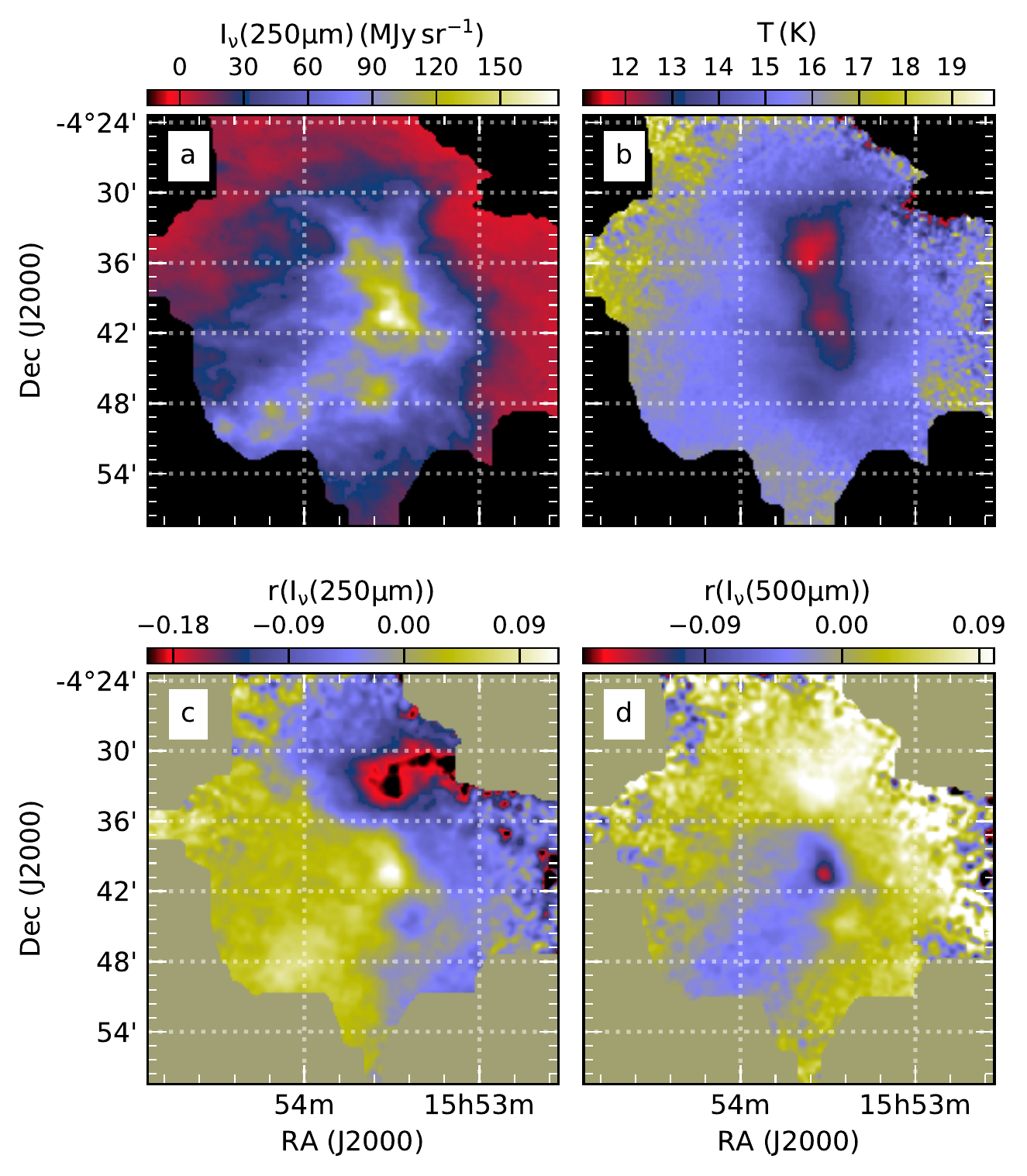}
\caption{
As Fig.~\ref{fig:G110_residuals} but showing the fit residuals for the field
G4.18+35.79.
}
\label{fig:G4_residuals}
\end{figure}

\begin{figure}
\includegraphics[width=8.8cm]{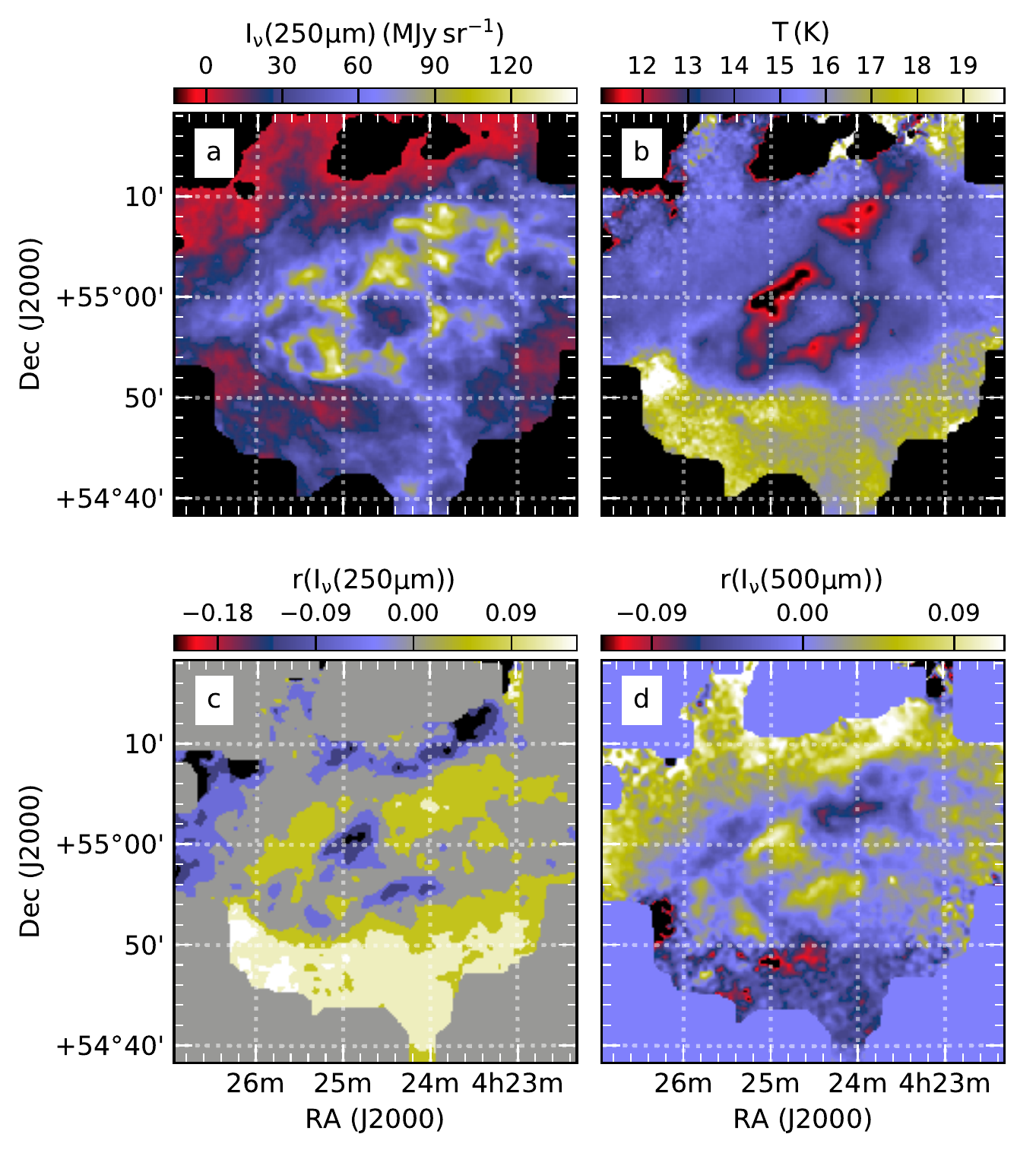}
\caption{
As Fig.~\ref{fig:G110_residuals} but showing fit residuals for the field G150.47+3.93.
}
\label{fig:G150_residuals}
\end{figure}

The radiative transfer modelling is described in Sect.~\ref{sect:RTmodels} with result
presented in Sect.~\ref{sect:RTresults}. In most cases the surface-brightness residuals
are of the order of 1\% and spatially uncorrelated. However, in some fields there are
significant, spatially correlated residuals that tend to be related to the properties of
the radiation field. The following Figures show three examples where the residuals are
caused either by a radiation sources within the field (Fig.~\ref{fig:G110_residuals}), a
strong anisotropy of the external field (Fig.~\ref{fig:G4_residuals}), or by an apparent
inconsistency between the heating of the dense and low-density regions
(Fig.~\ref{fig:G150_residuals}).

\section{Clump profile fits with 2D models}    \label{sect:2dfits}

The following Figures show the distributions of the 2D fits to a
sample of 85 clumps,
each with an area close to 10\,arcmin$^2$. 

\begin{figure*}
\includegraphics[width=18.2cm]{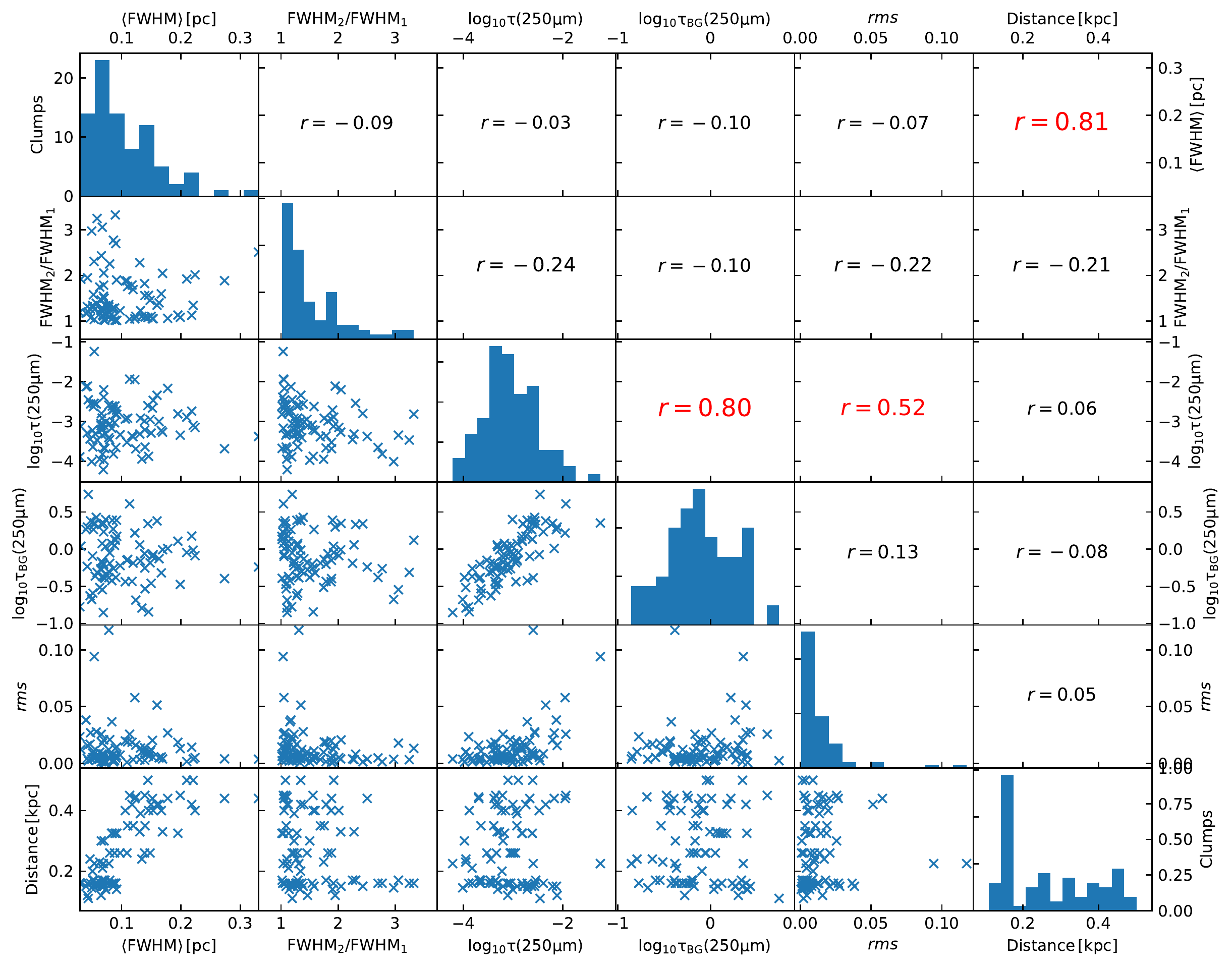}
\caption{
Distributions of the parameters of 2D Gaussian fits to the sub-sample of 85 clumps. The
frames in the lower left part of the Figure show scatterplots between the quantities:
clump size $\langle {\rm FWHM} \rangle$, clump elongation ${\rm FWHM_2/FWHM_1}$, peak optical
depth $\tau(250\mu{\rm m})$, background optical depth $\tau_{\rm BG}(250\mu{\rm m})$,
the relative rms error of the fit, and the distance. The frames on the diagonal show the
histograms of the individual quantities and each of the remaining frames indicates the
linear correlation coefficient between the corresponding pair of parameters.
}
\label{fig:monopeak_G}
\end{figure*}

Figure~\ref{fig:monopeak_G} shows the results for 2D Gaussian fits. Even this simple
model gives relatively good fits with rms residuals mostly below 10\%. Most pairs of
parameters do not show significant correlations. In particular, the angular size and the
shape of the clumps are independent of the $\tau(250\mu{\rm m})$ values of both the
background and the clump. Similarly, there is no dependence on the estimated distance,
apart from the expected correlation with the absolute size of the clumps, for which the
correlation coefficient is $r=0.81$. Even if the physical size were completely
uncorrelated with the distance, this amount of apparent correlation could be produced by
having 30\% uncertainty in the distance estimates. Statistically, there is no significant
difference between the parameter values obtained with and without the column density
corrections derived from the RT models ($N_{\rm cor}$ vs. $N_{\rm obs}$).

\begin{figure*}
\includegraphics[width=18.2cm]{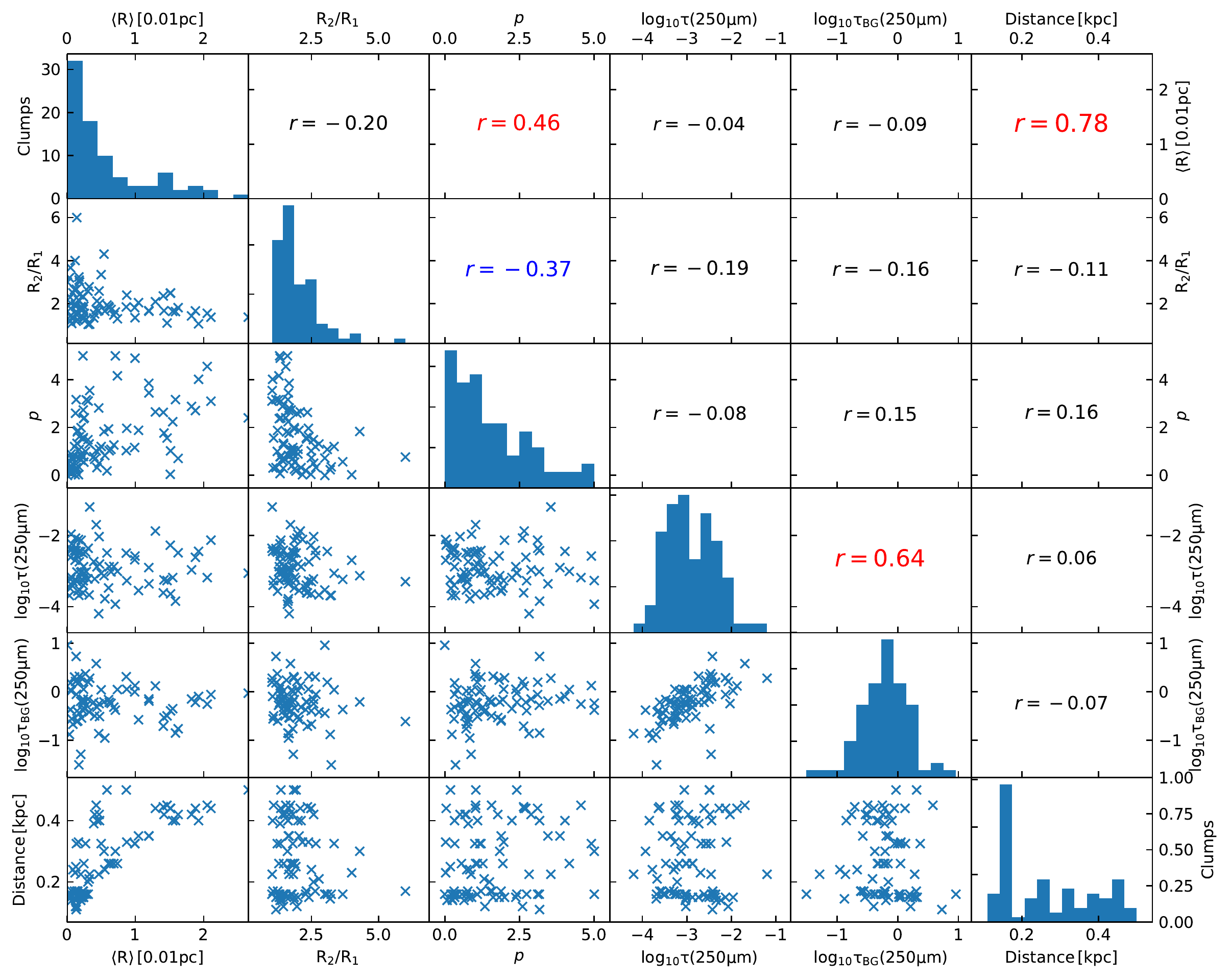}
\caption{
Distributions of the parameters of 2D Plummer profiles for the sub-sample of 85 clumps.
Scatterplots and correlations are shown between geometric mean of the flat radius values
$\langle R \rangle$, ratio of flat radius values ${\rm R_2/R_1}$, power-law exponent $p$,
peak optical depth $\tau(250\mu{\rm m})$, background optical depth $\tau_{\rm
BG}(250\mu{\rm m})$, and distance. The frames on the diagonal show the histograms of the
individual quantities and the frames in the upper right hand part of the Figure indicate
the linear correlation coefficients for the corresponding pairs of parameters.
}
\label{fig:monopeak_A}
\end{figure*}

In theory, to better characterise the radial profiles of the clumps,
one can fit the clumps with Plummer functions (see
Eq.~(\ref{eq:Plummer})). The results of these fits are shown in
Fig.~\ref{fig:monopeak_A}. However, because of the large number of
parameters and the dependence between the $R$ and $p$ parameters, the
fitted values show a large scatter. There is a significant positive
correlation between the optical depths of the target and the
background ($r=0.64$). There is a weaker negative correlation between
the clump elongation and the $p$ parameter, which is natural if the
source is reminiscent of a cylinder.

\section{Examples of cloud structure extracted with TM} \label{sect:example_TM}

\begin{figure*}
\includegraphics[width=18cm]{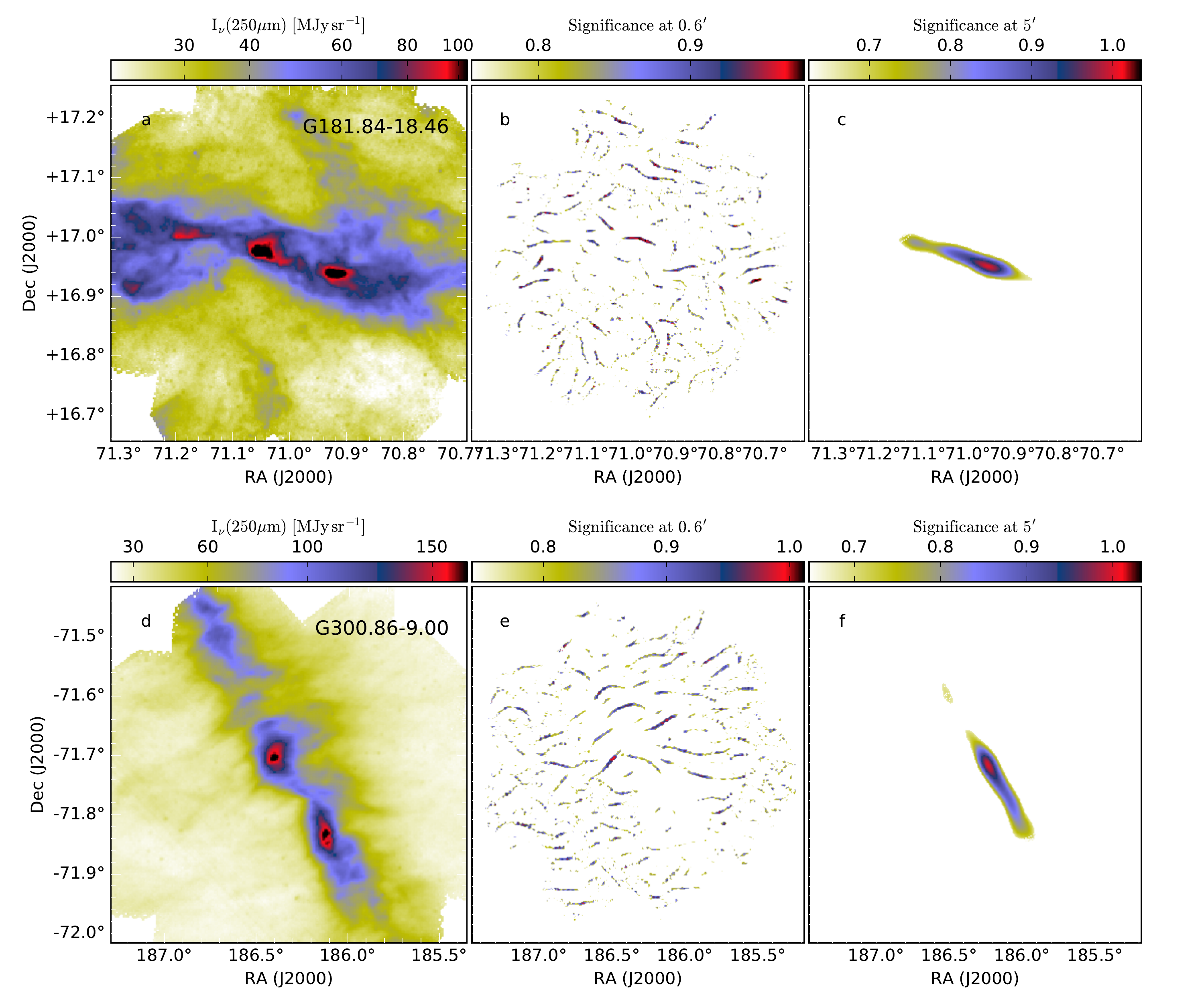}
\caption{
Examples of elongated structures extracted by the TM method. The upper row shows data for
the field G181.84-18.46 and the lower row for G300.86-9.00. The frames, from left to
right, are the 250\,$\mu$m surface brightness and the relative significance of the
aligned structures at the 0.6$\arcmin$ and 5$\arcmin$ scales. In the latter frames, pixels with
relative significance below the 90\% percentile are masked.
}
\label{fig:plot_TM_examples}
\end{figure*}

Figure~\ref{fig:plot_TM_examples} shows examples of structures extracted at the
0.6$\arcmin$ and 5$\arcmin$ scales. The fields G181.84-18.46 and G300.86-9.00 are both
dominated by a single filamentary structure. However, in Fig.~\ref{fig:oricor1}, they
represent extreme cases where the small- and large-scale structures are either aligned or
preferentially perpendicular.

\section{Angular dispersion functions} \label{sect:ADF}

The angular dispersion function
\begin{equation}
S(r, \delta) = \sqrt{ \frac{1}{N} \sum_{i=1}^{N} 
\left(  \theta(r) - \theta(r+\delta_i)  \right)^2  },
\label{eq:ADF}
\end{equation}
measures the coherence of angles $\theta$ around a position $r$ as the function of the
lag $delta$. The summation goes over all displacements $\delta_i$ of a given distance or
a distance range. For example, \citep{Planck_2015_XX} calculated the coherence of
polarisation angles using a wide annulus of [$\delta/2$, $3\delta/2$]. 

We apply $S$ to the TM results that characterise the structural anisotropy of the
surface-brightness images. The possible values of $\delta$ are limited by the typical
size of the images and, on the other hand, by the size of the template used in the TM
calculation. Depending on the TM parameters (the scale $F$ and the possible
normalisation, as described in Sect~\ref{sect:global}) result in the extraction of
structures of different size and would thus also change the subsequent estimates of $S$.
The position angles $\theta$ that are determined by TM are significant only for a small
subset of the map pixels. The sum in Eq.~(\ref{eq:ADF}) should only include pixels in
well-defined structures, but the number of such structures in a given environment can
become small, especially for small values of $\delta$.

\begin{figure}
\includegraphics[width=8cm]{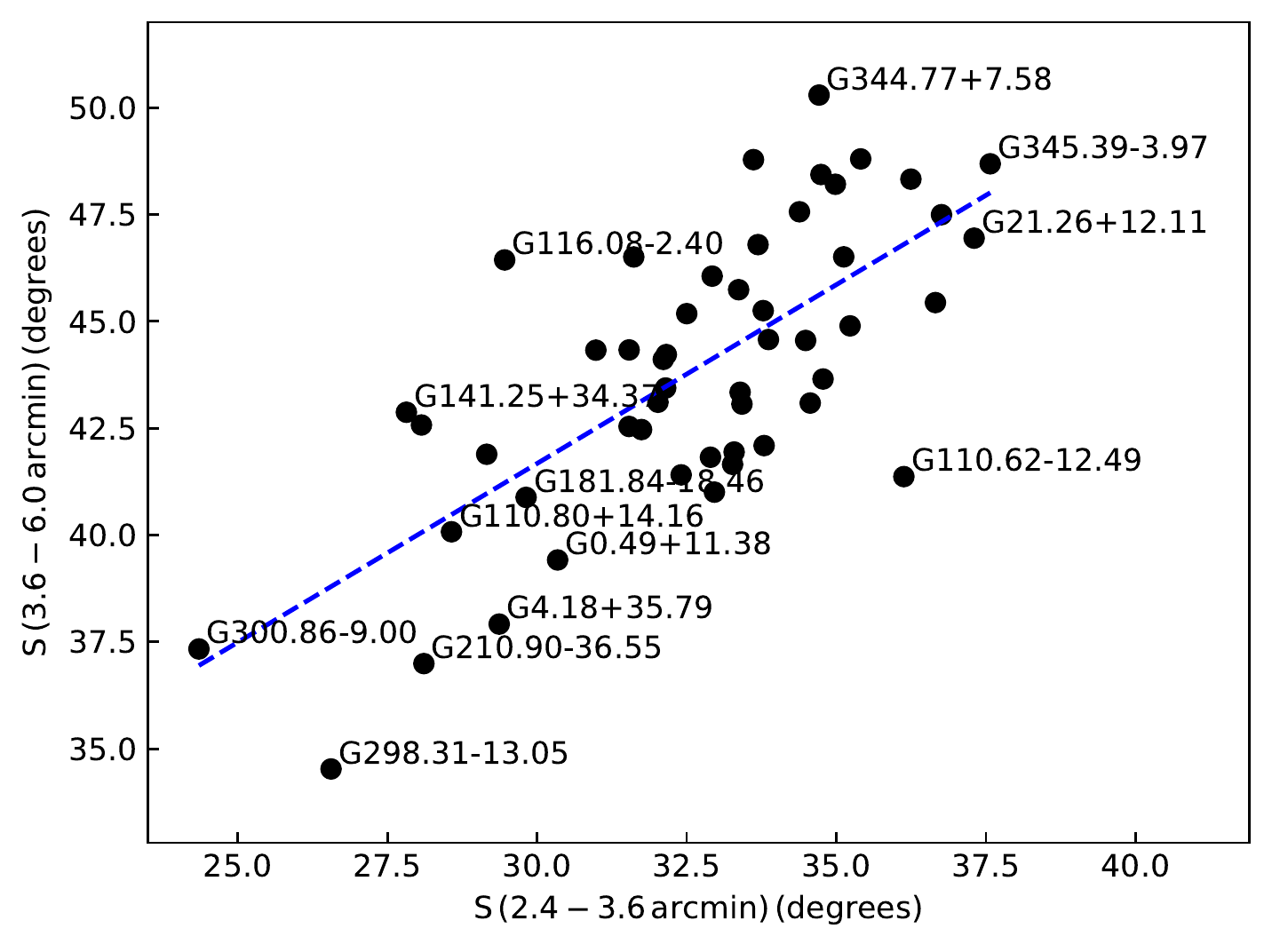}
\caption{
Angular dispersion function $S$ calculated from the position angle estimates of the TM
method. The plotted values are averages over individual fields, calculated using lags
$\delta$ within the ranges of 2.4-3.5$\arcmin$ and 3.6-6.0$\arcmin$. The dashed line
corresponds to the unweighted least squares fit.
}
\label{fig:ADF}
\end{figure}

As an example, we examine $S$ using the TM position angle estimates calculated with the scale
parameter $F=1.2\arcmin$ and using data normalisation (see Sect.~\ref{sect:global}) to
extract structures more uniformly over the whole map area. Figure~\ref{fig:ADF} compares the
map-averaged values of $S$ for two cases where the reference annulus extends over
$\theta$=2.4-3.5$\arcmin$ or over larger distances $\theta$=3.6-6.0$\arcmin$. We note that at a
distance of 2.4$\arcmin$ the $\theta$ estimates are not completely independent because of the
combined effect of the size of the template that was used to estimate the position angles
($2.4\arcmin \times 2.4\arcmin$) and the resolution of the underlying data (18$\arcsec$ for
the 250\,$\mu$m surface-brightness images).

As expected, the $S$ values at the different scales are correlated and larger lags tend
to correspond to larger dispersion (in Fig.~\ref{fig:ADF} on average by 34\%). The
smallest $S$ values are found for the nearby ($d<200$\,pc) fields G300.86-9.00 and
G298.31-13.05 that are dominated by one or two filamentary structures close to the scale
$F$ selected in TM. Small values are also found for the high-latitude Lynds clouds
G358.96+36.75, G6.03+36.73, and G4.18+35.79 (LDN~1780, LDN~183, and LDN~134,
respectively). Conversely, the angular dispersion is large in fields like G344.77+7.48
and G345.39-3.97, where most of the area is covered by low-column-density material that
has structure more consistent with generic turbulent fluctuations.

\section{Radial clump profiles} \label{app:radial}

The Figs.~\ref{fig:radial_first}-\ref{fig:radial_last} show azimuthally averaged column
density profiles for clumps with approximate size of 10\,arcmin$^2$. We have restricted
the sample to 85 clumps where the structures do not contain significant secondary 
peaks.

\begin{figure}
\includegraphics[width=8.2cm]{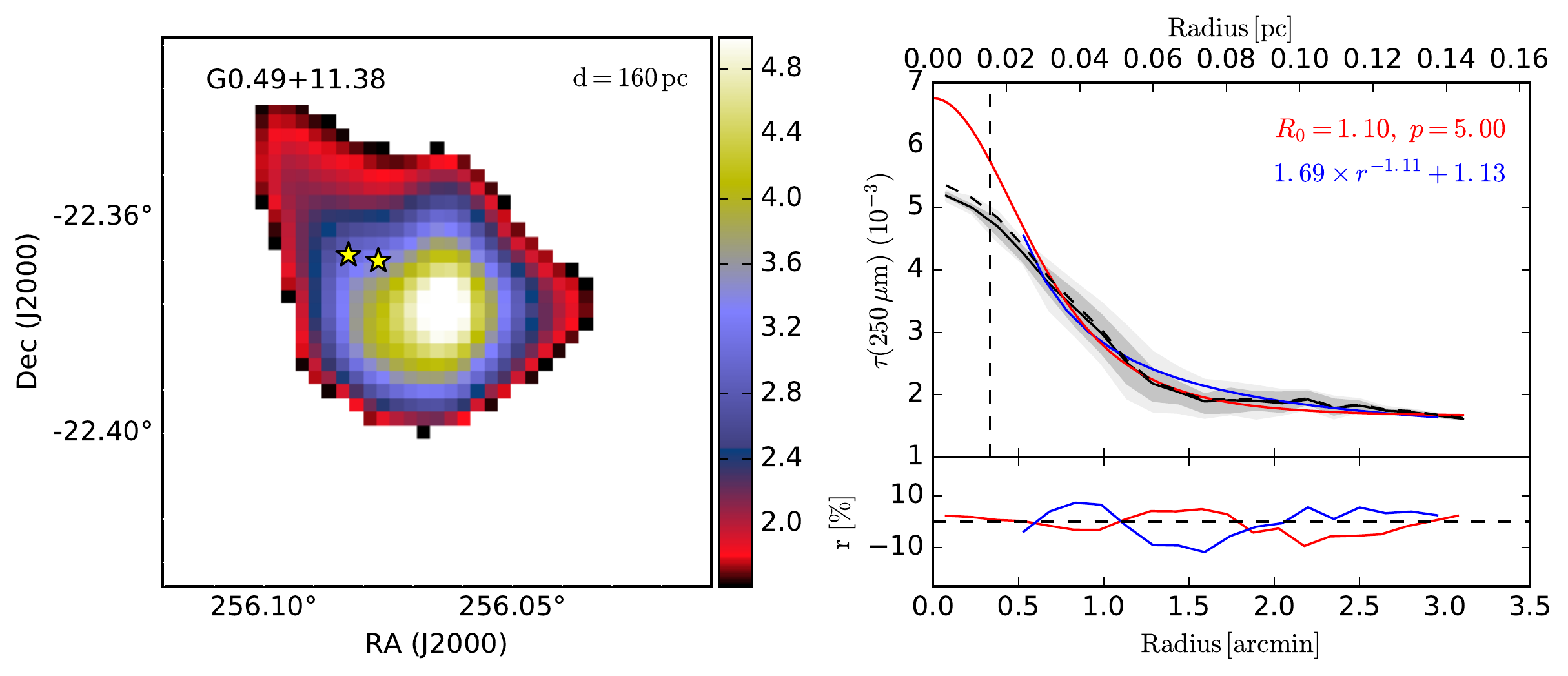}
\includegraphics[width=8.2cm]{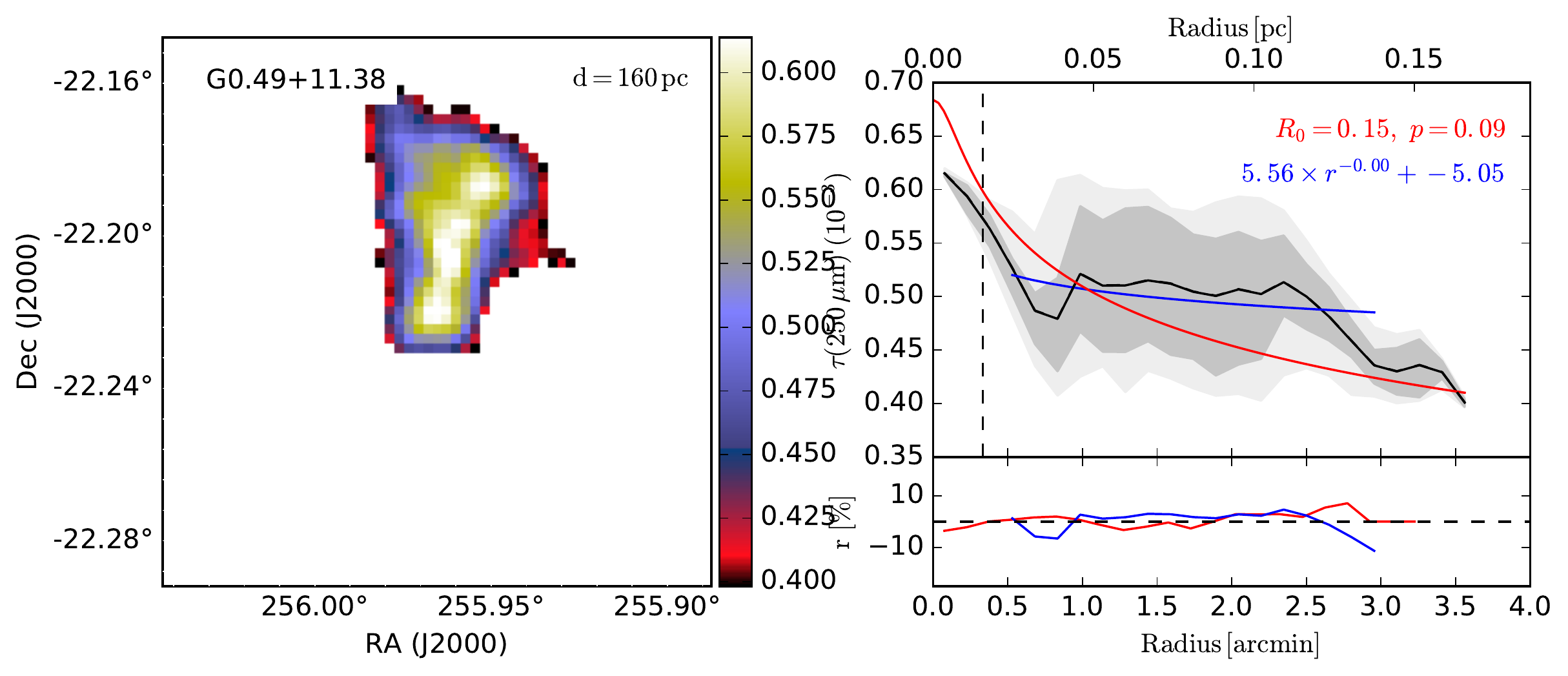}
\includegraphics[width=8.2cm]{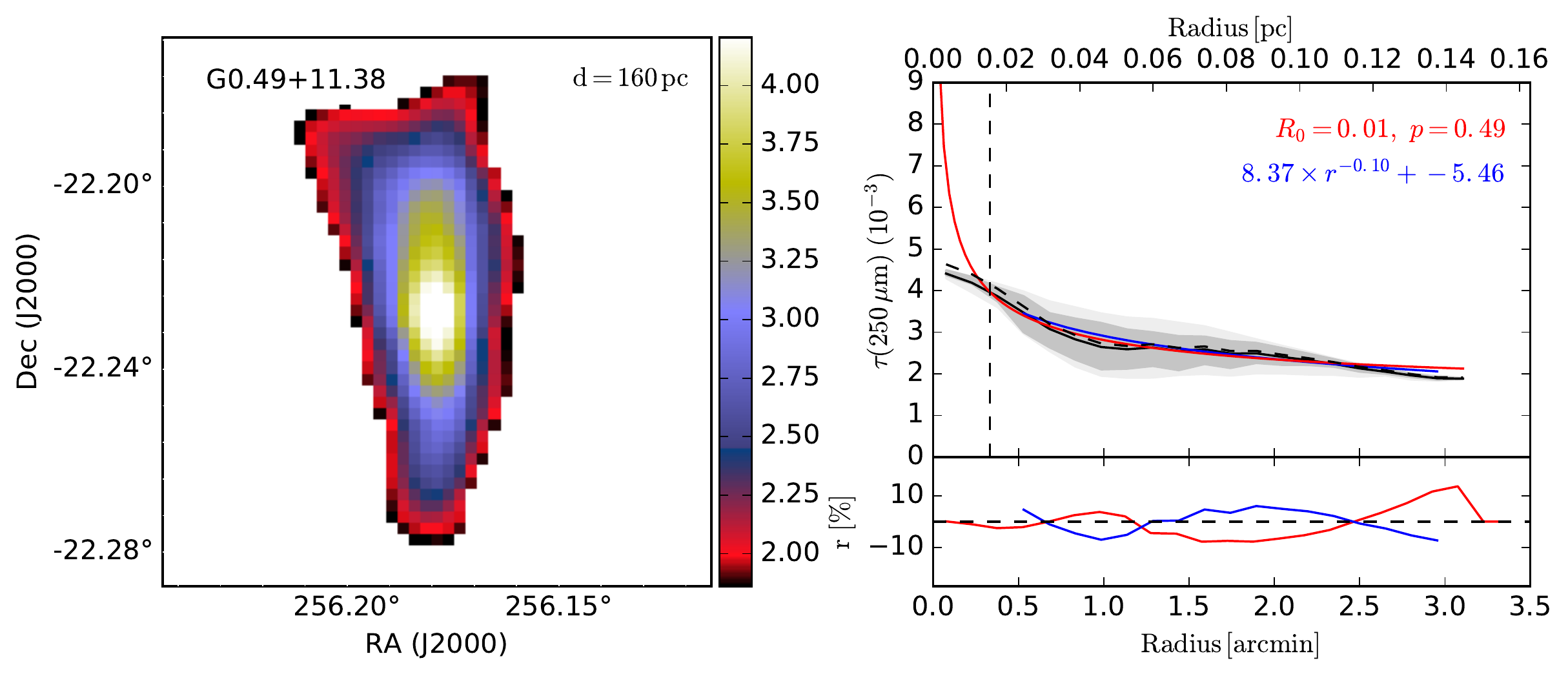}
\includegraphics[width=8.2cm]{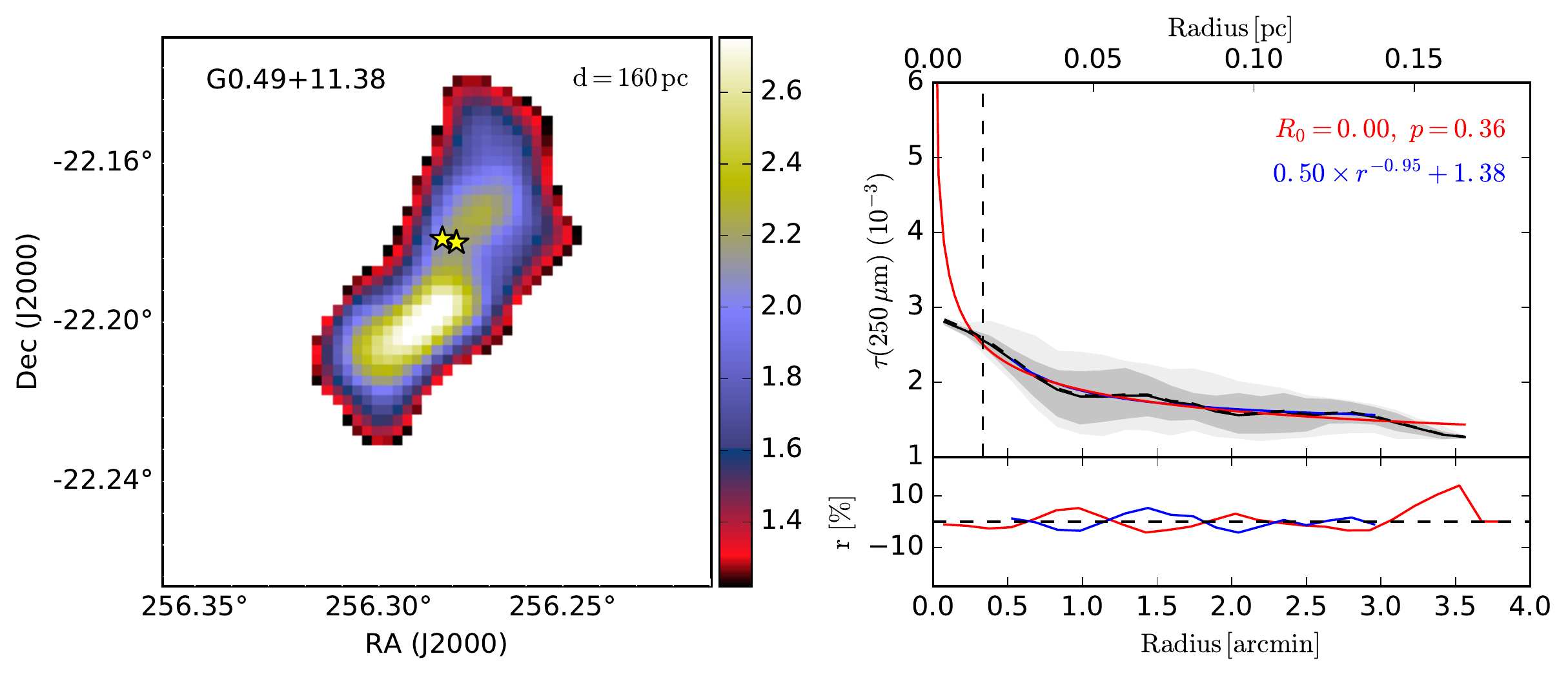}
\includegraphics[width=8.2cm]{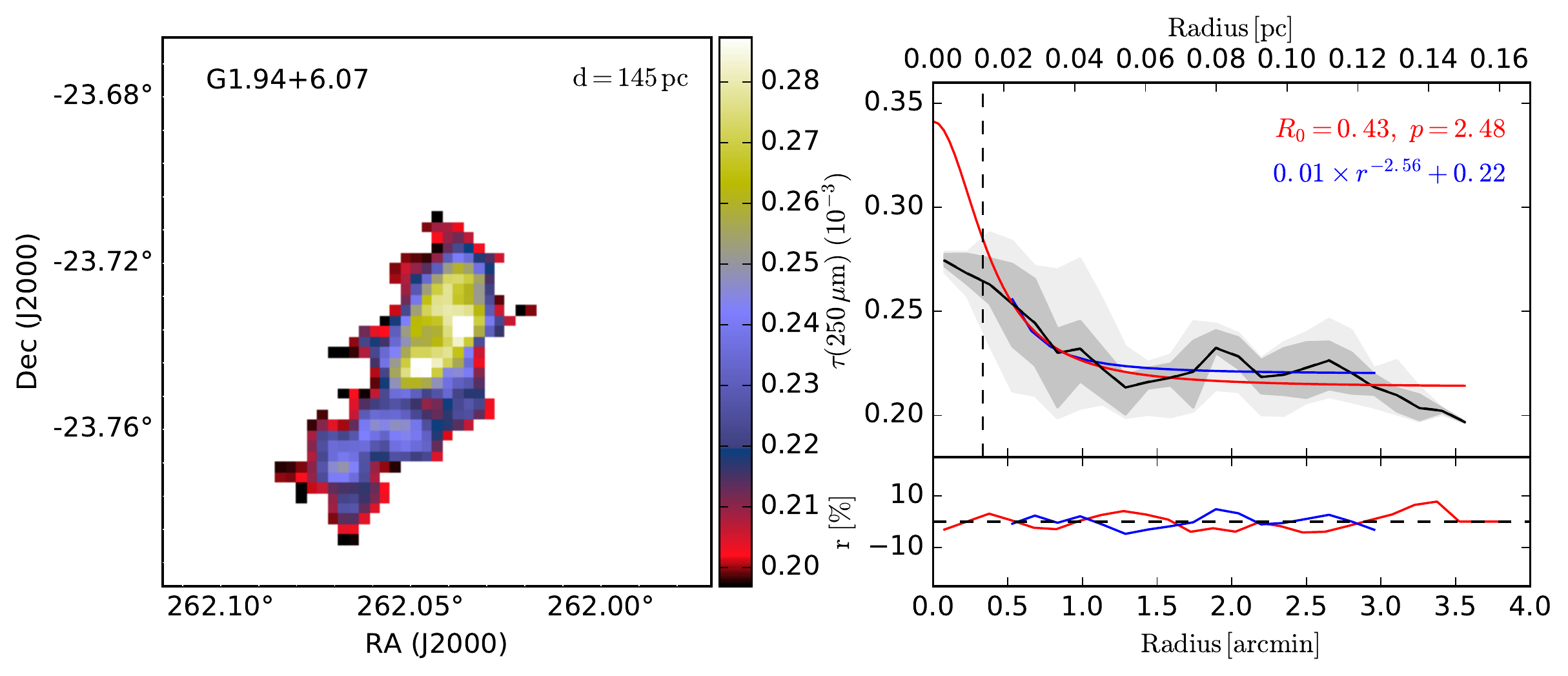}
\caption{
Maps and radial profiles of $10^{-3} \times \tau(250\mu {\rm m})$ for 10\,arcmin$^2$
clumps. On the map, YSO candidates \citep{GCC-IV} are indicated with stars. On the
right, the azimuthal averaged profiles are shown without (black solid line) and with
(dashed black line) corrections derived from RT models. The shaded areas show the
10-90\% and the 25-75\% intervals for the azimuthal variation. The red curve correspond
to the fitted, de-convolved Plummer profiles. The blue curve correspond to a powerlaw
fit to data outside 30$\arcsec$ radius (data corrected according to the RT models but
not deconvolved). The fit parameters are quoted in the frame. The vertical dashed line
at 20$\arcsec$ corresponds to the FWHM extent of a point source. The lower frame shows
the relative fit errors.
\label{fig:radial_first}
}
\end{figure}

\begin{figure}
\includegraphics[width=8.2cm]{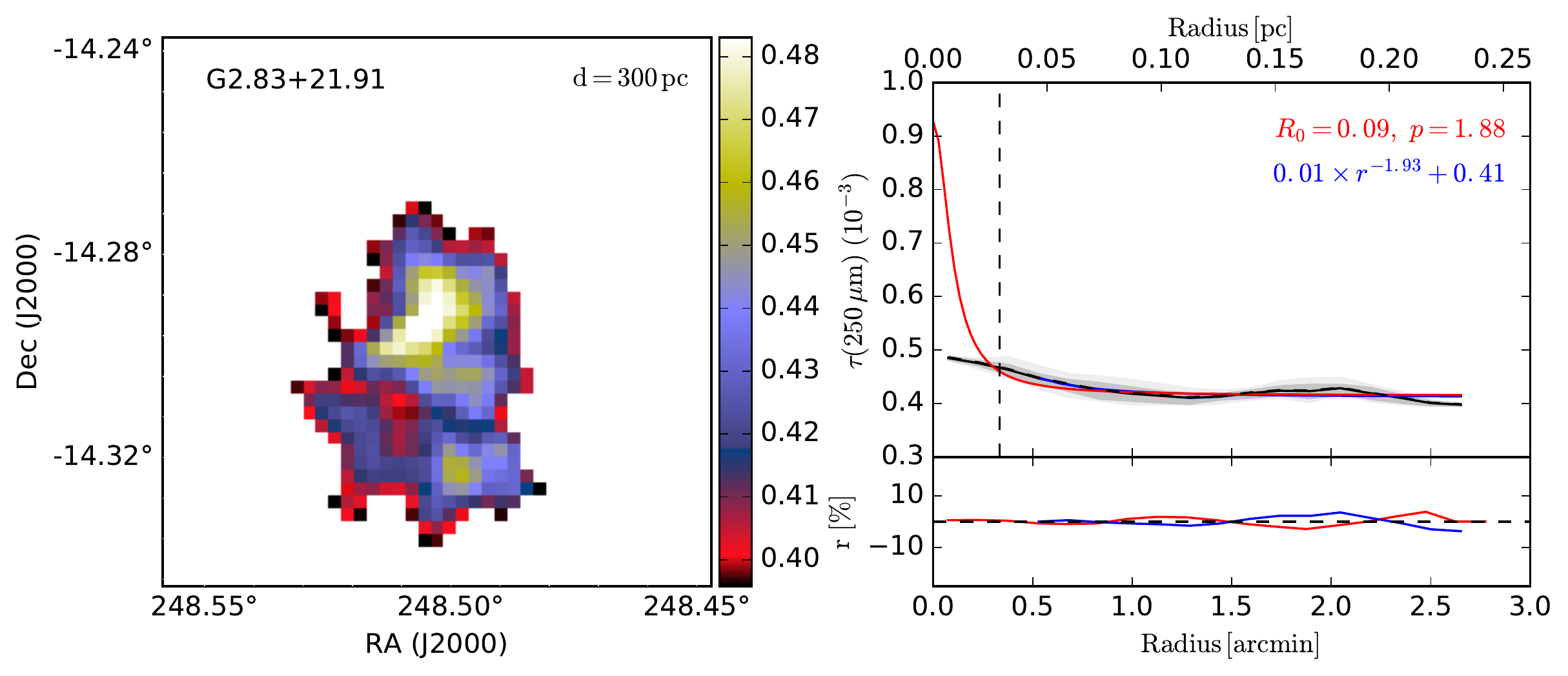}
\includegraphics[width=8.2cm]{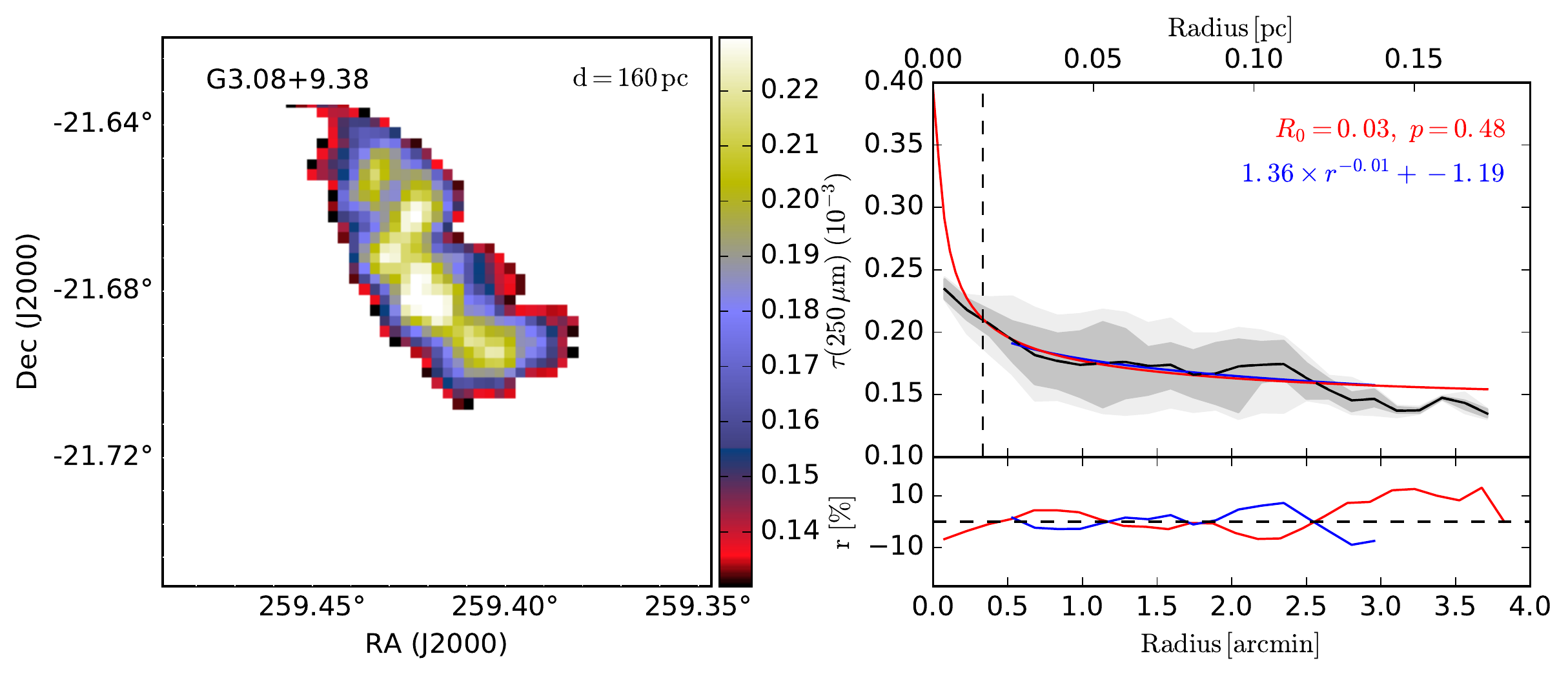}
\includegraphics[width=8.2cm]{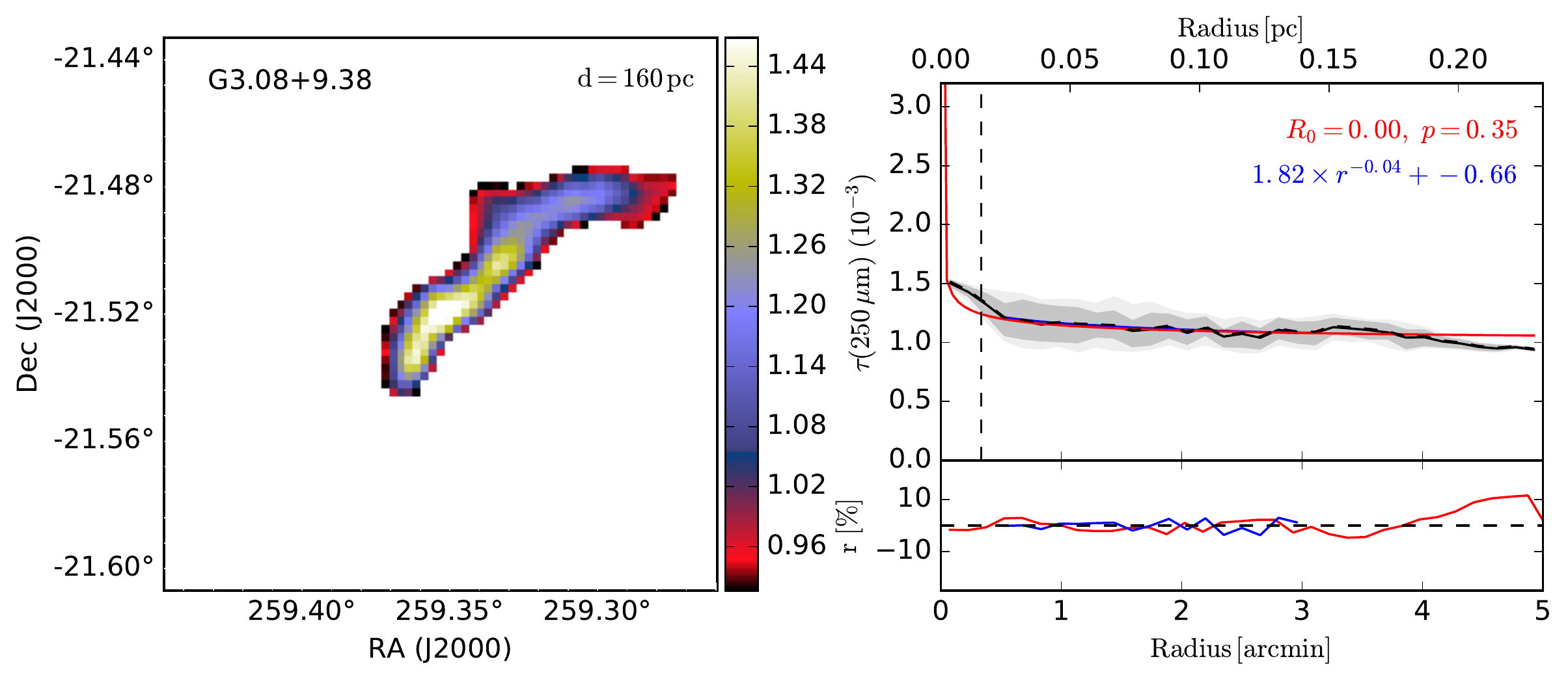}
\includegraphics[width=8.2cm]{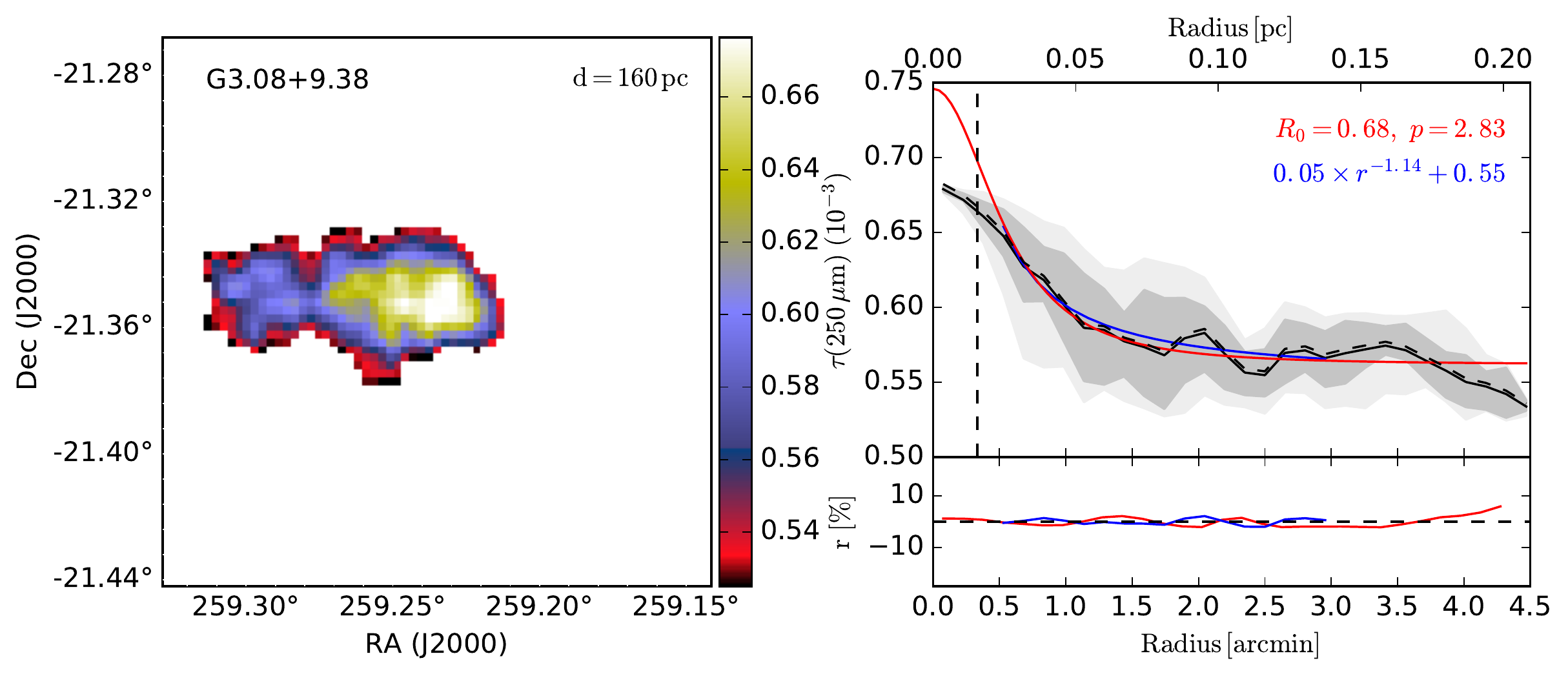}
\includegraphics[width=8.2cm]{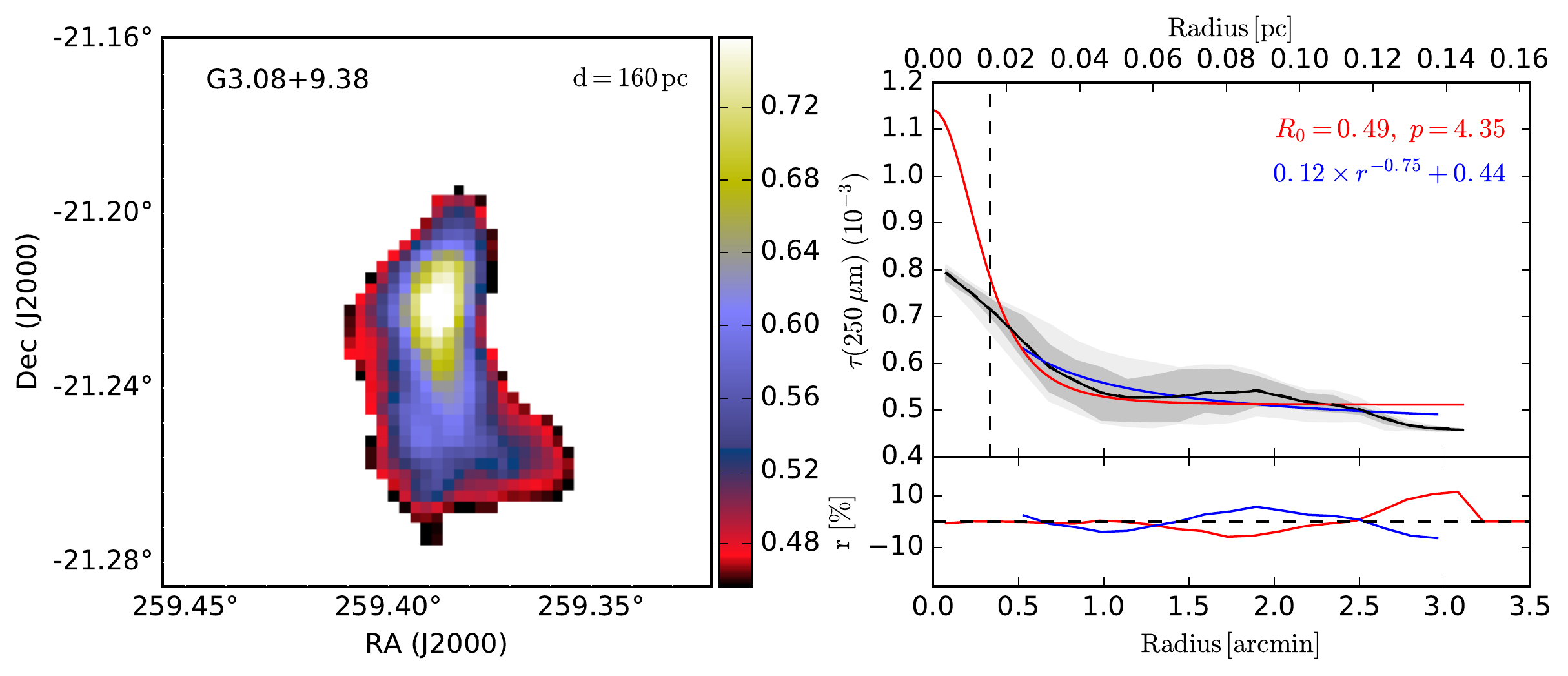}
\includegraphics[width=8.2cm]{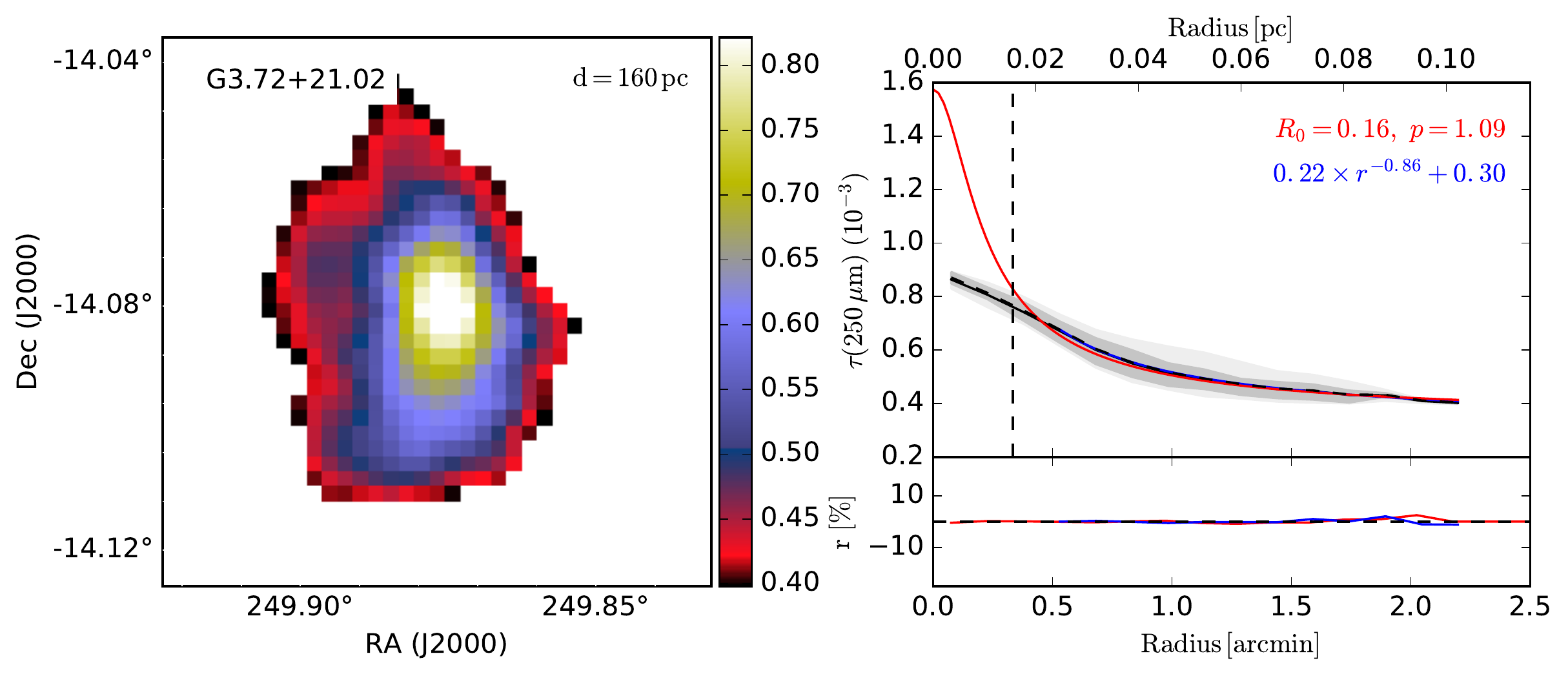}
\caption{continued.}
\end{figure}

\begin{figure}
\includegraphics[width=8.2cm]{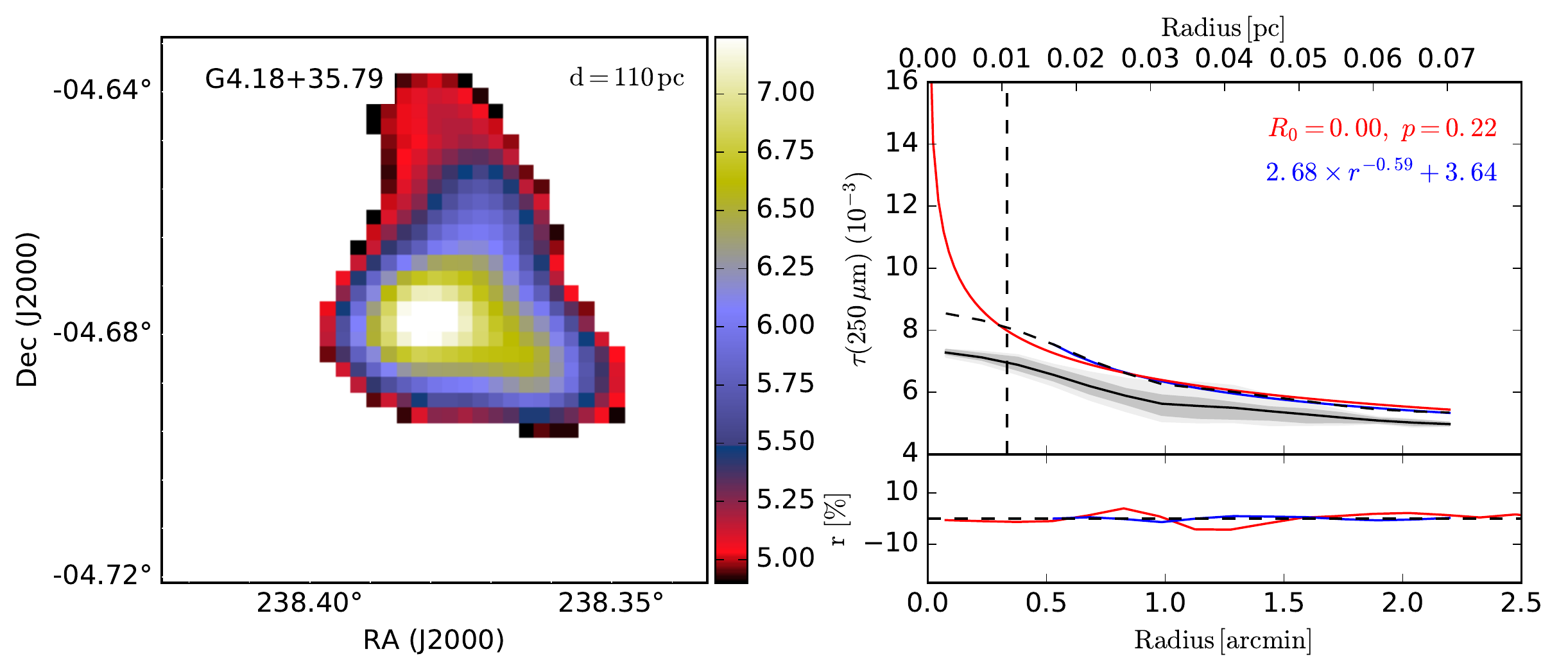}
\includegraphics[width=8.2cm]{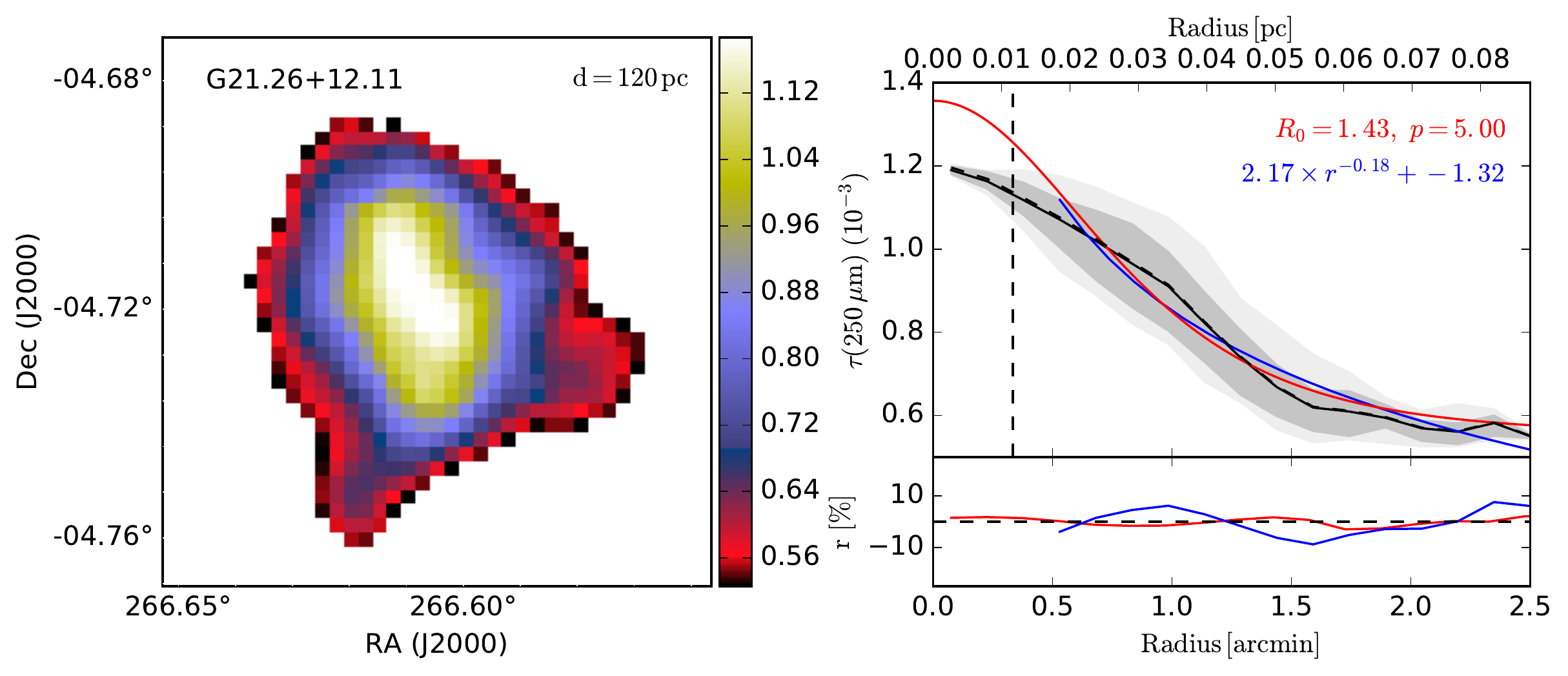}
\includegraphics[width=8.2cm]{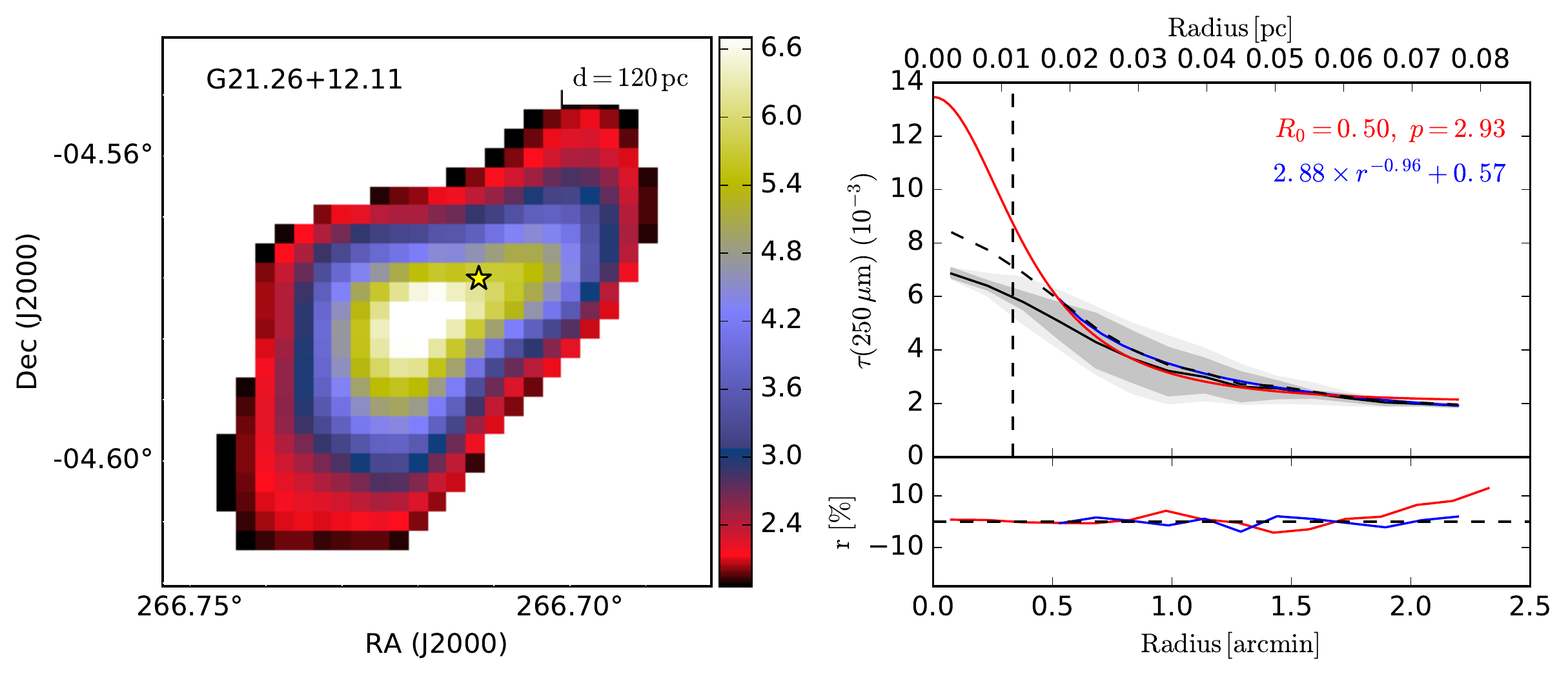}
\includegraphics[width=8.2cm]{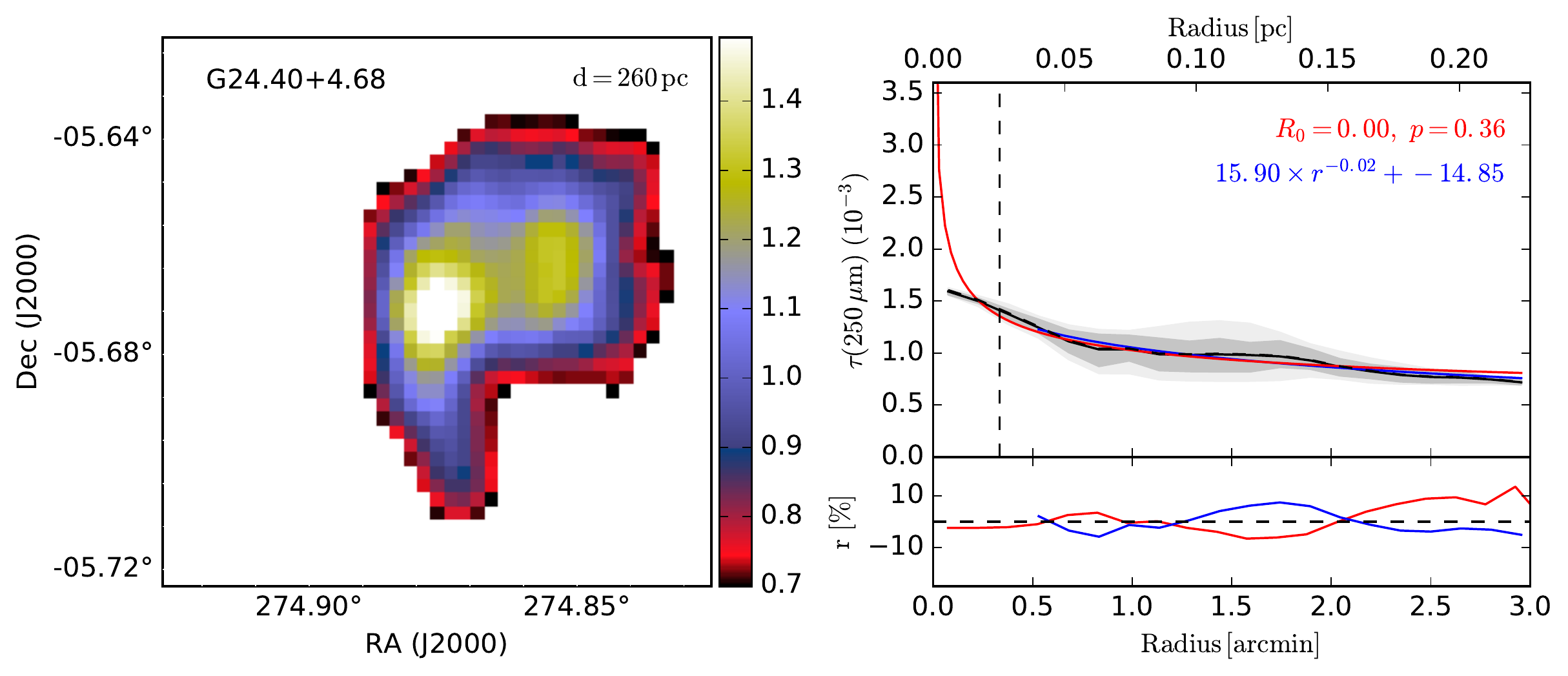}
\includegraphics[width=8.2cm]{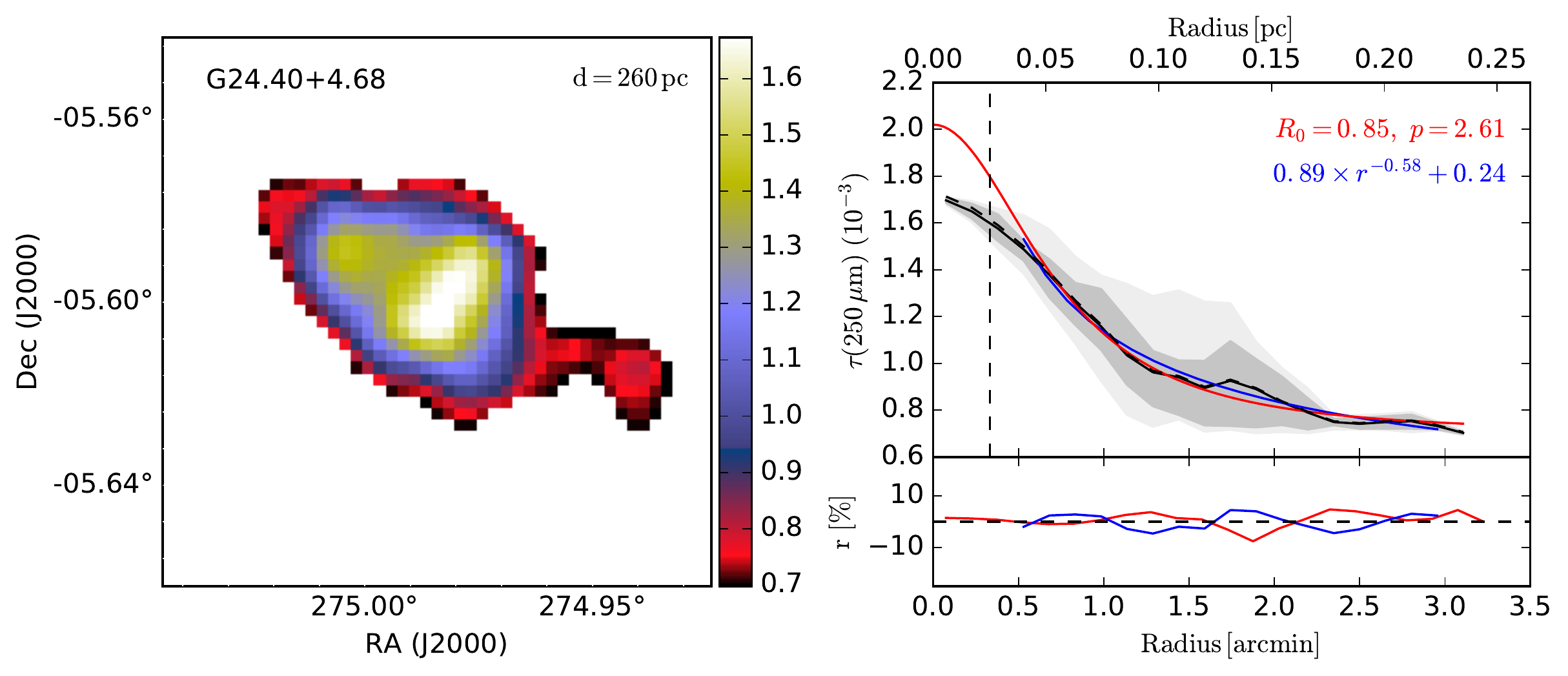}
\includegraphics[width=8.2cm]{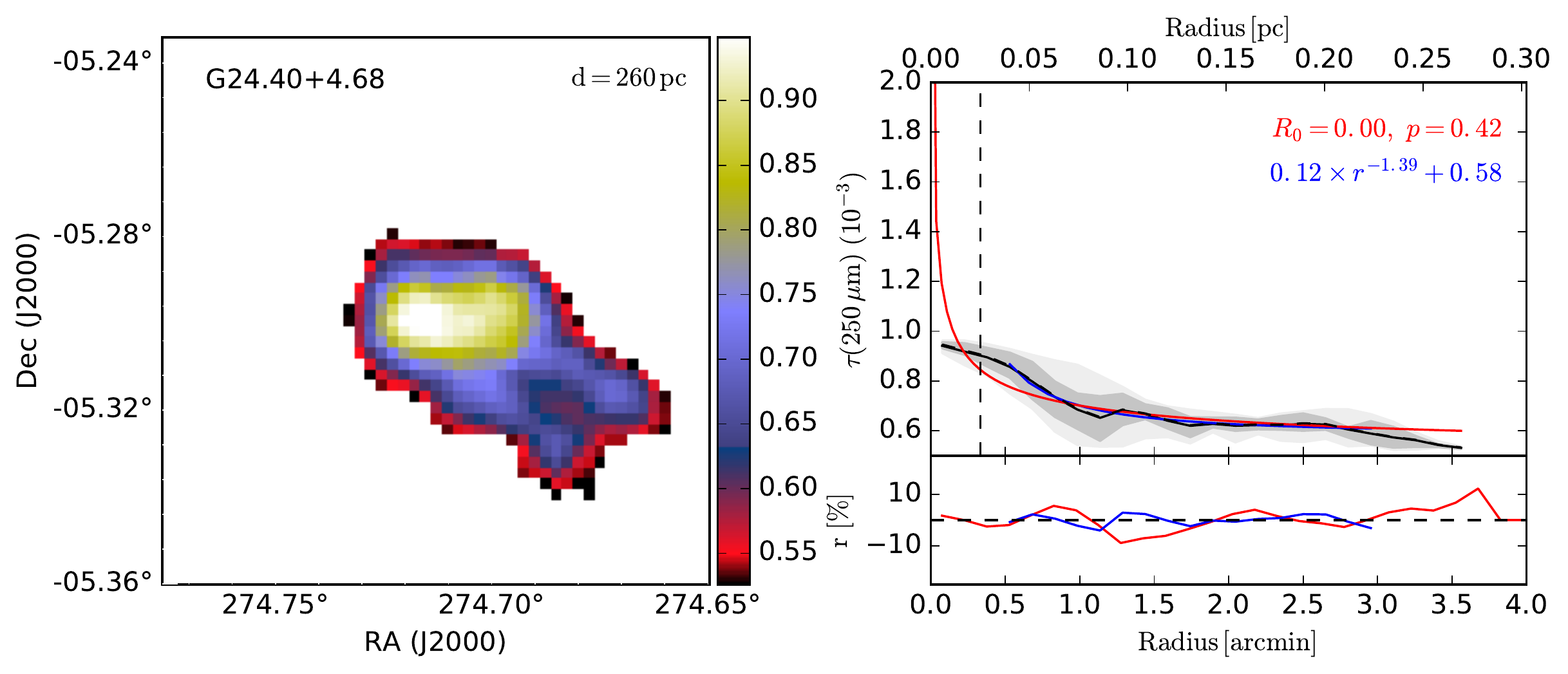}
\caption{continued.}
\end{figure}

\begin{figure}
\includegraphics[width=8.2cm]{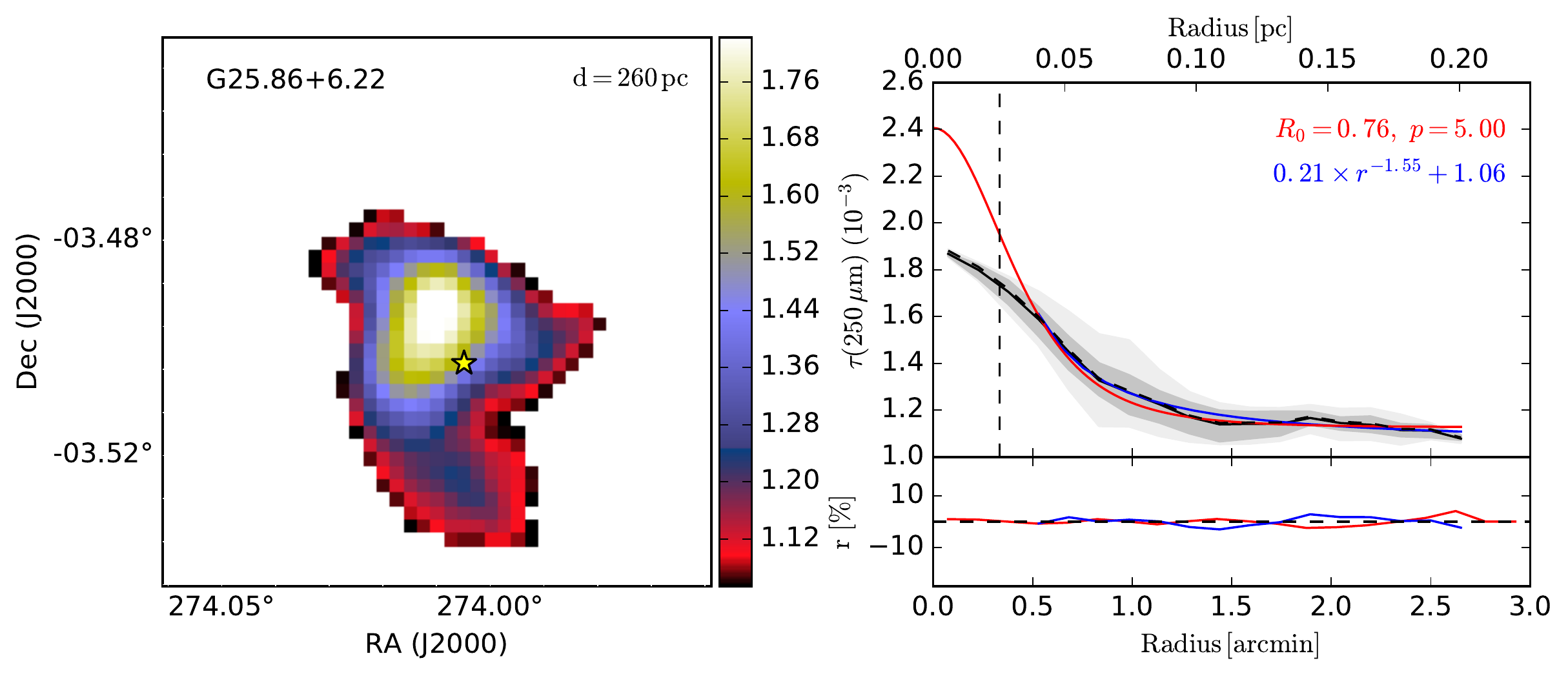}
\includegraphics[width=8.2cm]{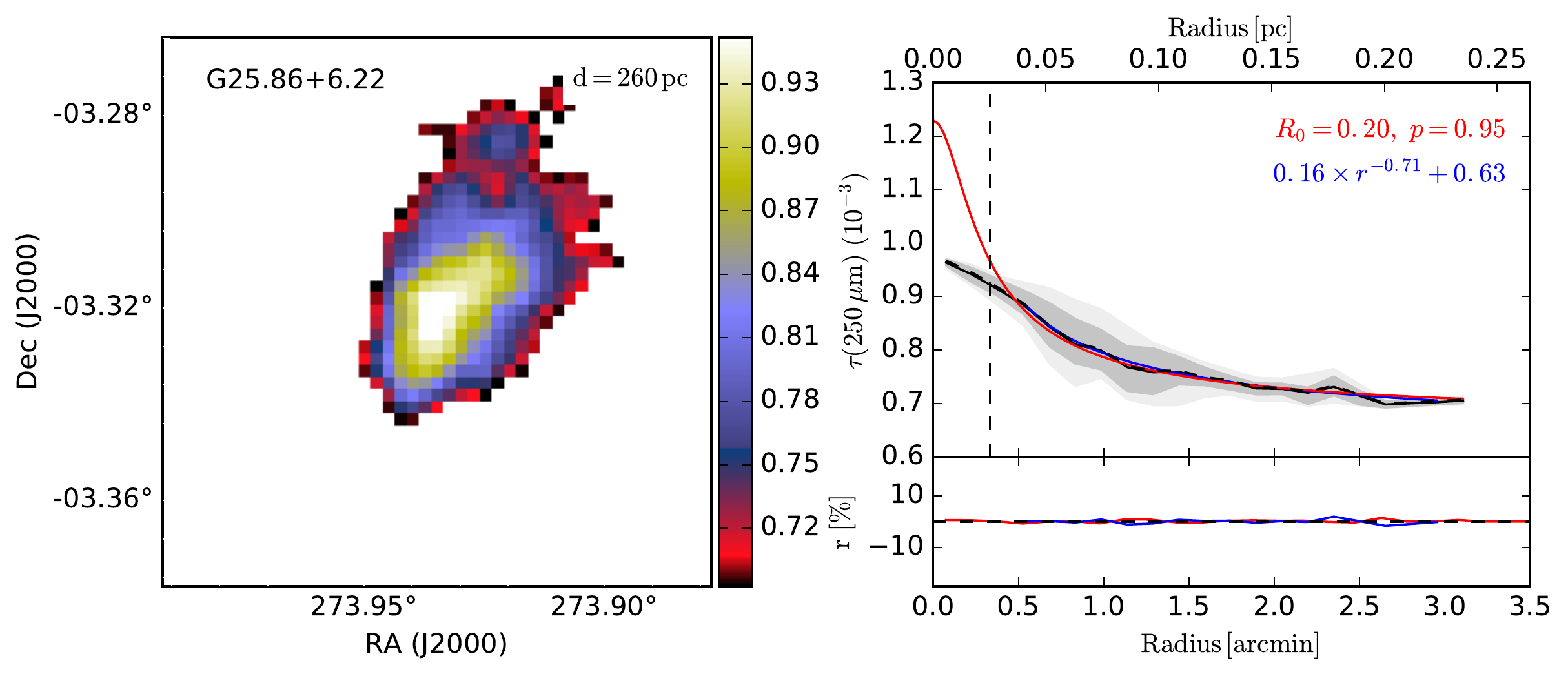}
\includegraphics[width=8.2cm]{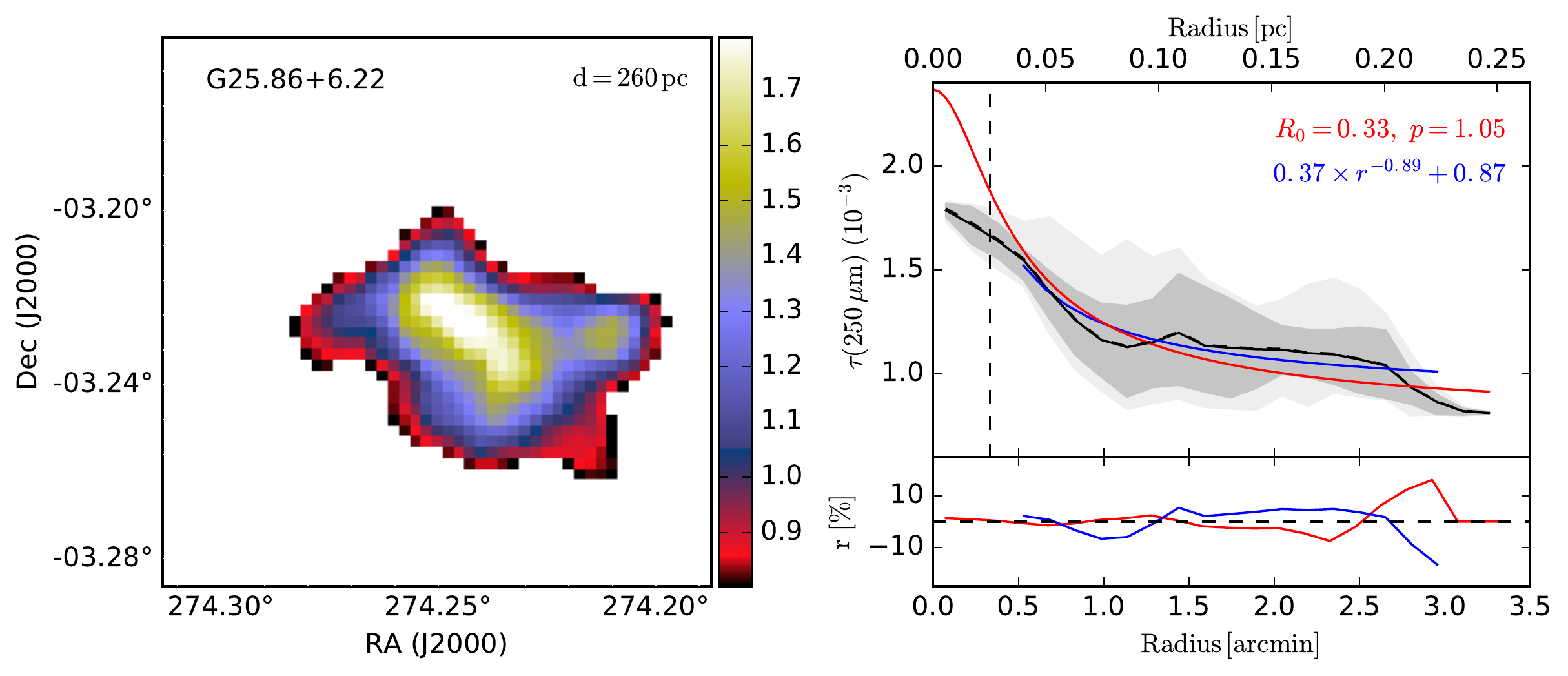}
\includegraphics[width=8.2cm]{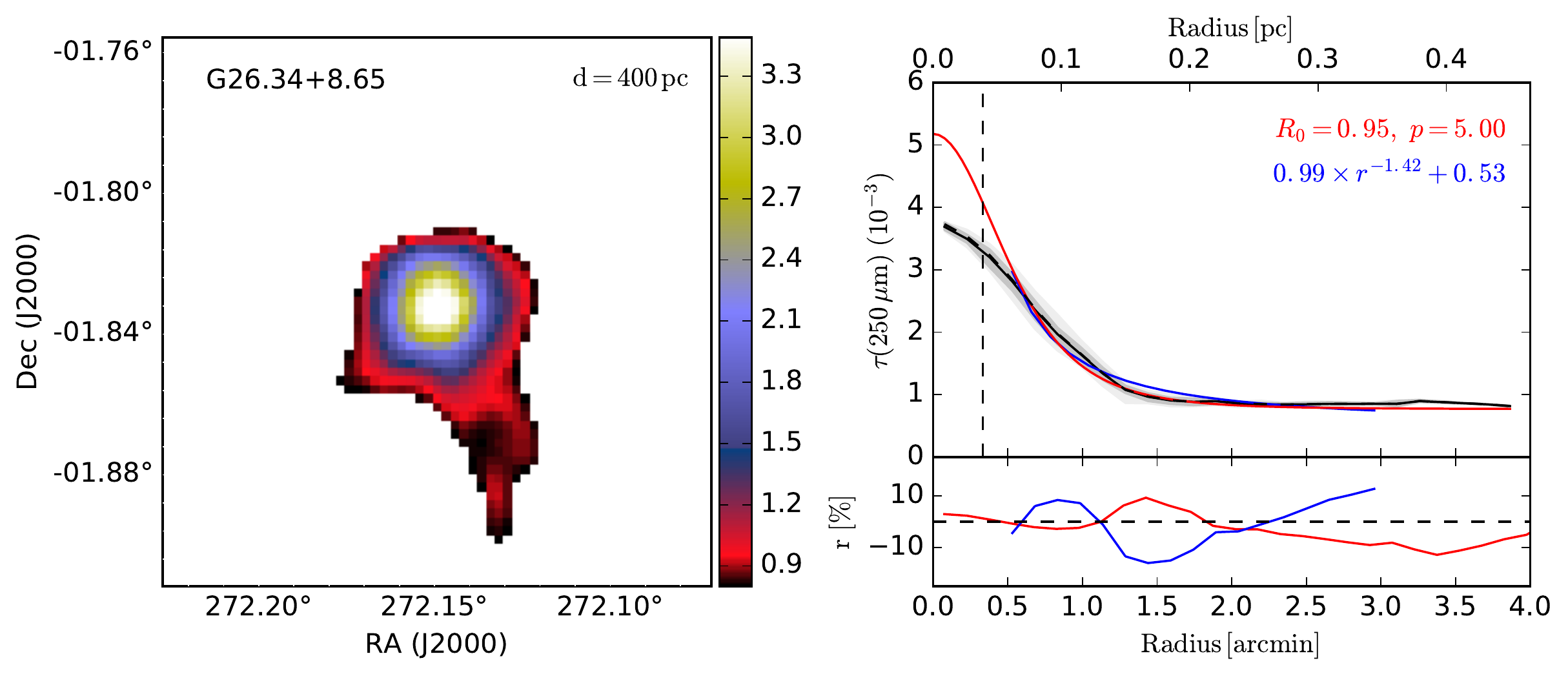}
\includegraphics[width=8.2cm]{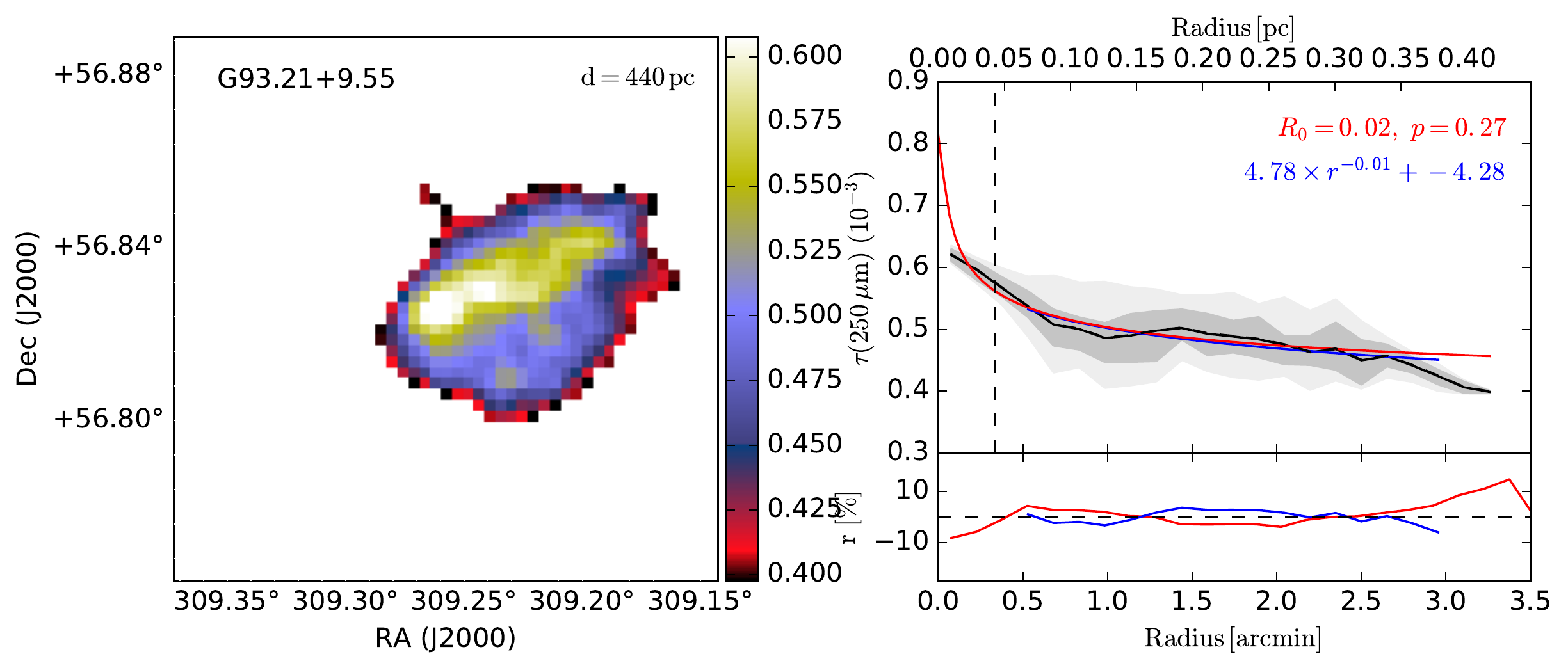}
\includegraphics[width=8.2cm]{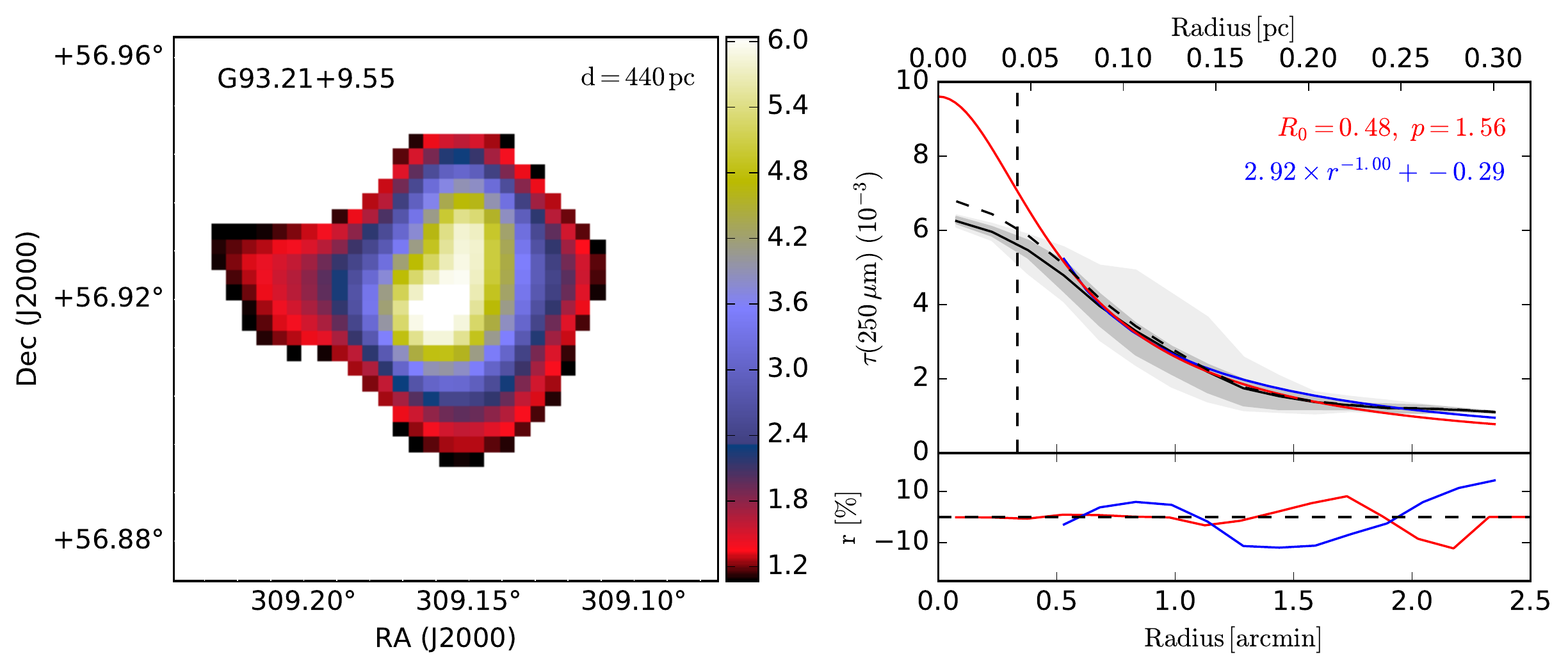}
\caption{continued.}
\end{figure}

\begin{figure}
\includegraphics[width=8.2cm]{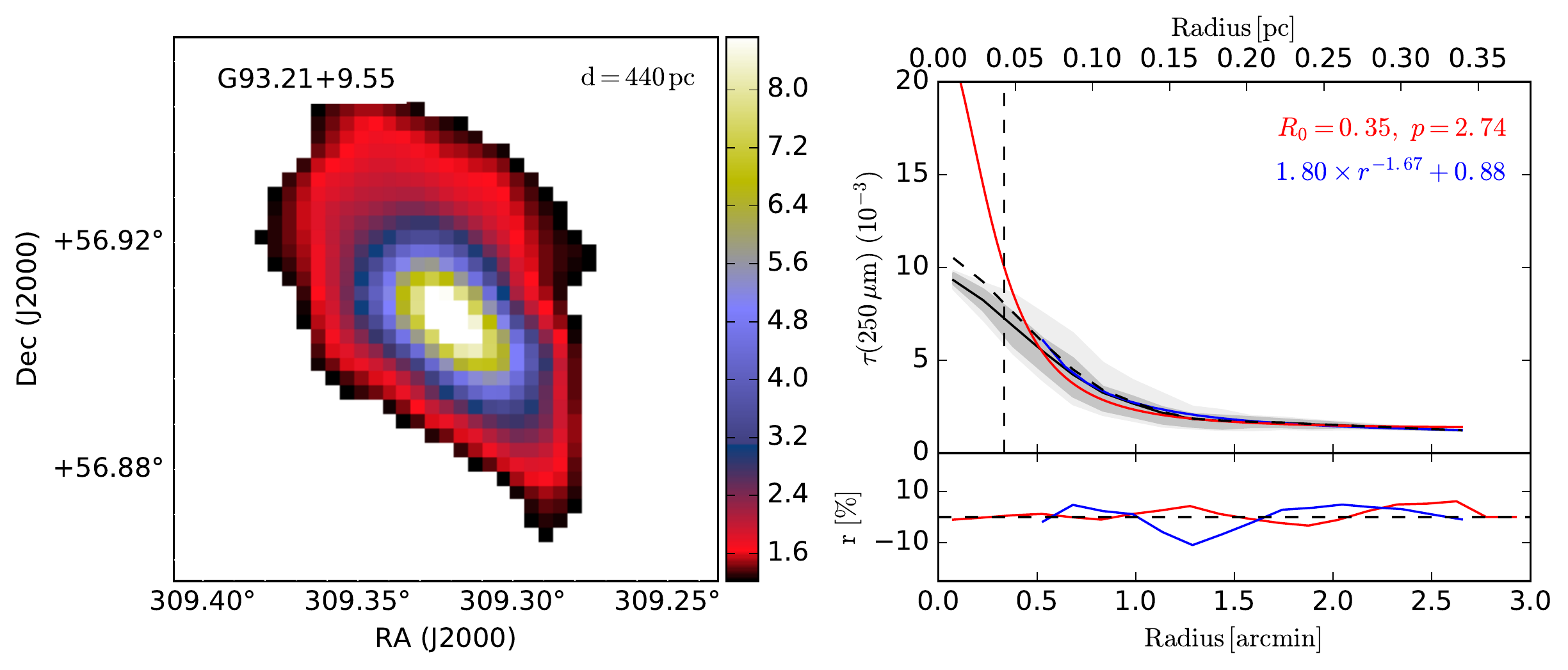}
\includegraphics[width=8.2cm]{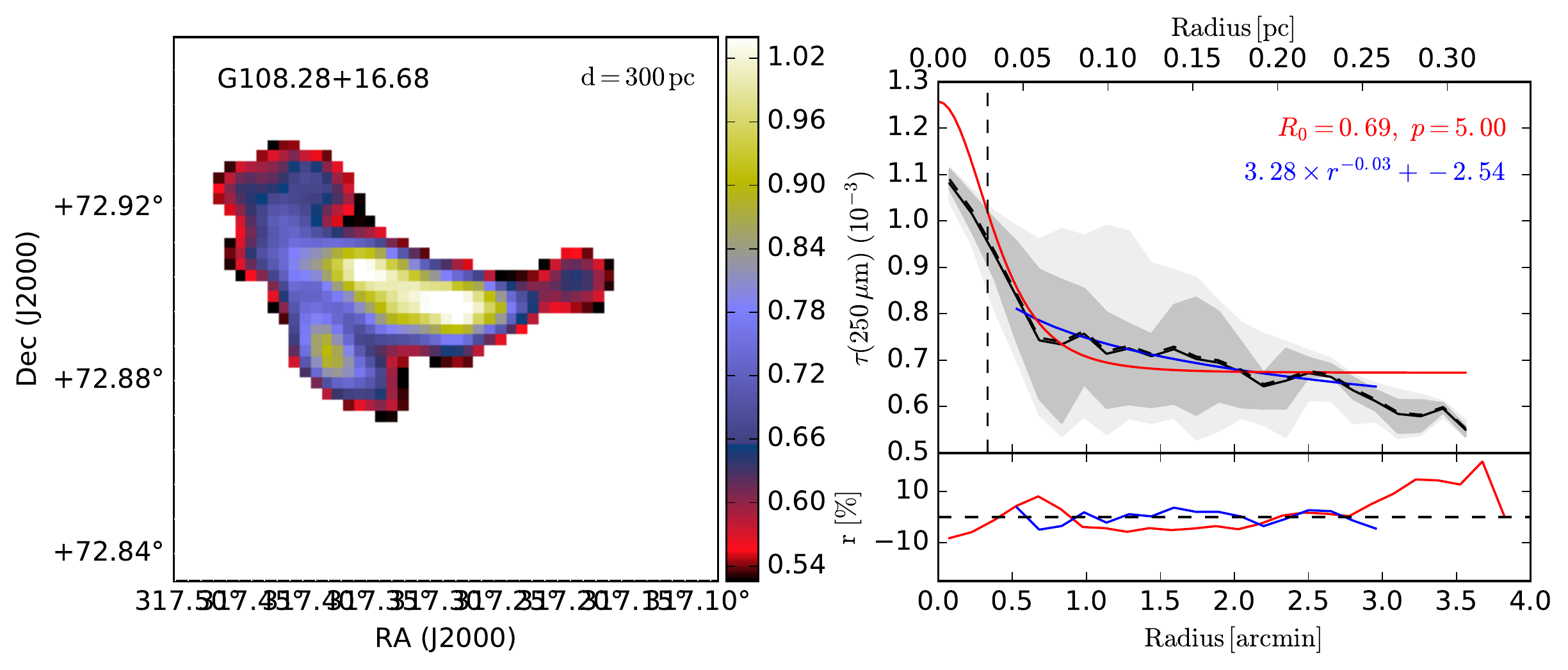}
\includegraphics[width=8.2cm]{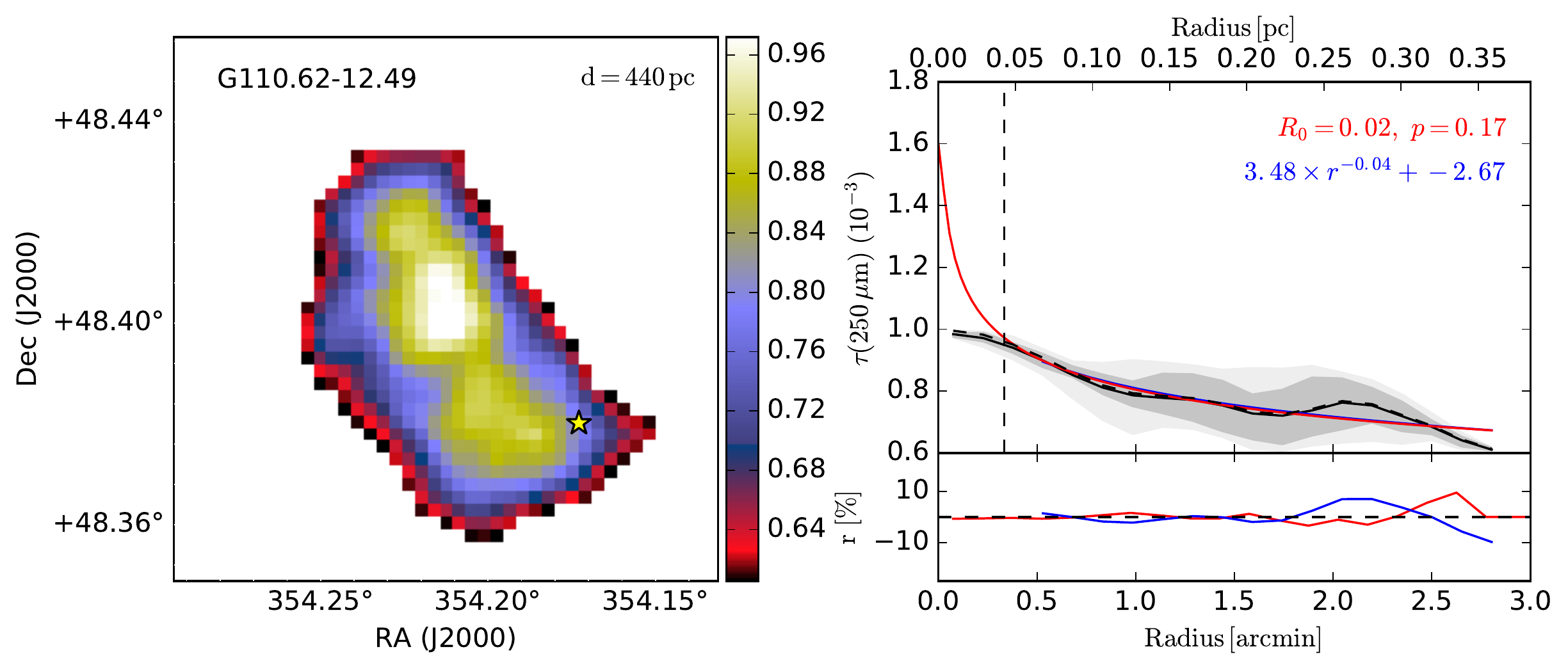}
\includegraphics[width=8.2cm]{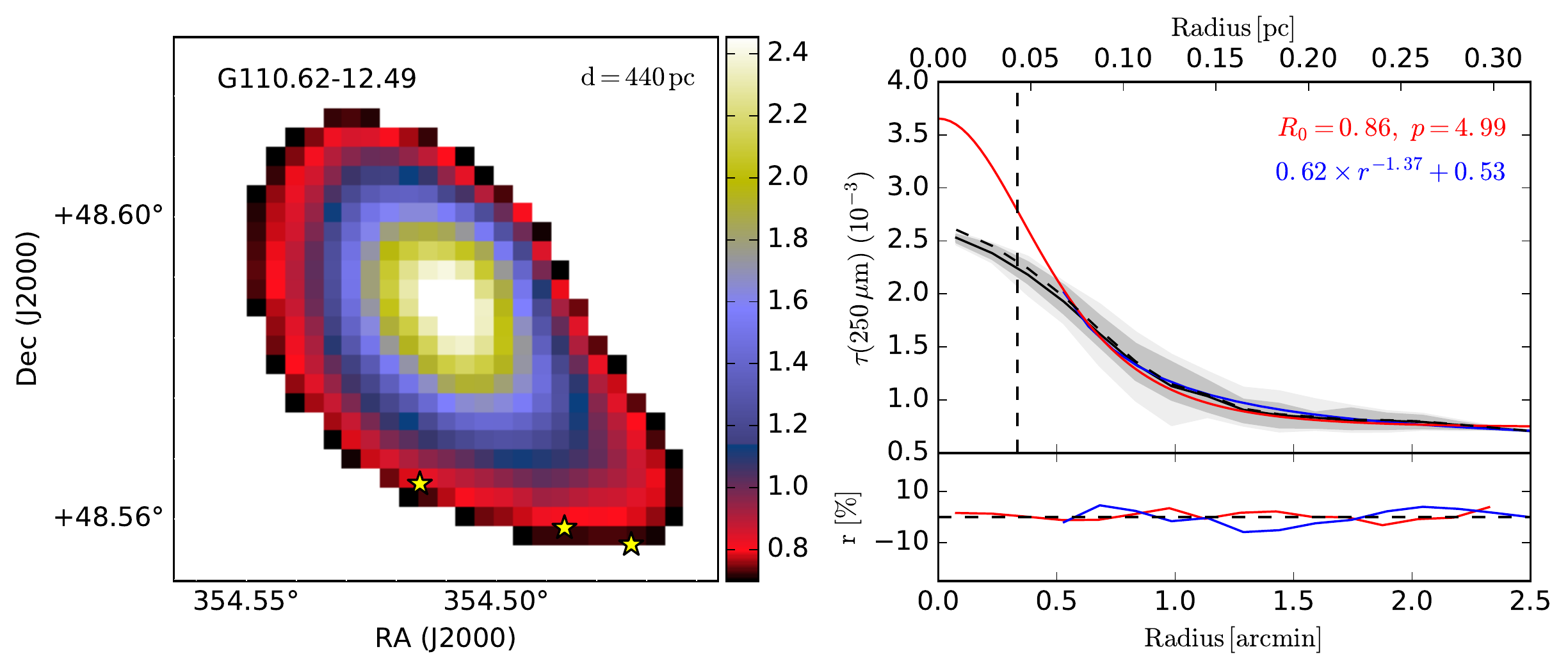}
\includegraphics[width=8.2cm]{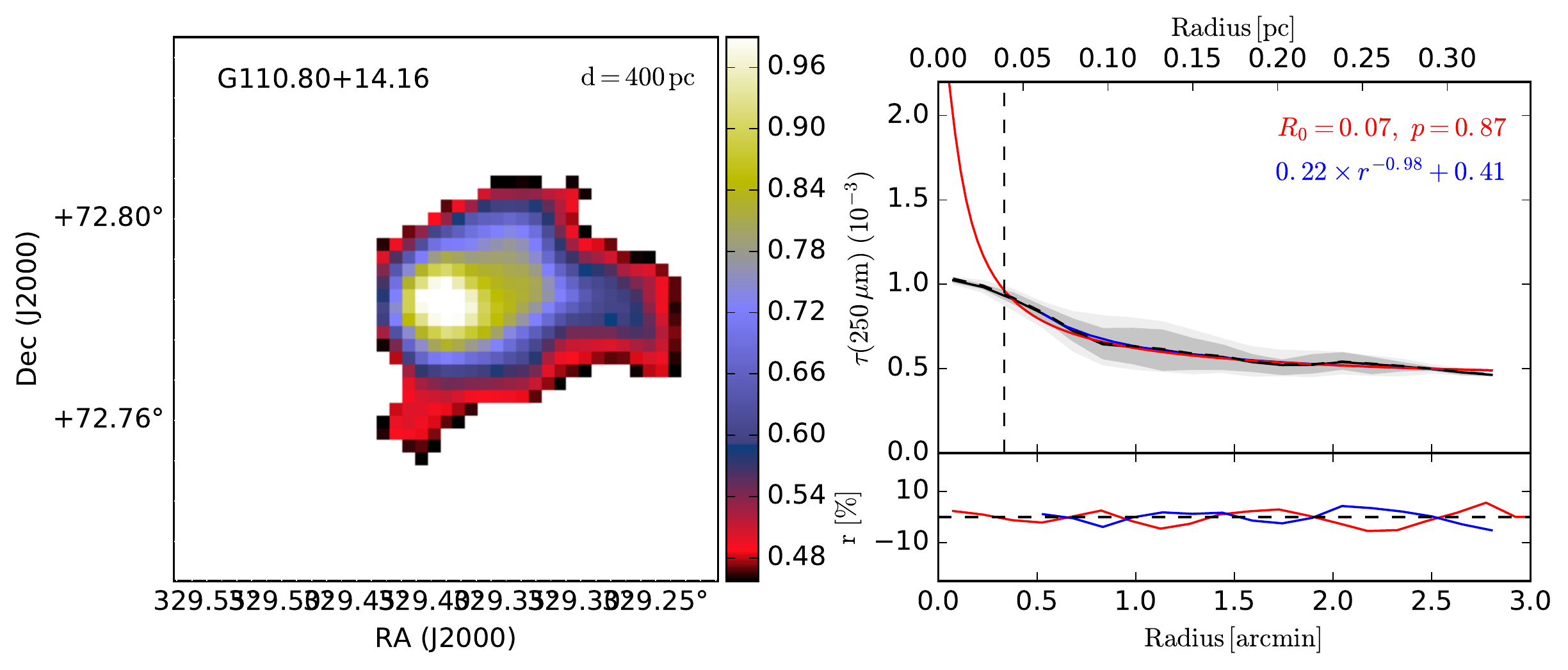}
\includegraphics[width=8.2cm]{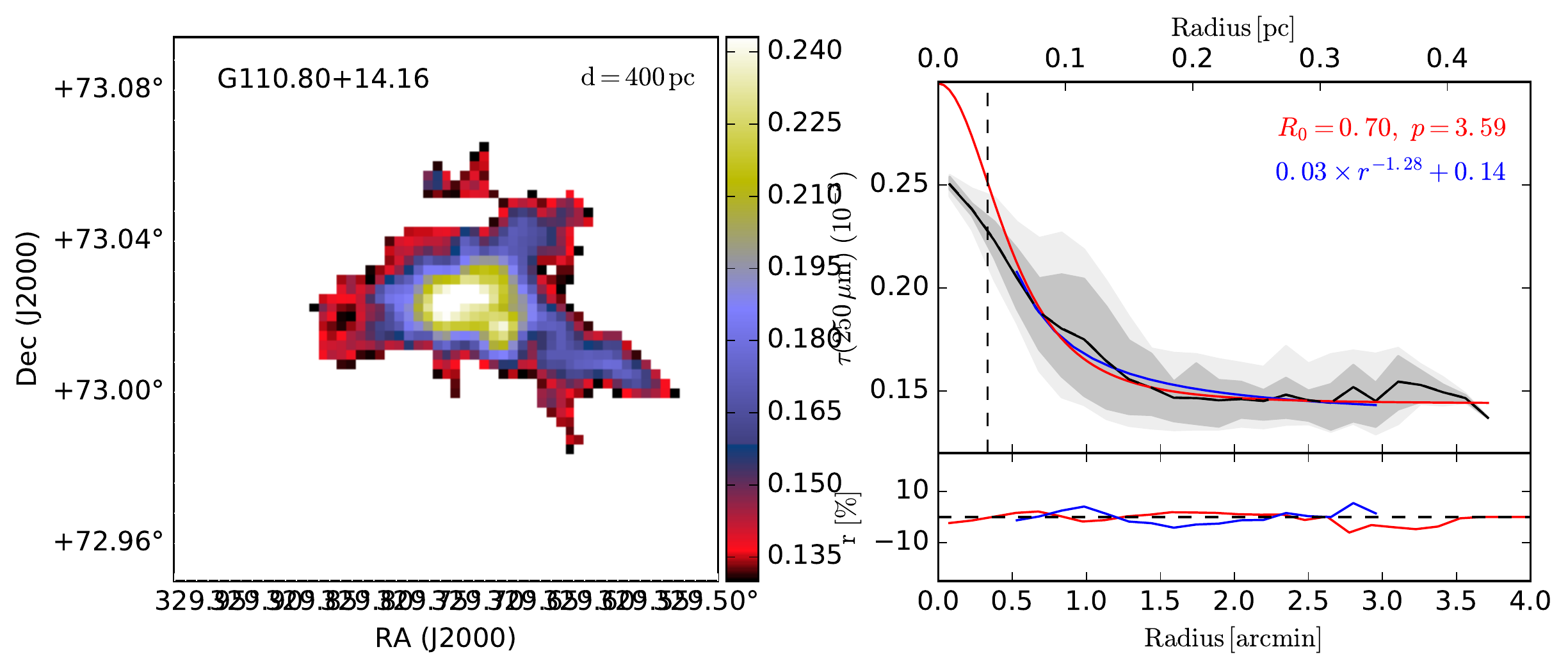}
\caption{continued.}
\end{figure}

\begin{figure}
\includegraphics[width=8.2cm]{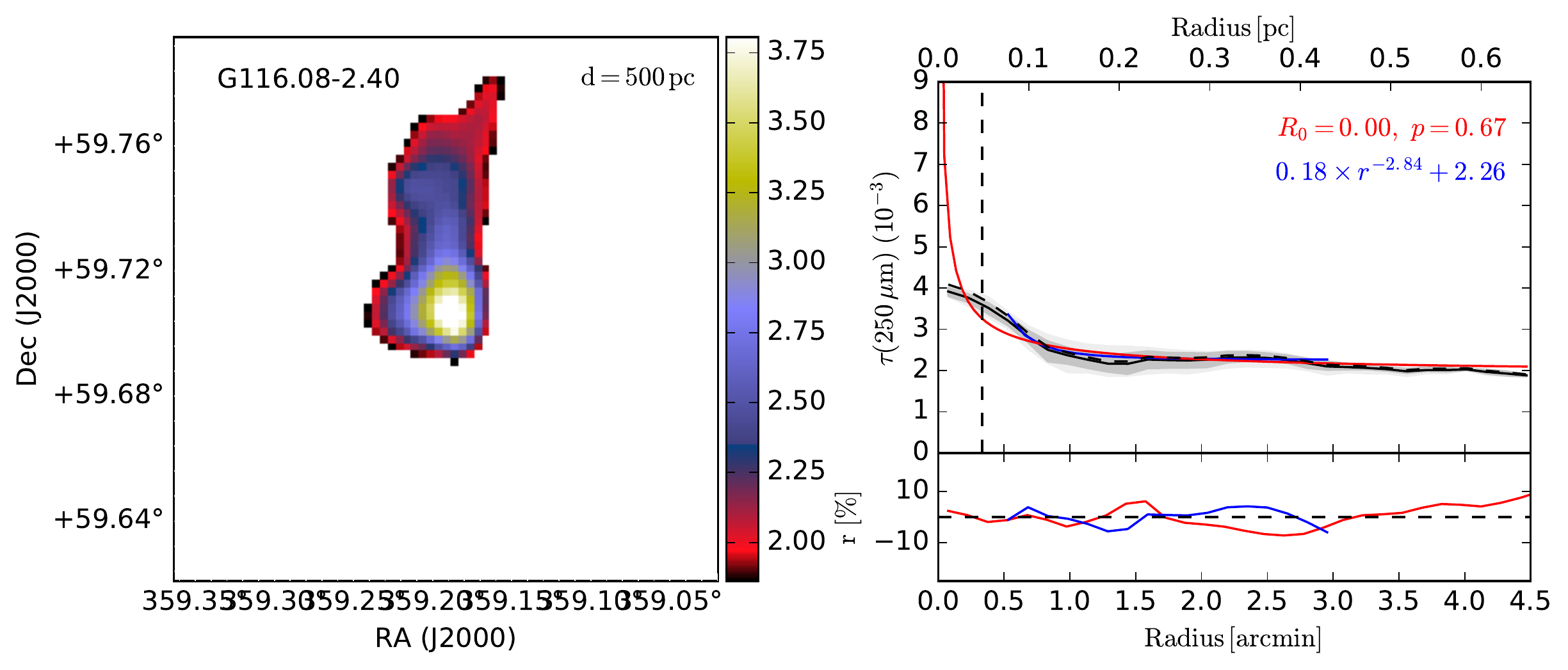}
\includegraphics[width=8.2cm]{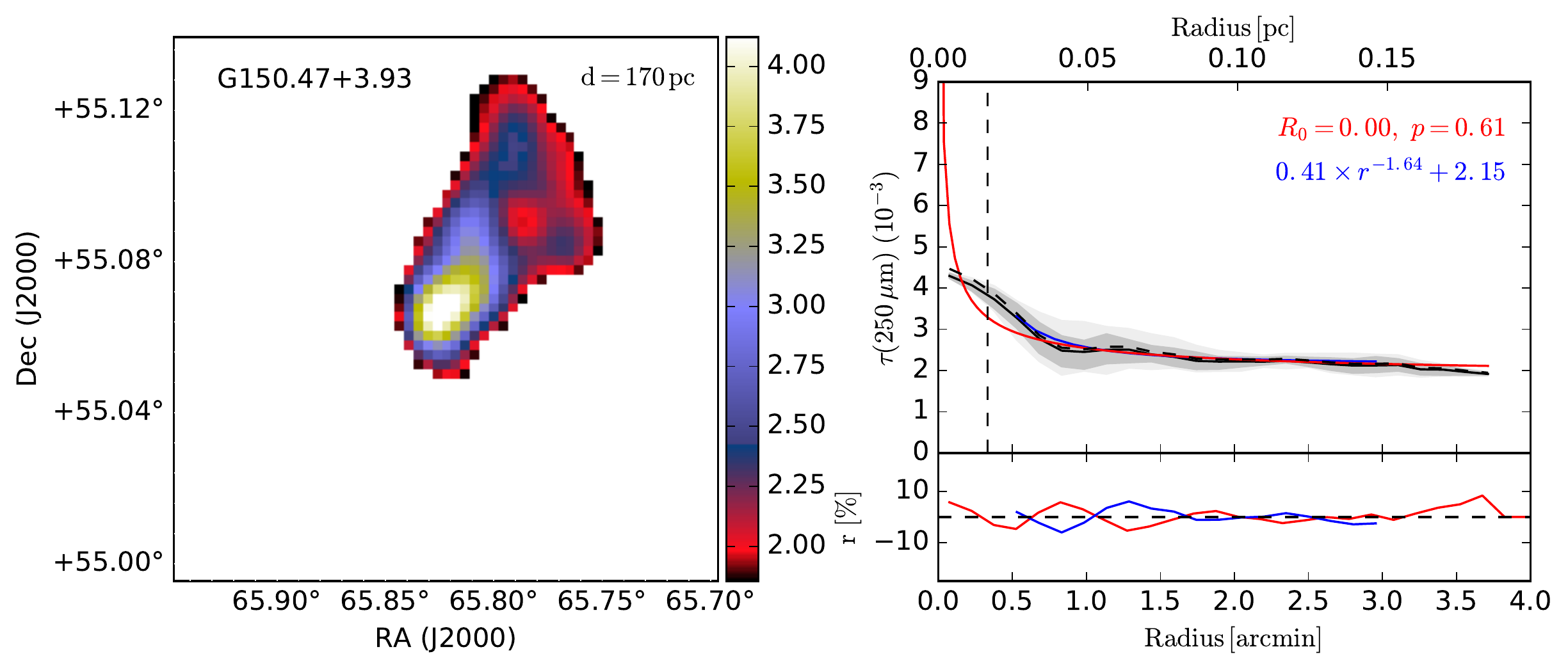}
\includegraphics[width=8.2cm]{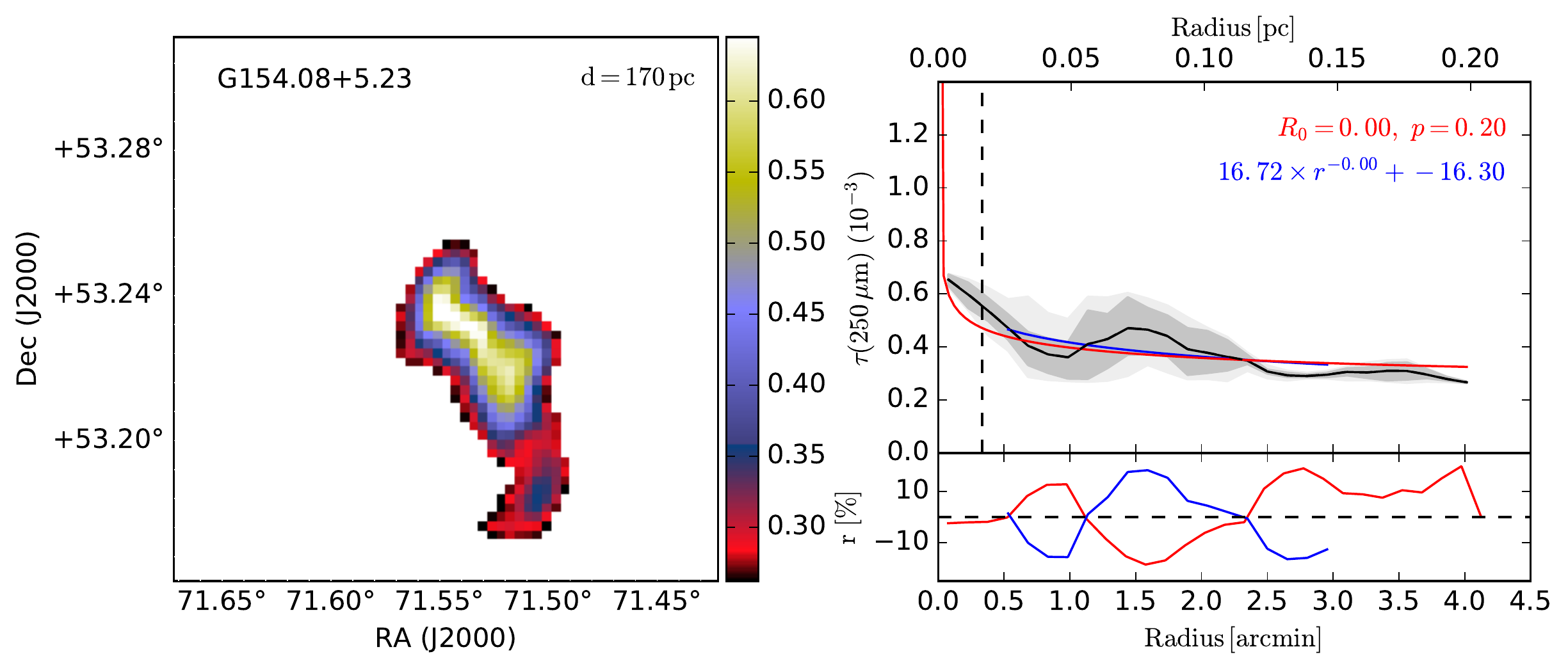}
\includegraphics[width=8.2cm]{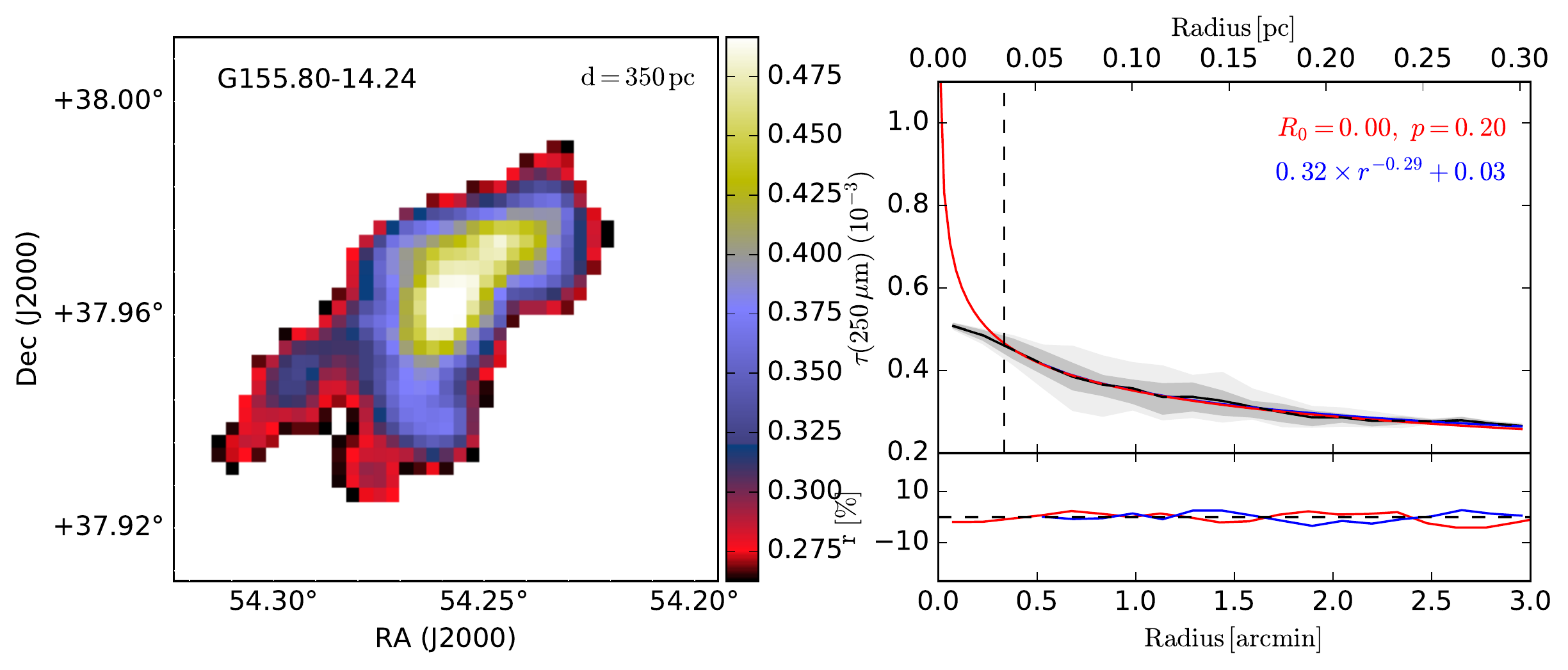}
\includegraphics[width=8.2cm]{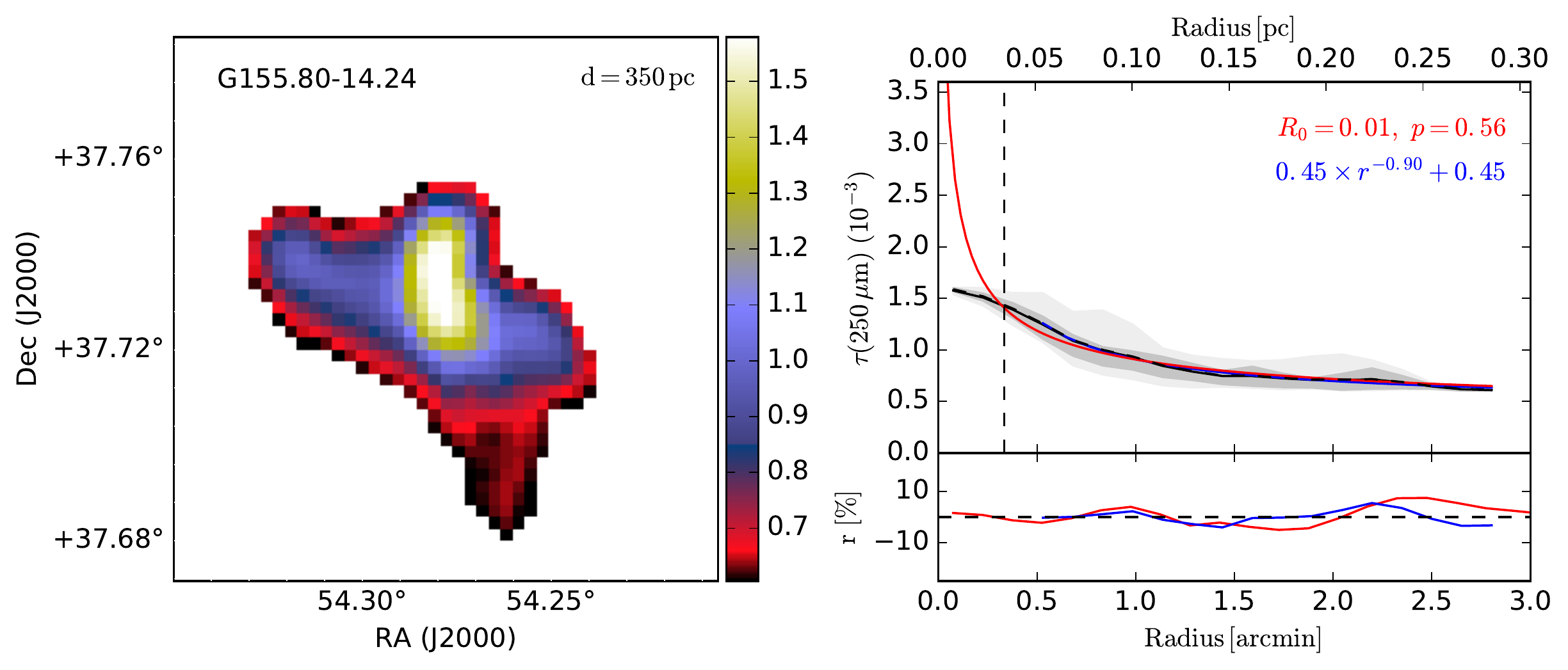}
\includegraphics[width=8.2cm]{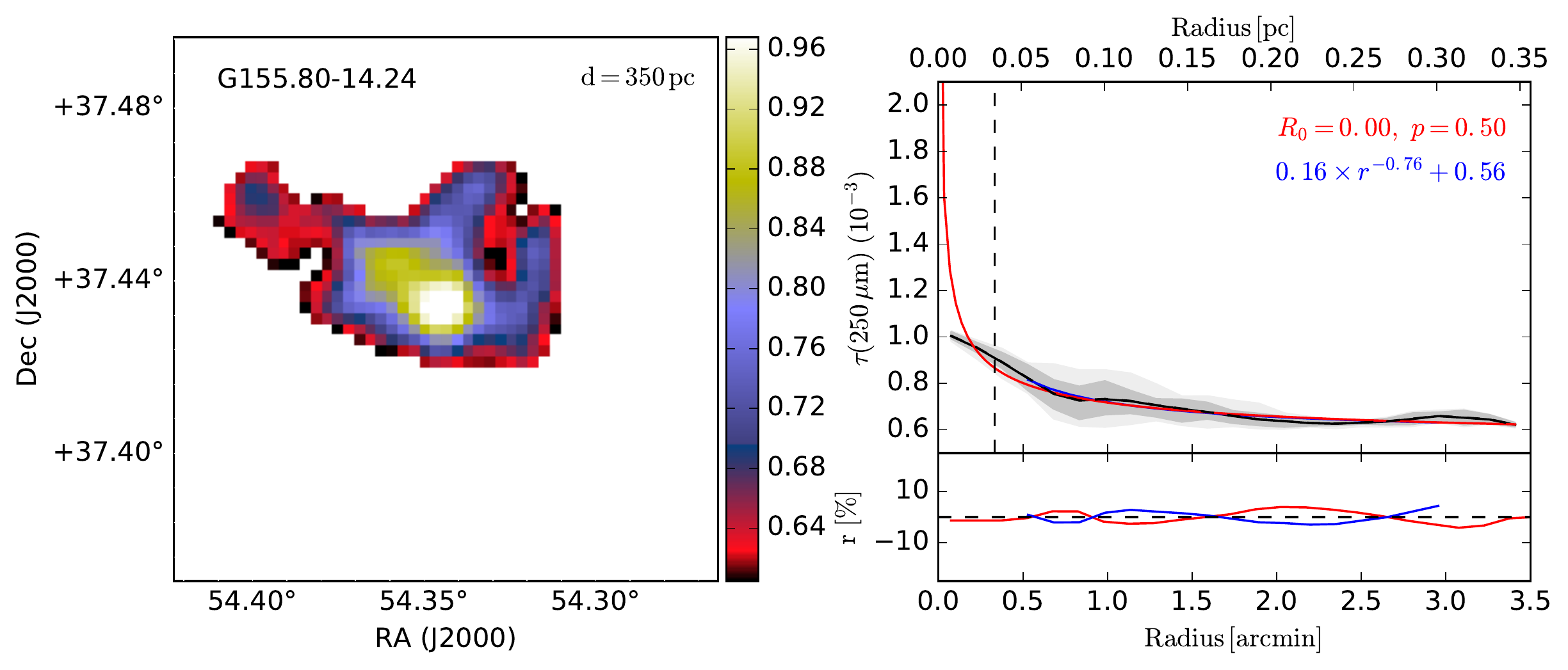}
\caption{continued.}
\end{figure}

\begin{figure}
\includegraphics[width=8.2cm]{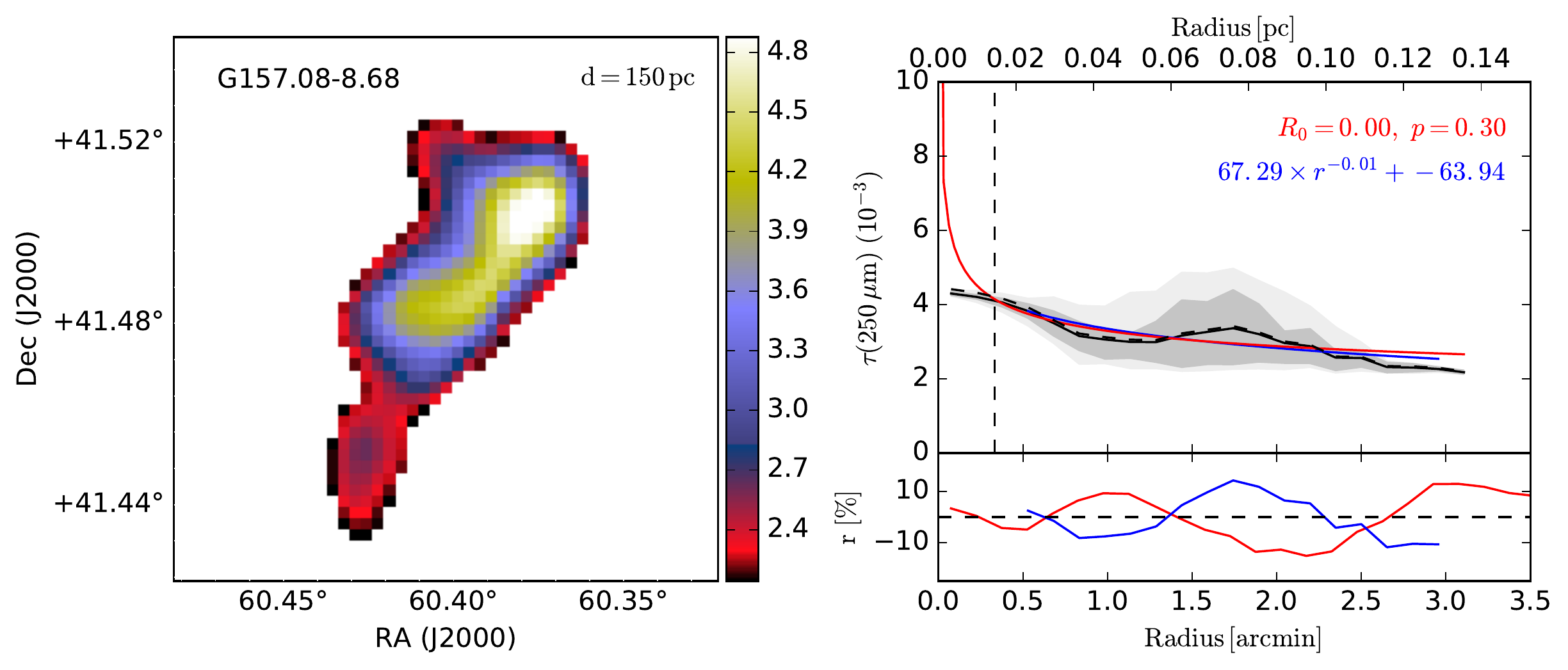}
\includegraphics[width=8.2cm]{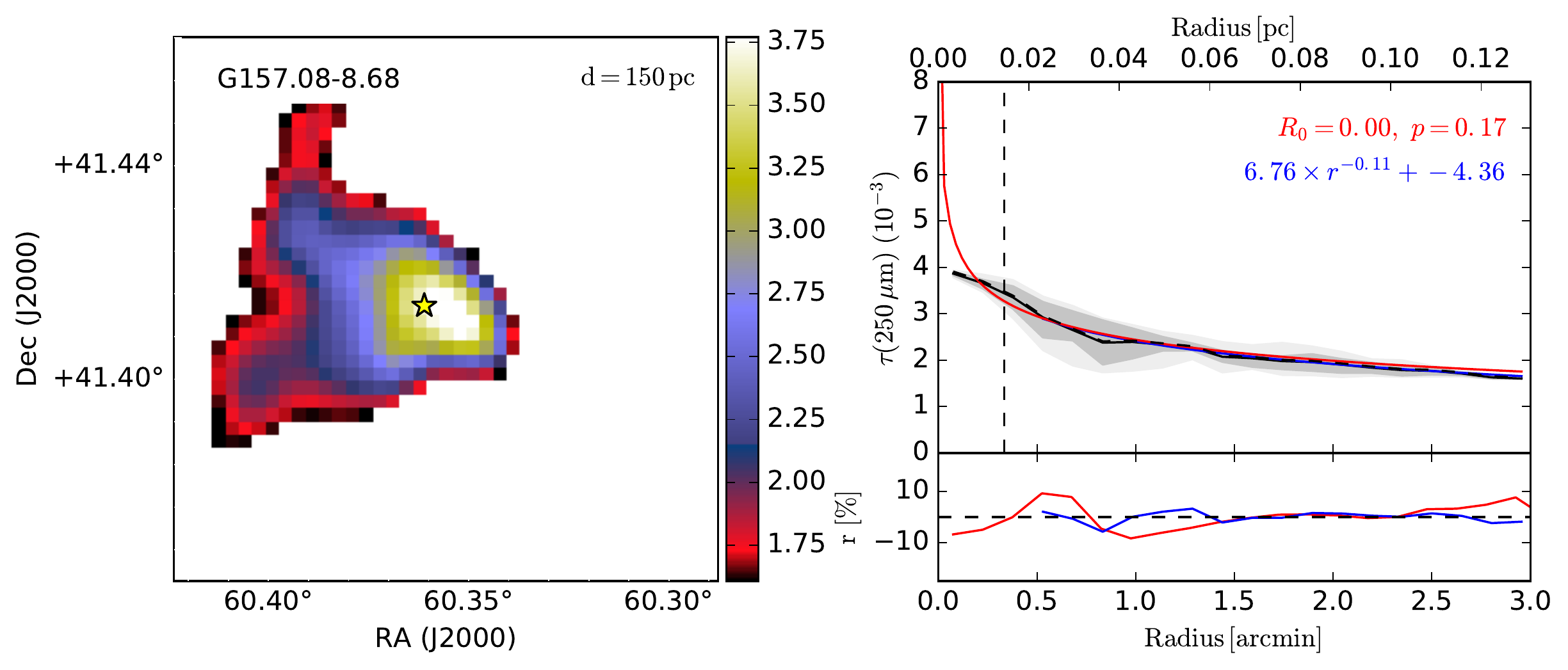}
\includegraphics[width=8.2cm]{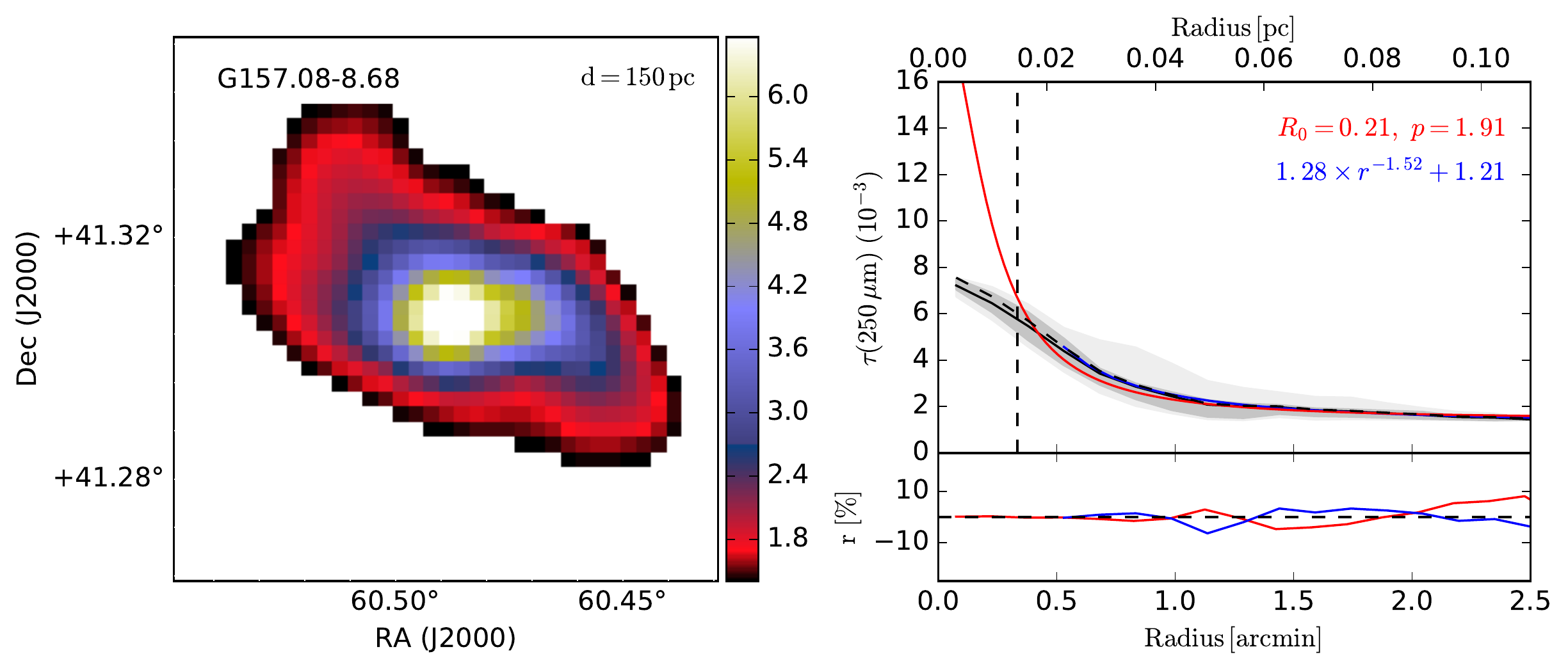}
\includegraphics[width=8.2cm]{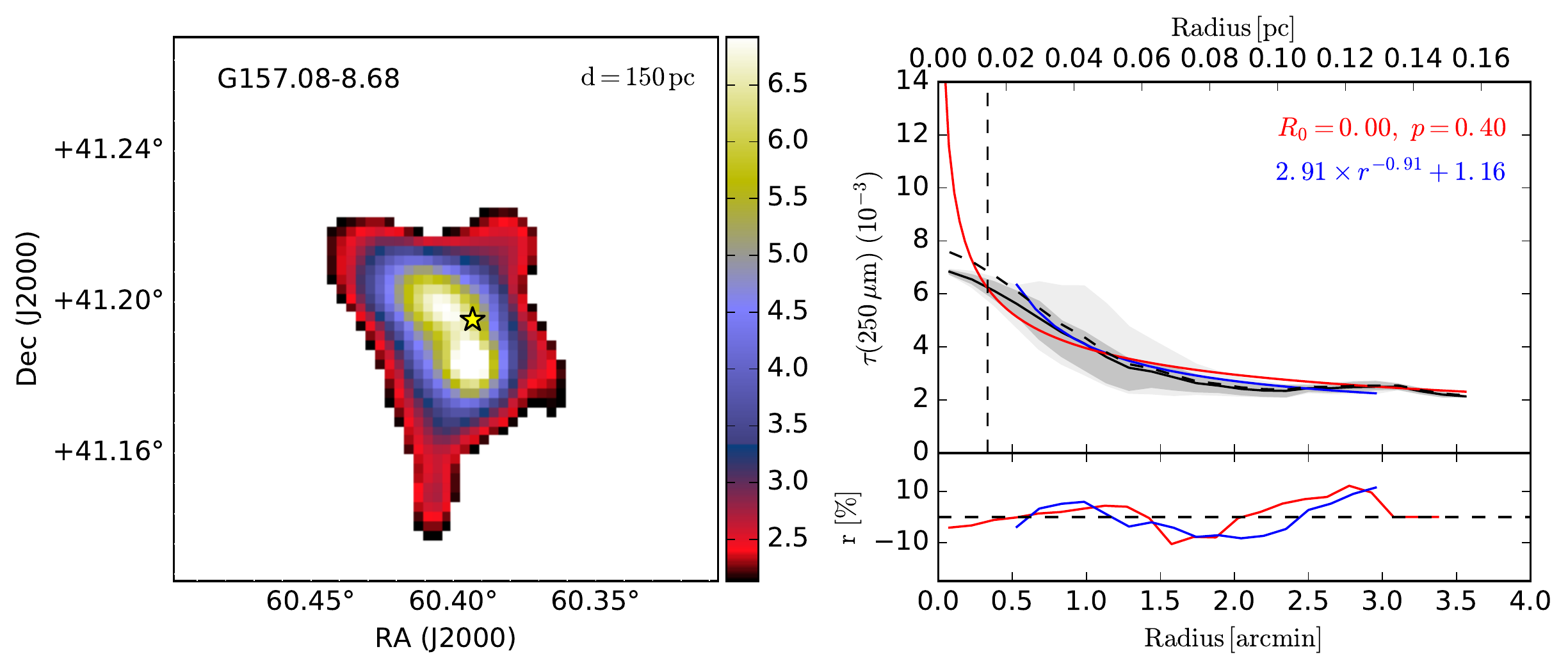}
\includegraphics[width=8.2cm]{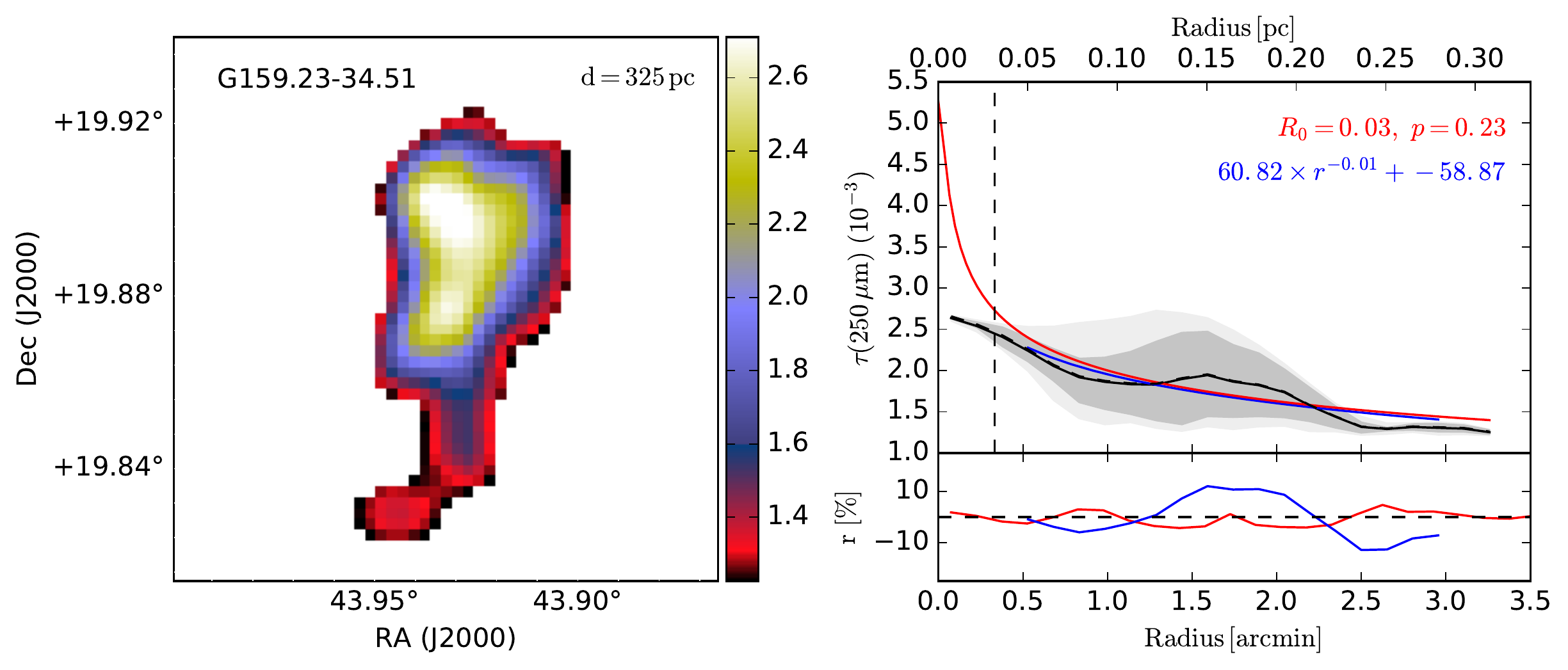}
\includegraphics[width=8.2cm]{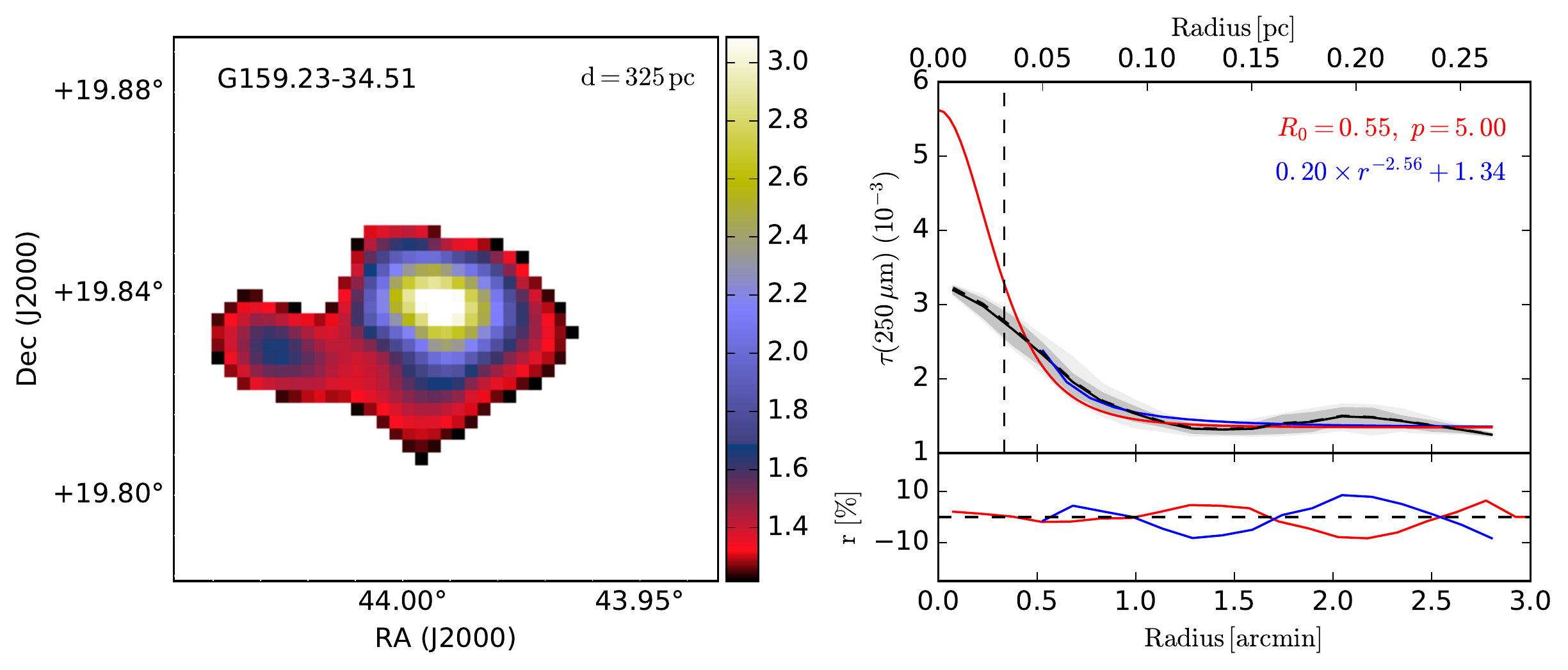}
\caption{continued.}
\end{figure}

\begin{figure}
\includegraphics[width=8.2cm]{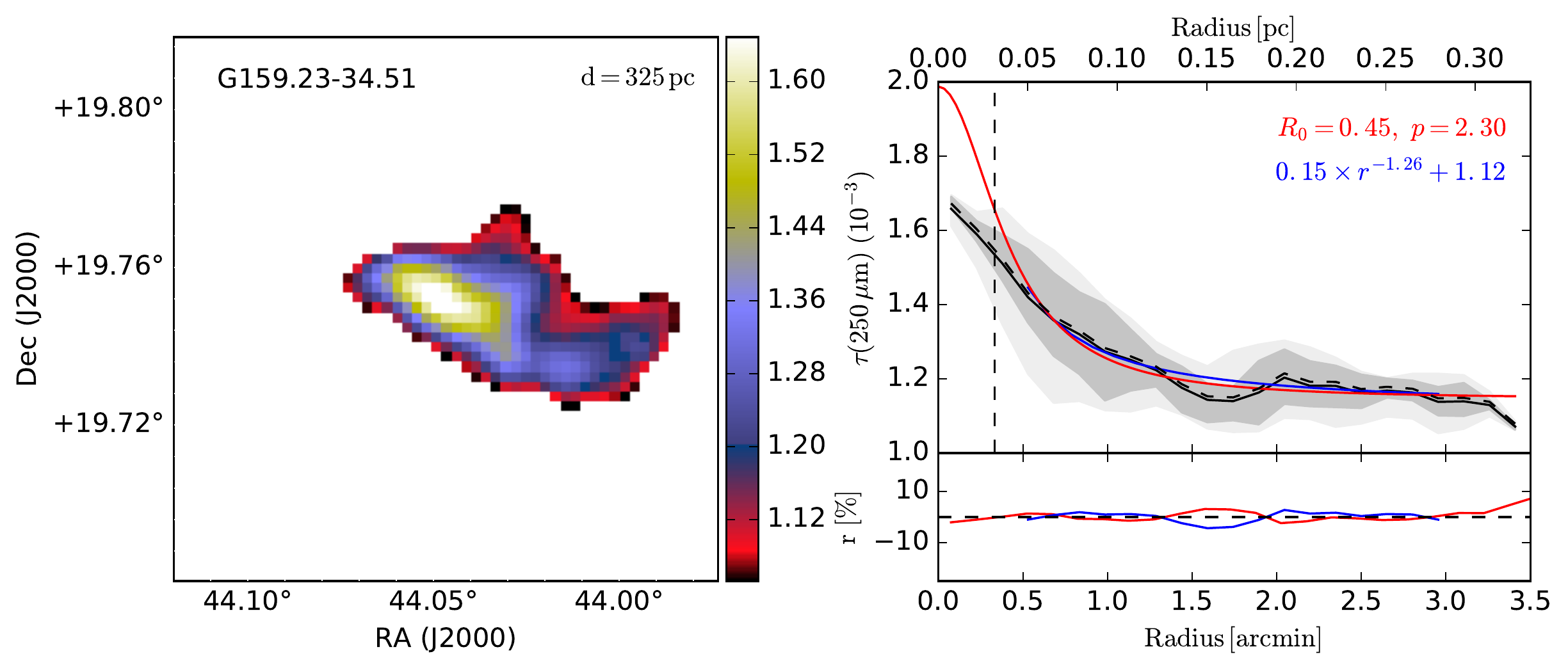}
\includegraphics[width=8.2cm]{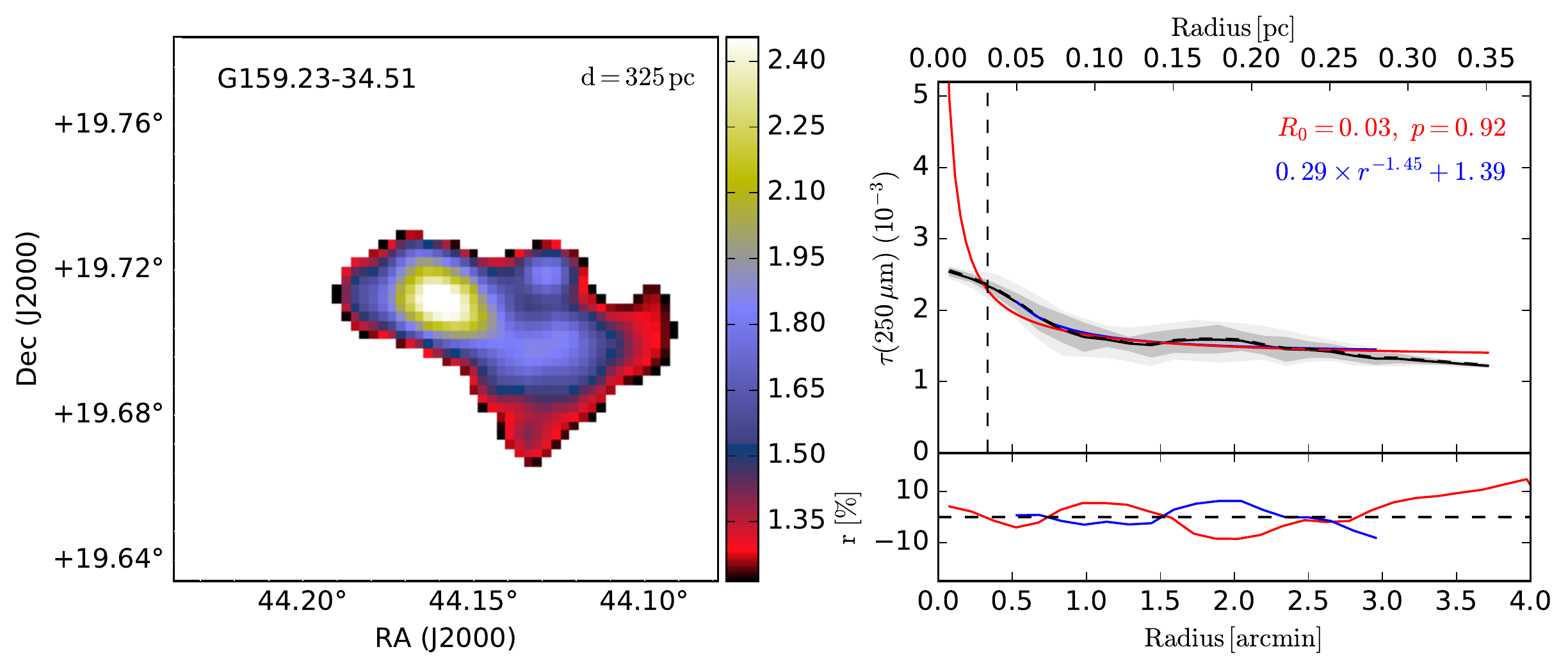}
\includegraphics[width=8.2cm]{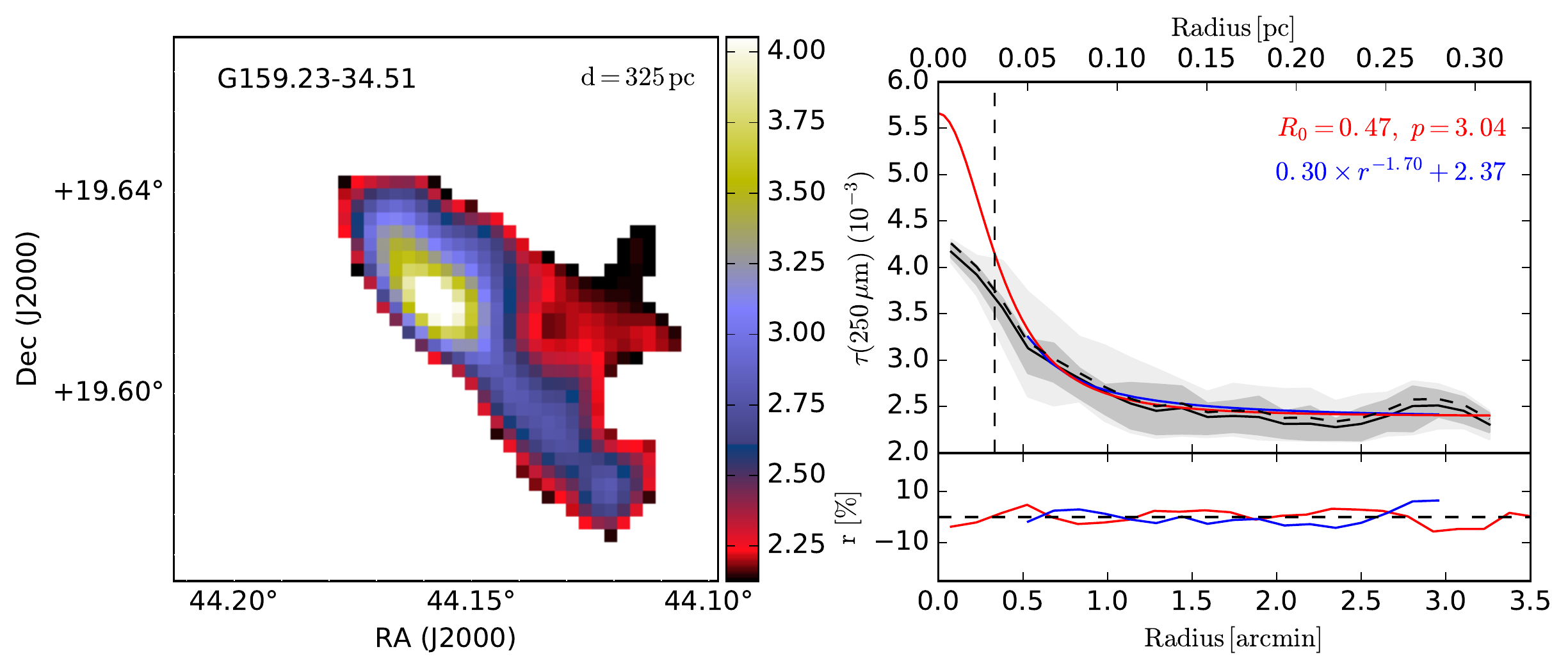}
\includegraphics[width=8.2cm]{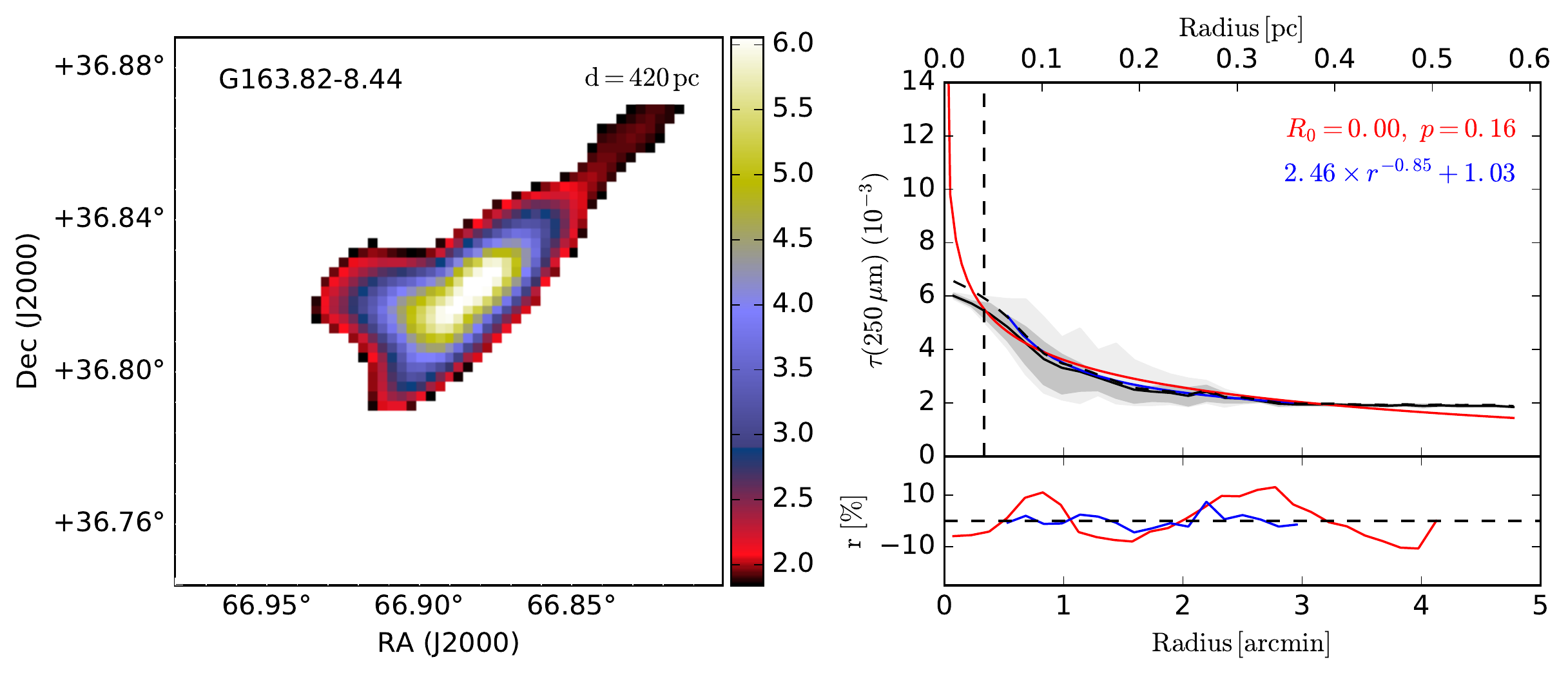}
\includegraphics[width=8.2cm]{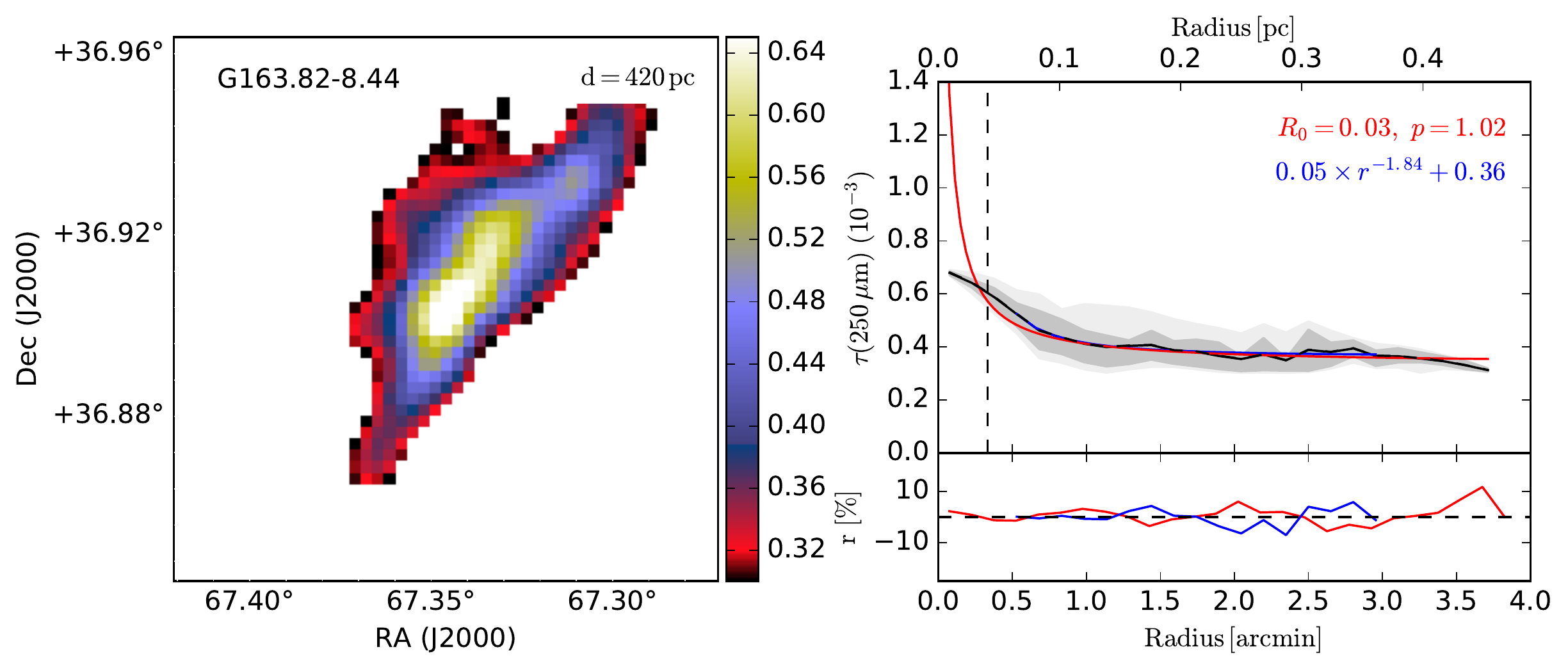}
\includegraphics[width=8.2cm]{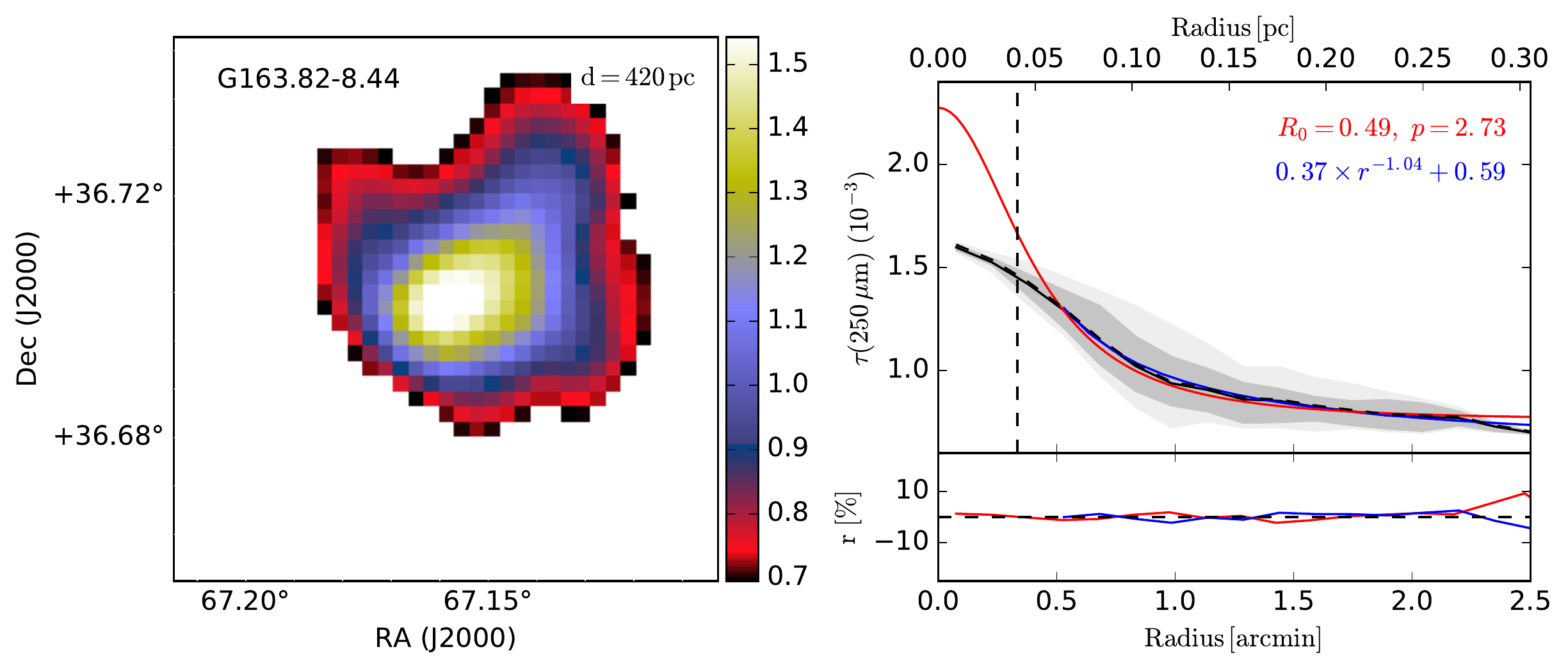}
\caption{continued.}
\end{figure}

\begin{figure}
\includegraphics[width=8.2cm]{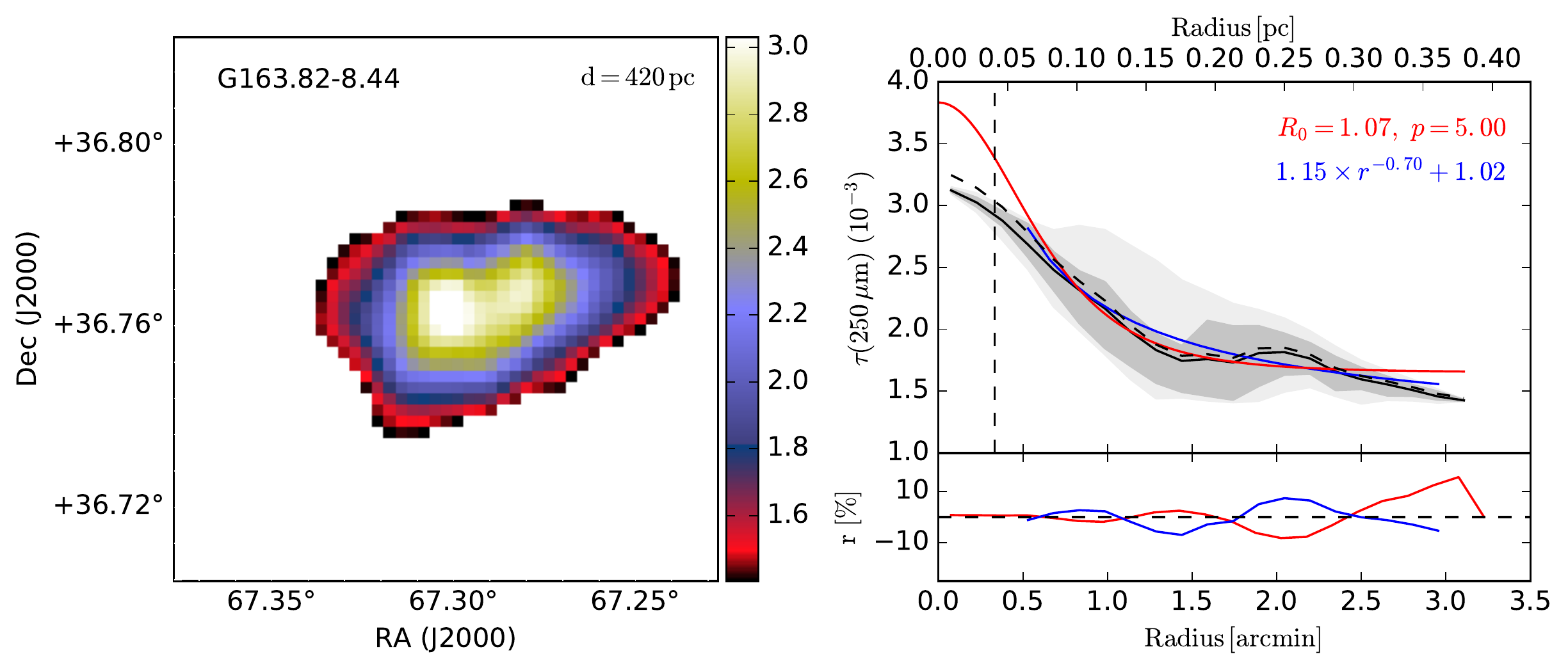}
\includegraphics[width=8.2cm]{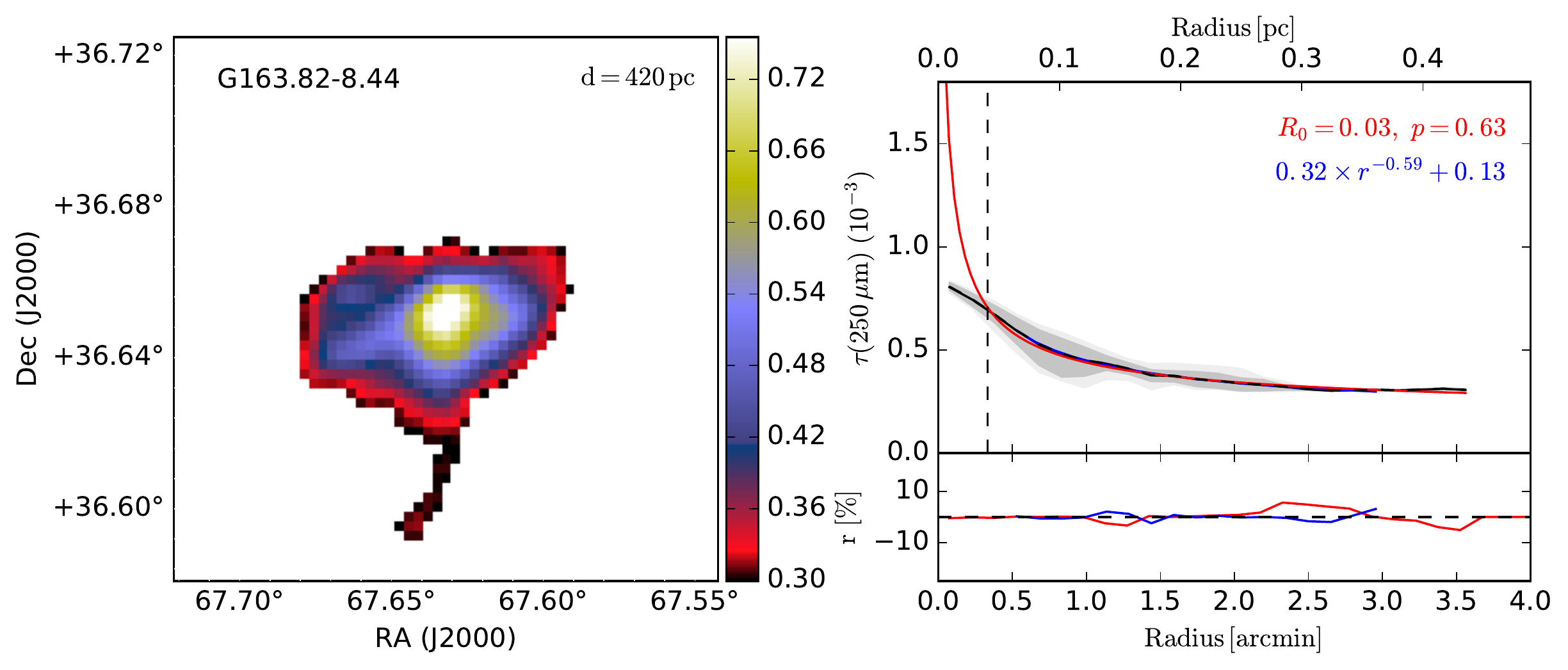}
\includegraphics[width=8.2cm]{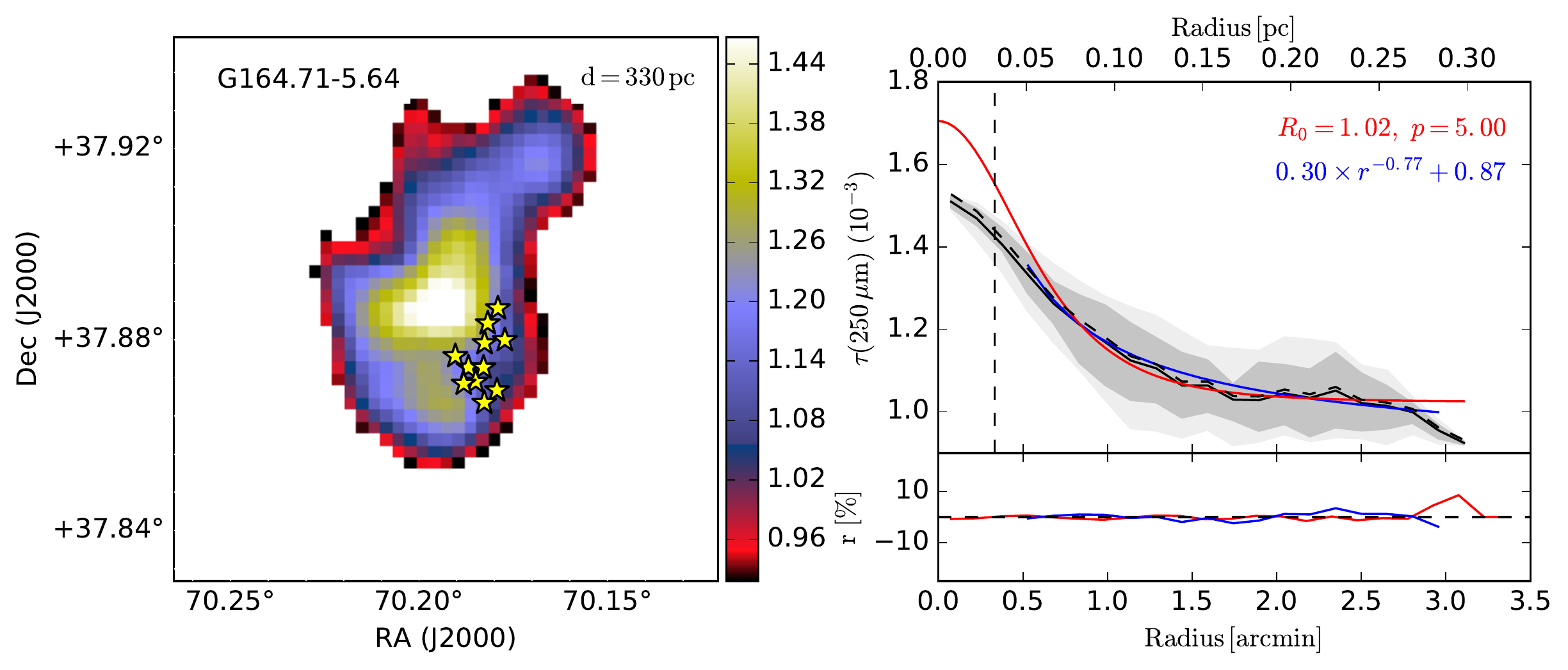}
\includegraphics[width=8.2cm]{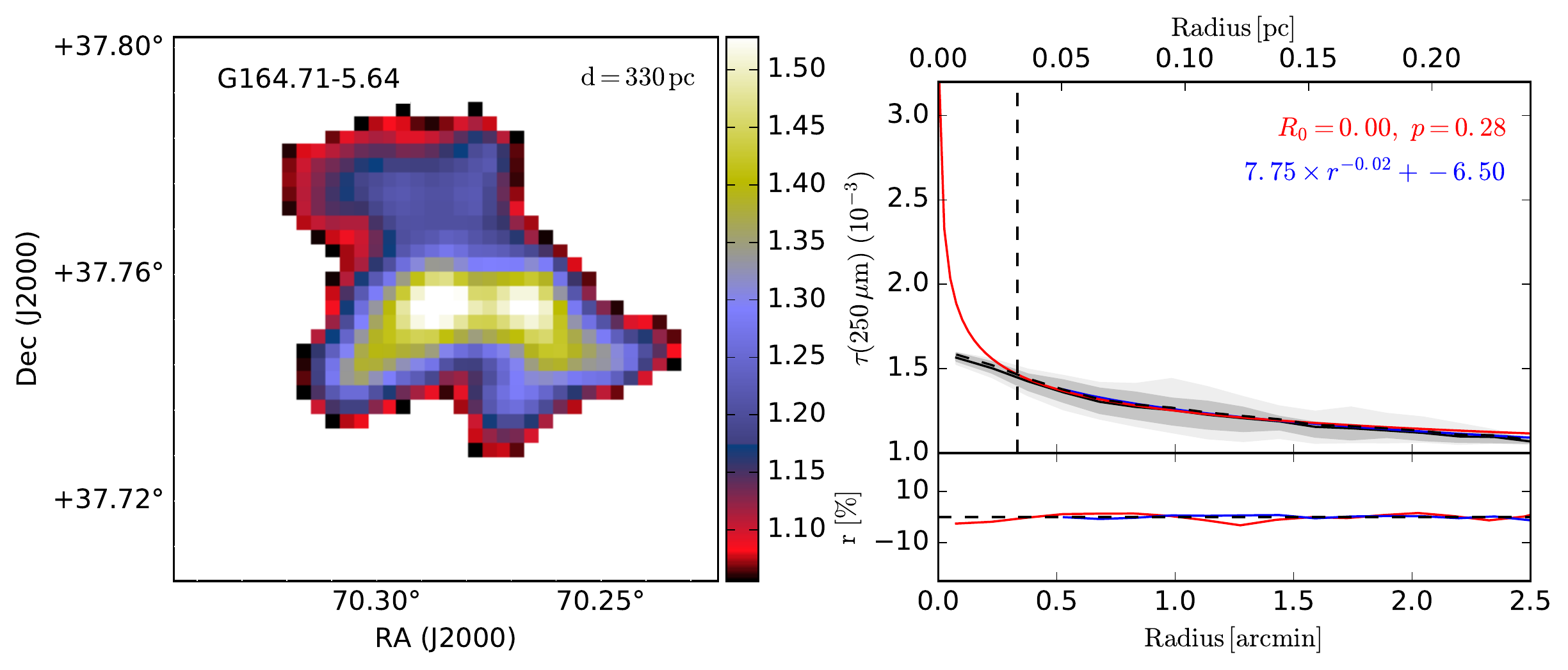}
\includegraphics[width=8.2cm]{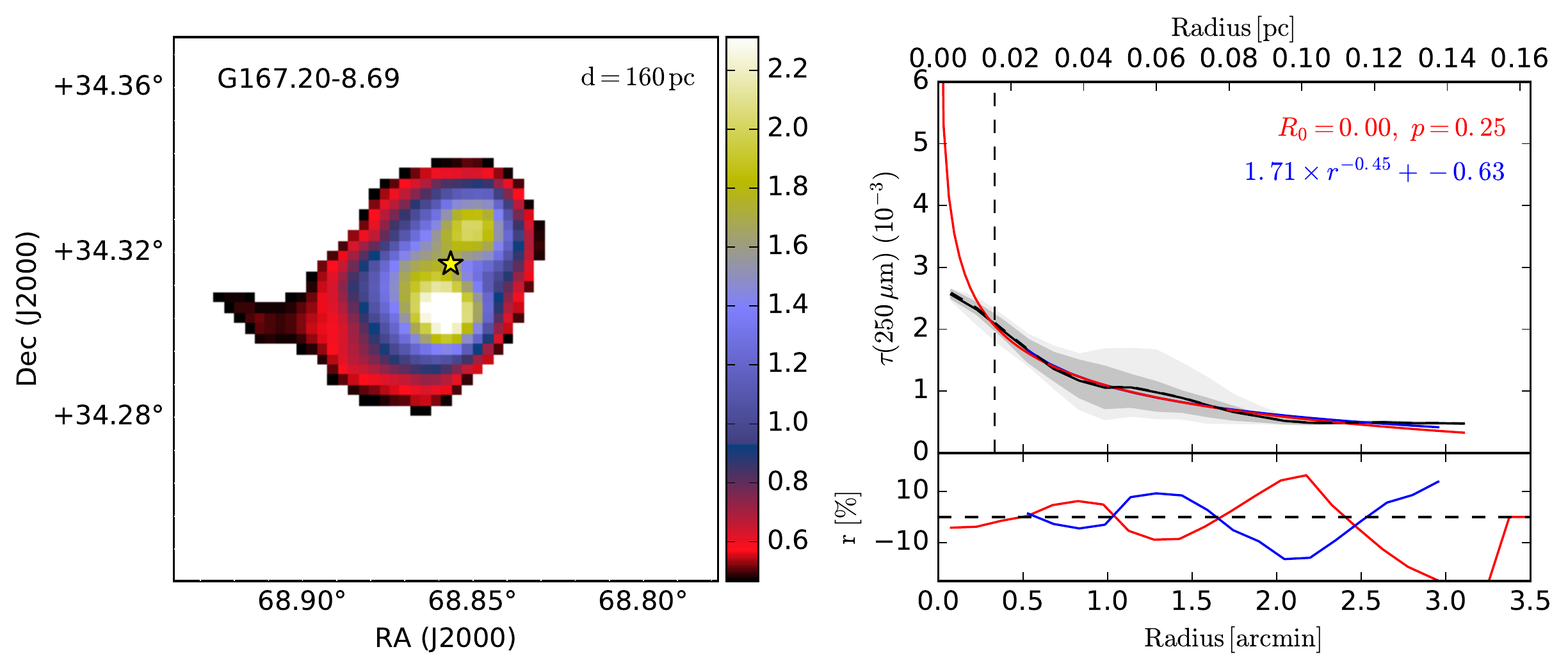}
\includegraphics[width=8.2cm]{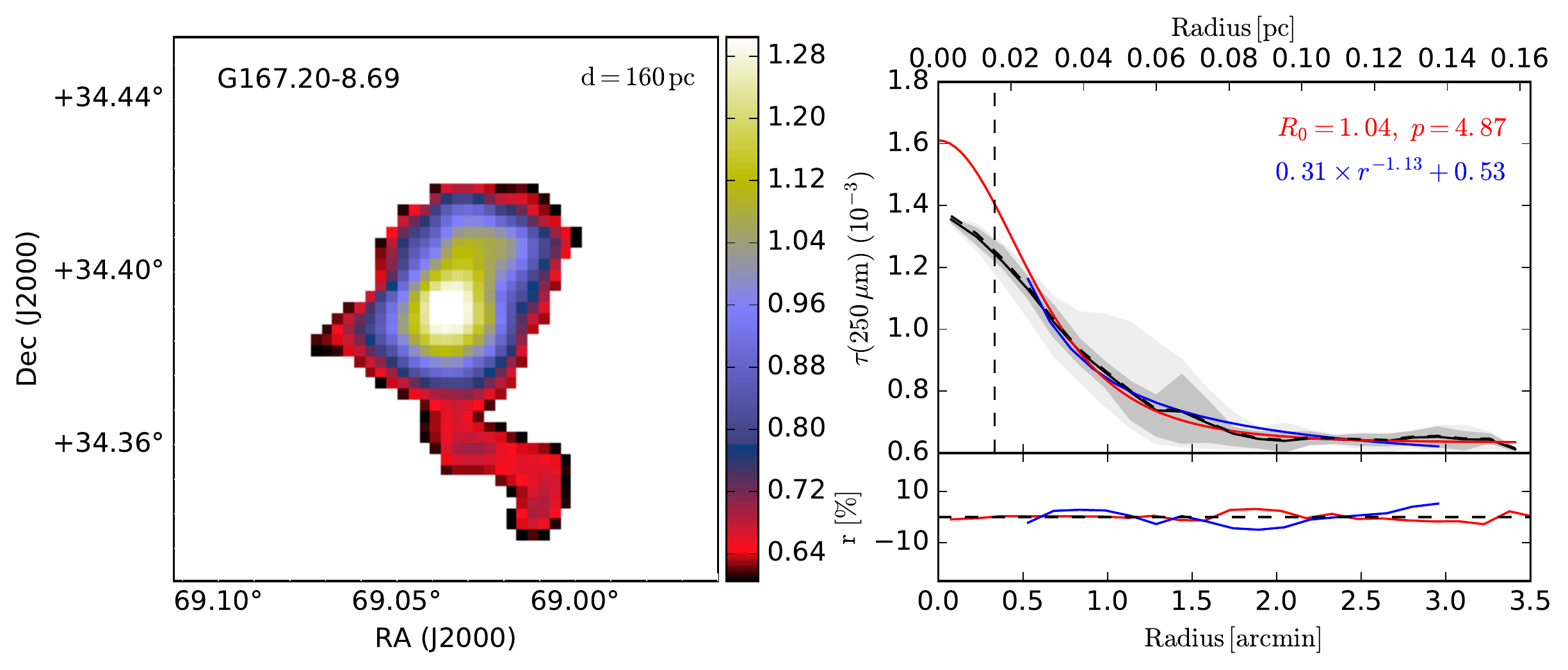}
\caption{continued.}
\end{figure}

\begin{figure}
\includegraphics[width=8.2cm]{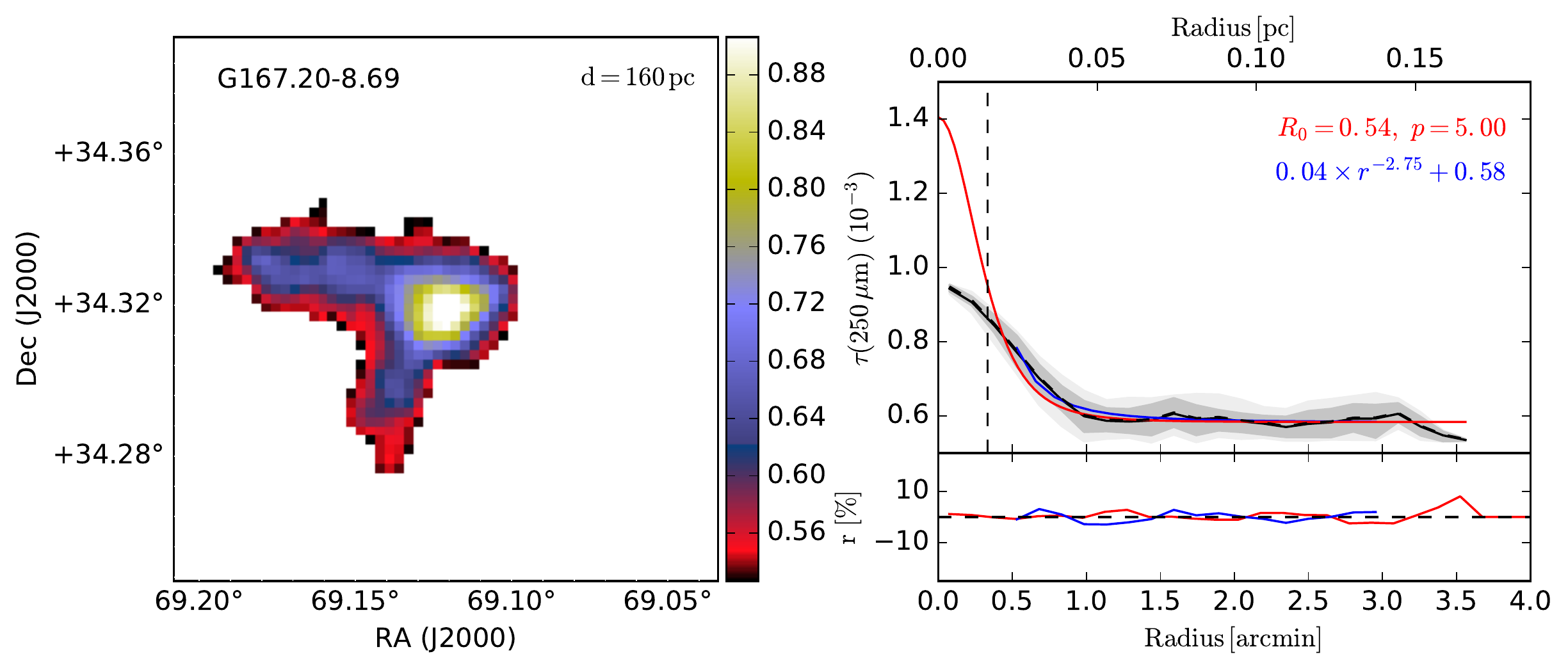}
\includegraphics[width=8.2cm]{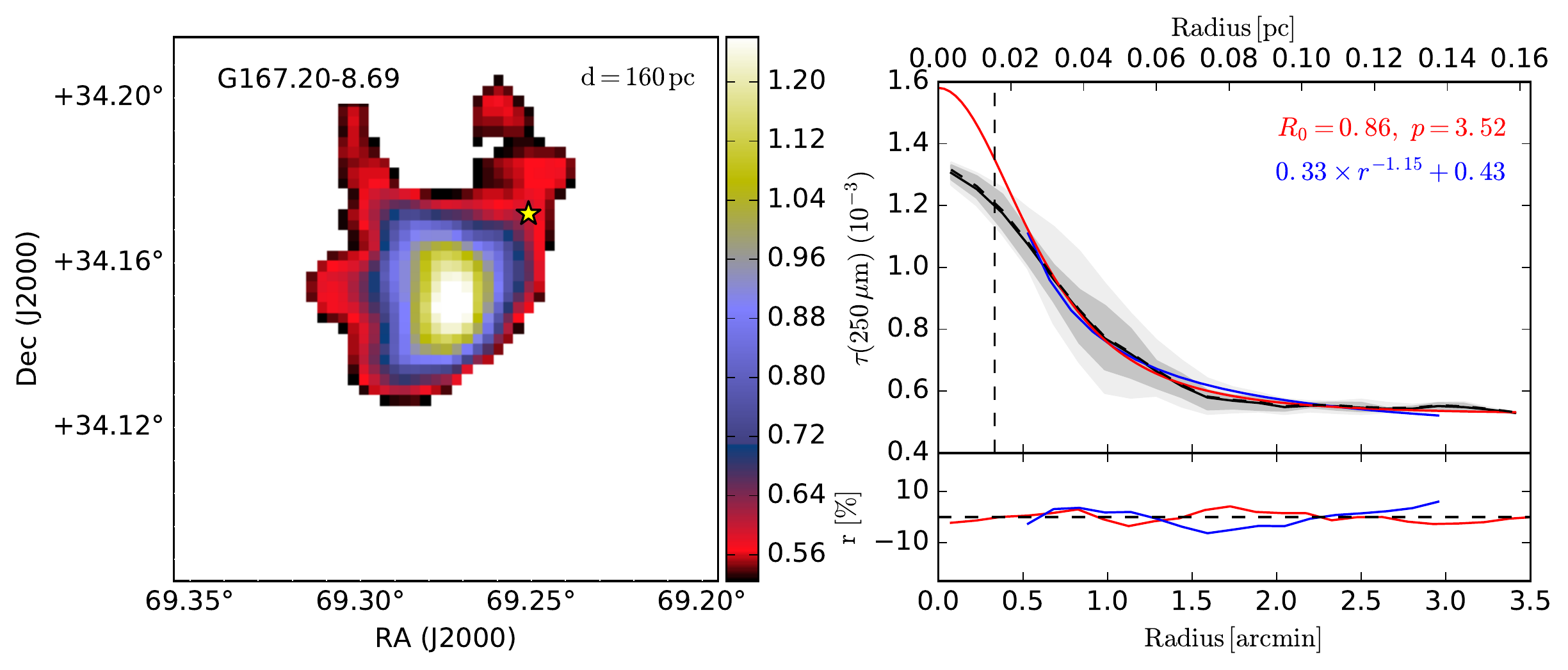}
\includegraphics[width=8.2cm]{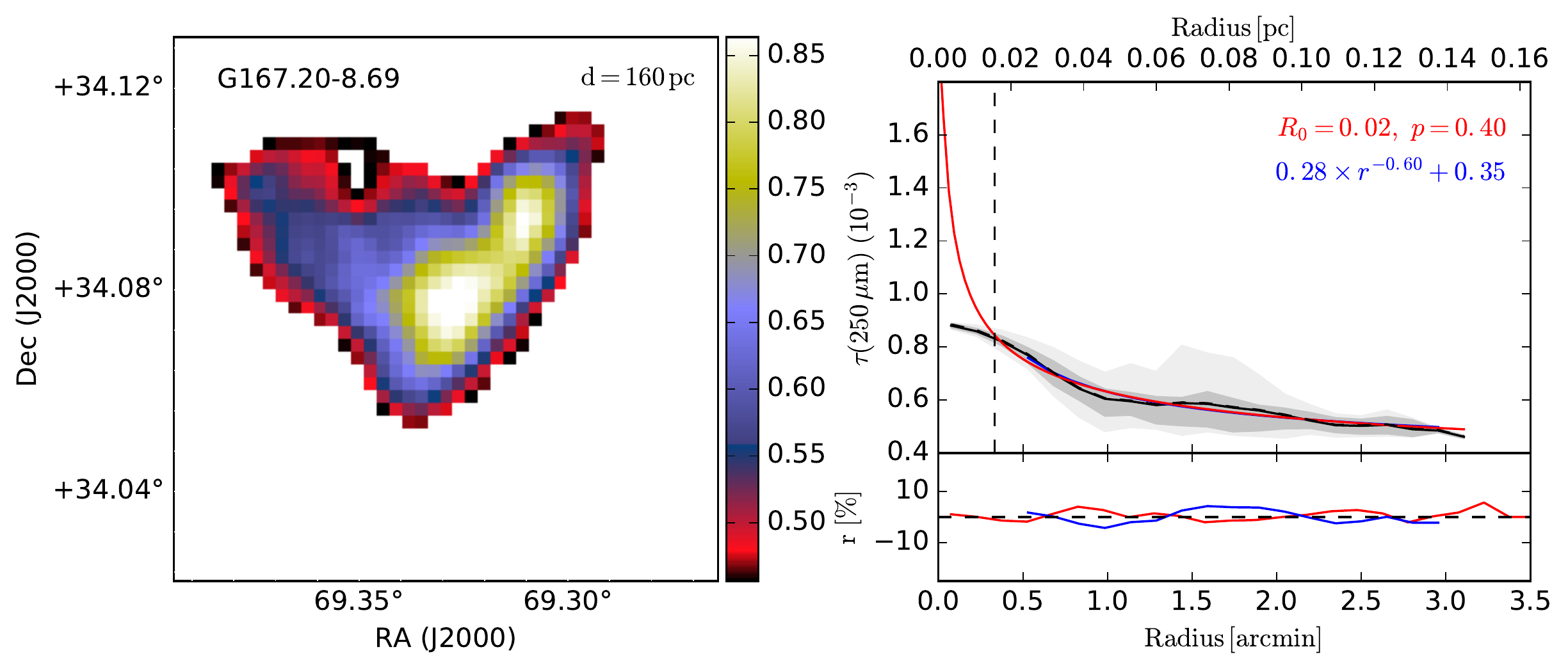}
\includegraphics[width=8.2cm]{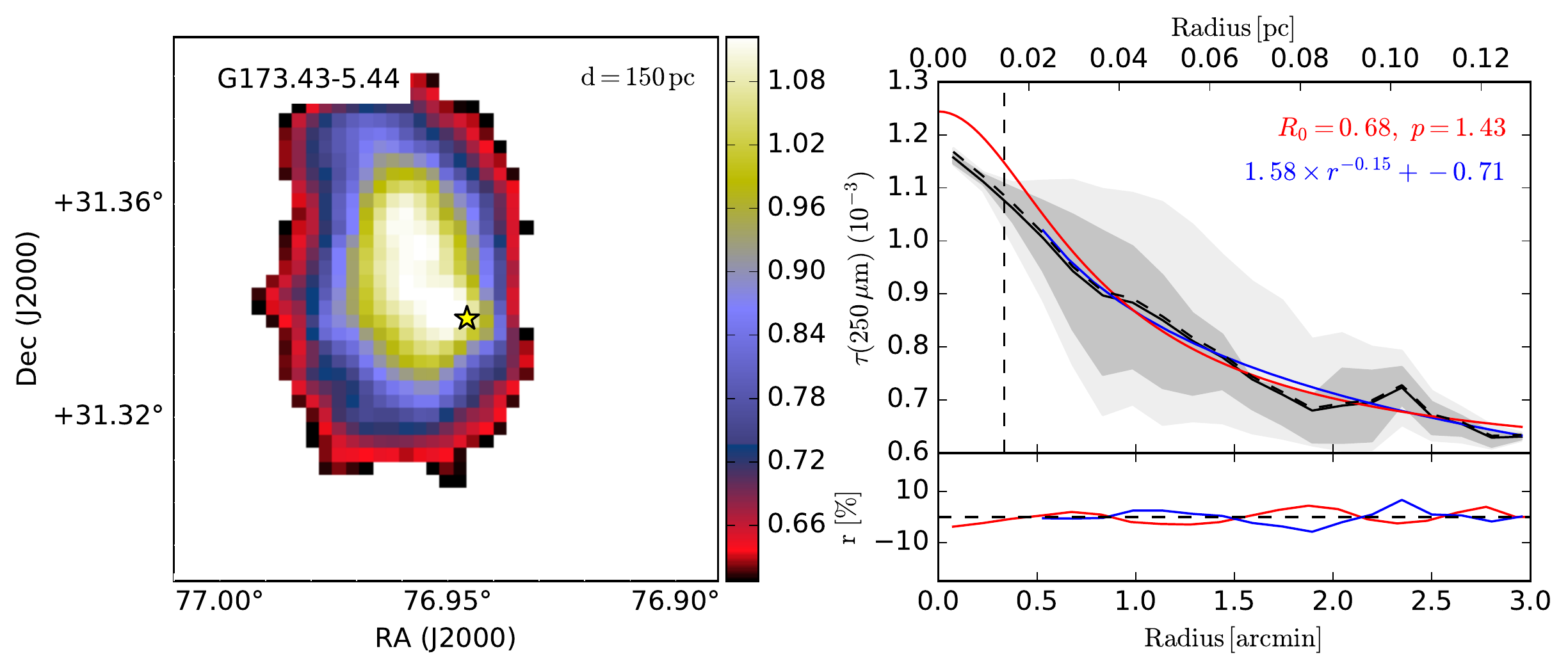}
\includegraphics[width=8.2cm]{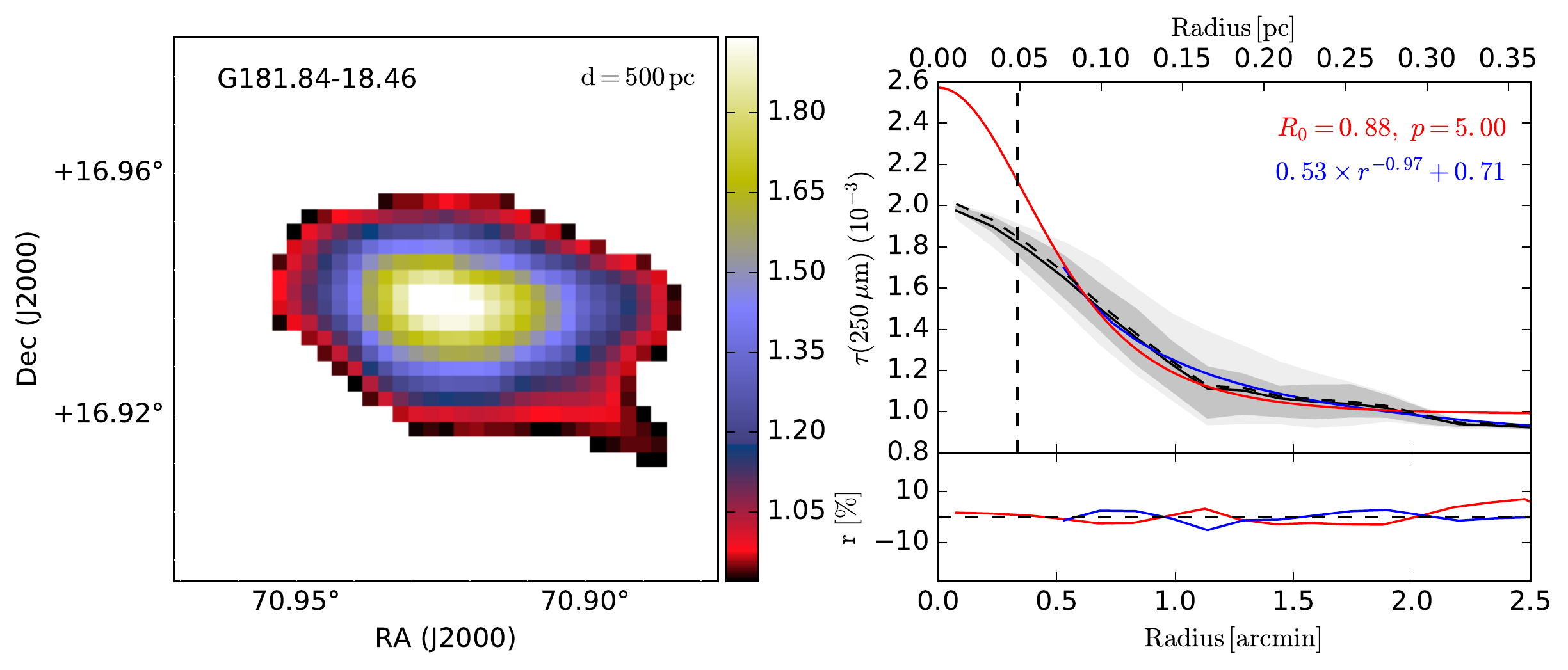}
\includegraphics[width=8.2cm]{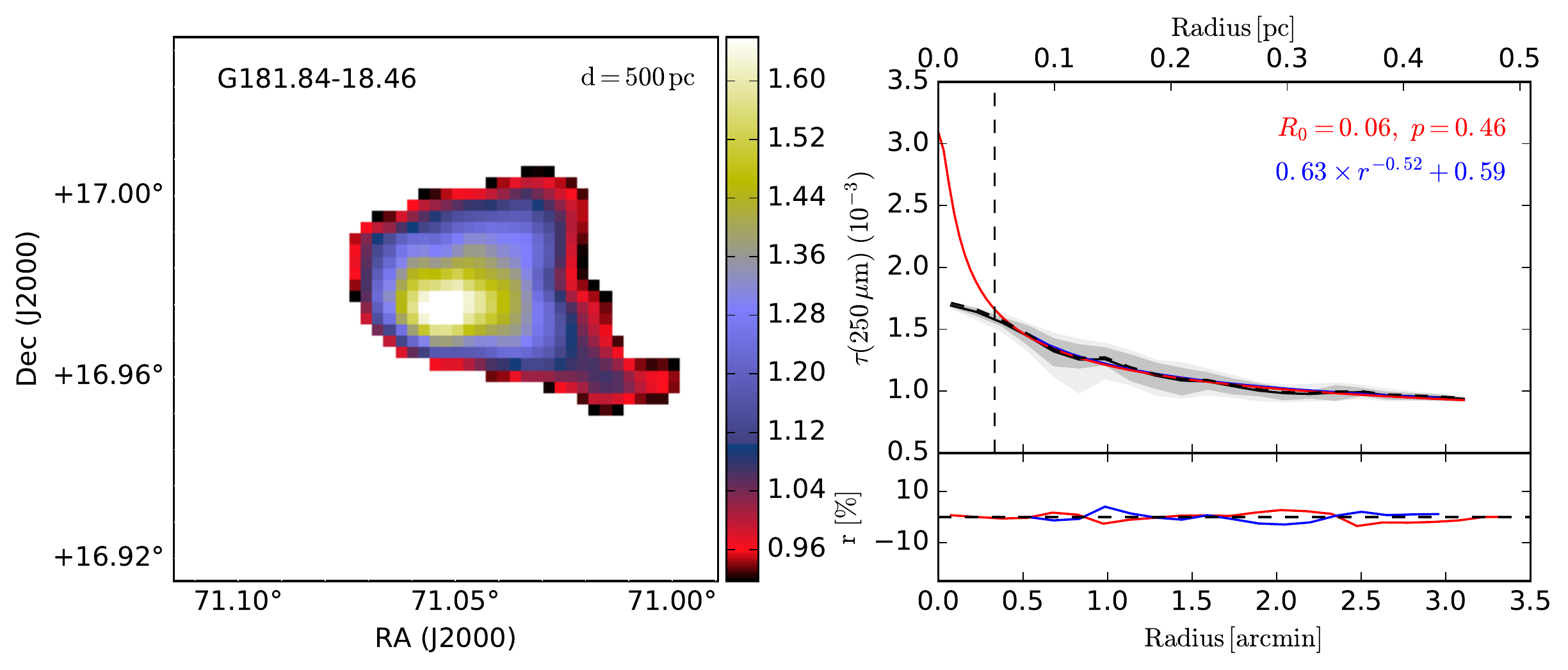}
\caption{continued.}
\end{figure}

\begin{figure}
\includegraphics[width=8.2cm]{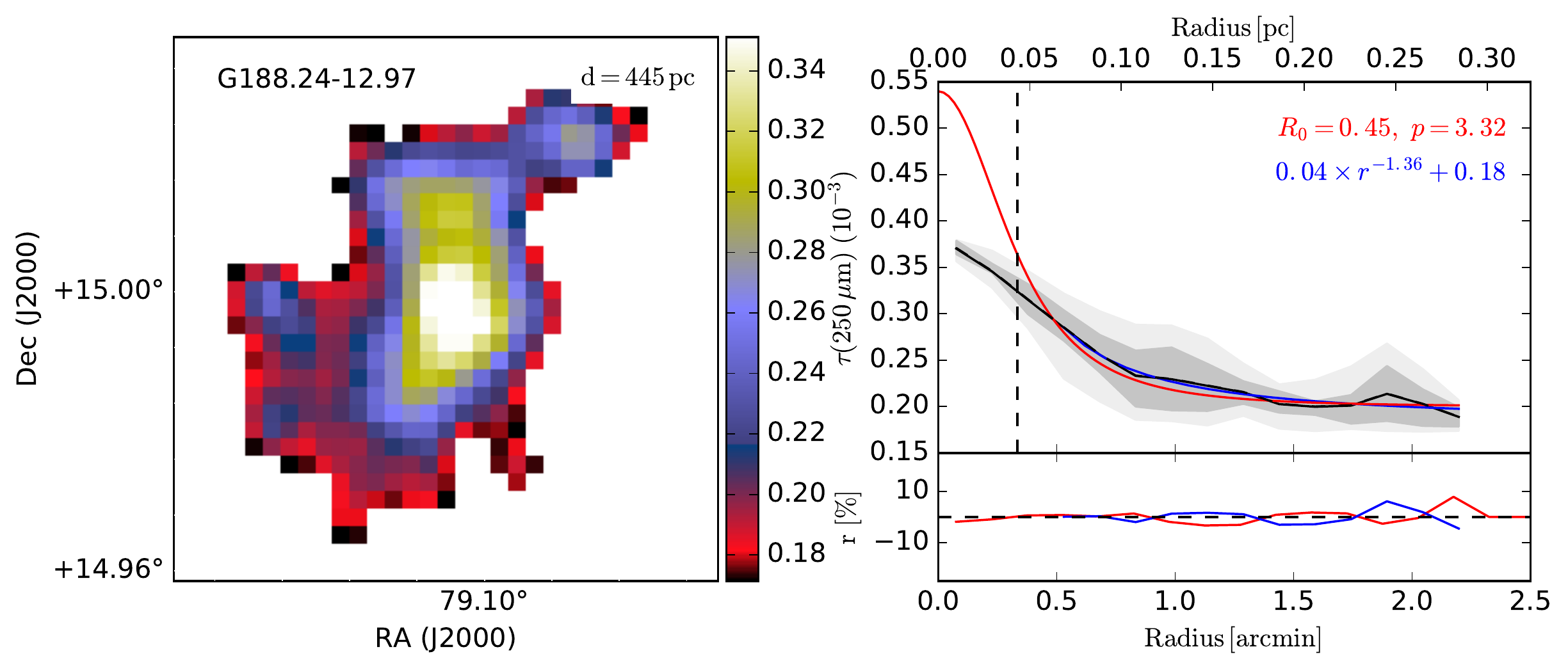}
\includegraphics[width=8.2cm]{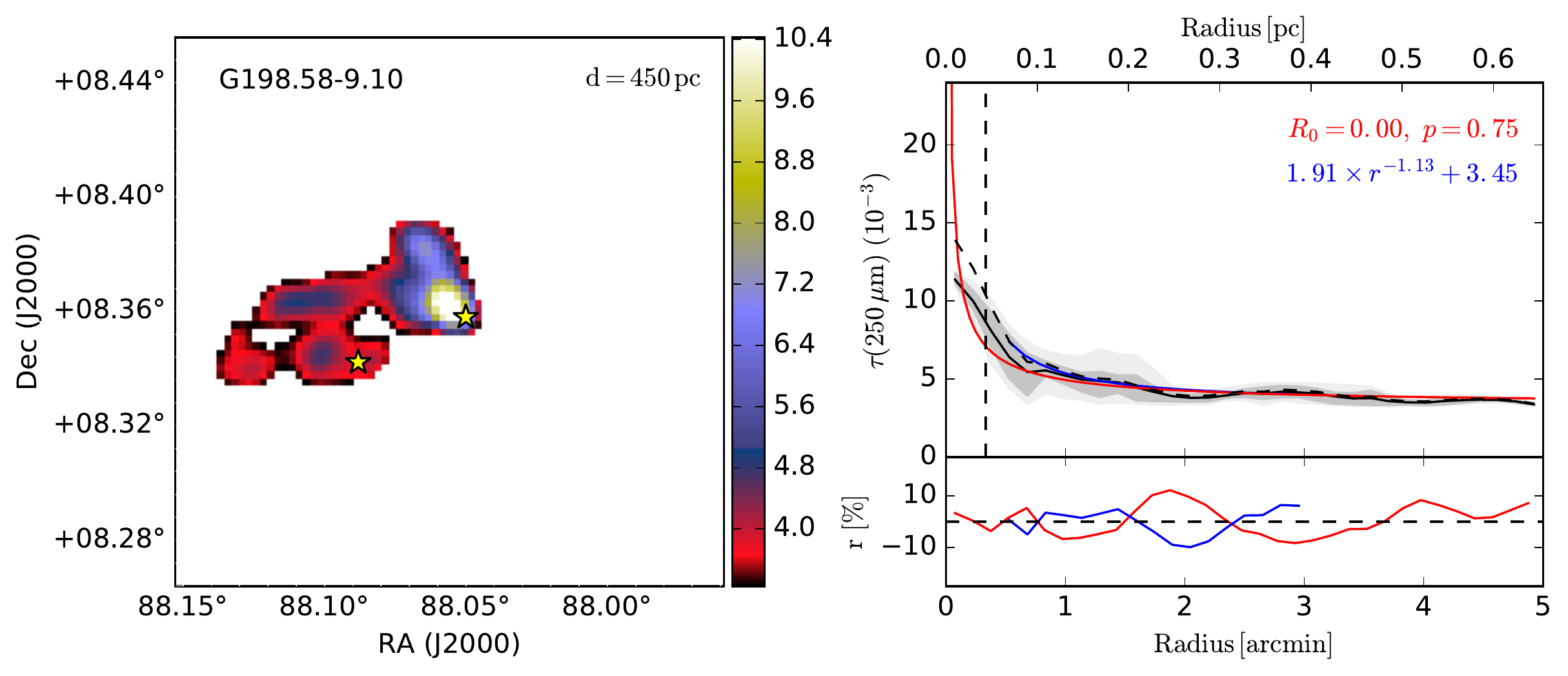}
\includegraphics[width=8.2cm]{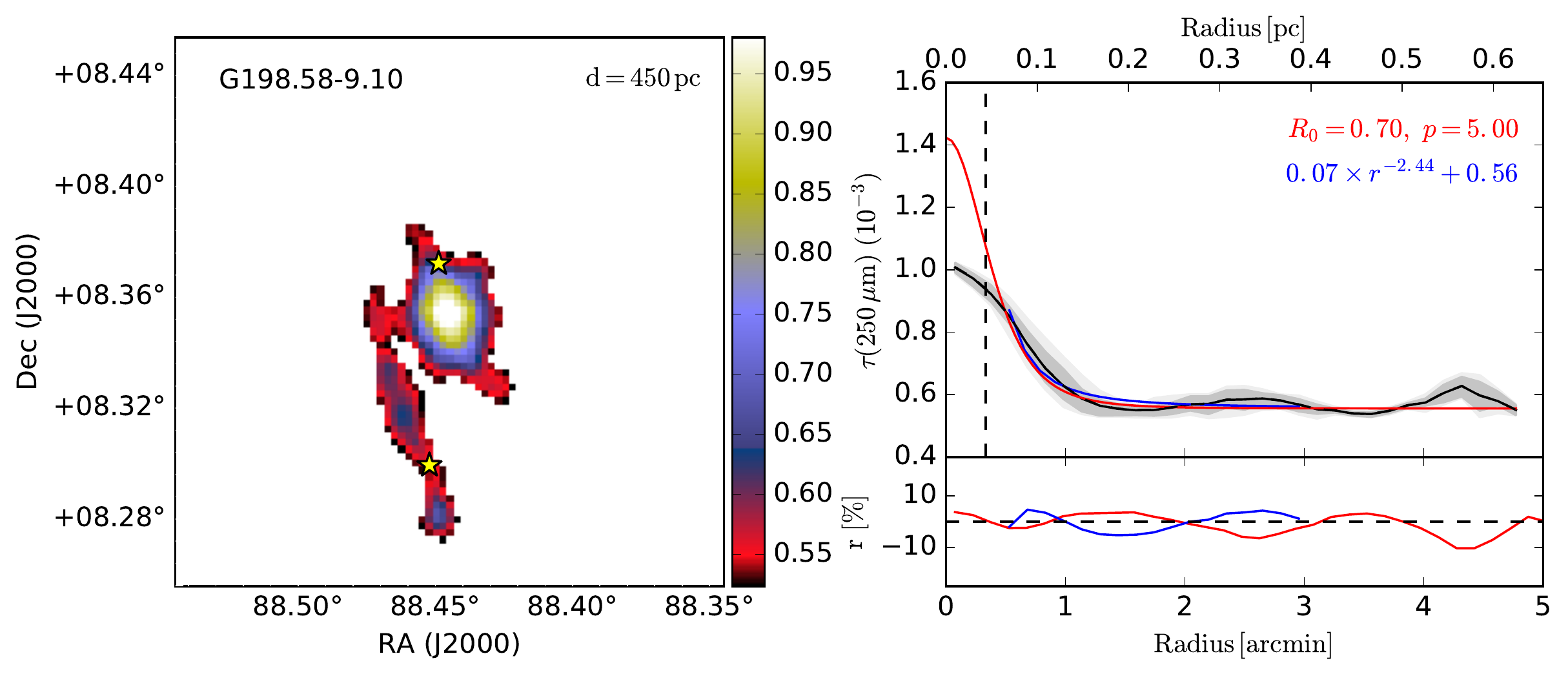}
\includegraphics[width=8.2cm]{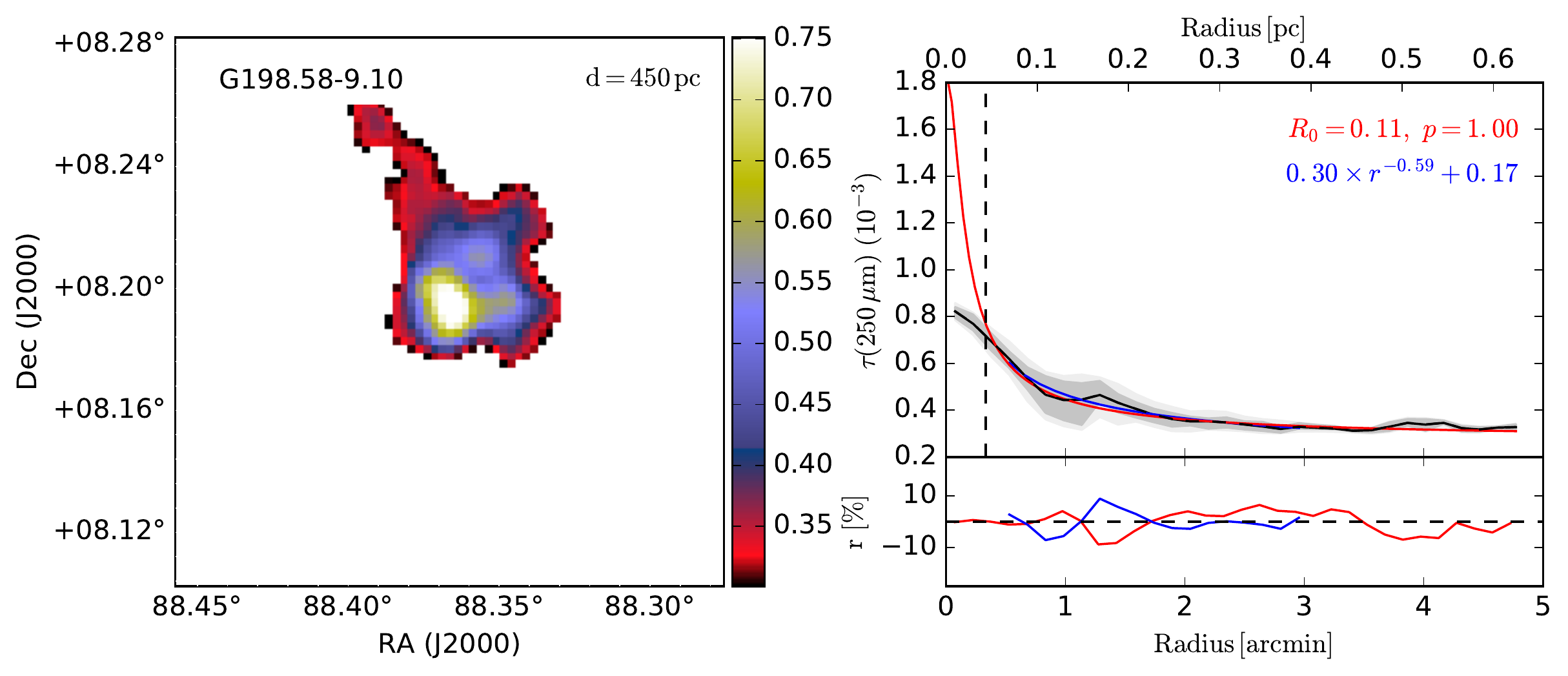}
\includegraphics[width=8.2cm]{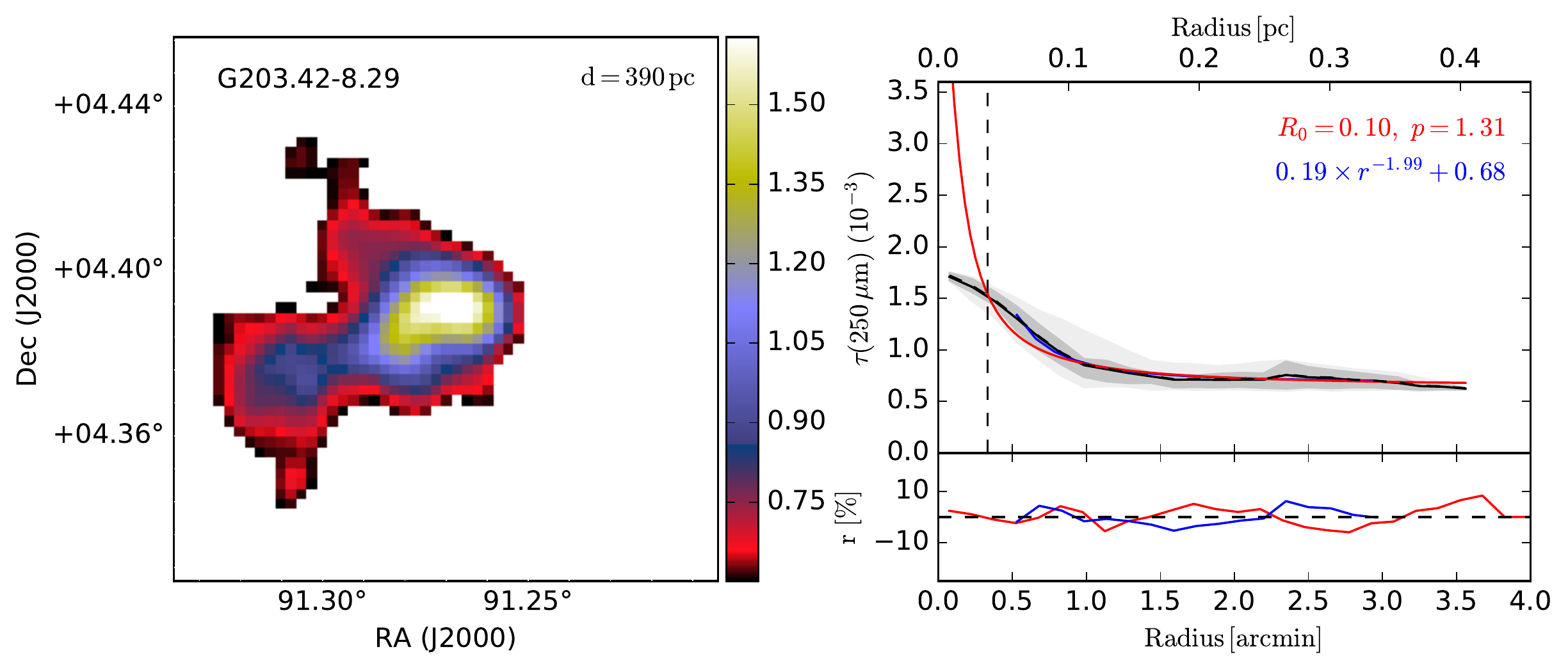}
\includegraphics[width=8.2cm]{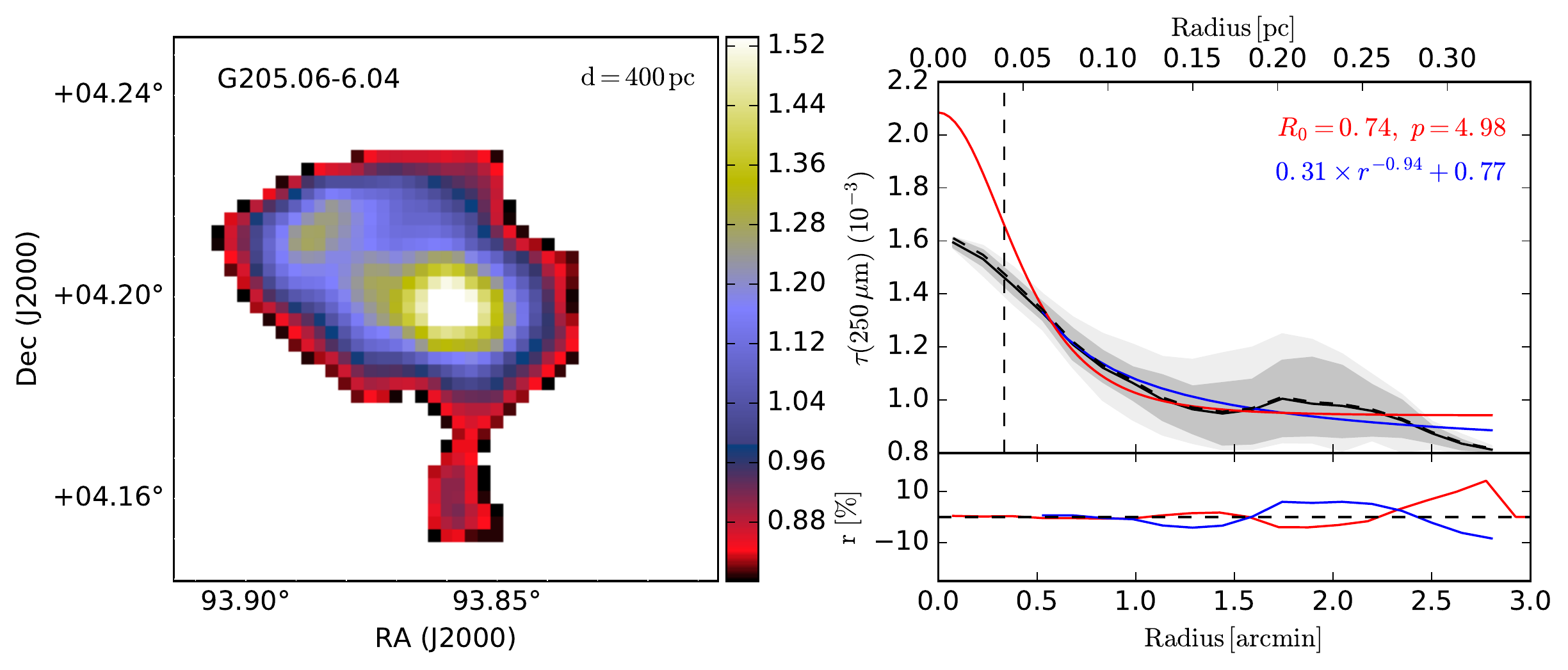}
\caption{continued.}
\end{figure}

\begin{figure}
\includegraphics[width=8.2cm]{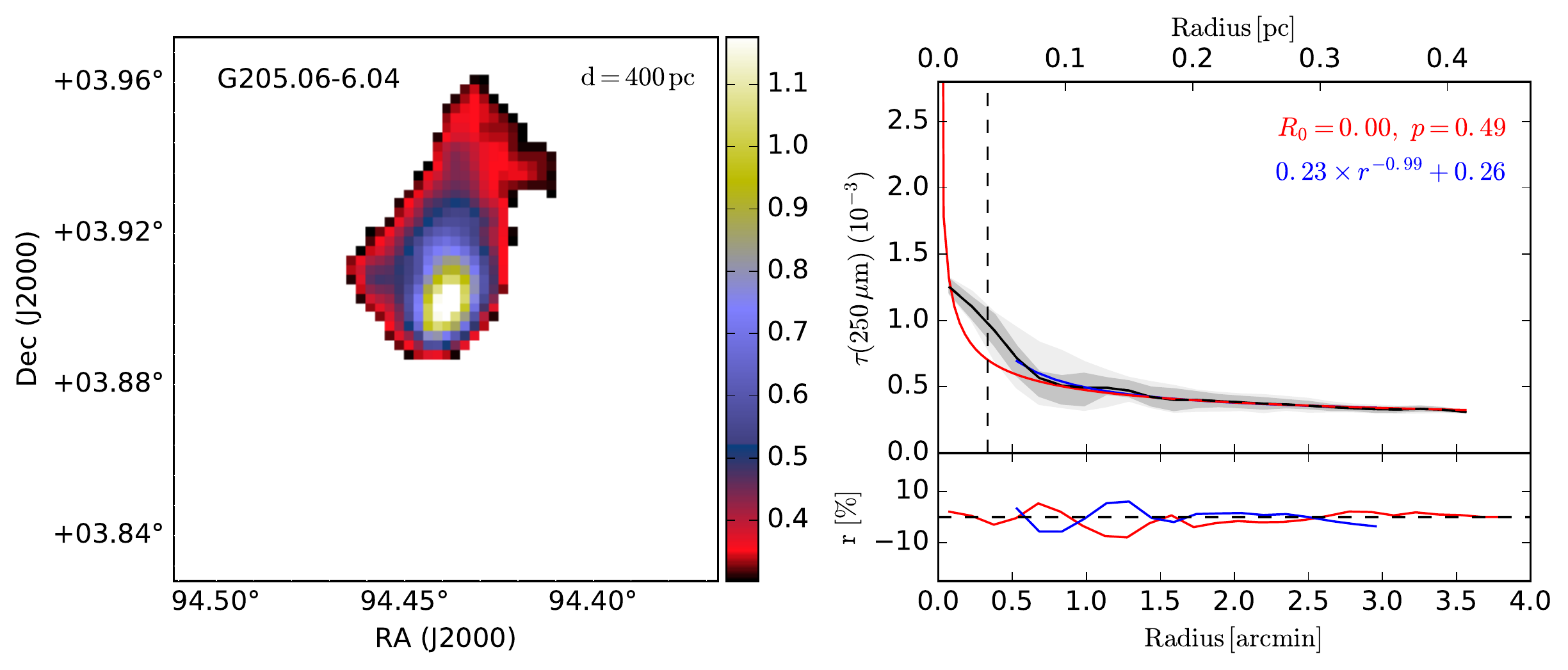}
\includegraphics[width=8.2cm]{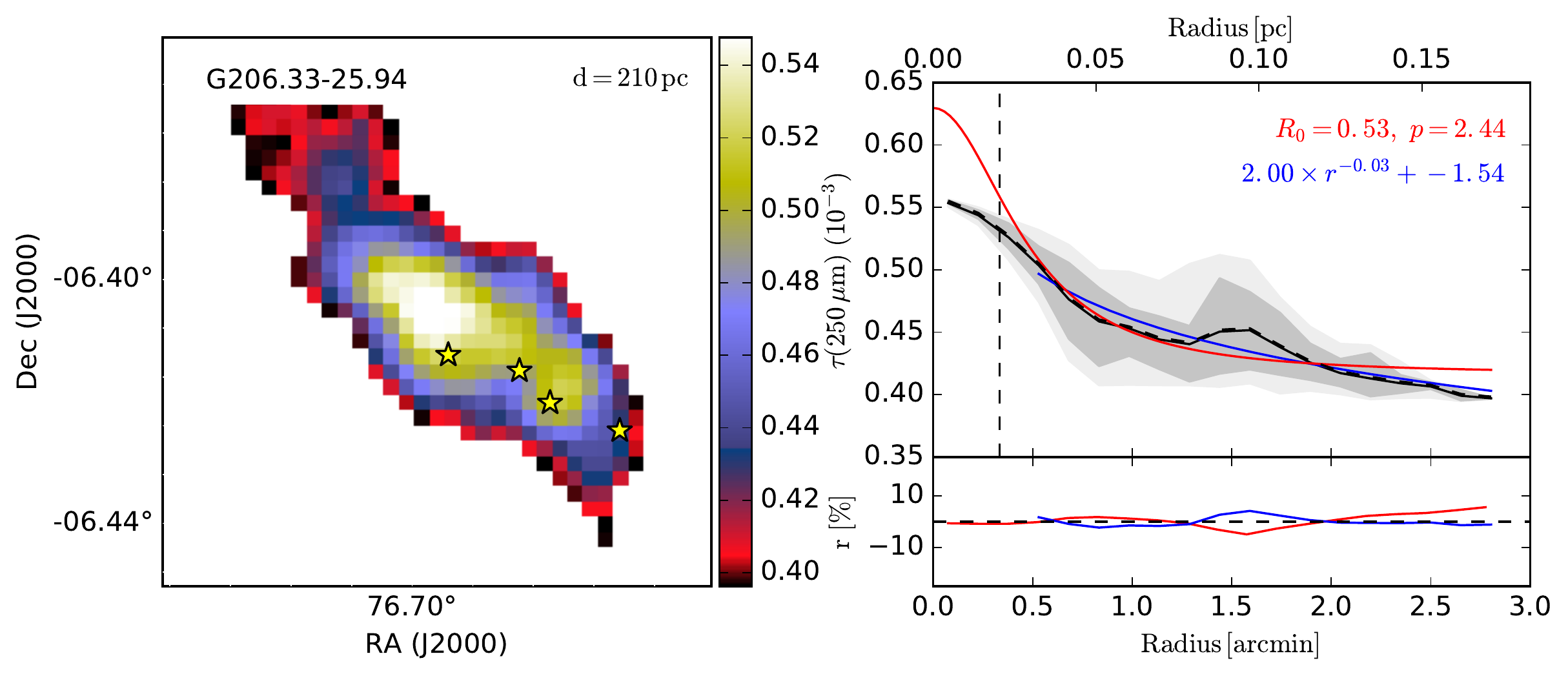}
\includegraphics[width=8.2cm]{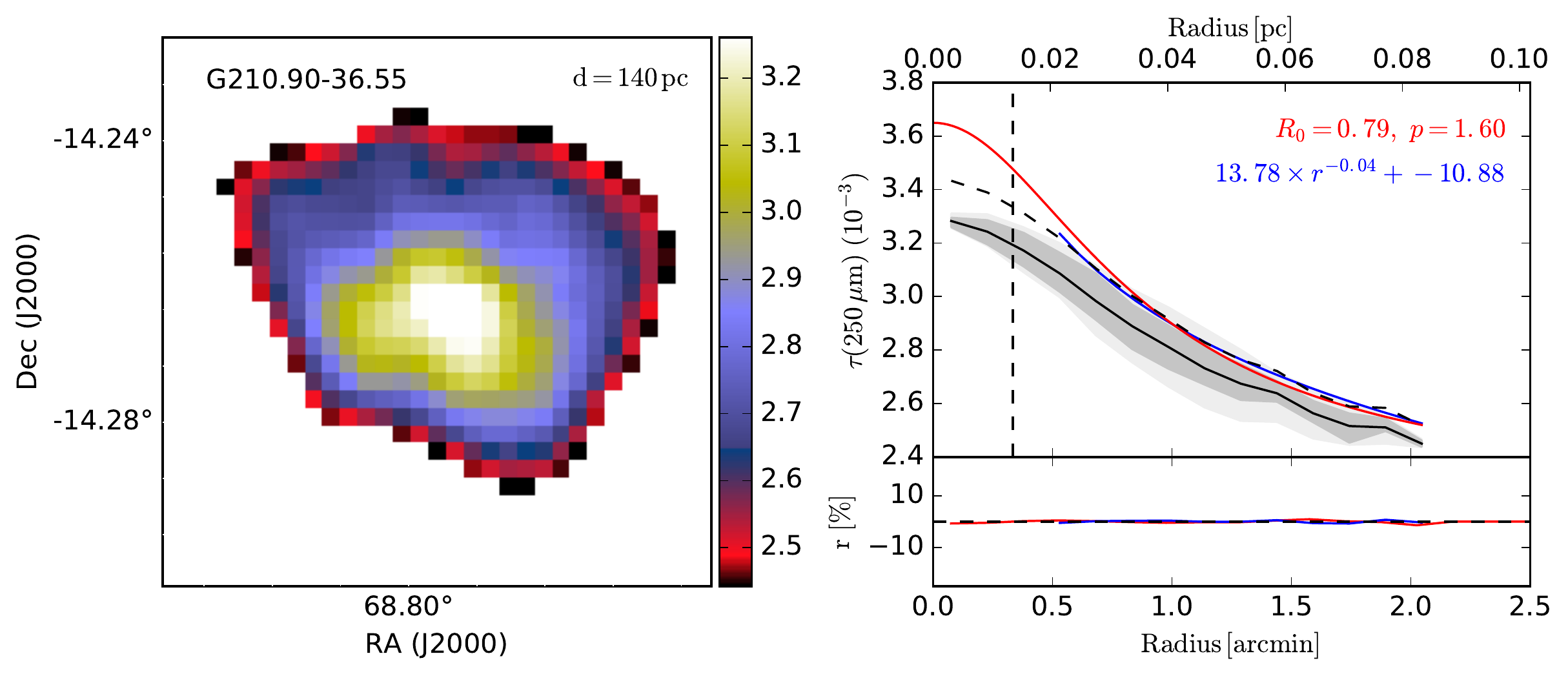}
\includegraphics[width=8.2cm]{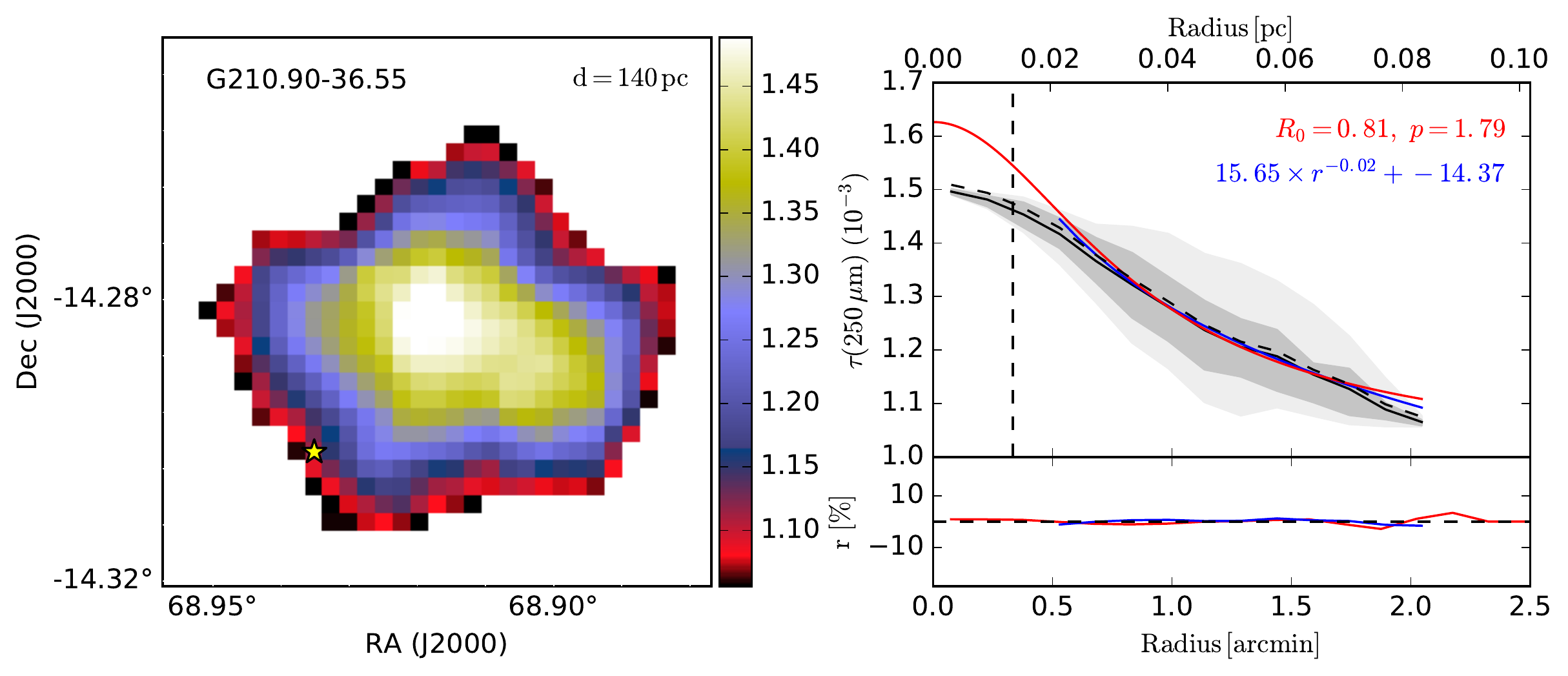}
\includegraphics[width=8.2cm]{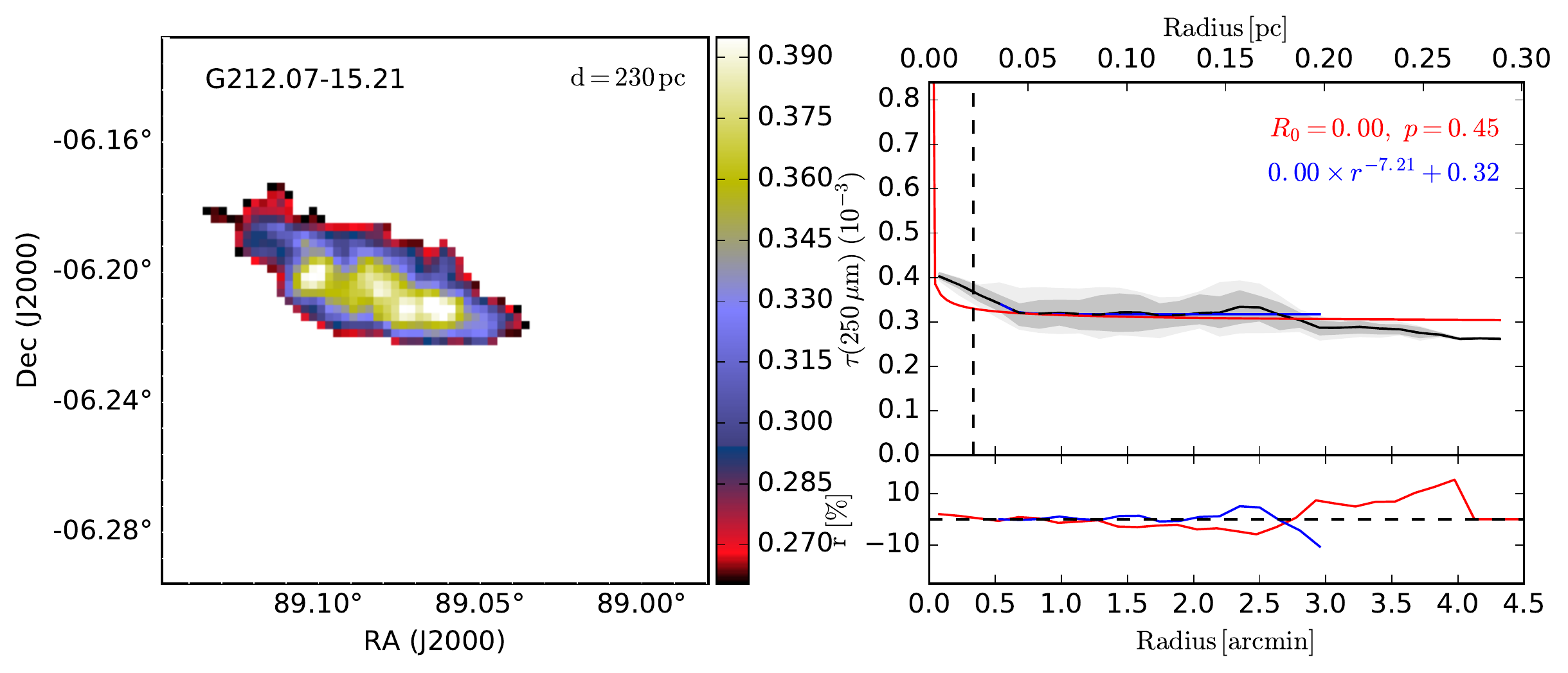}
\includegraphics[width=8.2cm]{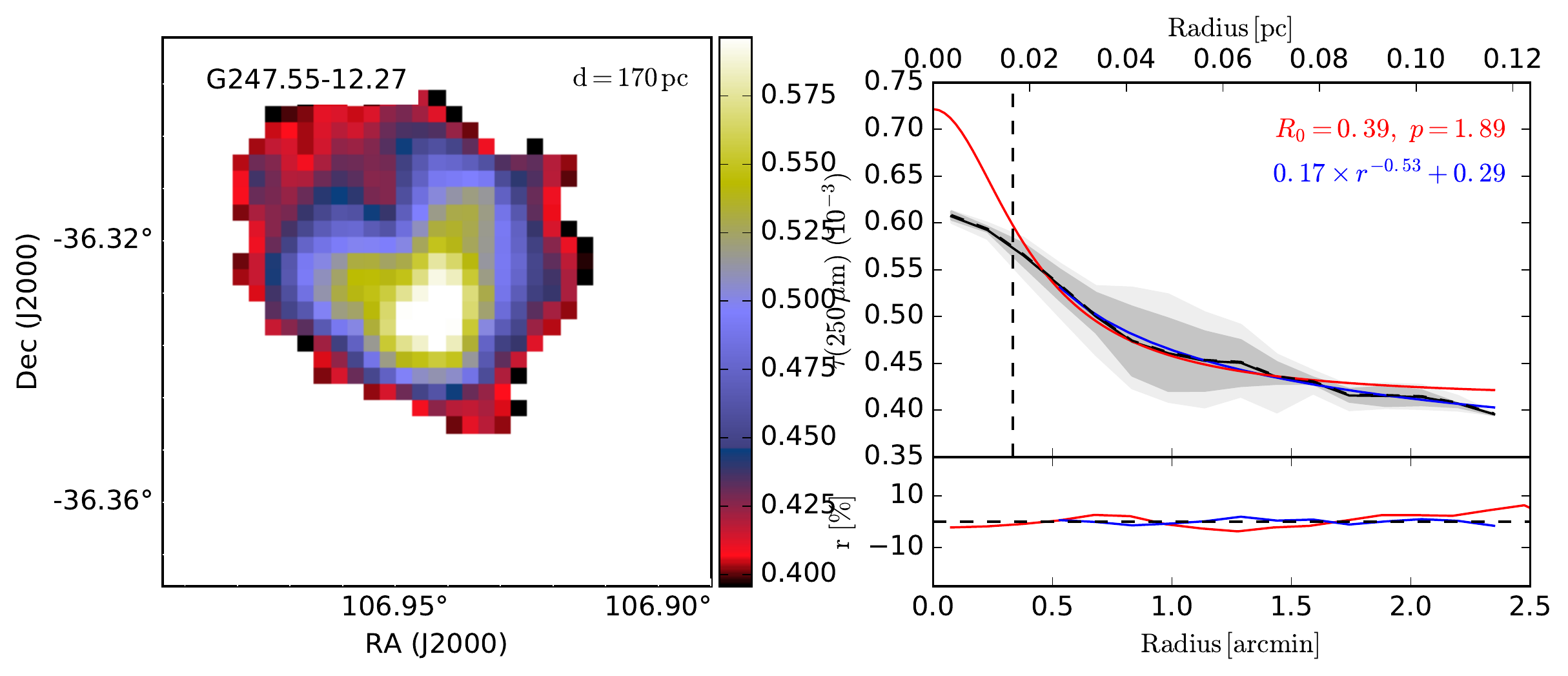}
\caption{continued.}
\end{figure}

\begin{figure}
\includegraphics[width=8.2cm]{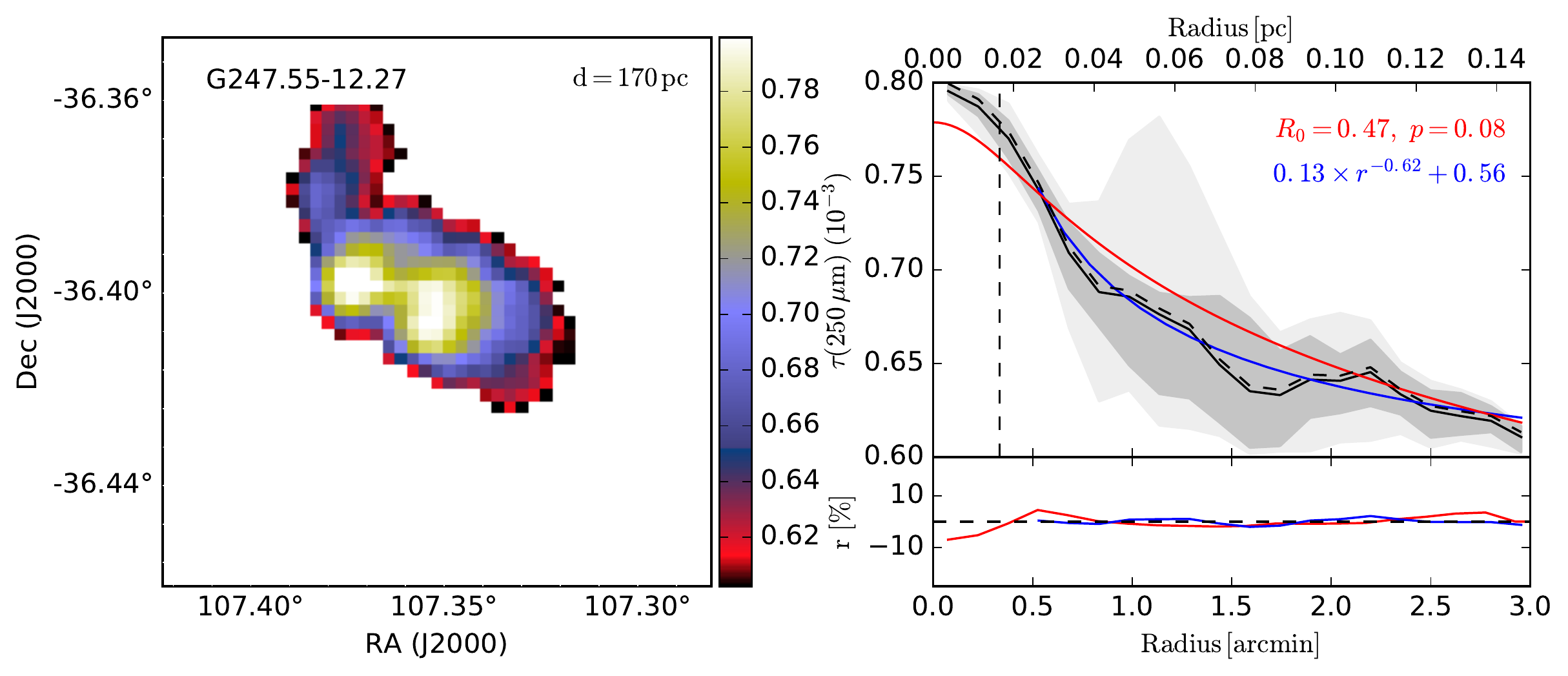}
\includegraphics[width=8.2cm]{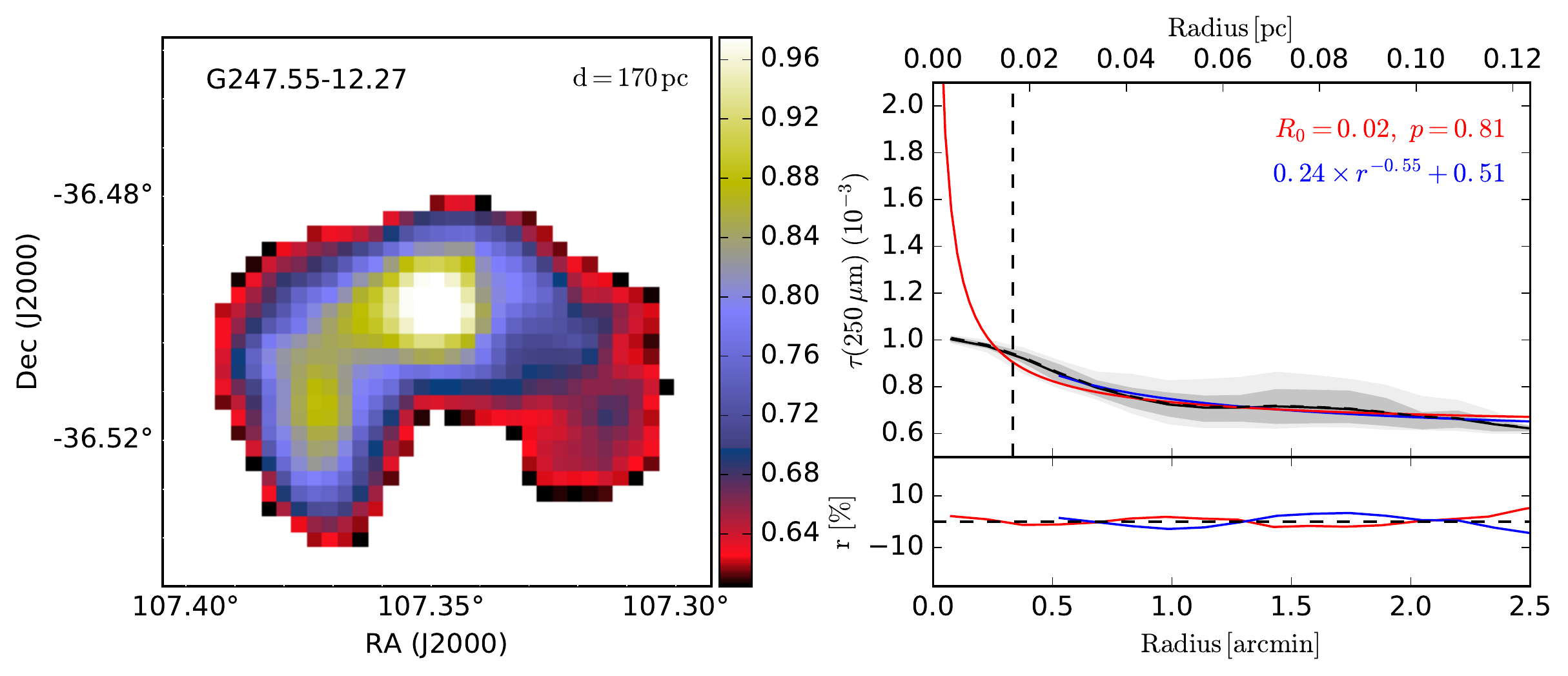}
\includegraphics[width=8.2cm]{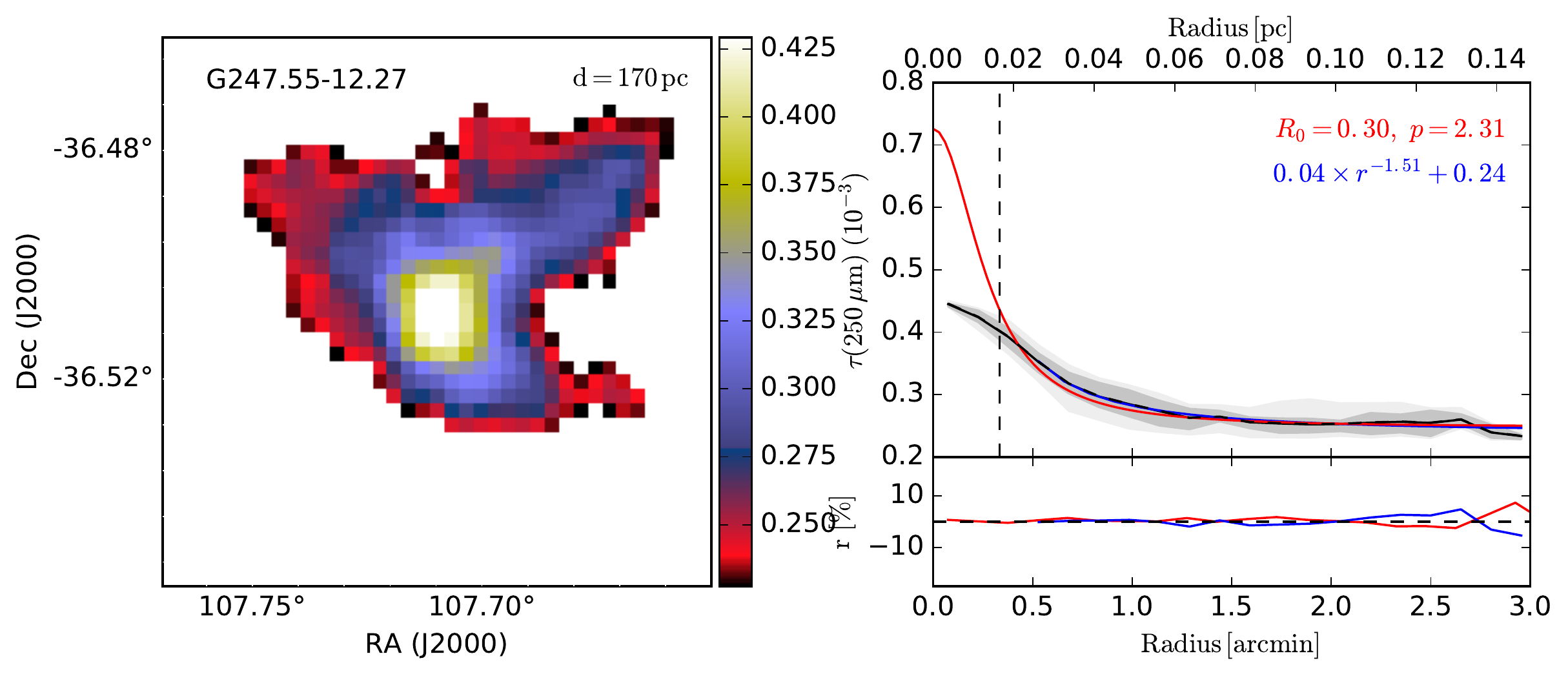}
\includegraphics[width=8.2cm]{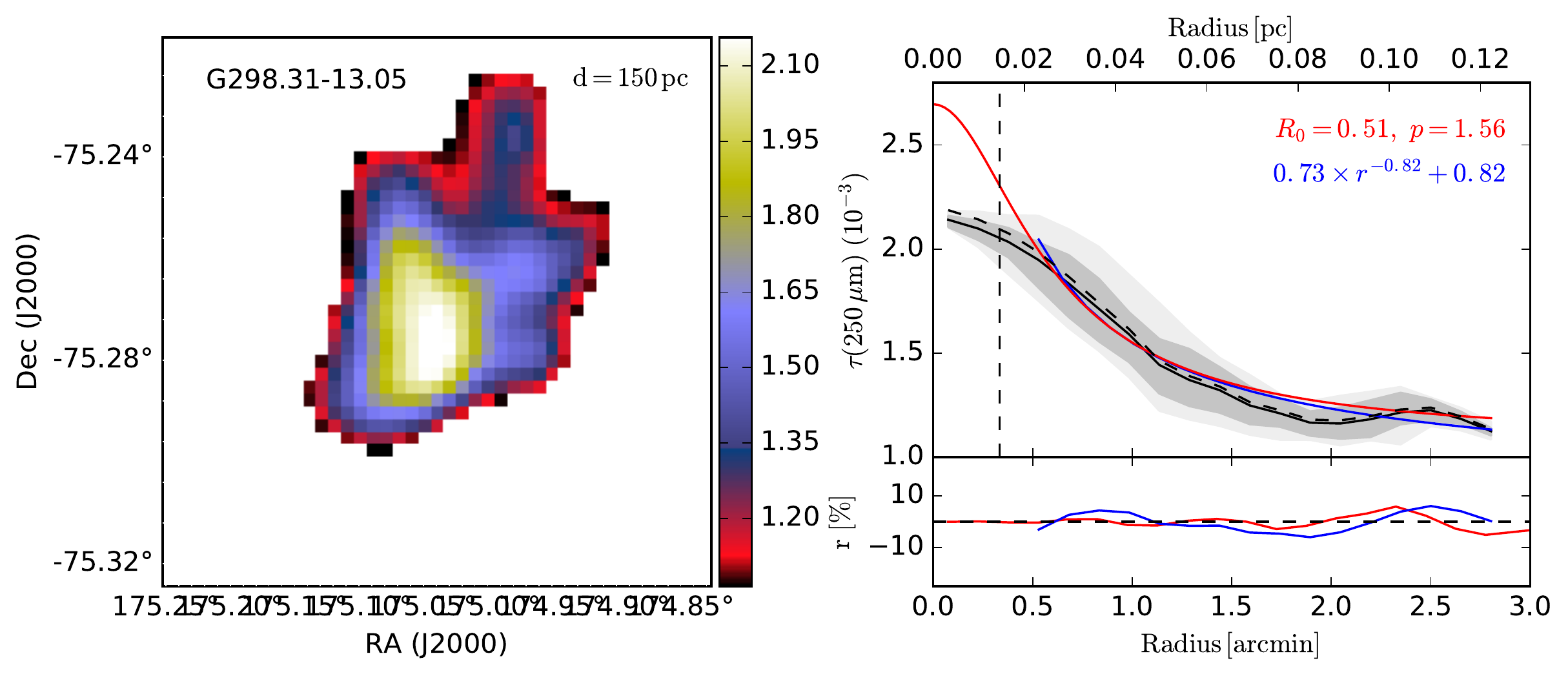}
\includegraphics[width=8.2cm]{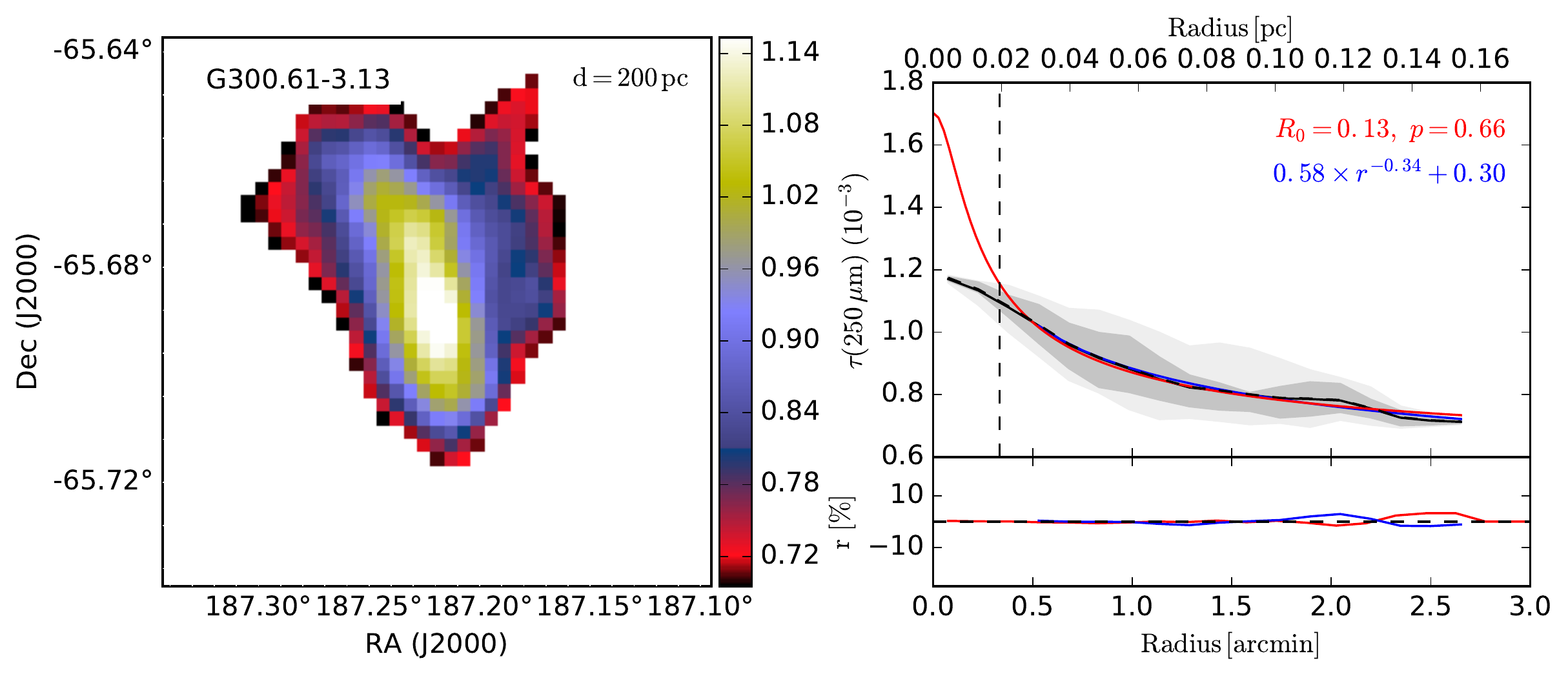}
\includegraphics[width=8.2cm]{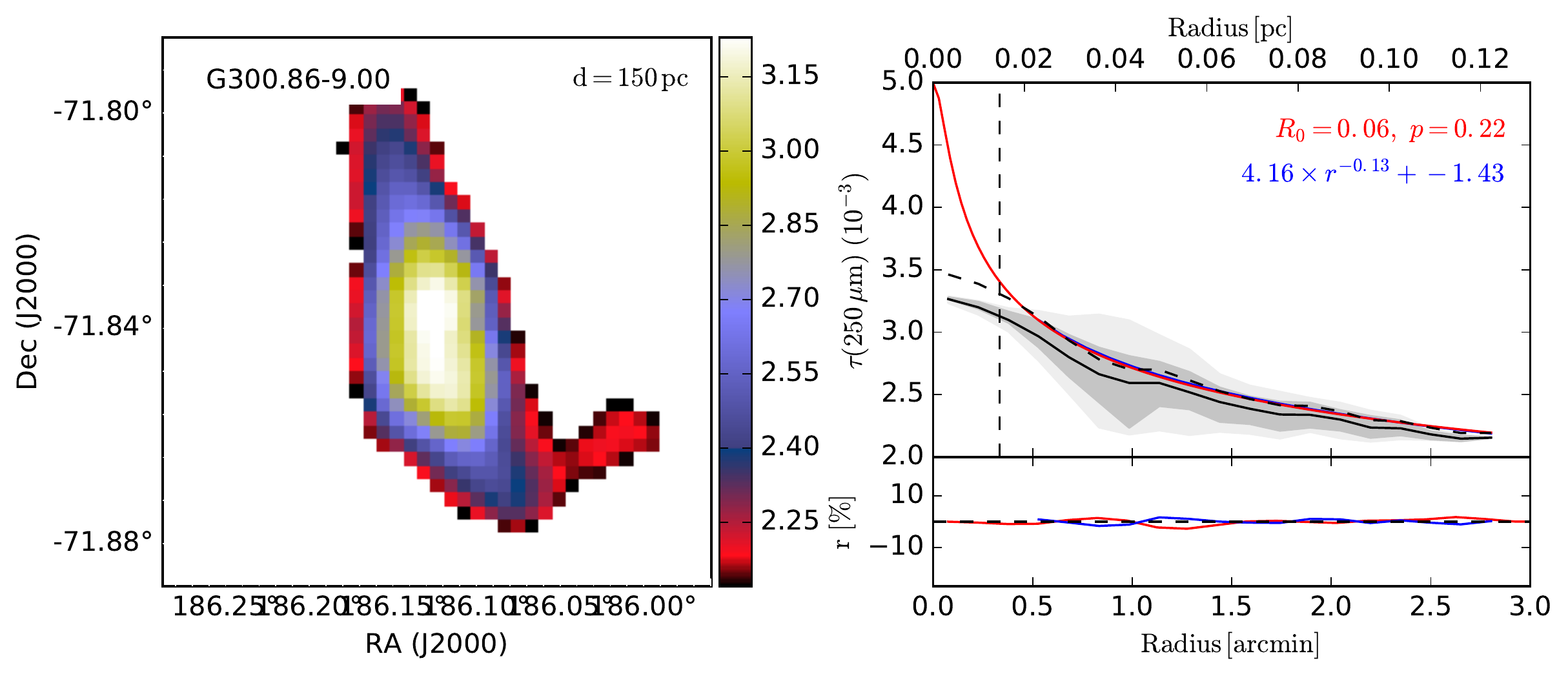}
\caption{continued.}
\end{figure}

\begin{figure}
\includegraphics[width=8.2cm]{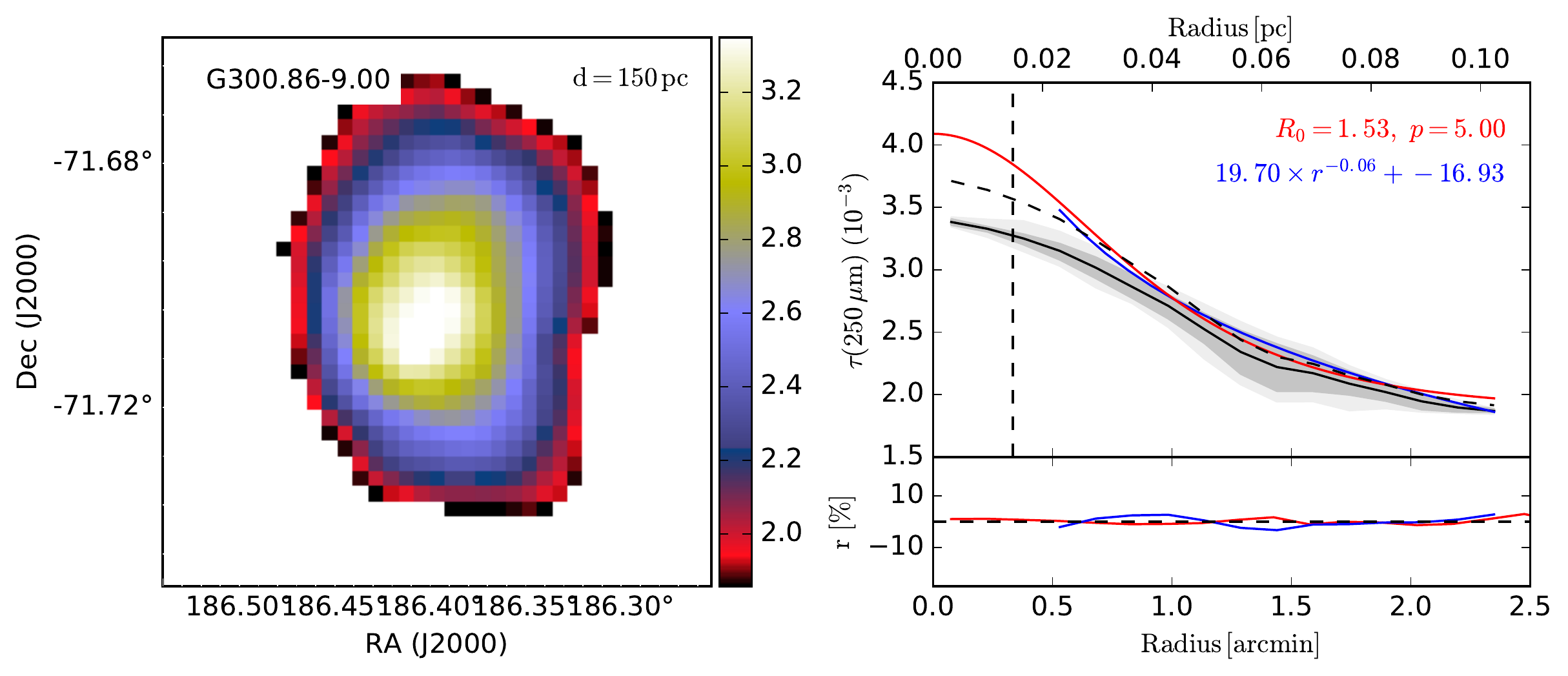}
\includegraphics[width=8.2cm]{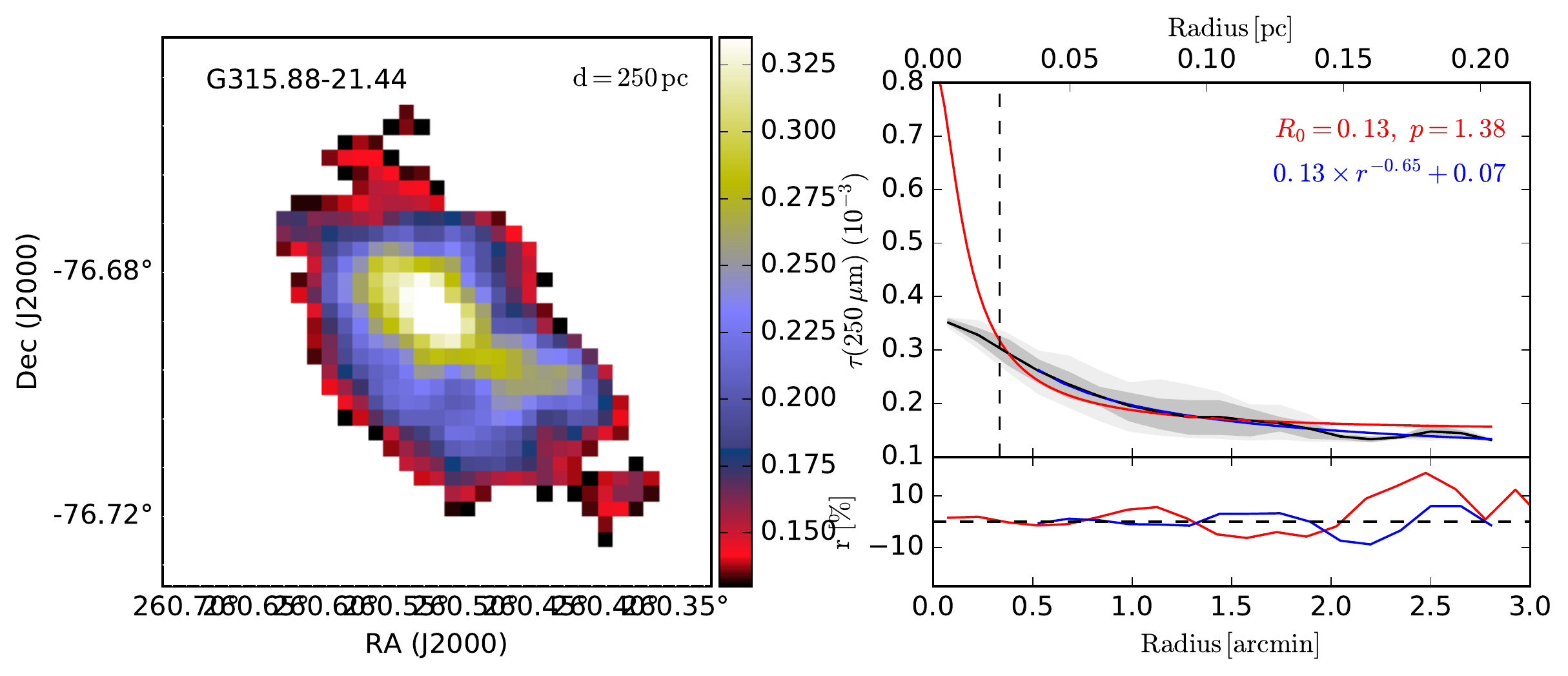}
\includegraphics[width=8.2cm]{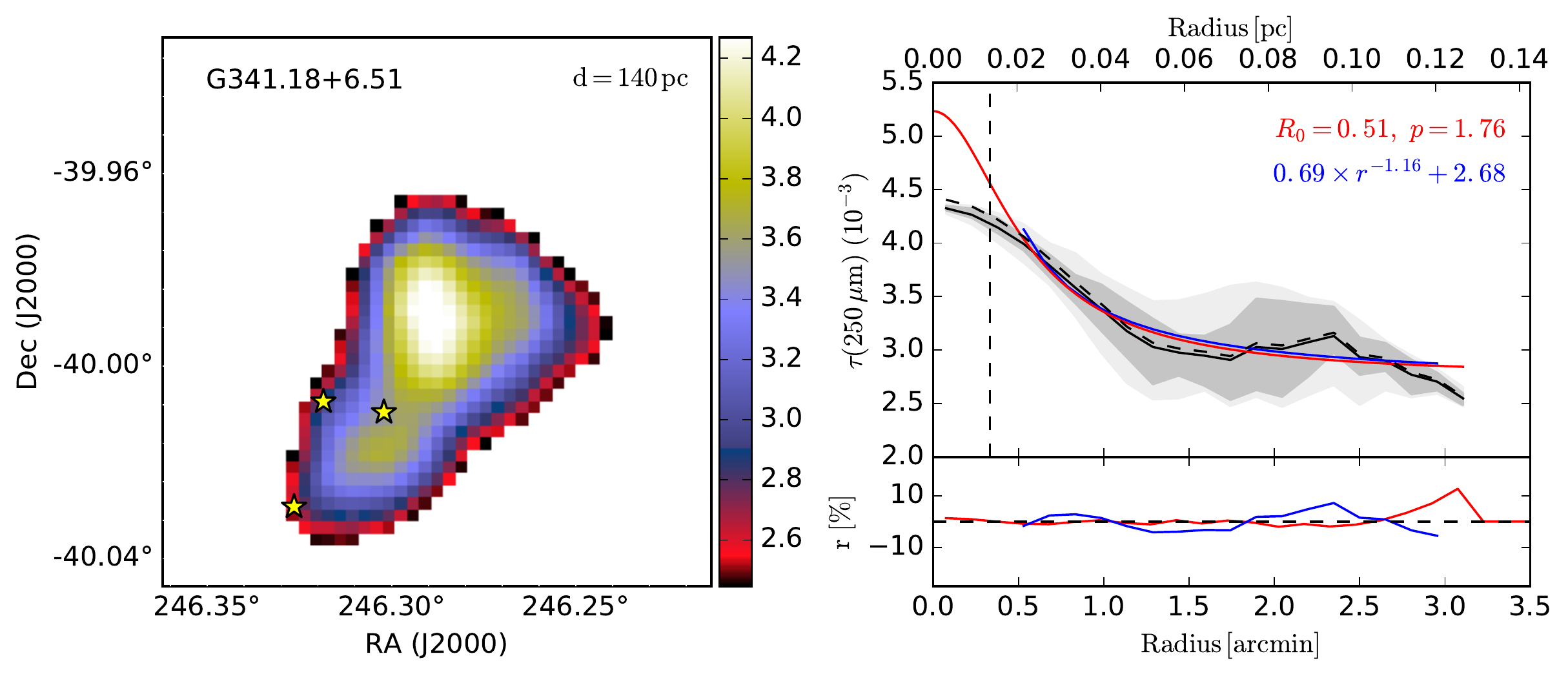}
\includegraphics[width=8.2cm]{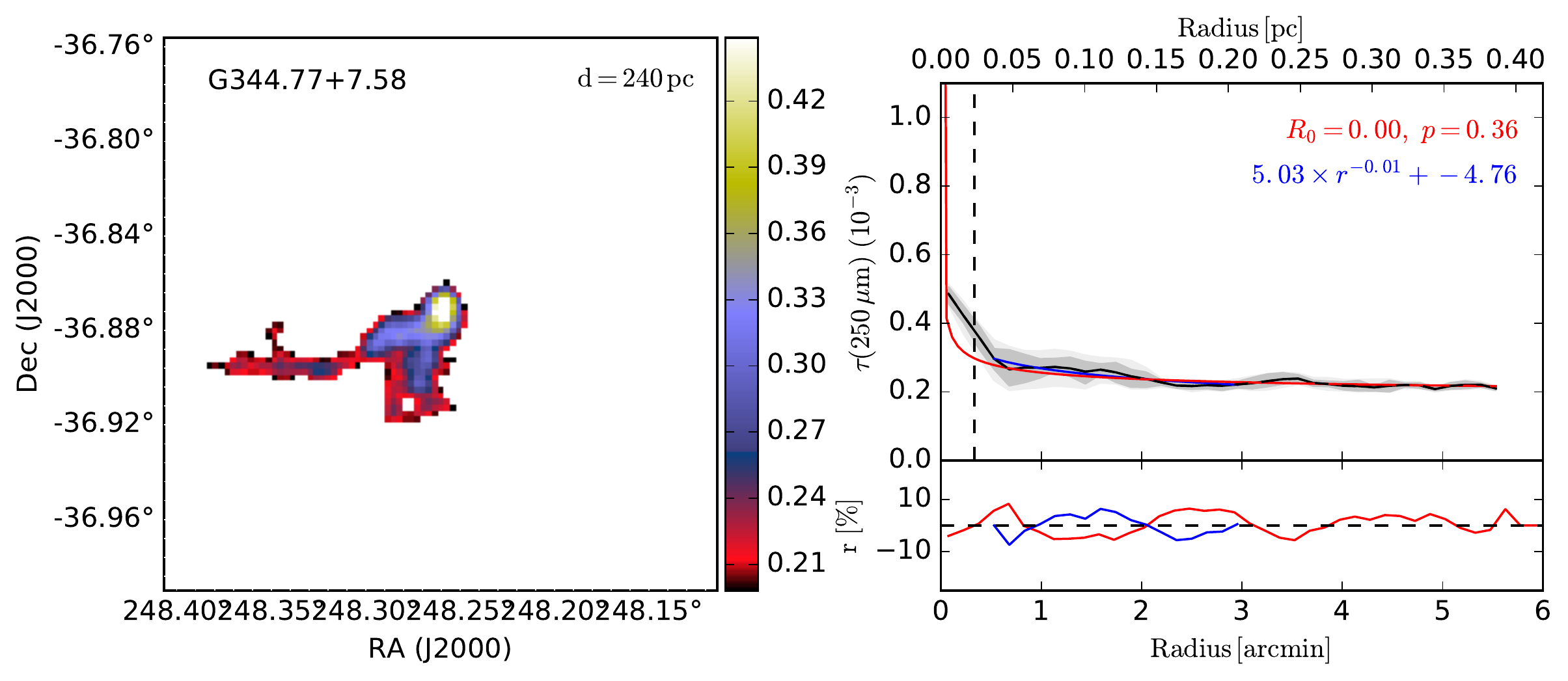}
\includegraphics[width=8.2cm]{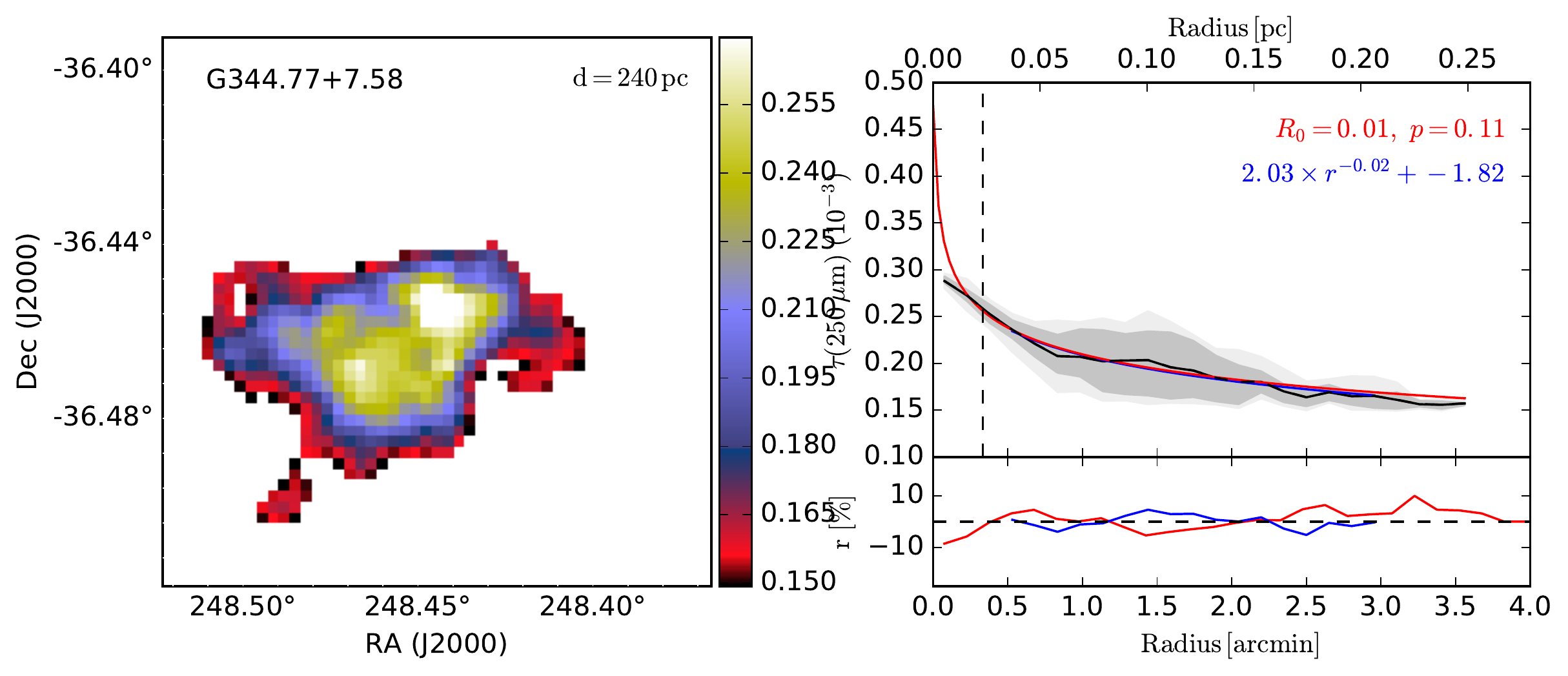}
\includegraphics[width=8.2cm]{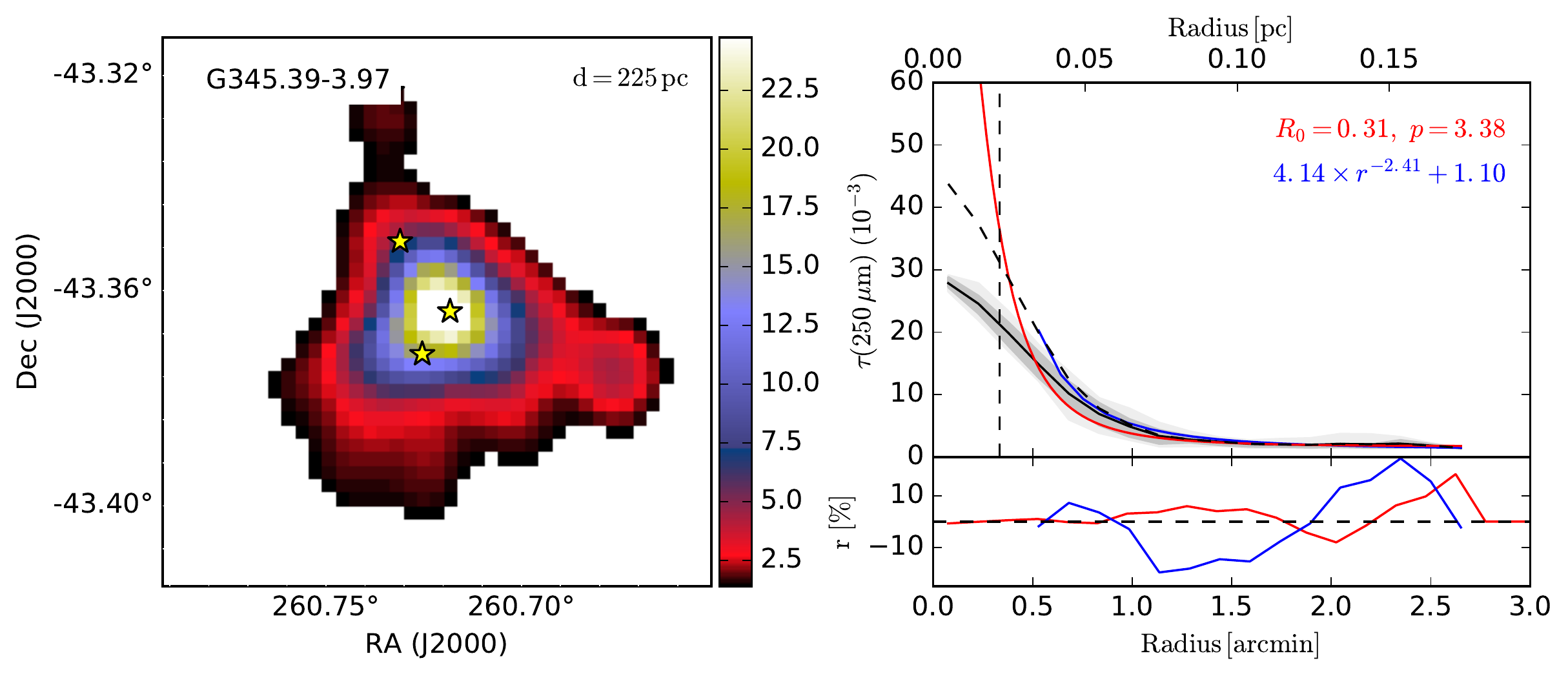}
\caption{continued.}
\end{figure}

\begin{figure}
\includegraphics[width=8.2cm]{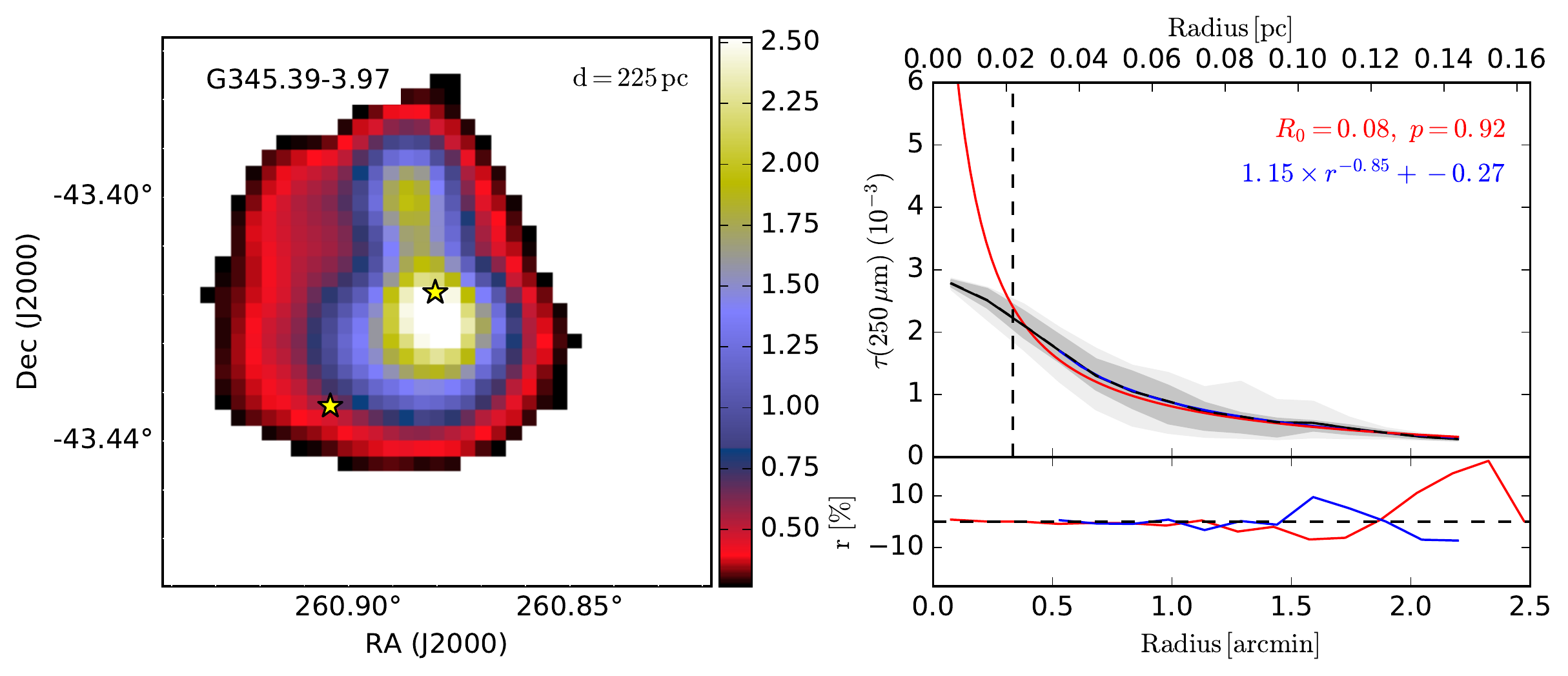}
\includegraphics[width=8.2cm]{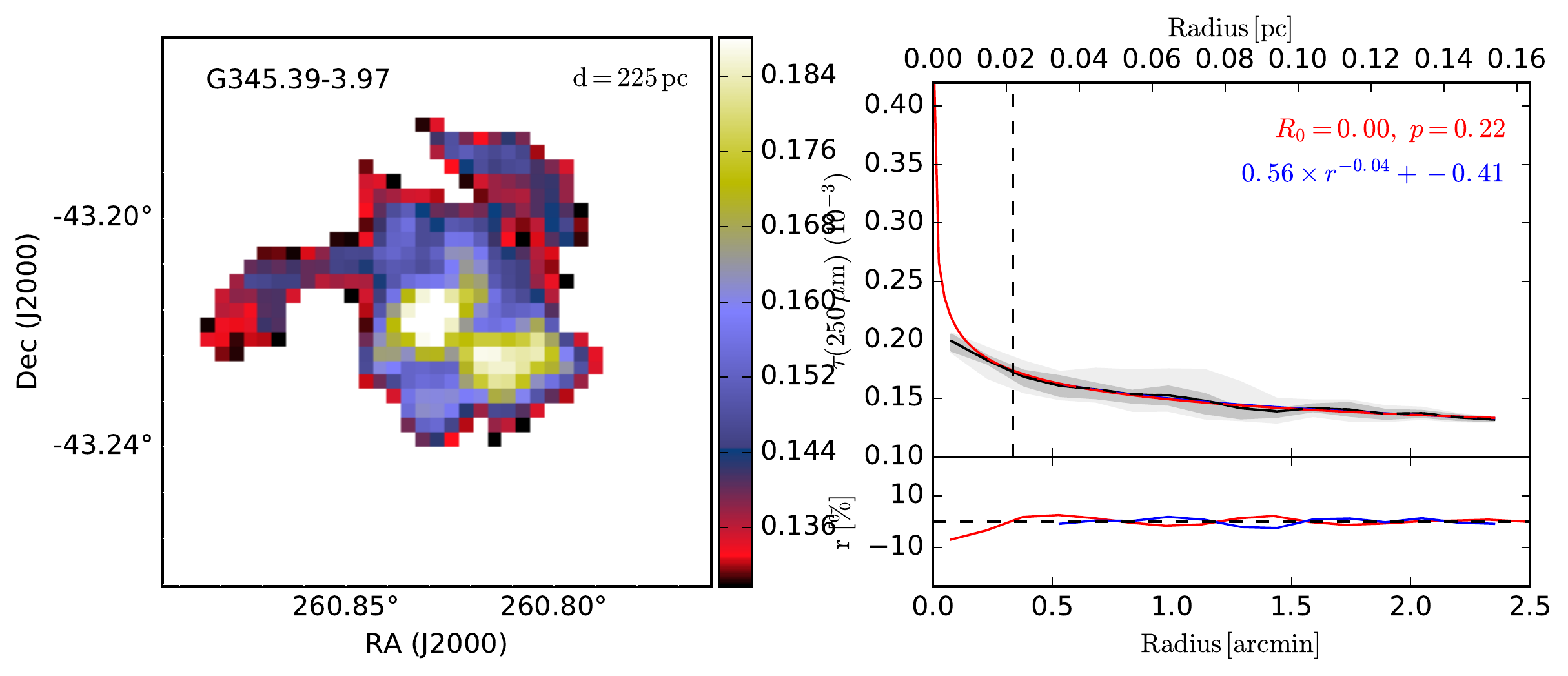}
\caption{continued.}
\label{fig:radial_last}
\end{figure}

\section{Column density PDFs of individual fields} \label{app:PDF}

We noted in Sect.~\ref{disc:PDF} the PDFs can sometimes reflect the
structure of individual objects rather than the general statistical
cloud properties. This is true especially because our observations
target pre-selected column density peaks.

Before background subtraction, the widest PDF is found in 
G206.33-25.94, a single cometary cloud observed towards an empty
high-latitude sky ($b=-26\degr$). The PDF directly reflects the mass
distribution with a dense cloud head and a progressively more diffuse
tail. In contrast, fields G20.72+7.07 and G1.94+6.07 have very narrow
PDFs. They do not have any particular morphological features and
simply happen to consist of structures of nearly uniform column
density.  

Fields G154.08+5.23, G198.58-9.10, and G345.39-3.97 exhibit
well-defined power-law tails towards high column densities. The latter
two are clear interface regions with significant column density
gradients and a dense boundary layer. Although the boundary layers may
contain some cores, the PDF shapes are connected to the particular
structure created by external forcing. The third field, G154.08+5.23,
looks more like a normal turbulent field. However, also here the PDF
tail is associated to elongated high-density regions. This could again
be a sharp cloud boundary that is seen more face-on than in the
previous examples. PDF shapes are of course also generally affected by
projection effects \citep{Schneider2015_LOS}.

There are a couple of fields where the PDF has a long tail towards
low column densities. G206.33-25.94 consist of a single cometary
cloud and the tail of the PDF directly corresponds to the tail of the
cloud. In G141.25+34.37, the PDF asymmetry is even more pronounced but
mainly reflects the density profile of the cloud. The average column
density of the G141.25+34.37 field is very low and the contribution of
the extragalactic background is noticeable. The tail at the low-$N$
side is, of course, very sensitive to the zero point of the quantity
used or how the background subtraction is carried out.

Differences between surface brightness and column density PDFs should
be correlated with high column density that makes large temperature
gradients possible. Anisotropic illumination can have a similar
effect. Therefore it may be significant that the difference between
surface brightness and column density PDFs is particularly large in
the aforementioned fields G198.58-9.10 and G345.39-3.97, both of which
exhibit sharp cloud edges.

\end{document}